\documentclass[12pt,a4paper]{article}
\usepackage[margin=2cm]{geometry}
\usepackage{arydshln}
\usepackage{stackengine,float}
\usepackage[T1]{fontenc} 
\usepackage[utf8]{inputenc} 
\usepackage[english]{babel}
\usepackage{array,epsfig,rotating}
\usepackage{epstopdf}
\usepackage[colorlinks=true, urlcolor=blue,  linkcolor=black, citecolor=cyan]{hyperref}
\usepackage{graphicx} 
\usepackage{subfigure}
\usepackage{xr} 
\usepackage{amsbsy}   
\usepackage{amsmath}
\usepackage{amssymb}
\usepackage{natbib}   
\usepackage{graphicx}   
\usepackage{longtable}
\usepackage{pdfpages}
\usepackage{multirow}
\usepackage{color}   
\usepackage{indentfirst}   

\usepackage{here}
\usepackage{appendix}
\usepackage{arydshln}  
\usepackage{dsfont}
\usepackage{multirow}
\usepackage{enumitem} 
\usepackage[yyyymmdd,hhmmss]{datetime}

\usepackage{hyperref}
\usepackage{authblk}

\usepackage{bold-extra}
\usepackage{listings}
\usepackage{setspace}

\title{Modeling the Evolution of Infectious Diseases with Functional Data Models:\\ The Case of COVID-19 in Brazil}
\author[1]{Julian A. A. Collazos} 
\author[2]{Ronaldo Dias}
\author[3]{Marcelo C. Medeiros} 
\affil[1]{New Granada Military University, Colombia} 
\affil[2]{State University of Campinas, Brazil} 
\affil[3]{Pontifical Catholic University of Rio de Janeiro, Brazil}
{
    \makeatletter
    \renewcommand\AB@affilsepx{: \protect\Affilfont}
    \makeatother

    \affil[ ]{Email}

    \makeatletter
    \renewcommand\AB@affilsepx{, \protect\Affilfont}
    \makeatother

    \affil[1]{julian.acuna@unimilitar.edu.co}
    \affil[2]{dias@unicamp.br}
    \affil[3]{mcm@econ.puc-rio.br}
}

\begin{document}
\date{}
\maketitle

\begin{abstract}
In this paper, we apply statistical methods for functional data to explain the heterogeneity in the evolution of number of deaths of Covid-19 over different regions. We treat the cumulative daily number of deaths in a specific region as a curve (functional data) such that the data comprise of a set of curves over a cross-section of locations. We start by using clustering methods for functional data to identify potential heterogeneity in the curves and their functional derivatives. This first stage is an unconditional descriptive analysis, as we do not use any covariate to estimate the clusters. The estimated clusters are analyzed as ``levels of alert'' to identify cities in a possible critical situation. In the second and final stage, we propose a functional quantile regression model of the death curves on a number of scalar socioeconomic and demographic indicators in order to investigate their functional effects at different levels of the cumulative number of deaths over time. The proposed model showed a superior predictive capacity by providing better curve fit at different levels of the cumulative number of deaths compared to the functional regression model based on ordinary least squares. 


\end{abstract}

\section{Introduction}
The fast spread of (new) diseases, like the Covid-19 pandemic, poses a number of challenges to scientists worldwide. One key research question is to understand which socioeconomic and demographic factors contribute to the spread and the lethality of the disease. The answer to this question not only shed light in explaining the heterogeneous patterns observed across different regions, but can also help governments to design more efficiently public policies related to health crises. Furthermore, from a statistical perspective, the full understanding of the heterogeneous behavior in the evolution of deaths can help the development of more precise forecasting and epidemiological models.

In this paper, we use methods for functional data analysis (FDA) to explain the heterogeneous patterns in the evolution of deaths of Covid-19 over different regions. FDA involves the analysis of data which units of observation are functions defined on some continuous domain, and the observed data consist of a sample of functions taken from some population, sampled over discrete intervals. In the present case, we treat the cumulative daily number of deaths in a specific region as a curve (functional data). Therefore, the data comprise of a set of curves over a cross-section of locations. We start by using clustering methods for functional data to identify potential heterogeneity in the curves. This first stage is an unconditional descriptive analysis, as we do not use any covariate to estimate the clusters. In the second and final stage, we estimate a functional quantile regression of the death curves on a number of socioeconomic and demographic indicators. This last analysis is conditional where the dependent variable is a function, but the regressors are scalars.

The advantages of using FDA instead of conventional clustering and regression techniques are threefold. First, and more importantly, by using FDA, we are able to explain heterogeneity in the pattern of the curves and not only over a summary statistic. For example, consider two regions with the same total number of Covid-19 cases but with completely different trajectories. Standard regression methods where the dependent variable is the total number of deaths will not be able to grasp any difference between the two locations. On the other hand, methods that explore the functional nature of the data will take the differences in trajectories into account. Second, functional regression allows covariates to have different relative importance during different stages of the evolution of the disease in each location. This is an important advance over panel data methods, where the regression coefficients are usually fixed over the time dimension. Finally, functional clustering methods are able to group locations with the same evolution pattern for the disease.

\subsection{Main Takeaways}
We use deaths data across Brazilian municipalities to show how the socioeconomic and demographic heterogeneity in Brazil explains the differences in the evolution and spread of Covid-19 over the country. The methods proposed here can be adopted by any country where data for different regions are available. The methods can also be applied to larger regions as well. Brazil is a nice ``laboratory'' to test the methodology considered in this paper due to many reasons. First, it is a country with continental dimensions and very heterogeneous in terms of social, economic, health, and demographic indicators. Second, and mostly unfortunate, Brazil is a country where the disease has been considered of being out of control for a long period, with many waves and new variants of the virus hitting the region. Third, the vaccination of the population has been conducted at a very slow pace.

Whilst vaccination of the population over the Globe is taking place and the pandemic may be close to a final outcome, we still believe that our results are useful. First, new variants of the disease are being discovered and there are still uncertainty about the effectiveness of the available vaccines and the duration of the immunization. At the same time, large countries like Brazil are still struggling to control the evolution of new cases and deaths. Third, and probably more importantly, the results presented here can shed light on how infectious diseases evolve in large and heterogeneous countries/regions like Brazil. Uncovering correlations between the disease dynamics and social, economic, and demographic factors may help policy makers and health agents to better manage crisis like the Covid-19 one in the future. Finally, the paper illustrates a nice combination  of unsupervised as well as supervised  statistical methods for FDA that can applied more generally.

Although it is known that countries with continental proportions, such as Brazil, India and the United States present high heterogeneity, it is worth to verify whether there are groups of locations that show some homogeneity among the death count curves over time. That is, we perform a clustering functional data analysis. There are several techniques for clustering functional data. In particular, we use \cite{Zambom19}. Using this technique, it is possible to group not only the original data curves, but also their first and second functional derivatives. Thus, it is possible to group the velocity and acceleration of the death count curves as function of time. In fact, it can be seen that after the clustering, some state capitals of Brazil have more similarities in the levels, the velocity and acceleration curves than others and, therefore, were grouped accordingly. In addition, we perform a functional quantile regression model with variable selection based on LASSO (Least Absolute Shrinkage and Selection Operator) penalty. For this functional quantile regression, the explanatory variables are scalar socioeconomic covariates. The idea is to verify whether these socioeconomic covariates can explain death count curves over the time.

\subsection{A Brief Comparison to the Literature}

Not many papers consider the use of FDA methods to disclosure the relationship between the curves of death counts and social, economic, and demographic indicators. The most notable exception is \citet{Carroll20}. However, there are a number of differences between their paper and ours. First, the authors use FDA tools to analyze the case dynamics and not deaths. Second, they consider only functional regression and, here, we not only put more emphasis on the dynamics of death counts, but we also provide a new functional clustering analysis. Third, their set of covariates is more limited than ours. In terms of social and demographic factors, the authors consider only population density, the proportion of the population over age 65 years, the log cumulative case counts per million, and the lagged decrease in workplace mobility. The dataset in the present paper also includes a larger number of social and economic indicators and a richer set of demographic/climatic variables, such as geographic location and temperature.

The other papers in this recent literature usually take a different route and present only a descriptive analysis of the relation between cases or deaths and social, economic, and demographic variables. For instance, \citet{Ortiz2020} evaluated the spread of Covid-19 through Latin America and the Caribbean region by means of a correlation study between climate and air pollution indicators. \citet{Dowd2020} consider a descriptive analysis to show how age structure among population helps to explain the differences in fatality rates across countries. Their work is extended by \citet{nepomuceno2020} who also include other variables in the study. See also \citet{fortaleza2020}.

\subsection{Organization of the Paper}
In addition to this Introduction, this paper is organized in the following sections. Section \ref{Data} describes the dataset used in our analysis. The statistical methodology is presented in Section \ref{Statistical}. Section \ref{strategy} outlines the empirical strategies adopted to obtain relevant results. The functional clustering analysis for capital and non-capital cities is presented in Section \ref{funclustanalysis}. Section \ref{FOSQR} exhibits the main results of the proposed functional quantile regression model to estimate and predict the cumulative death curves of Covid-19. Some concluding remarks are given in Section \ref{conclusions}.

\section{Data}\label{Data}

The dataset examined in this work is based on the informative reports of daily deaths from Covid-19 from Brazilian cities. The dataset is publicly available and can be found at https://brasil.io/dataset/covid19/caso/.

Our dependent functional variable is the cumulative daily death counts per 100,000 inhabitants in each municipality in Brazil. The dataset is organized as a panel in epidemiological time, i.e, $Y_i(t)$ represents the cumulative deaths in municipality $i$, $t$ days after the 240th case and with at least 5 cases of death by Covid-19. The time spam goes until May 20, 2021. A total of 1,921 cities are considered in this study. As explanatory variables we consider a number of socioeconomic, demographic variables, and epidemiological variables as described below.

\begin{figure}
\centering
\stackunder[5pt]{\includegraphics[height=4cm,width=4cm]{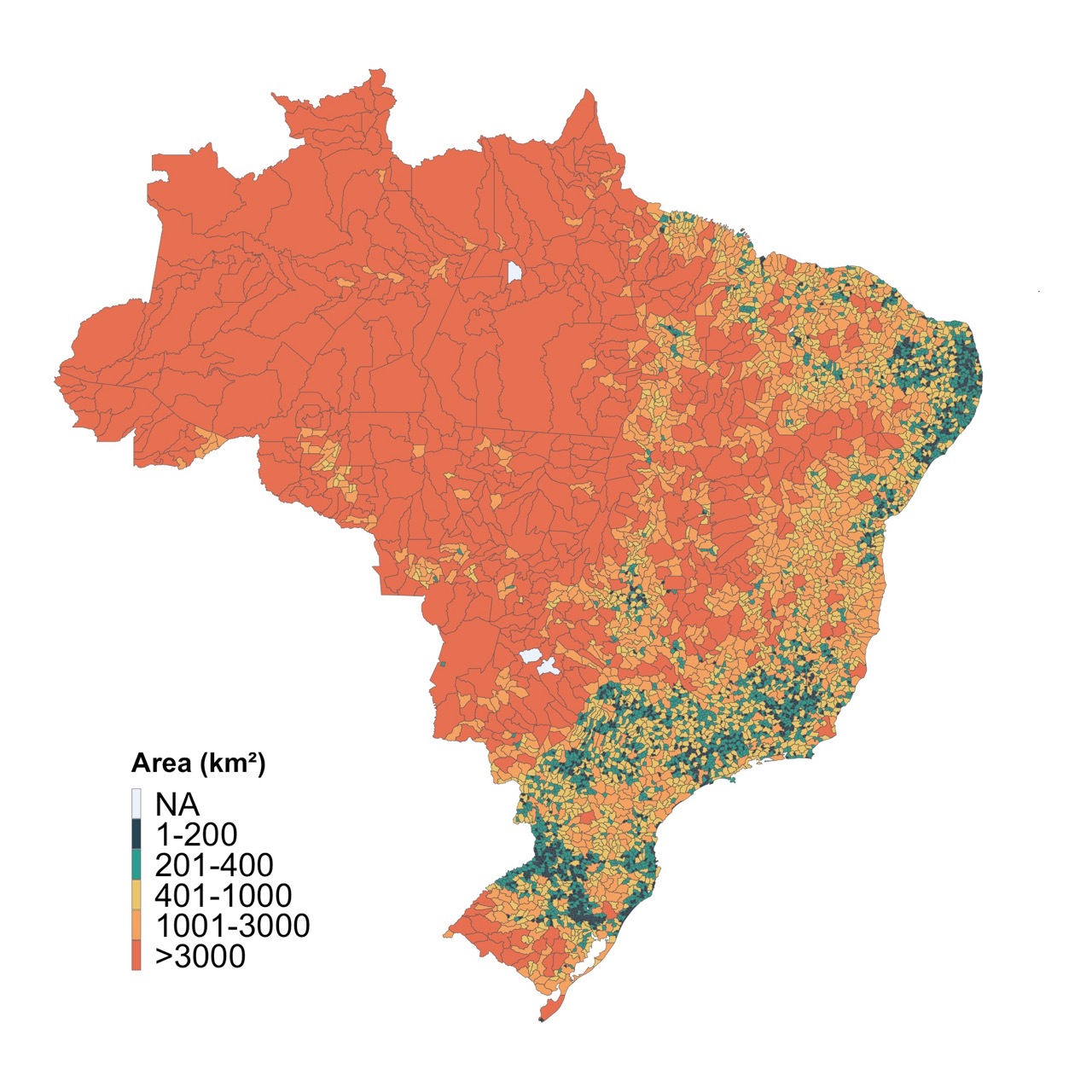}}{(a) Area}\quad
\stackunder[5pt]{\includegraphics[height=4cm,width=4cm]{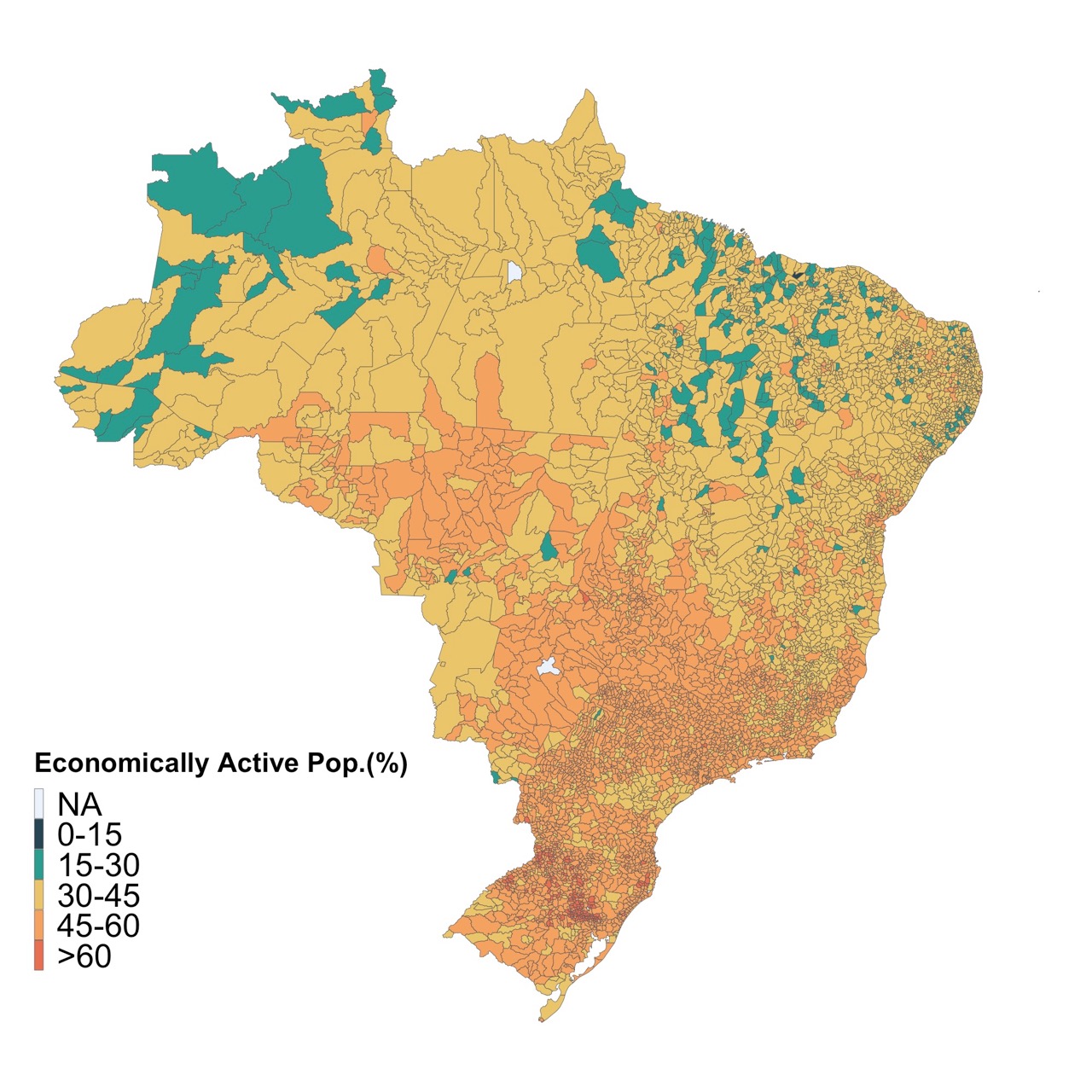}}{(b) Econ. Act. Pop.}\quad
\stackunder[5pt]{\includegraphics[height=4cm,width=4cm]{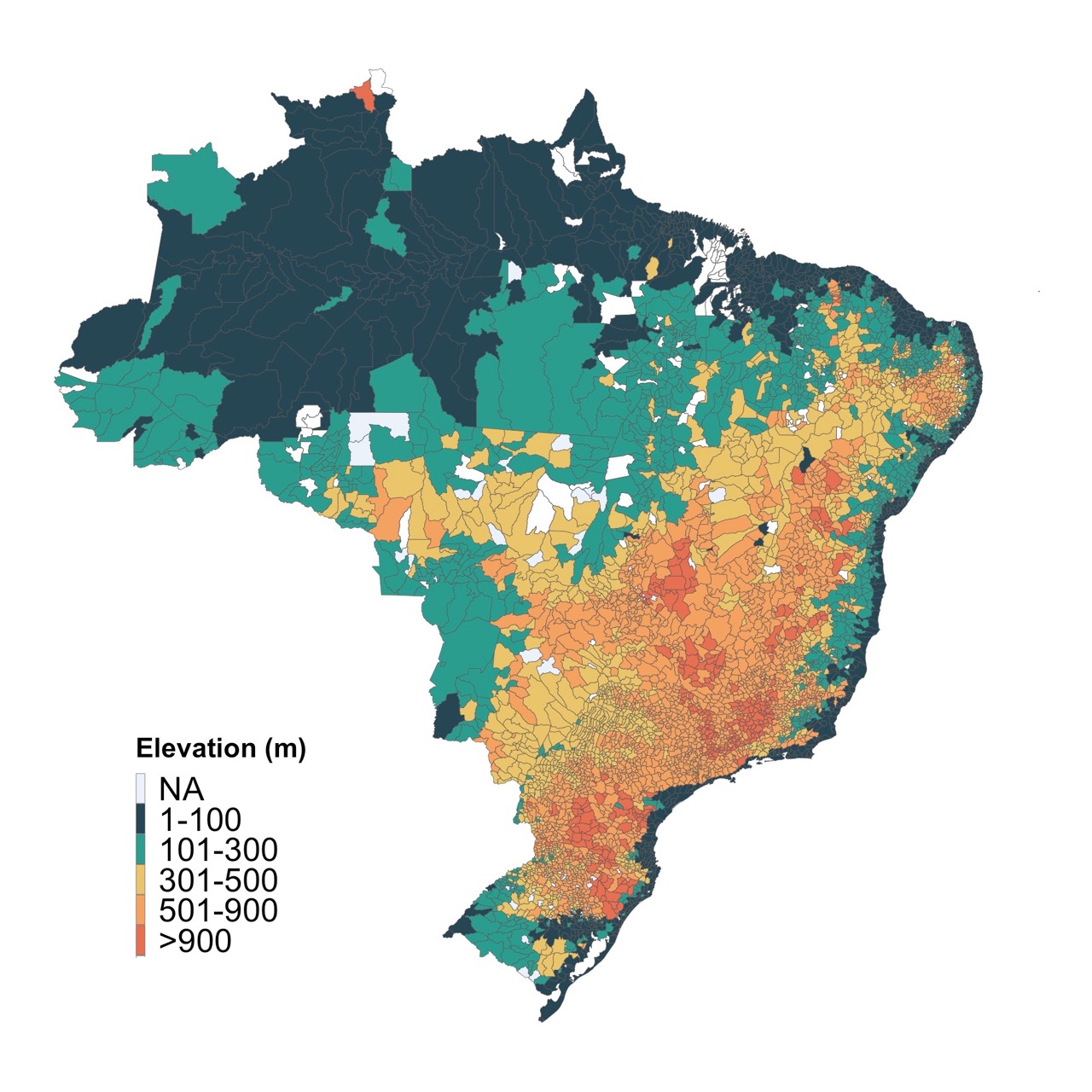}}{(c) Elevation}\\
\stackunder[5pt]{\includegraphics[height=4cm,width=4cm]{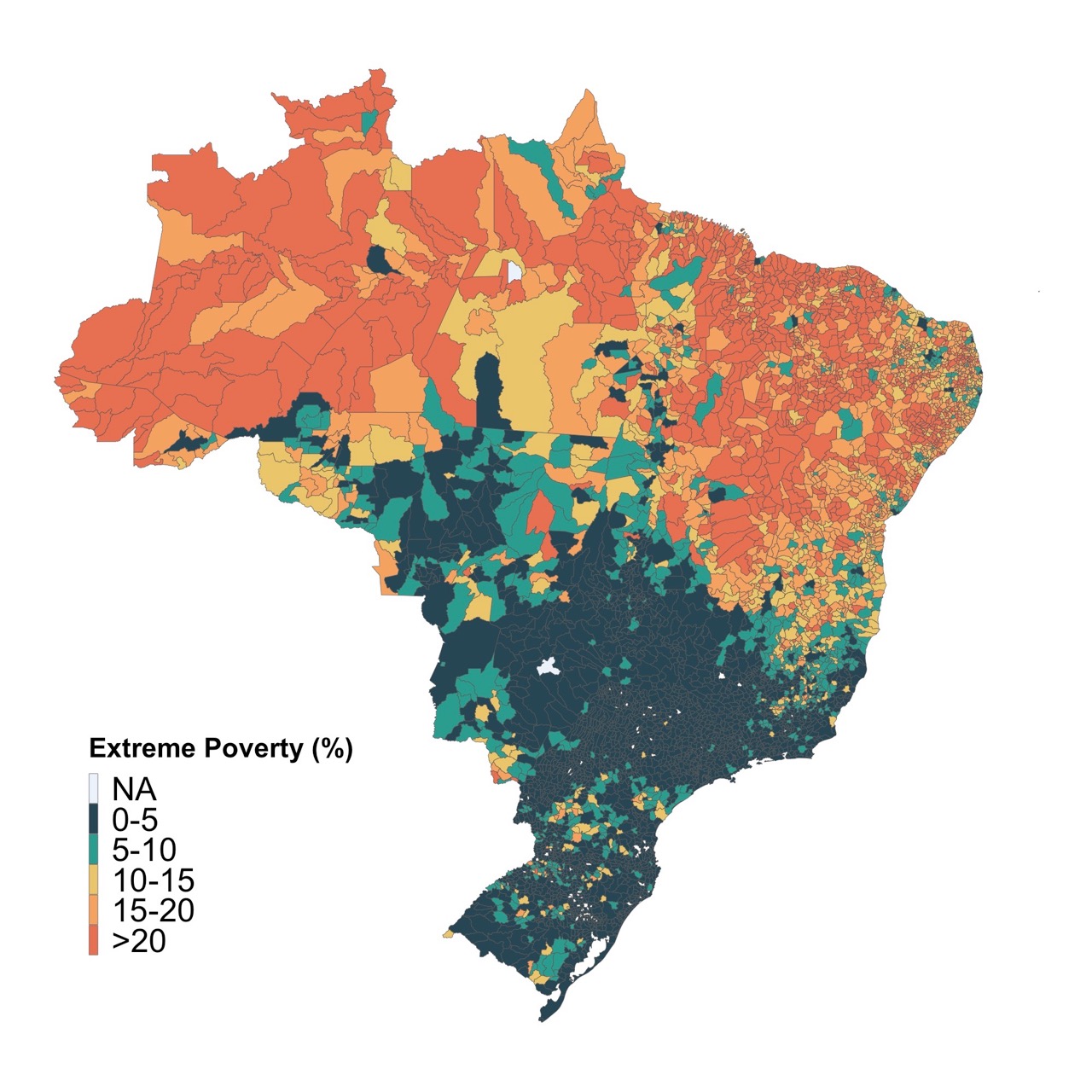}}{(d) Ext. Poverty}\quad
\stackunder[5pt]{\includegraphics[height=4cm,width=4cm]{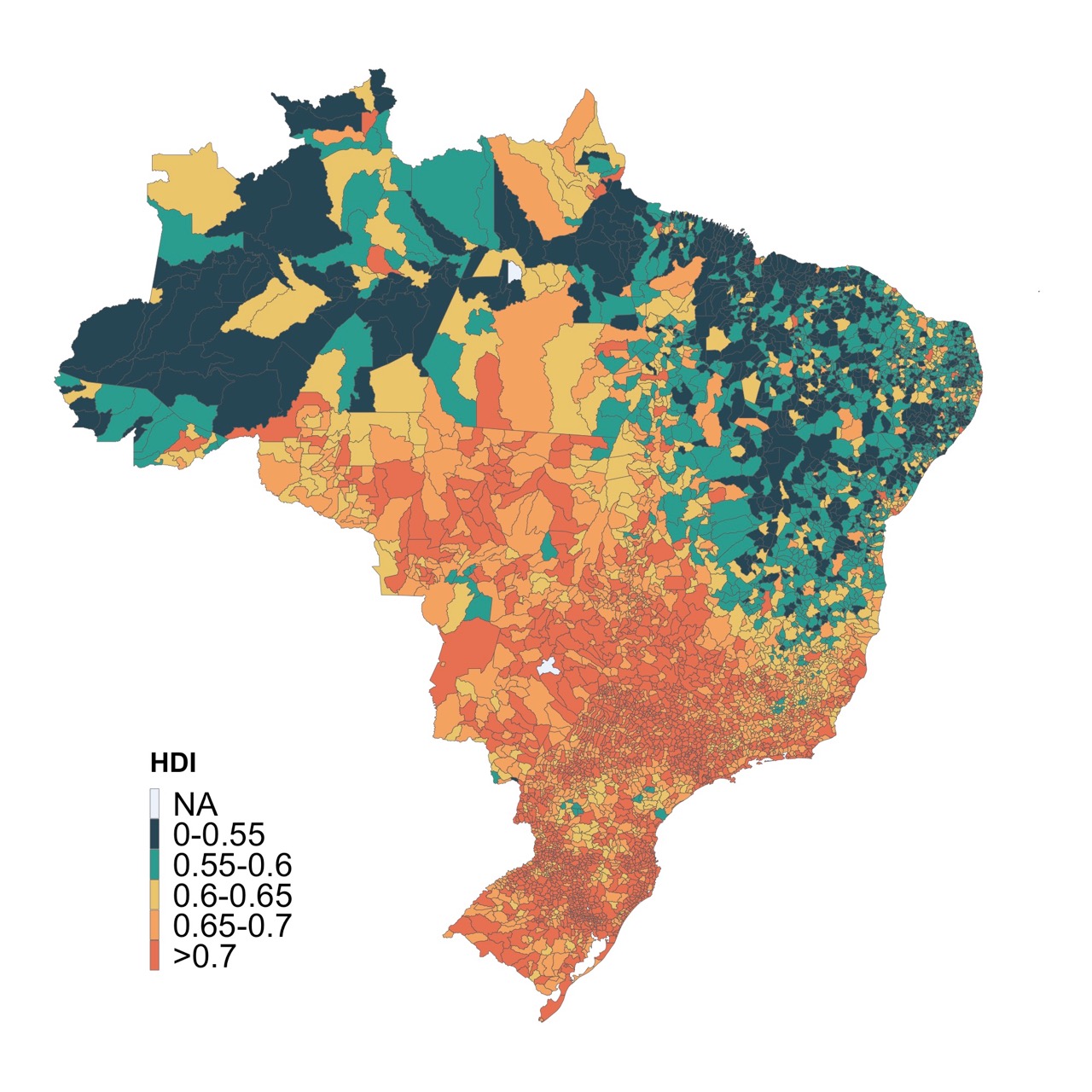}}{(e) HDI}\quad
\stackunder[5pt]{\includegraphics[height=4cm,width=4cm]{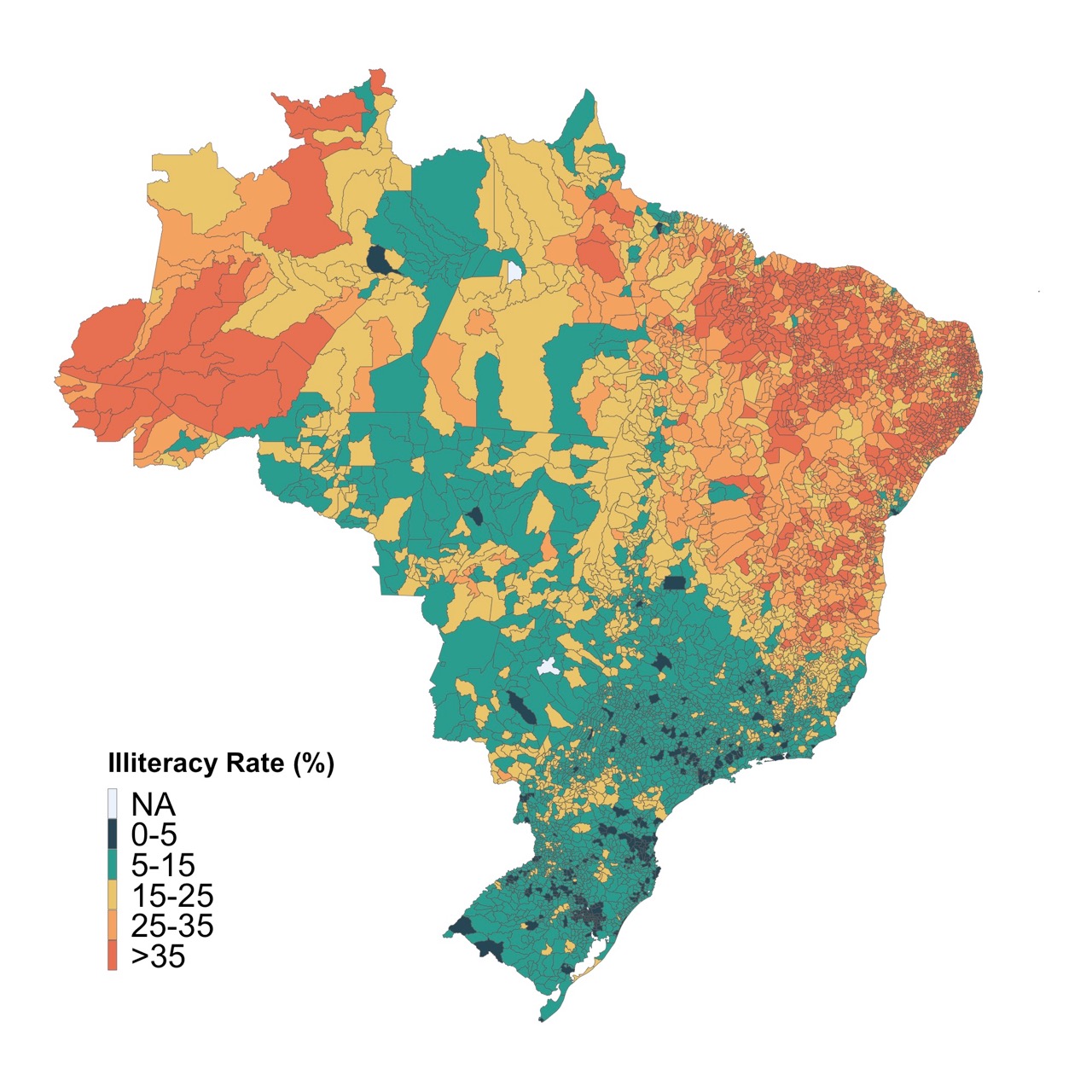}}{(f) Illiteracy Rate}\\
\stackunder[5pt]{\includegraphics[height=4cm,width=4cm]{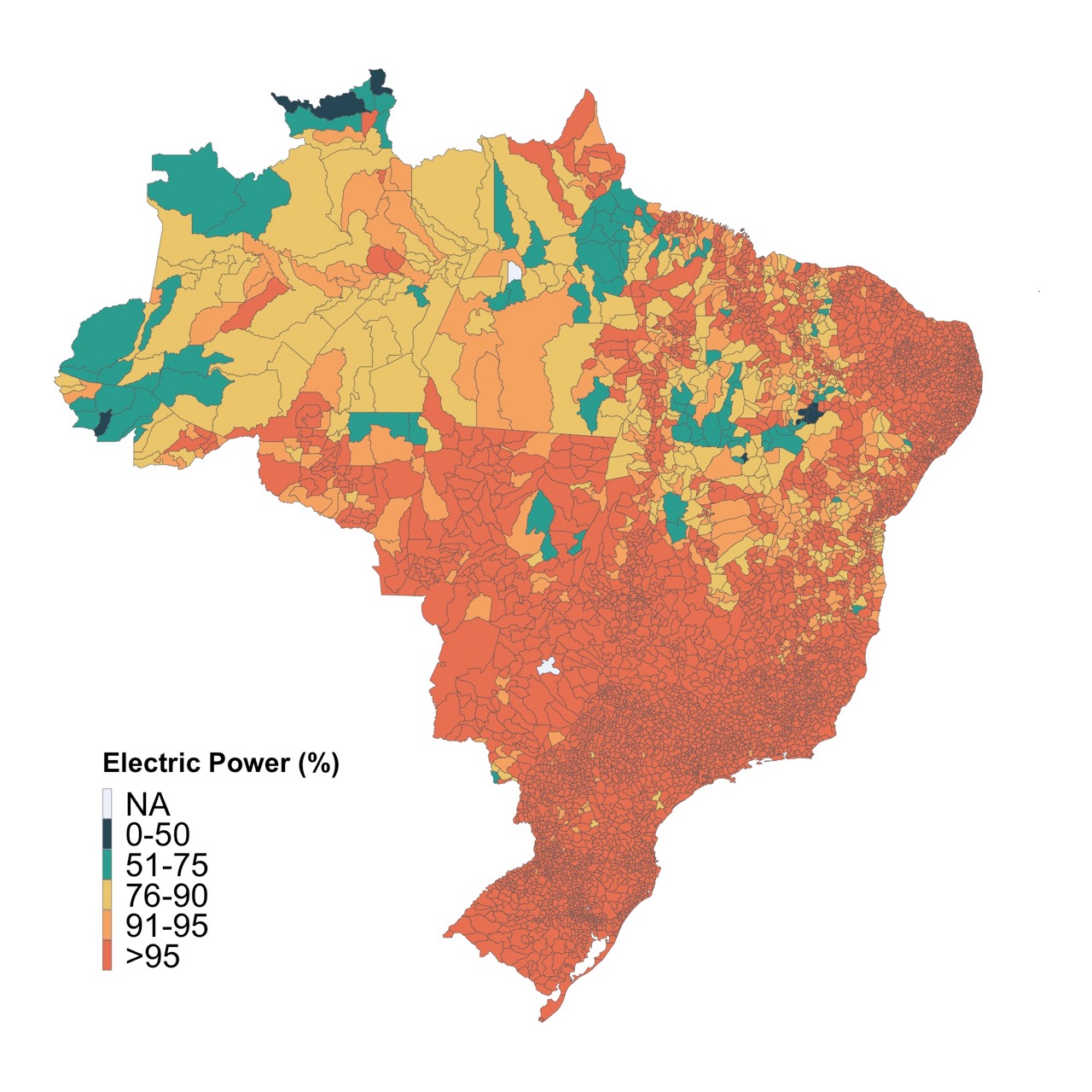}}{(g) Electric Power}\quad
\stackunder[5pt]{\includegraphics[height=4cm,width=4cm]{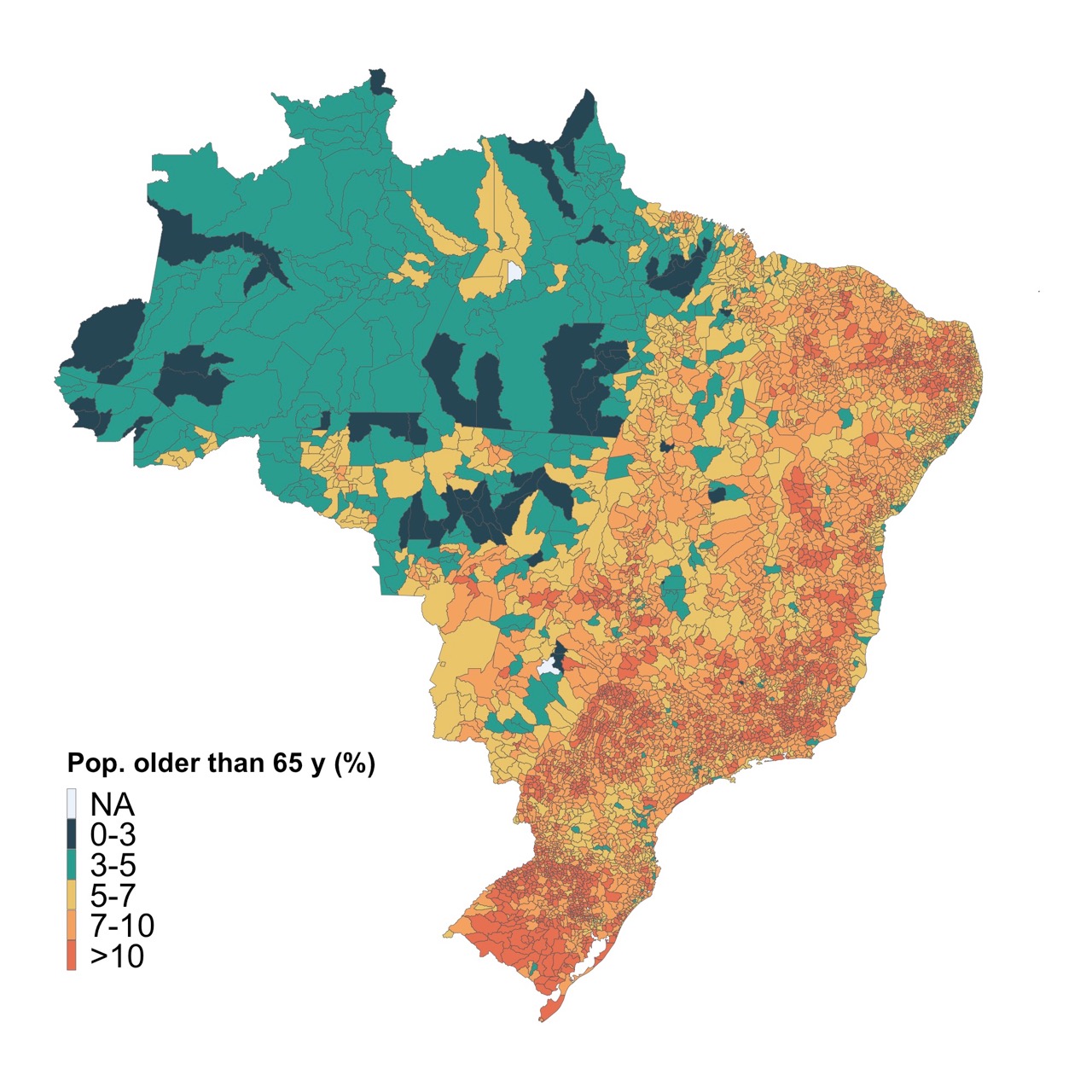}}{(h) Pop. Older than 65 Yr.}\quad
\stackunder[5pt]{\includegraphics[height=4cm,width=4cm]{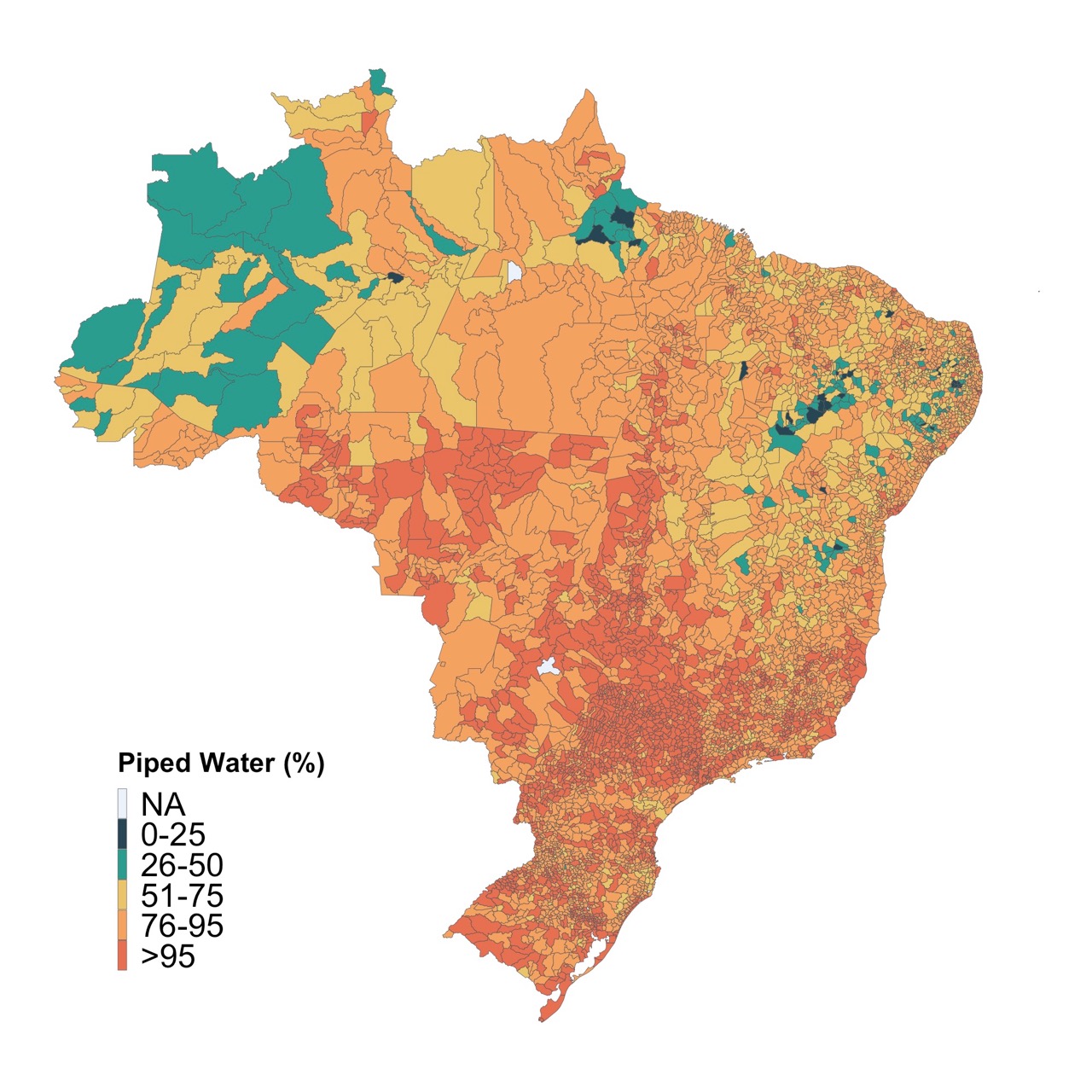}}{(i) Piper Water}\\
\stackunder[5pt]{\includegraphics[height=4cm,width=4cm]{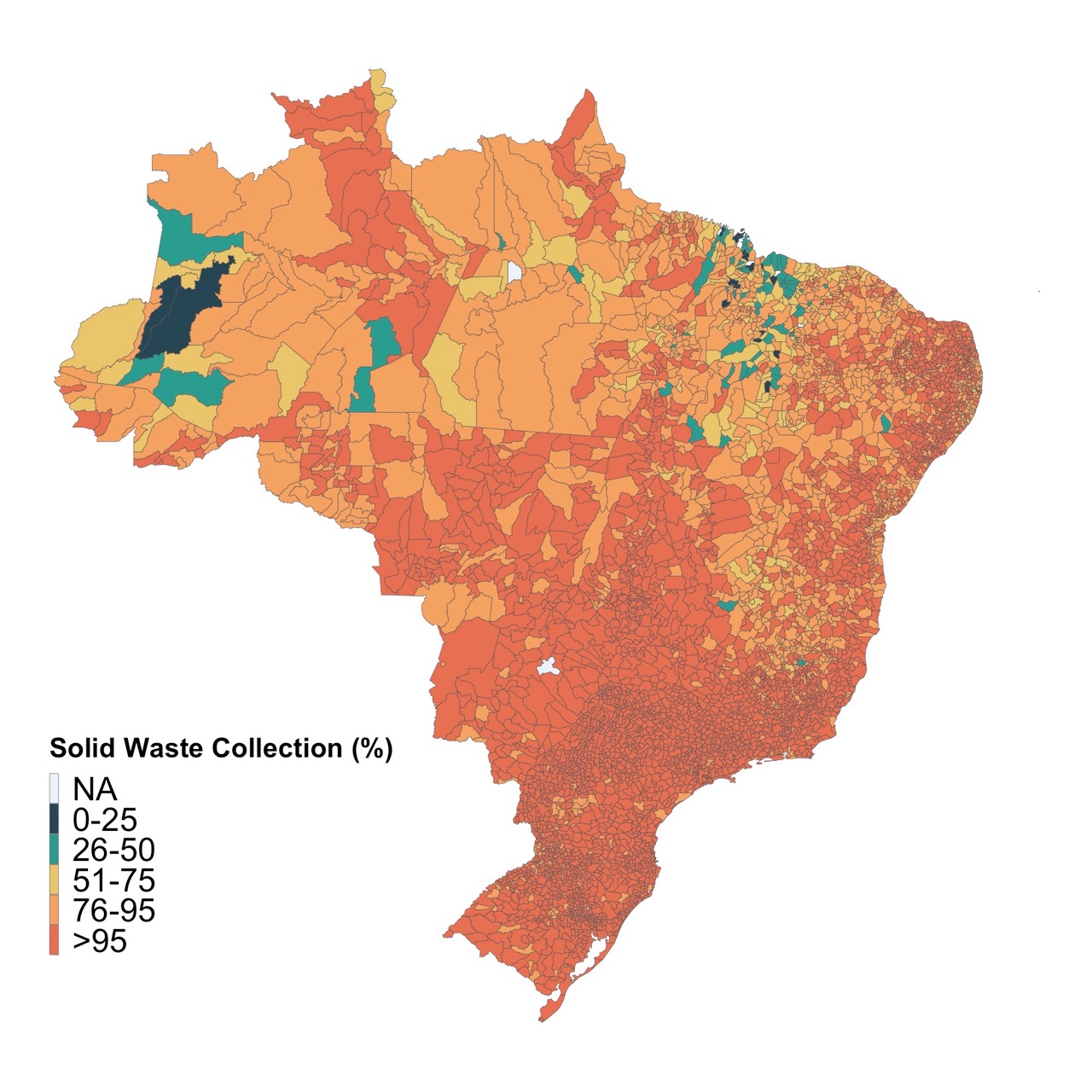}}{(j) Solid Waste Col.}\quad
\stackunder[5pt]{\includegraphics[height=4cm,width=4cm]{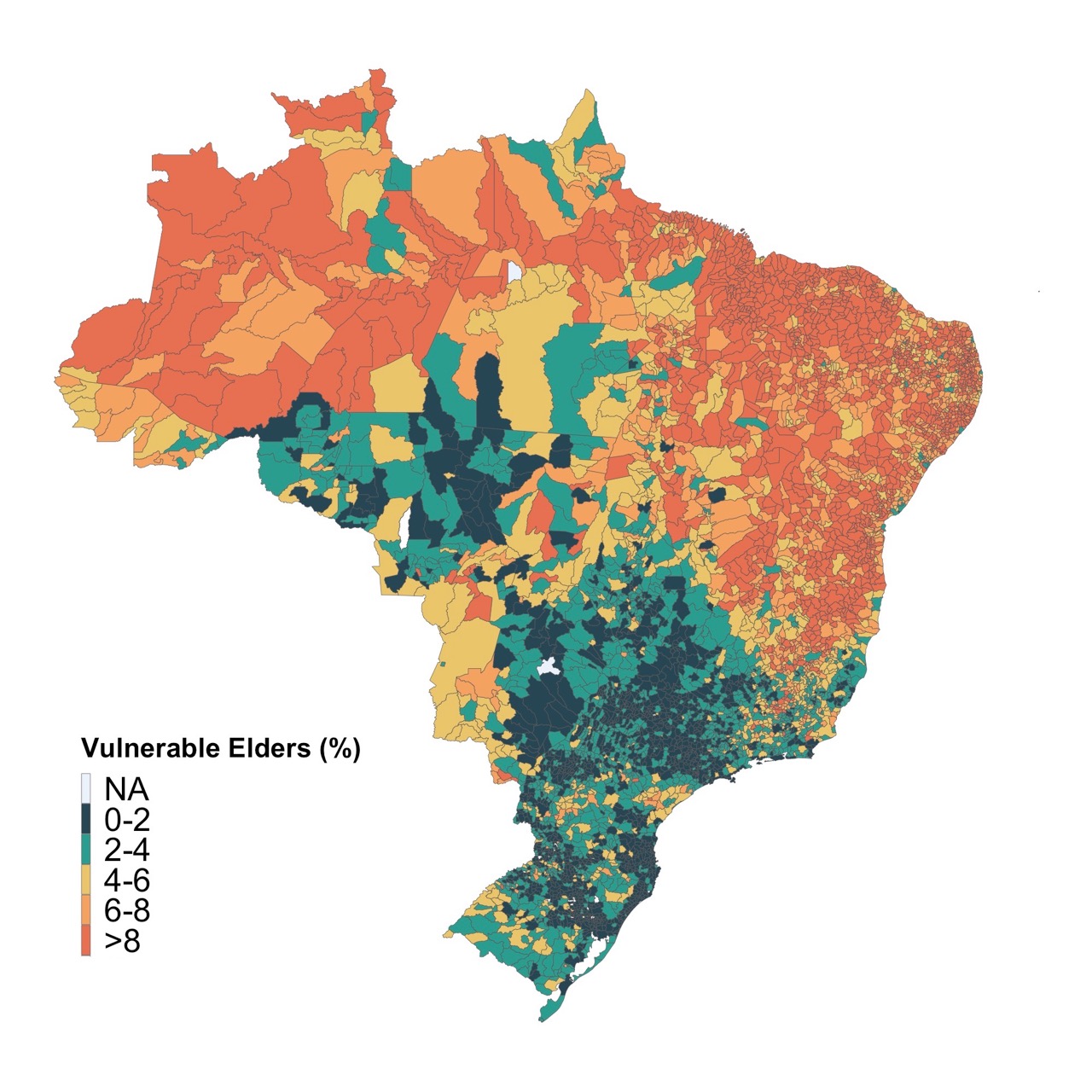}}{(k) Vulnerable Eldery}
\caption{Geographic Dispersion of Social and Economic Indicators.}
\begin{minipage}{1\linewidth}
\begin{footnotesize}
The maps in the figure illustrate the geographic dispersion of a number of social, economic and demographic indicators across Brazilian municipalities. Map (a) illustrates the dispersion in terms of the area of the municipalities. The remaining maps shows the dispersion of the following indicators: (b) economic active population; (c) elevation of the municipality (meters above sea level); (d) fraction of the total population in extreme poverty; (e) human development index; (f) fraction of the population that is illiterate; (g) fraction of the population with access to electric power; (h) population over 65 years old; (i) fraction of the population with piped water; (j) fraction of the population with access to solid state collection; and (k) proportion of vulnerable elder people in the population.  
\end{footnotesize}
\end{minipage}
\label{fig:maps}
\end{figure}

Brazil is very heterogeneous country. In order to illustrate the social, economic and demographic heterogeneity of Brazil, Figure \ref{fig:maps} presents several maps. The maps in the figure illustrate the geographic dispersion of a number of social, economic and demographic indicators across Brazilian municipalities. Map (a) illustrates the dispersion in terms of the area of the municipalities. The remaining maps shows the dispersion of the following indicators: (b) economic active population; (c) elevation of the municipality (meters above sea level); (d) fraction of the total population in extreme poverty; (e) human development index; (f) fraction of the population that is illiterate; (g) fraction of the population with access to electric power; (h) population over 65 years old; (i) fraction of the population with piped water; (j) fraction of the population with access to solid state collection; and (k) proportion of vulnerable elder people in the population.

\section{Statistical Methodology}\label{Statistical}


The statistical methodology is based on the analysis of functional data which studies the behavior of functional data (curves) and the relationships between them. In our case, the data on the cumulative daily number of deaths from Covid-19 has a curve structure for each city or municipality analyzed. The methodology is based on three stages. The first stage consists of the pre-processing of the curves. Since the death curves have different lengths, it is necessary to represent them in the same dimension to be able to analyze them in the next stages. This representation can be carried out using B-splines basis functions in order to obtain all the curves with the same dimension. Through the B-splines basis functions, the curves of the first and second functional derivatives can be calculated to observe variations in the velocity and acceleration of the cumulative death curves. In the second stage, the curves represented are used in a functional clustering analysis to analyze the heterogeneity of the death curves and their functional derivatives through the estimation of functional clusters. These functional clusters define three alert levels (low, moderate, and high) in the cumulative number of deaths by cities. The estimated clusters on the curves of the first and second functional derivatives will determine if the curve of the number of deaths in a city changes alert level in the velocity and acceleration of cumulative deaths. In the final stage, we use a functional quantile regression model with LASSO penalty. We estimate the functional effects of scalar socioeconomic indicators in a given quantile on the cumulative death curves. This way, we can estimate the quantile curve of the fitted curves of the model that most closely approximate the mean curves of the estimated clusters as alert levels. Therefore, this approach will allow estimating the functional effects of scalar socioeconomic indicators for each alert level of the cumulative death curves.

\subsection{Curve Representation via B-splines}

\begin{figure}[!htbp]
     \begin{center}
        \subfigure[]{%
            \label{bscaps}
            \includegraphics[width=8cm,height=8cm]{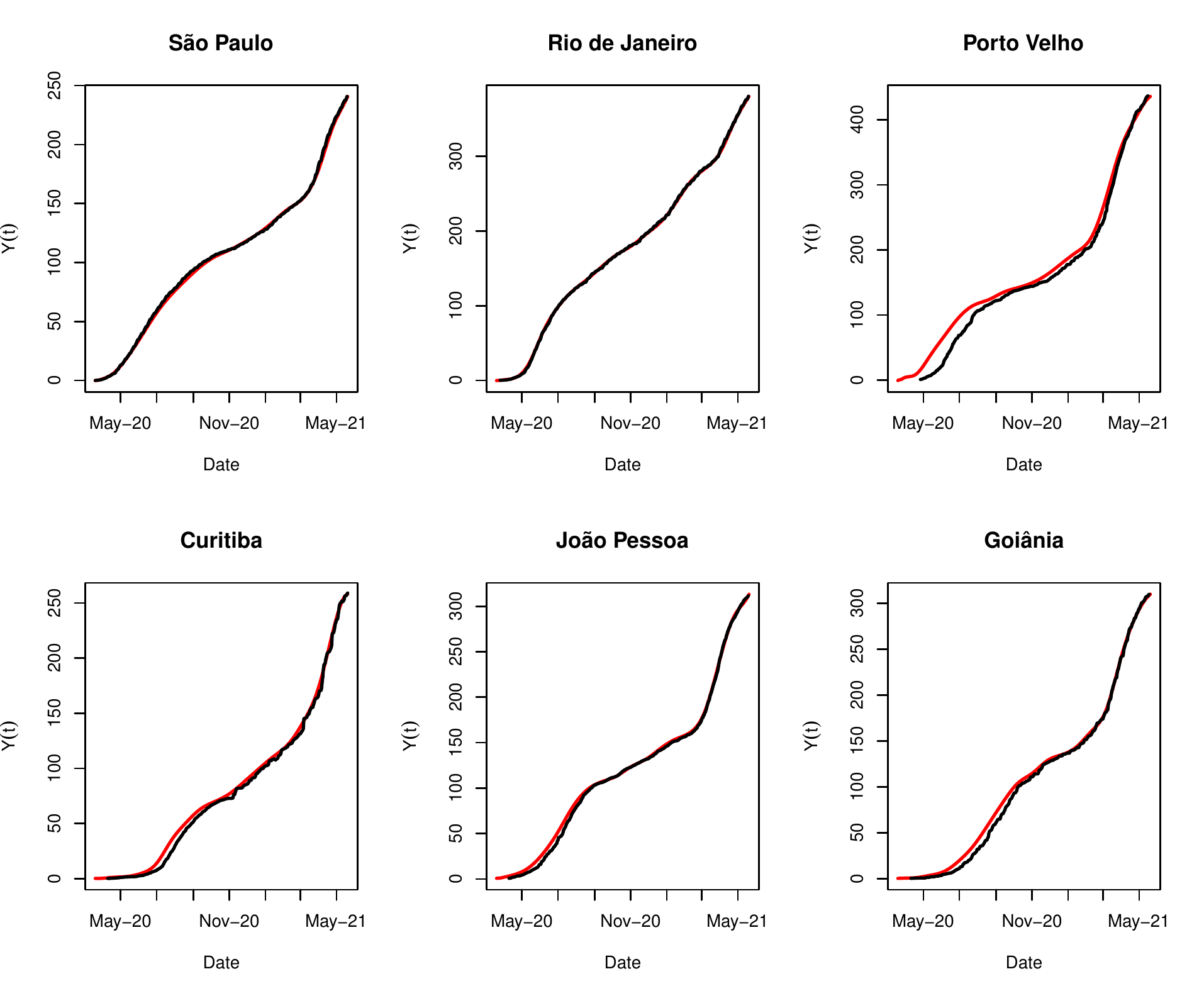}
        }
        \subfigure[]{%
           \label{bsnocaps}
           \includegraphics[width=8cm,height=8cm]{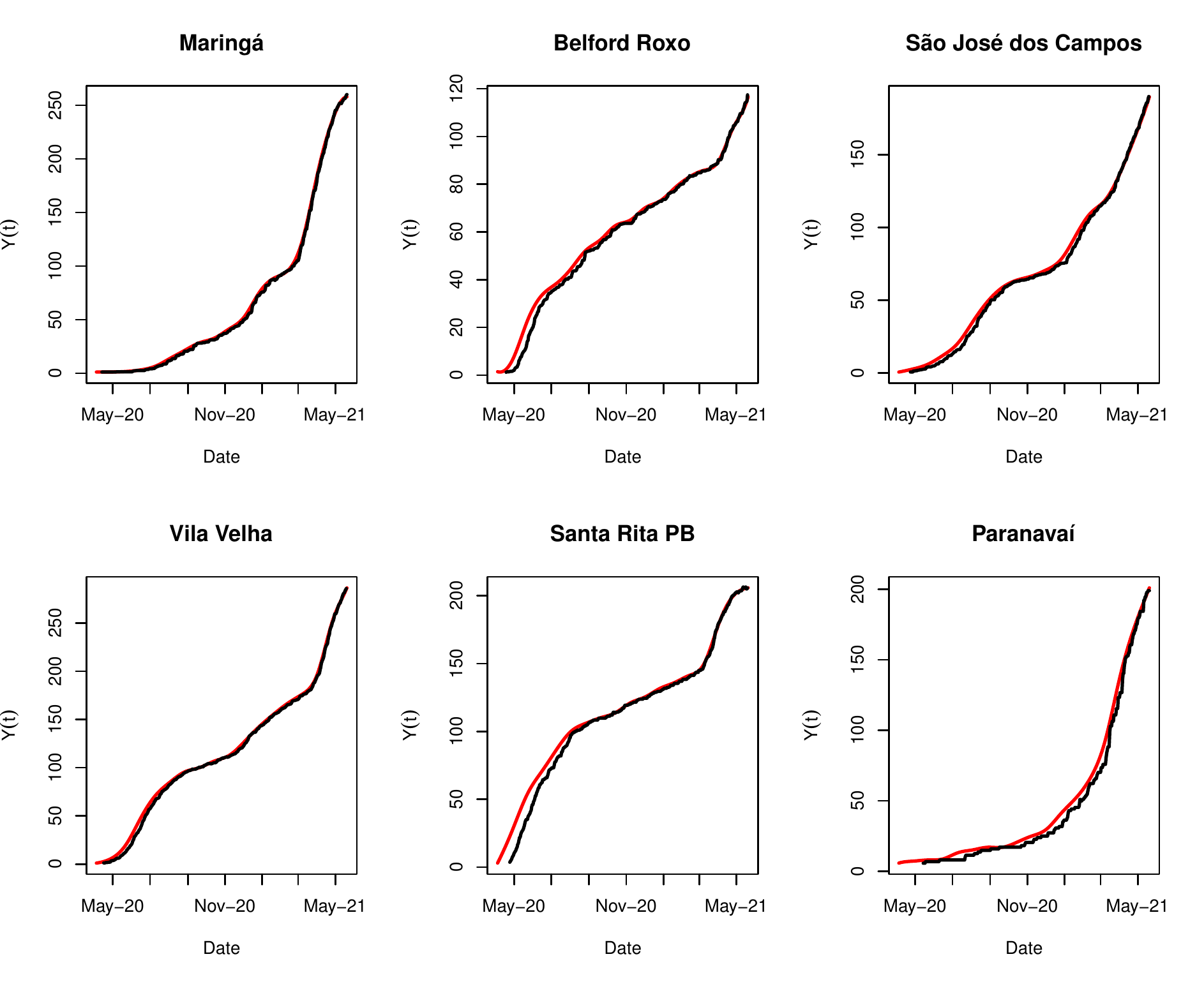}
        }
    \end{center}
    \caption{%
        Raw curves (black lines) and estimated curves (red lines) for (a) capital cities and (b) non-capital cities.}
   \label{rawcurves}
\end{figure}

Covid-19 death count curves are not measured at the same time (days) for all cities, as outbreaks and subsequent deaths were not reported at the same times for a given location. The curves, in this case, are called irregular curves in the FDA literature. Figure \ref{rawcurves}, exhibits the irregular structure of the raw curves (black lines) of some randomly chosen capital (\ref{bscaps}) and non-capital cities (\ref{bsnocaps}). In Figure \ref{rawcurves}, we can see the black lines as the raw curves of some randomly chosen capital (\ref{bscaps}) and non-capital cities (\ref{bsnocaps}) where their irregular structure can be seen.

To represent these curves, we make use of the well known B-splines basis function, equation (\ref{bs}). For the death count curves of the cities with more than 240 days of recorded data with the total number of $n = 1,921$ cities, we take $K=20$ basis functions to smooth and represent the  curves. Figure \ref{rawcurves}, shows the fitted curves (red lines) and the raw curves (black lines). More specifically, we have the following low rank representations of every functional data considered in this study
\begin{eqnarray}\label{bs}
Y_i(t)\approx \sum_{k=1}^{K} \omega_{ik}\phi_{k}(t) = \boldsymbol\omega_i'\boldsymbol\phi(t), 
\end{eqnarray}
where $\boldsymbol\omega$ is a ($K\times 1$) vector of weight coefficients and $\boldsymbol\phi(t)=(\phi_{1}(t),..., \phi_{K}(t))'$ is a vector of cubic B-spline basis functions. In general, the vector $\boldsymbol\omega$ is obtained as the solution of Penalized Least Square optimization problem.


Once the cumulative death curves are represented according to equation (1), we can compute the first and second functional derivatives of these curves applying a linear differential operator using the same number of B-splines basis functions. This can be done by the library \emph{fda} of the R Project software which allows to compute both functional derivatives. Thus, it is possible to obtain the functional derivative of order $\ell$ of the curves as follows:
\begin{eqnarray}\label{bsderiv}
D^{\ell}Y_i(t)= \sum_{k=1}^{K} \omega_{ik}D^{\ell}\phi_{k}(t) = \boldsymbol\omega_i'D^{\ell}\boldsymbol\phi(t). 
\end{eqnarray}


\subsection{Clustering Functional Data}


Countries like Brazil with continental dimension present high heterogeneity in many different aspects, mainly characterized by a large gap between social classes. This can be drastically aggravated in situations such as the global pandemic, where not only public health is affected, but also a country's economy. So it is important to try to understand how (or if) socioeconomic variables can explain the cases and the number of deaths over time. However, a group of cities/towns may have a similar dynamic behavior in terms of their case and death curves. After identifying these groups, it would be interesting to propose a Functional Regression model with the socioeconomic variables as predictors of the count death curves for each group.

In clustering functional data, one seeks to identify homogeneous subsets of observations according to a random component in order to describe certain patterns in the data.
Clustering Functional Data is still a difficult problem mainly due to the obstacles encountered when working in spaces of infinite dimension. However, an approximation in a finite dimensional space can decrease complexity and facilitate estimation.
The most well known approach for cluster analysis is based on distance measurements. Classical methods compute distances between data points and identify clusters according to a partition algorithm. Techniques such as the hierarchical clustering and the $k$-means clustering can be found in (\cite{ward:1963}) and (\cite{hartigan1975clust}; \cite{hartigan1979algorithm}) respectively.
Several adaptations of the $k$-means algorithm for functional data have been proposed in the recent literature, such as \cite{abra:corn:lobe:moli:2003}, \cite{Ieva:paga:pigo:vite:2013}, \cite{Yamamoto14} and  \cite{tarpey2003clustering}. For hierarchical clustering, a few versions for functional data can be found in 
\cite{ferra:vieu:2004} \cite{ferra:vieu:2006} and \cite{boulle2014nonparametric}. Differently than cluster analysis based on distance, model-based clustering assumes that the data comes from a mixture of probability distributions and their parameters are, in general, estimate  by an EM algorithm. The number of clusters and model selection are based on criteria such as Akaike information criterion (AIC) or Bayesian information criterion (BIC), or, in some cases, by LASSO algorithm and its variants. Some examples of this approach for functional data can be found in \cite{bouveyron2011model}, \cite{JacquesPreda2014}, \cite{yao2005functional}, who use functional principal components analysis (FPCA); \cite{james2003clustering}, \cite{giacofci2013wavelet} where the infinite
dimensional space is approximated by a finite basis expansion and its coefficients are used to compute the distance measurements; \cite{wang2008nonparametric} who proposes a nonparametric version; and \cite{Serban12}, who investigate clustering methods for multilevel functional data and not only use mixture models but also use multilevel functional principal components. In this paper we make use of the  approach for clustering functional data based on hypothesis testing (see details in \cite{Zambom19}). There, the authors developed a procedure that, instead of applying the classical k-means algorithm by estimating the basis coefficients to compute the distance between curves, the procedure, at each step of the $k$-means algorithm, decides to which cluster a curve belongs based on a combination of two hypothesis test statistics. The advantage of this algorithm is that it is not restricted to fixed smoothing levels for all curves, since test statistics are used instead of basis coefficients.

\subsection{Function-on-Scalar Quantile Regression Model}

Recently, with the advancement of technology, more data can be recorded over a period of time (or space). In particular, one can easily obtain samples of curves or trajectories recorded individually. A Functional Linear Model is essentially a generalization of a linear model. There are several types  of functional linear model. For example,  scalar response variable with functional covariates (scalar-on-function), functional response variable  with scalar covariates (function-on-scalar) and the more general model, functional response with functional covariates (function-on function).  For instance, let $Y_i(t)$ be the cumulative daily number of deaths per 100k inhabitants for every location represented by B-splines according to the equation (\ref{bs}), for $i \in\{i=1,\dots,n\}$, taking values in the set $\{0,1,\ldots,N\}$. A functional linear model, known as function-on-scalar regression (\cite{chen2016variable,fan2000two}), is given by
\begin{eqnarray}\label{eq1}
Y_i(t)= \beta_{0}(t)+\sum_{j=1}^{p} X_{i,j}\beta_{j} (t)+\epsilon_{i}(t)
\end{eqnarray} 
for all $t \in [0,\infty)$ where $X_{ij}$ are the scalar covariates such as social-economic indicators with $j=1,..., p$, $\beta_j(t)$ are the functional coefficients and $\epsilon_{i}(t) \sim N (0, \Sigma)$ 
function drawn from a continuous stochastic process with expectation zero and covariance function $\boldsymbol{\Sigma}(s, t)=\text{cov}(\epsilon_{j}(s), \epsilon_{j}(t))$ with $s, t \in [0,\infty)$.
For a good introduction to Functional Data Analysis and Function-on-Scalar Regression see: \cite{chen2016variable} \cite{rams:1991}, \cite{rams:silv:2005} and \cite{Kokoszka17}. For a more advanced reading related to this work see, \cite{ferra:vieu:2006}, \cite{Aneiros15} and \cite{Gerth13}. Moreover, is not uncommon to perform variable selection in order to identify the most significant covariates that explains the data in the simplest way. This reduction of dimensionality can be achieved by different methods. The most common ones are via LASSO and Functional Principal components, see \cite{James2000}, \cite{matsui11}, \cite{romo20}.
Particularly, in this work we consider a functional quantile linear model with functional response and scalar covariates. Given the heterogeneity in the cumulative death curves, we consider using the functional quantile regression model instead of the functional linear model due to its robustness in the presence of extreme or atypical curves. Also the idea of using this model is also because it is possible to estimate different quantile values on the cumulative death curves that can be approximated to the mean curves of the clusters estimated as alert levels in the cumulative number of deaths by cities. For example, quantiles from 5\% to 25\% for curves at a low level, quantile 50\% to estimate the median curve or quantiles from 75\% to 95\% for curves at a high level of the cumulative number of deaths. Based on the approach of \cite{fan2000two}, let $Q_{\tau}(Y_{i}(t)|X_{i})$ be the conditional $\tau th$ quantile function of the functional response $Y_{i}(t)$ given the true scalar covariates $X_{i,j}$ with $\tau \in (0, 1)$, for $i=1,..., n$ and $j=1,.., p$. The function-on-scalar quantile regression model is assumed as
\begin{eqnarray}\label{eq1_1}
Q_{\tau}(Y_{i}(t)|X_{i}) = \beta_{0,\tau}(t)+\sum_{j=1}^{p} X_{i,j}\beta_{j,\tau}(t),
\end{eqnarray} 
where $\beta_{0,\tau}(t)$ is the functional quantile-level varying intercept and $\beta_{j,\tau}(t)$ is the functional quantile-level regression coefficient and is the main target of interest. The estimation process is based on two steps. First, a quantile regression model with LASSO penalty is fitted separately for each point \emph{t} along the functional response variable $Y(t)$ together with the scalar covariates. The quantile regression LASSO for this step is then given by the minimization problem 
\begin{eqnarray}\label{eq1_1_lasso}
\boldsymbol{\beta}_{\tau} = \arg\min_{\boldsymbol{\beta}\in R^{p}}\sum_{i=1}^{n}\rho_{\tau}(Y_{i}-\boldsymbol{X}_{i}^{'}\boldsymbol{\beta}_{\tau})+n\lambda\sum_{j=1}^{p}|\beta_{j}|, \:\:\: j=1,..., p, \:\:\: i=1,2,..., n,
\end{eqnarray}
where $\lambda$ is the tuning parameter that shrinks some beta coefficients towards zero and the second term is the LASSO penalty term which uses the $L_{1}$ norm. The minimization problem of equation (\ref{eq1_1_lasso}) can be solved by linear programming algorithms such as the Barrodale and Roberts algorithm (see \cite{koenker_2005}). Secondly, the resulting estimated coefficients are smoothed using B-splines basis functions to obtain the estimated functional coefficients as follows
\begin{eqnarray}\label{funcbetas}
\widehat{\beta}_{j,\tau}(t) \approx \sum_{k=1}^{K} \widehat{\beta}_{j,\tau,k}\phi_{k}(t) = \boldsymbol{\widehat{\beta}_\tau'}\boldsymbol\phi(t), \:\:\: j=1,..., p.
\end{eqnarray}
For each quantile used in the functional quantile regression model, a set of estimated functional coefficients are obtained. To the best of our knowledge, there are still no published research works in the literature on functional quantile regression models with functional response and scalar predictors. Some published works on functional quantile regression consider scalar response and functional predictors (\cite{kato2012estimation,cardot2005quantile,li2016inference}), also using a functional partially linear predictor (\cite{lu2014functional,yao2017regularized}) and for predictors with ultra-high dimensions as in \cite{ma2019quantile}. 

\section{Empirical Strategy}\label{strategy}
\subsection{Functional Clustering and Quantile Regression Model Strategies}
The first strategy is based on clustering the curves of the capital cities and the municipalities of the Brazilian states. For this, it can be taken into account that FDA models are typically non-parametric models which, in turn, provide a more flexible modeling process. Hence, smoothing plays an important rule for the statistical analysis of functional data (see for example \cite{rams:silv:2002}). The smoothing is then performed by assuming that every curve can be approximate by a linear combination of basis functions, in particular, B-splines, as described in equation (\ref{bs}) and Figure \ref{rawcurves} show the result of the smoothing procedure. The first and second functional derivatives of the represented curves by B-splines basis functions are calculated and viewed as the velocity and acceleration curves of the cumulative number of deaths. After smoothing all the curves, the methodology developed in \cite{zamb:colla:dias:2016a} is applied to clustering the them.

The second strategy for the analysis of the cumulative death curves per 100k inhabitants was developed by estimating the function-on-scalar quantile regression model in Equation (\ref{eq1_1}). This model is proposed in order to estimate the quantile curves that most closely approximate the mean curves estimated in the previous strategy. It is also used to know the effect of scalar socioeconomic covariates on the cumulative death curves at each alert level according to the quantile used in the estimation of the model. The model is based on the approach of \cite{fan2000two} where first a separate quantile regression model with penalty LASSO is fitted at each point along the functional response and then the estimated coefficients are smoothed to obtain the estimated functional coefficients. Thus, the estimated quantile curves for some quantile values could give a closer approximation to the mean curves estimated in the functional clustering strategy. For this reason we use the functional quantile regression model than the functional linear regression model.

\subsection{Preliminary Results}
\begin{figure}[!htbp]
     \begin{center}
        \subfigure[]{%
            \label{fig:histcases}
            \includegraphics[width=8cm,height=7.5cm]{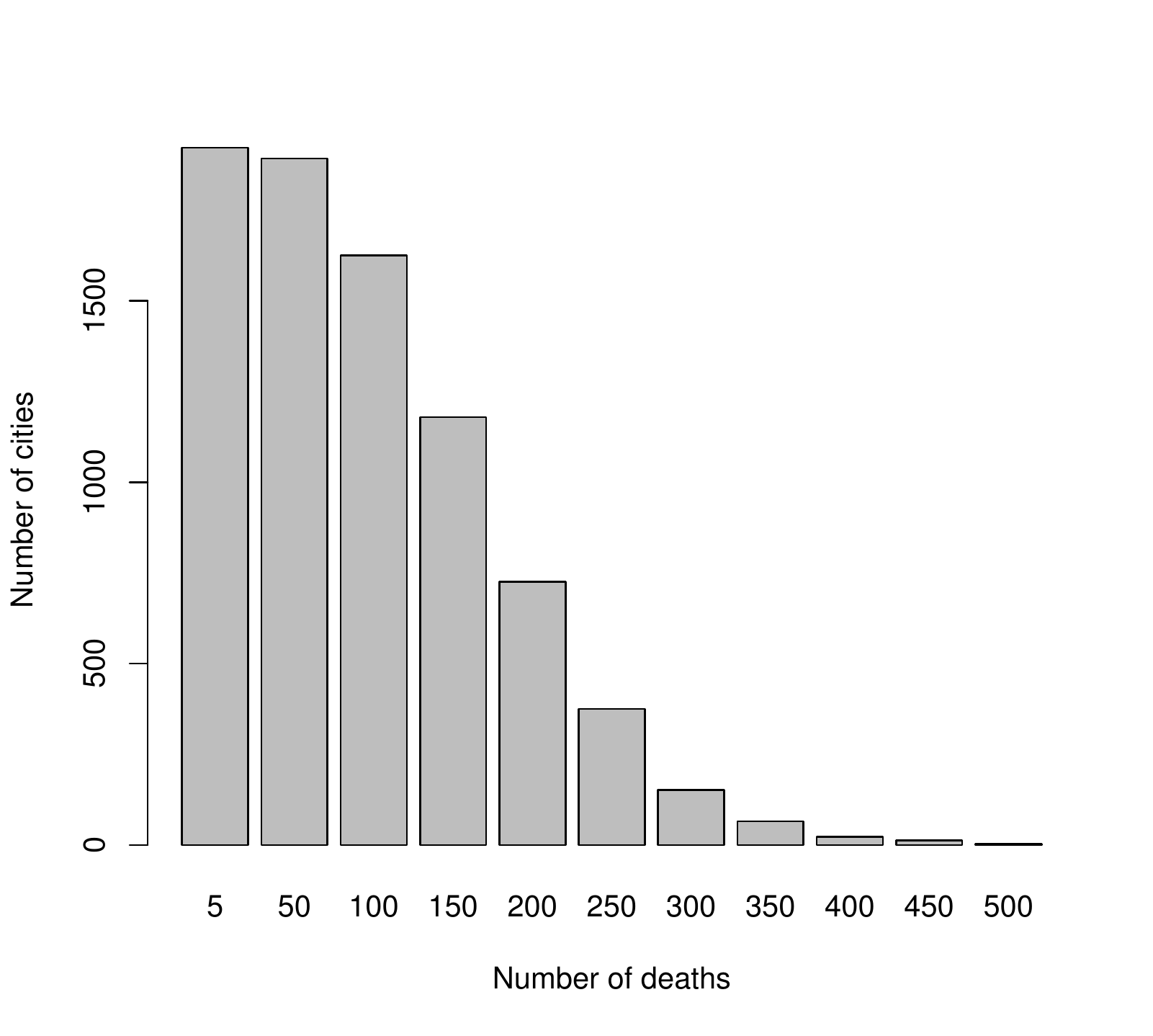}
        }
        \subfigure[]{%
           \label{fig:corrmatrix}
           \includegraphics[width=8cm,height=7.5cm]{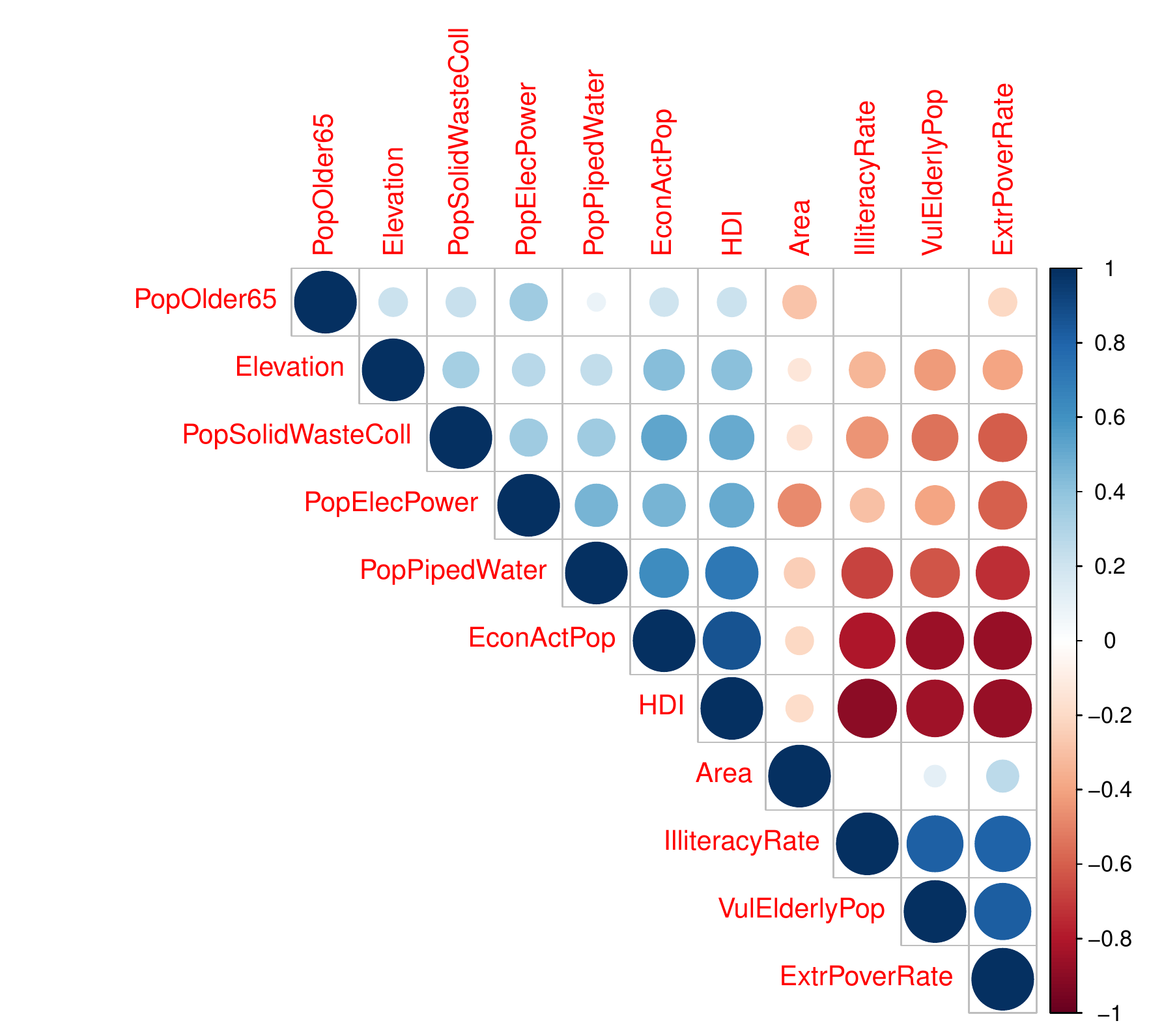}
        }
    \end{center}
    \caption{%
        (a) Distribution of cumulative deaths by number of cities  (b) Correlation plot chart of scalar covariates}
   \label{preliminar}
\end{figure}

\begin{figure}[!htbp]
     \begin{center}
        \subfigure[]{%
           \label{meancaps}
           \includegraphics[width=7cm,height=5.5cm]{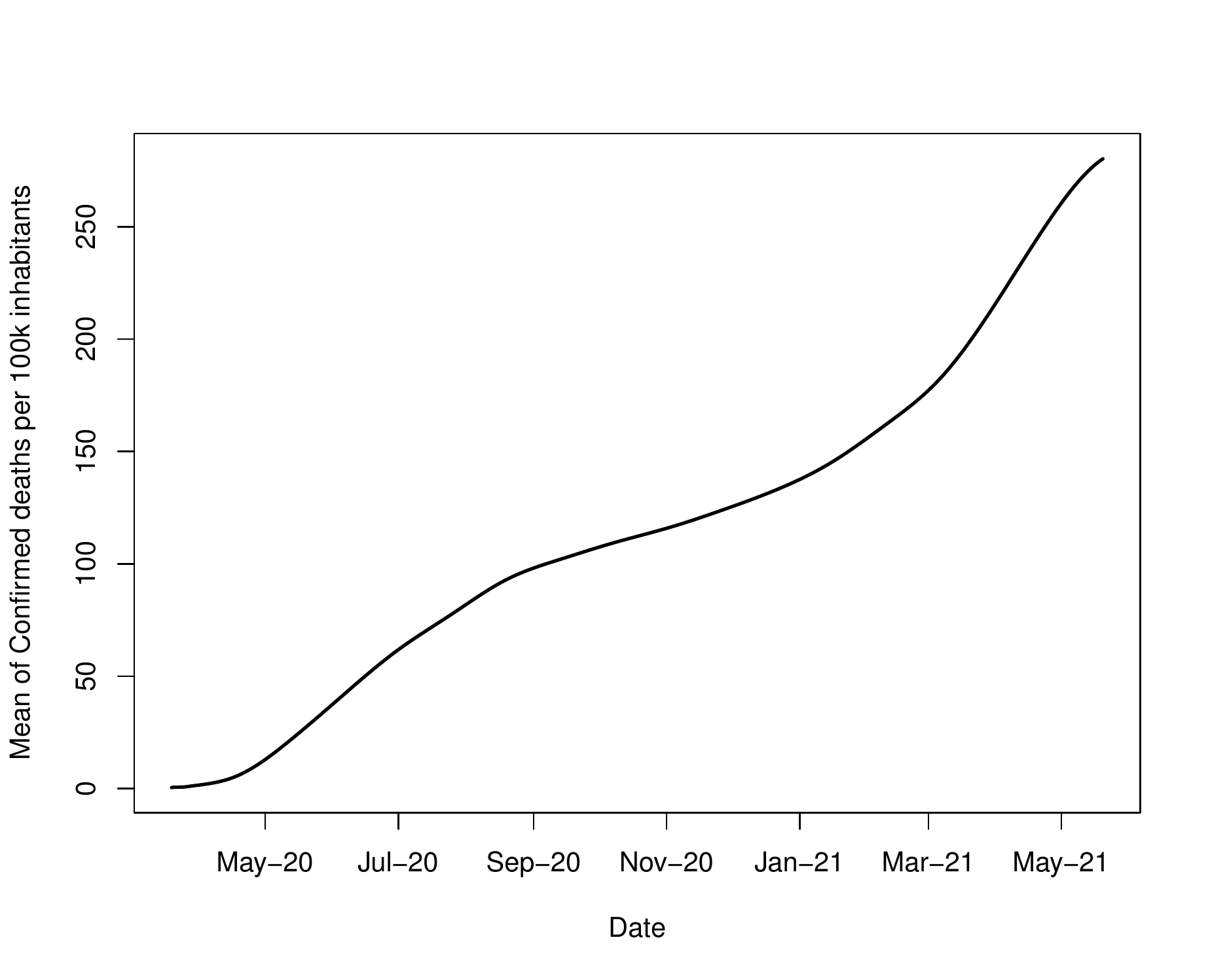}
        }
        \subfigure[]{%
           \label{sdcaps}
           \includegraphics[width=7cm,height=5.5cm]{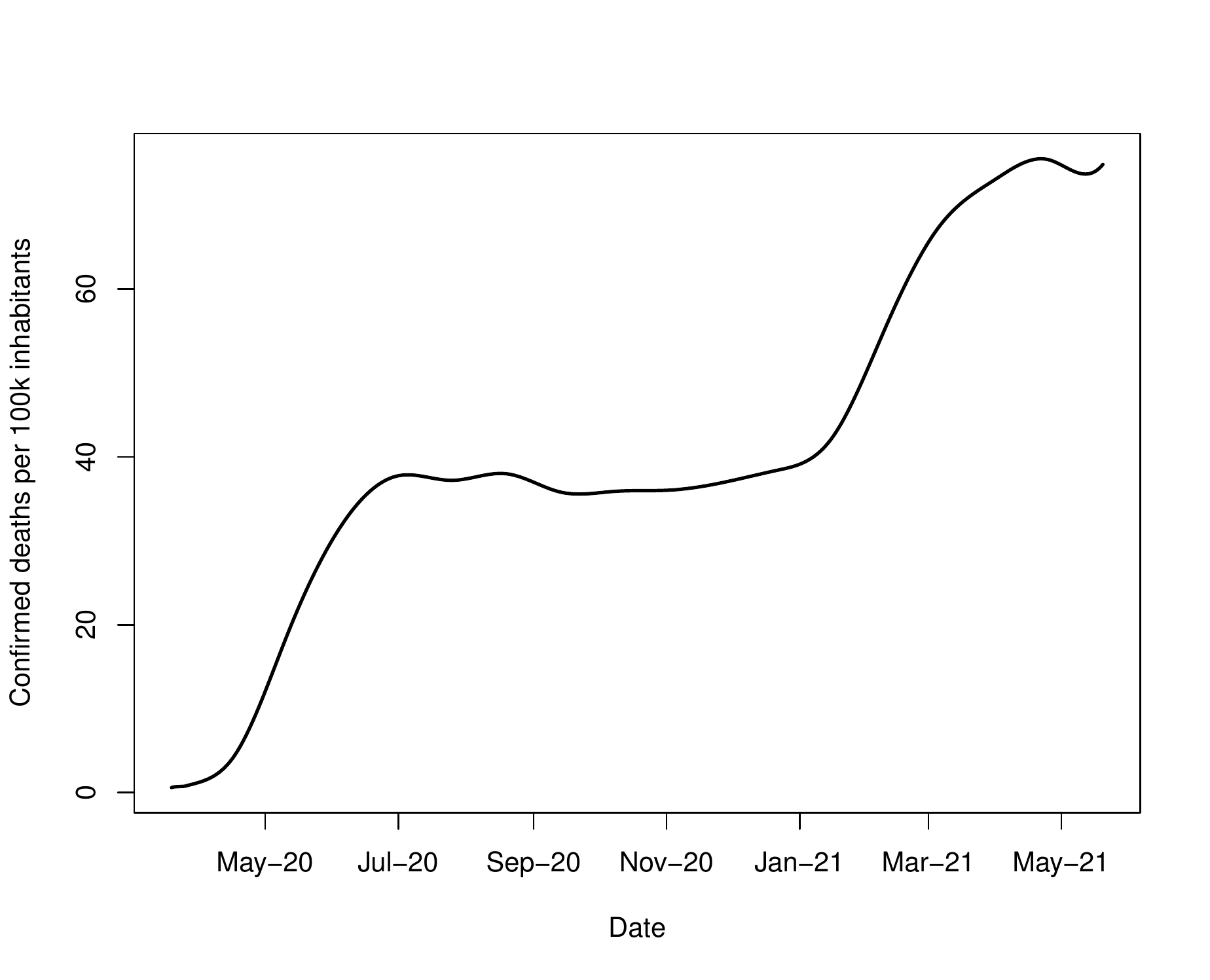}
        }\\
        \subfigure[]{%
           \label{meannocaps}
           \includegraphics[width=7cm,height=5.5cm]{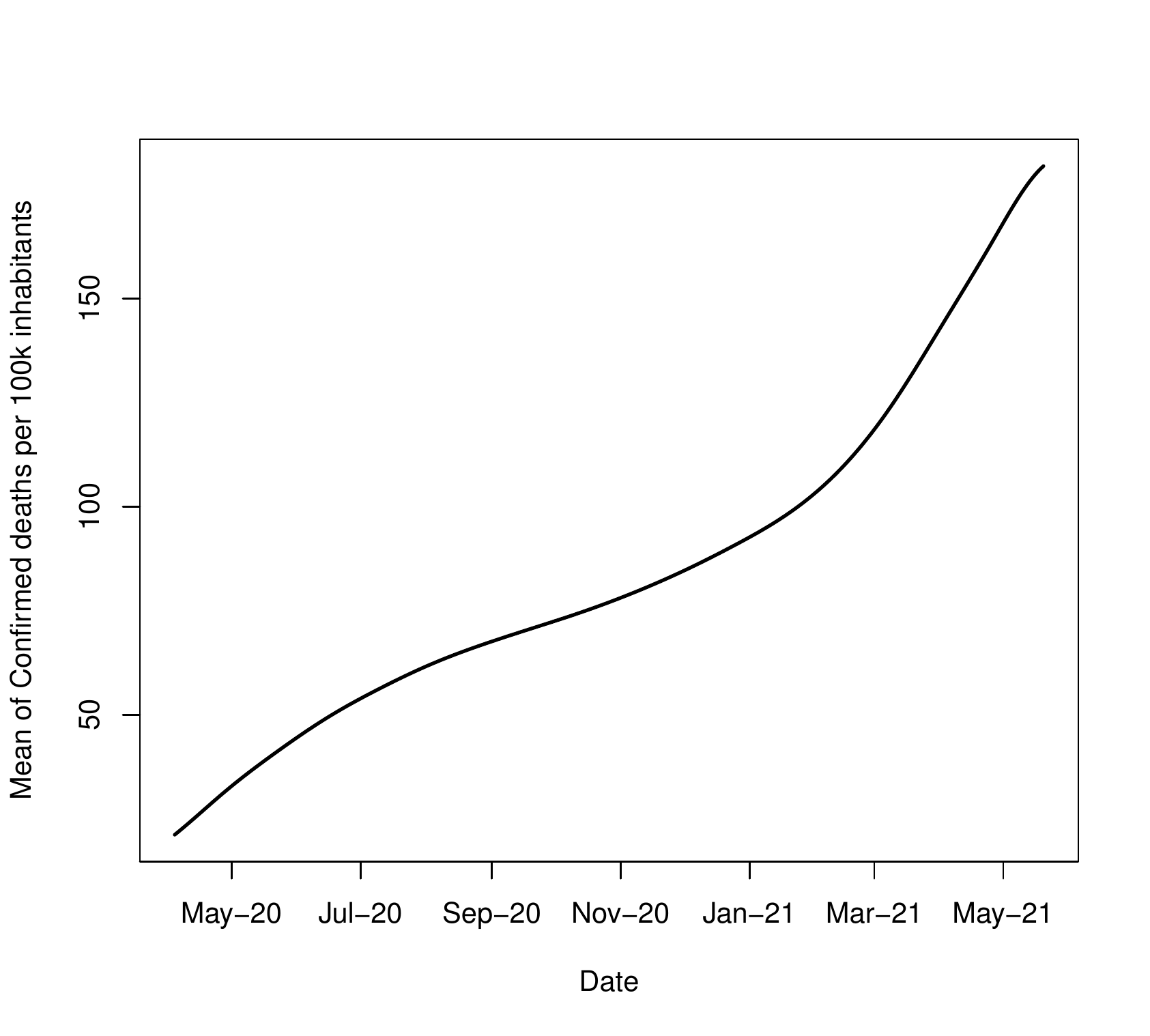}
        }
        \subfigure[]{%
           \label{sdnocaps}
           \includegraphics[width=7cm,height=5.5cm]{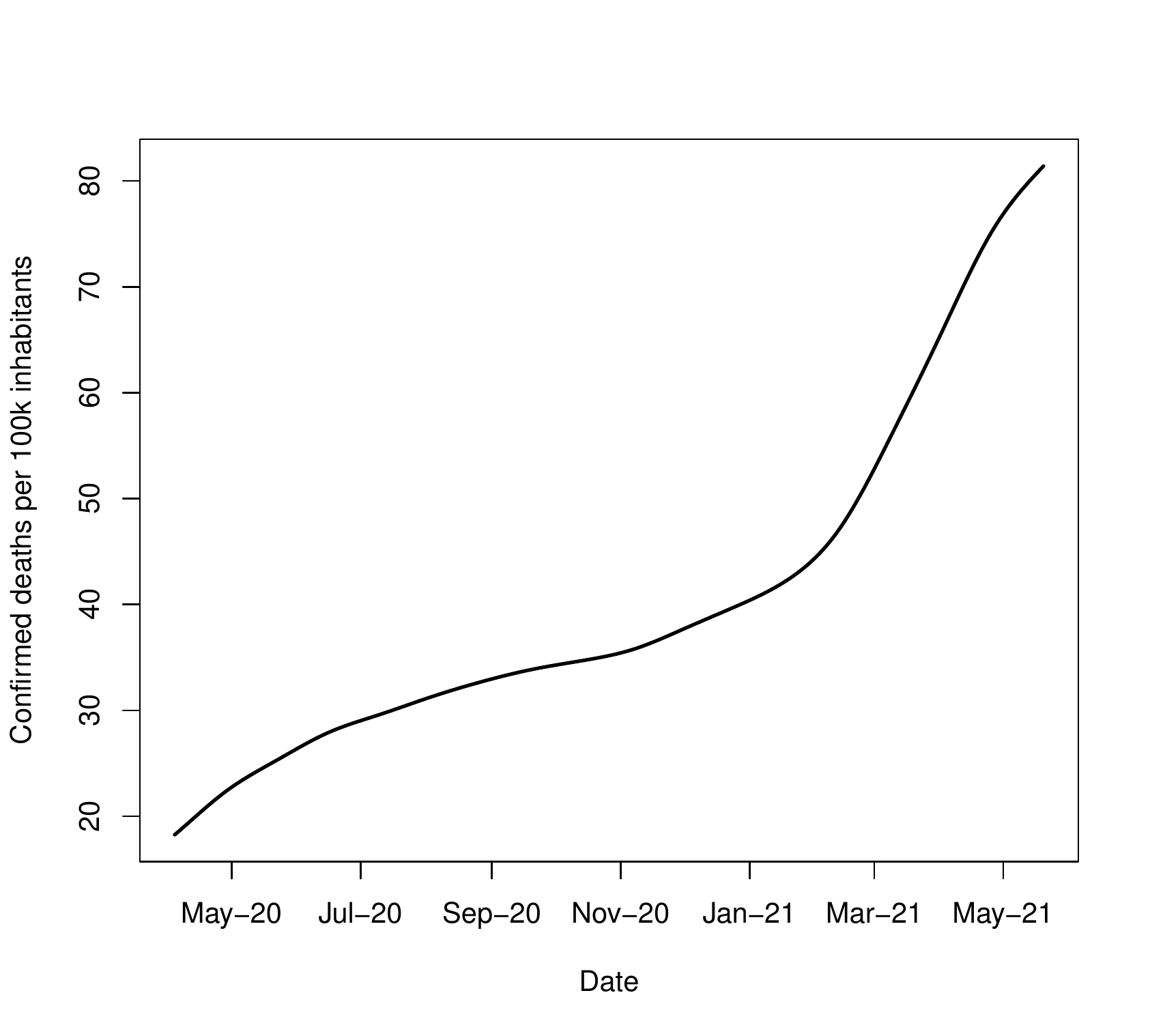}
        }
    \end{center}
    \caption{%
        Functional descriptive measures for cumulative death curves per 100k inhabitants. First line: functional (a) mean and (b) standard deviation for capital cities. Second line: functional (c) mean and (d) standard deviation for non-capital cities.}
   \label{fdescript}
\end{figure}

A total of 1,921 cities were considered as explained in Section \ref{Data}. Figure \ref{fig:histcases} shows the distribution of cities by cumulative number of deaths. We can observe that most cities have a lower number of deaths and few cities have a high number of deaths. For these cities, we use some indicators as scalar covariates in the functional quantile regression model of equation (\ref{eq1_1}). The correlations between all covariates and their functional relation can be seen in Figure \ref{fig:corrmatrix}. Some covariates are positively associated while others are negatively associated. For example, the covariates HDI (Human Development Index), population with piped water and economically active population are negatively correlated with vulnerable elderly population and rates of illiteracy and extreme poverty.
For capital cities, the mean curve (Figure \ref{meancaps}) of the cumulative number of deaths per 100k inhabitants increased considerably at the beginning of the pandemic until August 2020 and also in the beginning of 2021 until May 2021. The curve of the standard deviation (Figure \ref{sdcaps}) shows a great dispersion during the same dates due to the uncontrolled daily increase with large outbreaks of new confirmed cases. The mean death curve for non-capital cities (Figure \ref{meannocaps}) shows a behavior similar to that of capital cities in terms of the same dates of increases in deaths, however the dispersion curve (Figure \ref{sdnocaps}) of the cumulative number of deaths during the pandemic is presents more slowly than in capital cities. The increase in cumulative deaths from the beginning of 2021 to the end of April is due to the appearance of new variants of Covid-19 that affected more young people than older adults. Given the differences in the heterogeneity of the curves, in the next section the curves of the cumulative number of deaths per 100k inhabitants of capital and non-capital cities will be analyzed separately.

\newpage
\section{Functional clustering analysis}\label{funclustanalysis}
This section presents a clustering functional analysis for two types of datasets: cumulative death curves per 100k inhabitants for capital and non-capital cities. Both datasets are grouped into 3 clusters in order to identify and locate cities at three alert levels (low, moderate, high). The estimated functional clusters are presented for the death curves represented by B-splines basis functions and also for their first two functional derivatives that show the velocity and acceleration curves of the cumulative daily number of deaths. The dynamic of the mean death curves and their locations are shown in subsections \ref{subsec:capital} and \ref{subsec:noncapital}.

\subsection{Functional clustering for capital cities}\label{subsec:capital}
\begin{figure}[!htbp]
     \begin{center}
         \subfigure[]{%
           \label{fig:clustercapitals}
           \includegraphics[width=8cm,height=8cm]{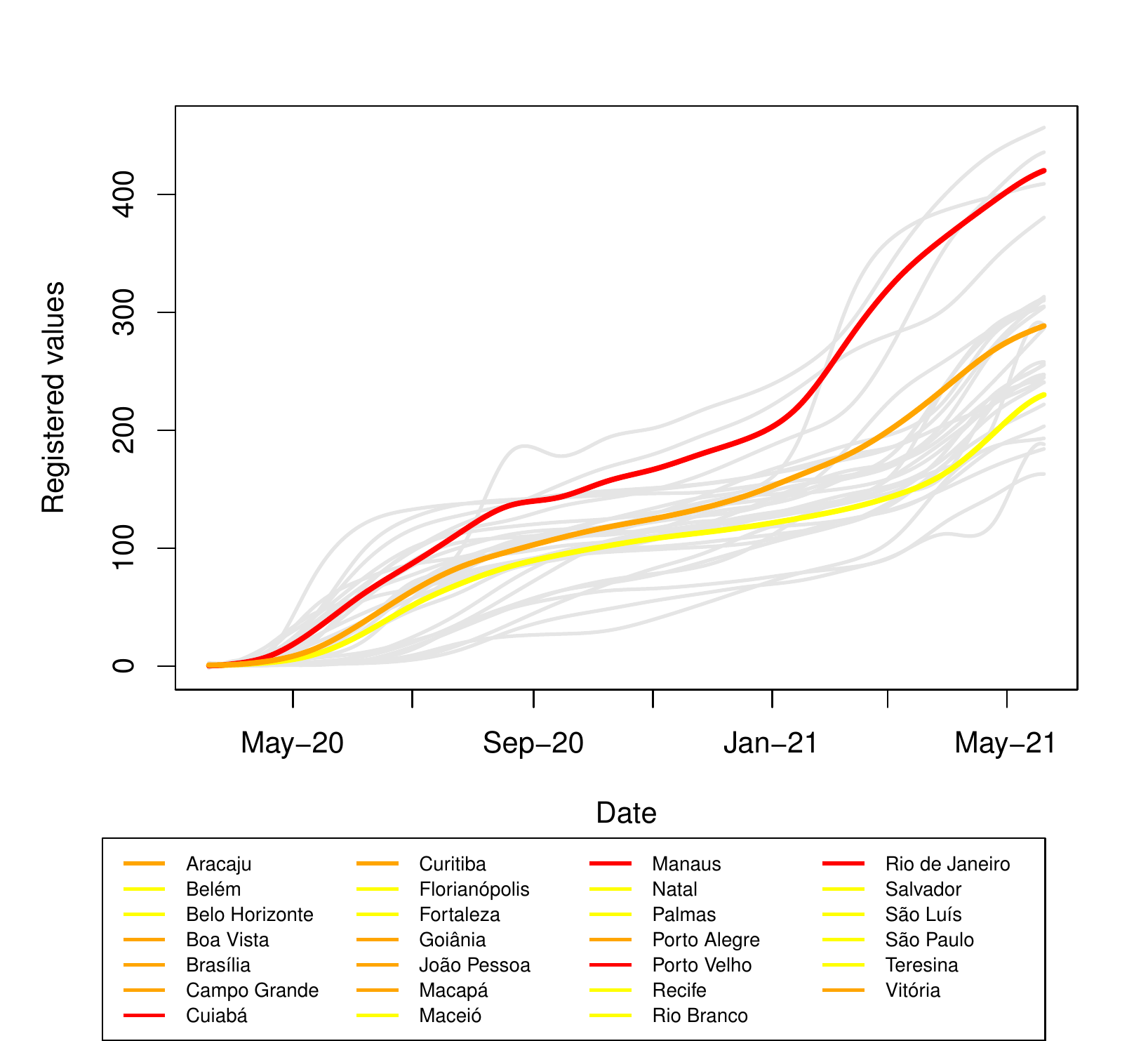}
        }
          \subfigure[]{%
           \label{mapcapK3-2}
           \includegraphics[width=8cm,height=8cm]{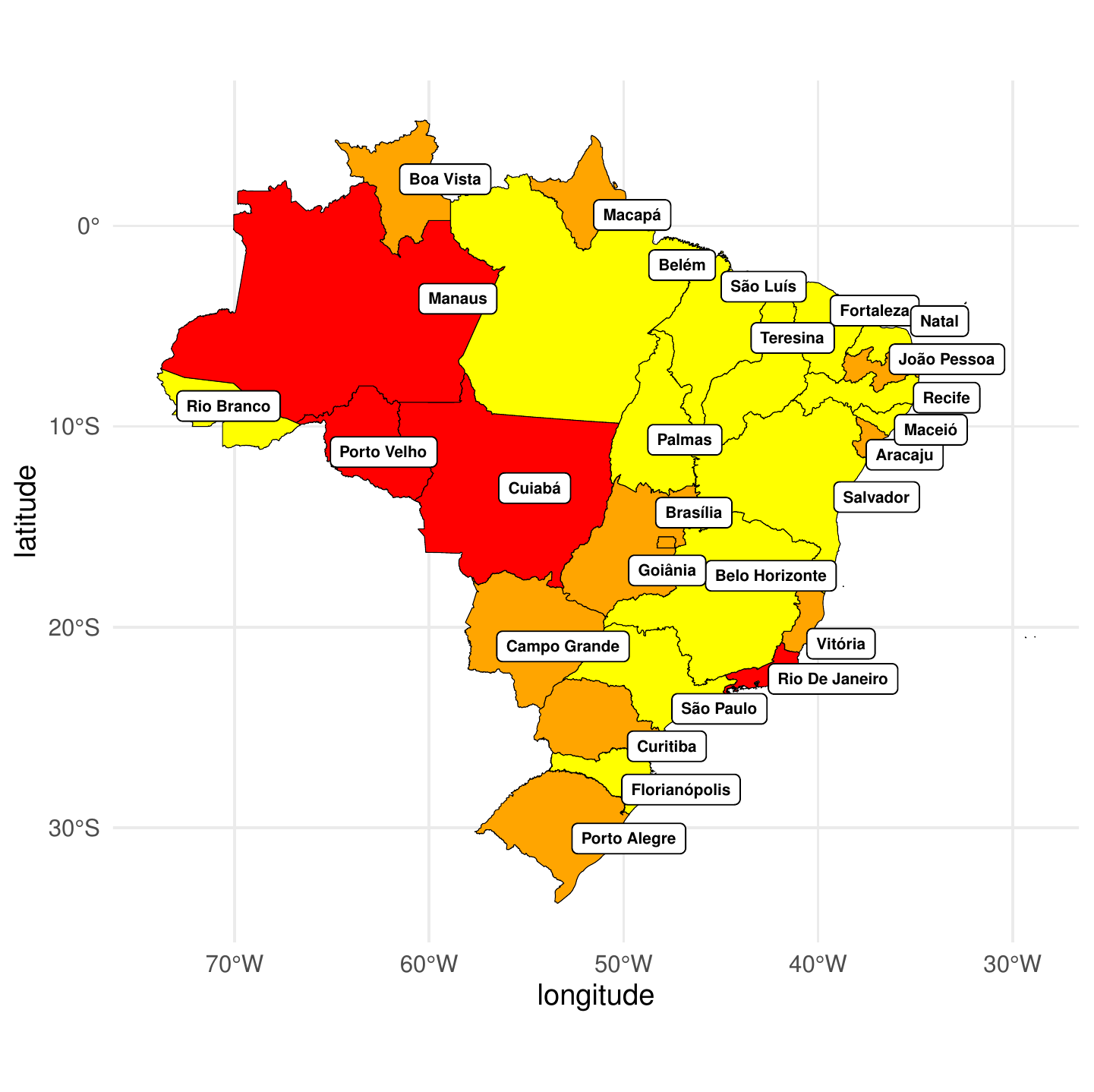}
        }
    \end{center}
    \caption{%
        Represented death curves per 100k inhabitants of the capital cities: (a) estimated functional clusters with K = 3 and (b) spatial location according to estimated functional clusters.}
   \label{fig:fpcacurves}
\end{figure}
Figure \ref{fig:fpcacurves}\subref{fig:clustercapitals} exhibits the result of clustering the curves for the capital cities and the respective mean curves of each cluster. For better visualization, 
Figure \ref{fig:fpcacurves}\subref{mapcapK3-2} 
locates on the map the states according to their capitals and clusters to which they belong.
\begin{figure}[!htbp]
     \begin{center}
       \subfigure[]{%
           \label{fig:clustercapitalsD1}
           \includegraphics[width=8cm,height=8cm]{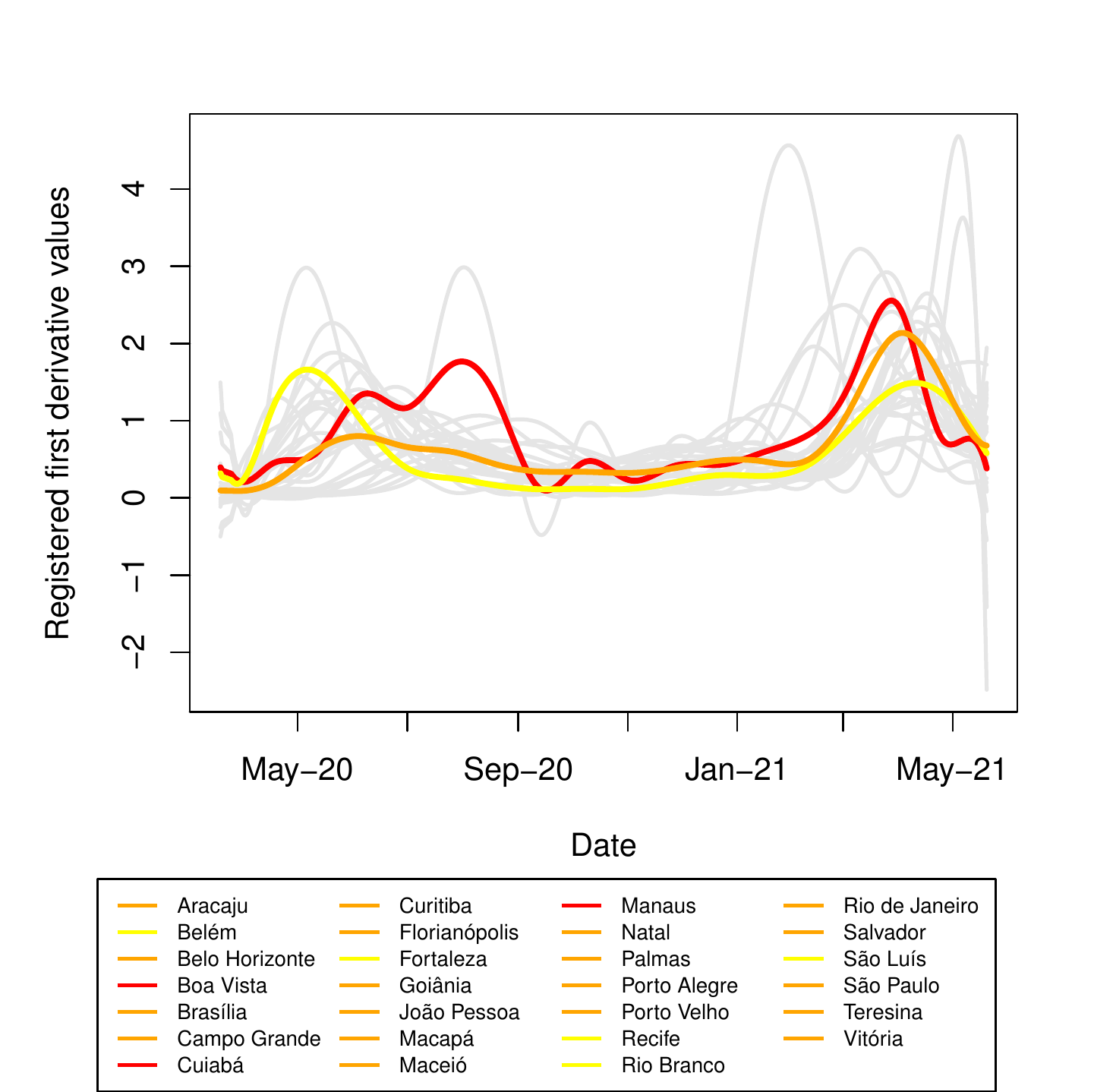}
        }
        \subfigure[]{%
           \label{mapcapK3D1-2}
           \includegraphics[width=8cm,height=8cm]{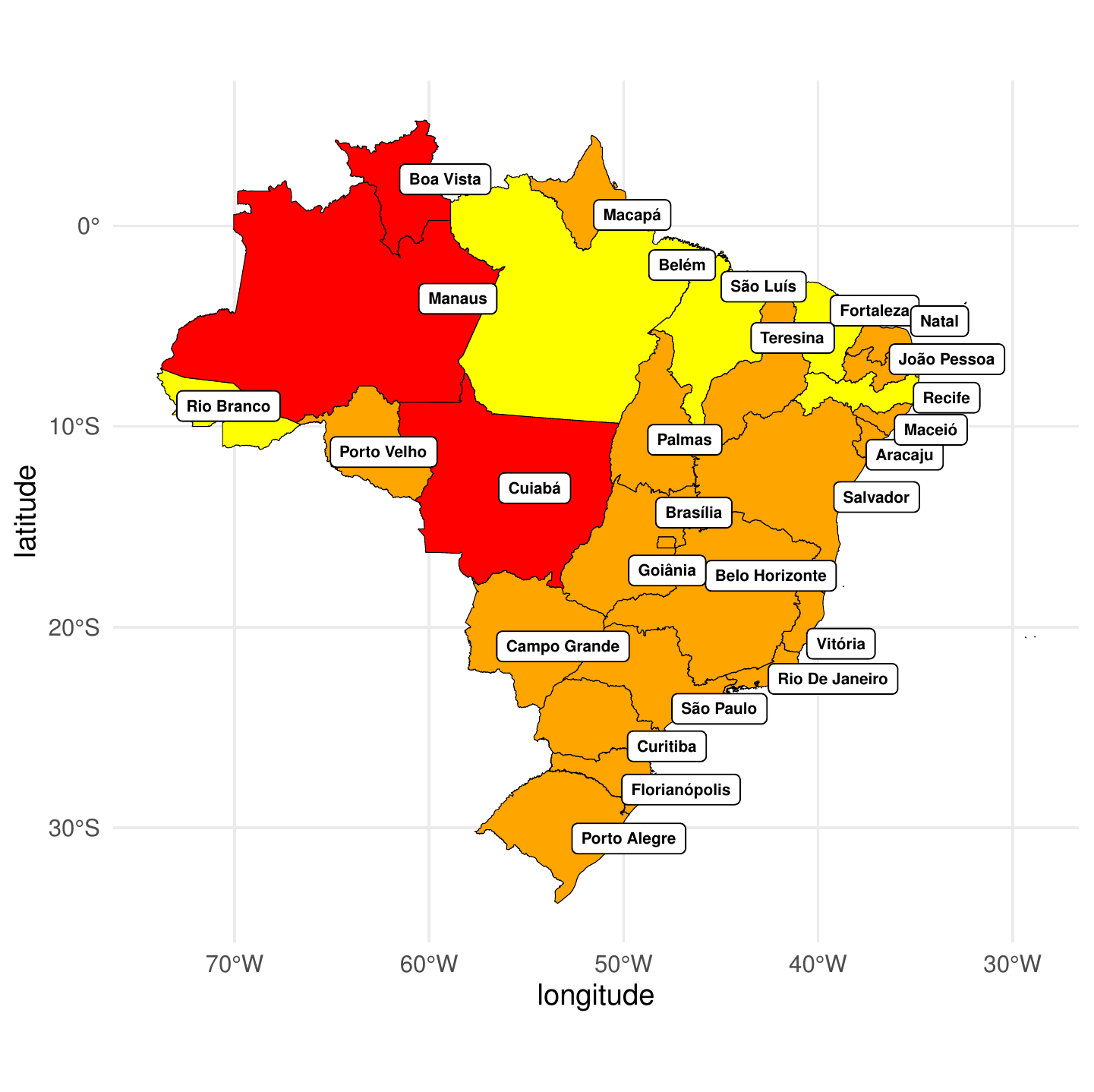}
        }
    \end{center}
    \caption{%
        First functional derivative of death curves per 100k inhabitants of the capital cities: (a) estimated functional clusters with K = 3 and (b) spatial location according to estimated functional clusters.}
   \label{fig:fpcaderiv1}
\end{figure}
Figure \ref{fig:clustercapitalsD1} one can see the dynamic of the mean curves of their clusters. To see the clusters on the map, Figure \ref{mapcapK3D1-2} display the effects of this clustering procedures on the Brazil map. Comparing Figures \ref{mapcapK3D1-2} and \ref{mapcapK3-2}, one may notice that some states capitals that were originally allocated as low number of deaths (yellow) have moved to the category of moderate number (orange) while other states have moved from moderate number (orange) to high number of deaths (red) when considering the velocity of the curves. Similar conclusion might be seen when looking  acceleration in Figure \ref{fig:fpcaderiv2}. For example, the city of Manaus in the northern state of Amazonas, in the represented (smoothed) curves as in the velocity and acceleration curves remains at the same high alert level, which means that the velocity and acceleration of the cumulative number of deaths increases considerably causing that this city could have another high peak of deaths in its population. The city of Rio de Janeiro would evidence a possible improvement situation when going from a high alert level (red) to low level (yellow) through the three scenarios.
\begin{figure}[!htbp]
     \begin{center}
        \subfigure[]{%
           \label{fig:clustercapitalsD2}
           \includegraphics[width=8cm,height=8cm]{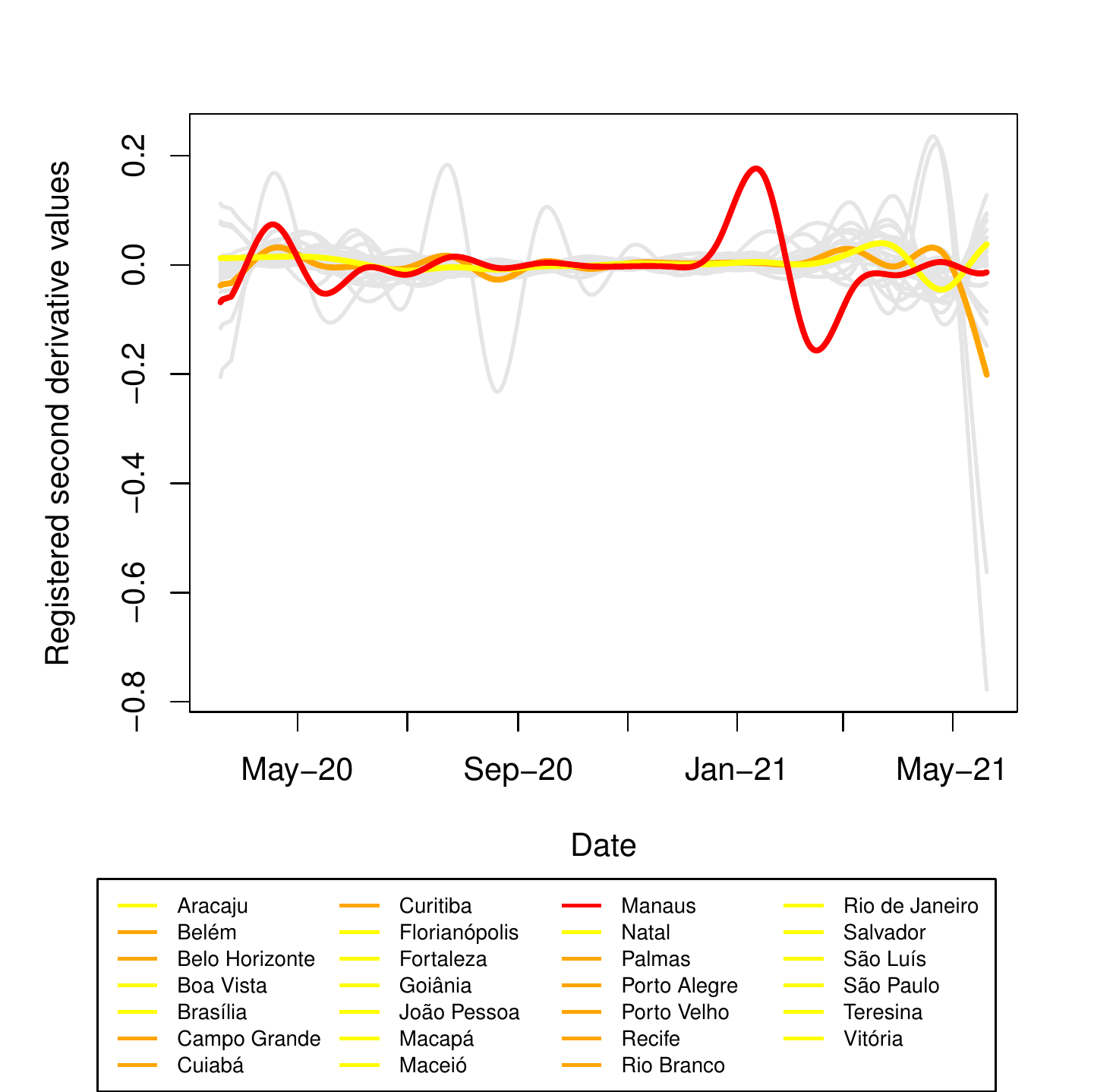}
        }
        \subfigure[]{%
           \label{mapcapK3D2-2}
           \includegraphics[width=8cm,height=8cm]{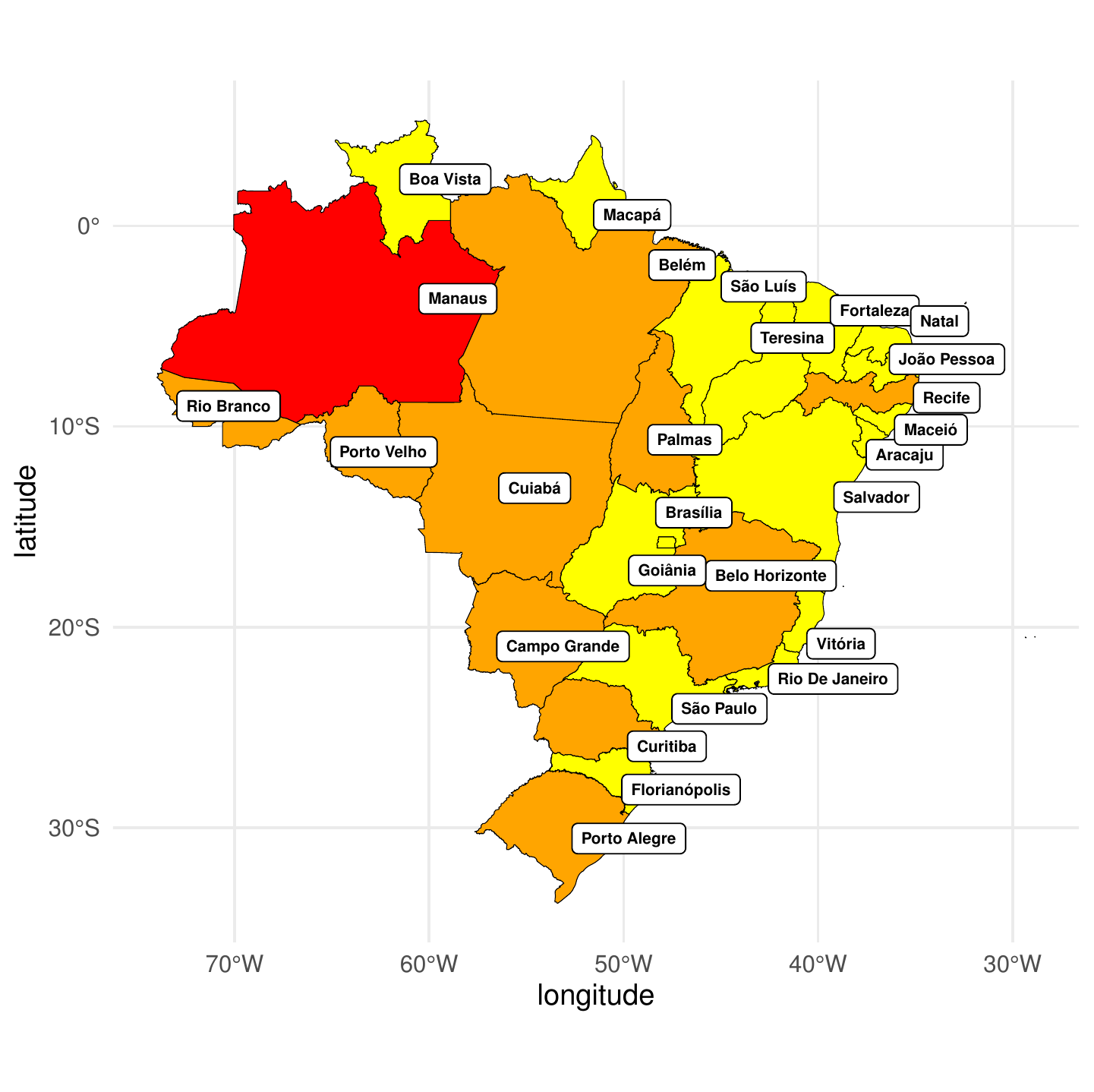}
        }
    \end{center}
    \caption{%
        Second functional derivative of death curves per 100k inhabitants of the capital cities: (a) estimated functional clusters with K = 3 and (b) spatial location according to estimated functional clusters.}
   \label{fig:fpcaderiv2}
\end{figure}

\newpage
Increasing or decreasing the alert level through the three scenarios can reveal a possible situation that generates critical events of severity or improvement in the cumulative number of deaths in a given city. It is important to mention that this type analysis could have helped policymakers to act properly avoiding a possible increase in number of deaths in case they had also the information about velocity and acceleration of the deaths count curves.

\subsection{Functional clustering for non-capital cities}\label{subsec:noncapital}
\begin{figure}[!htbp]
     \begin{center}
        \subfigure[]{%
            \label{fig:noncap}
            \includegraphics[width=5.3cm,height=8cm]{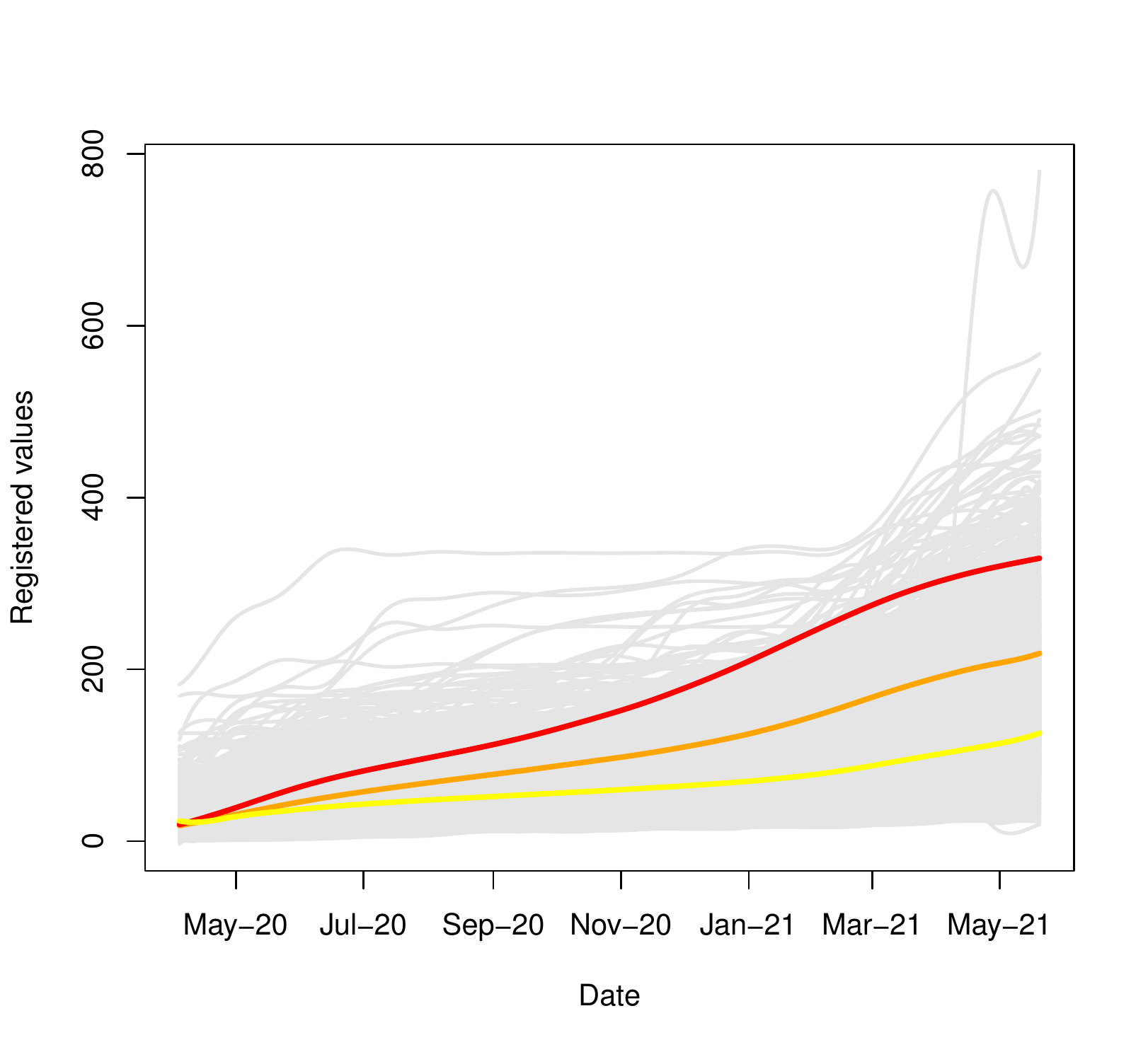}
        }
        \subfigure[]{%
           \label{fig:noncap_1-deriv}
           \includegraphics[width=5.3cm,height=8cm]{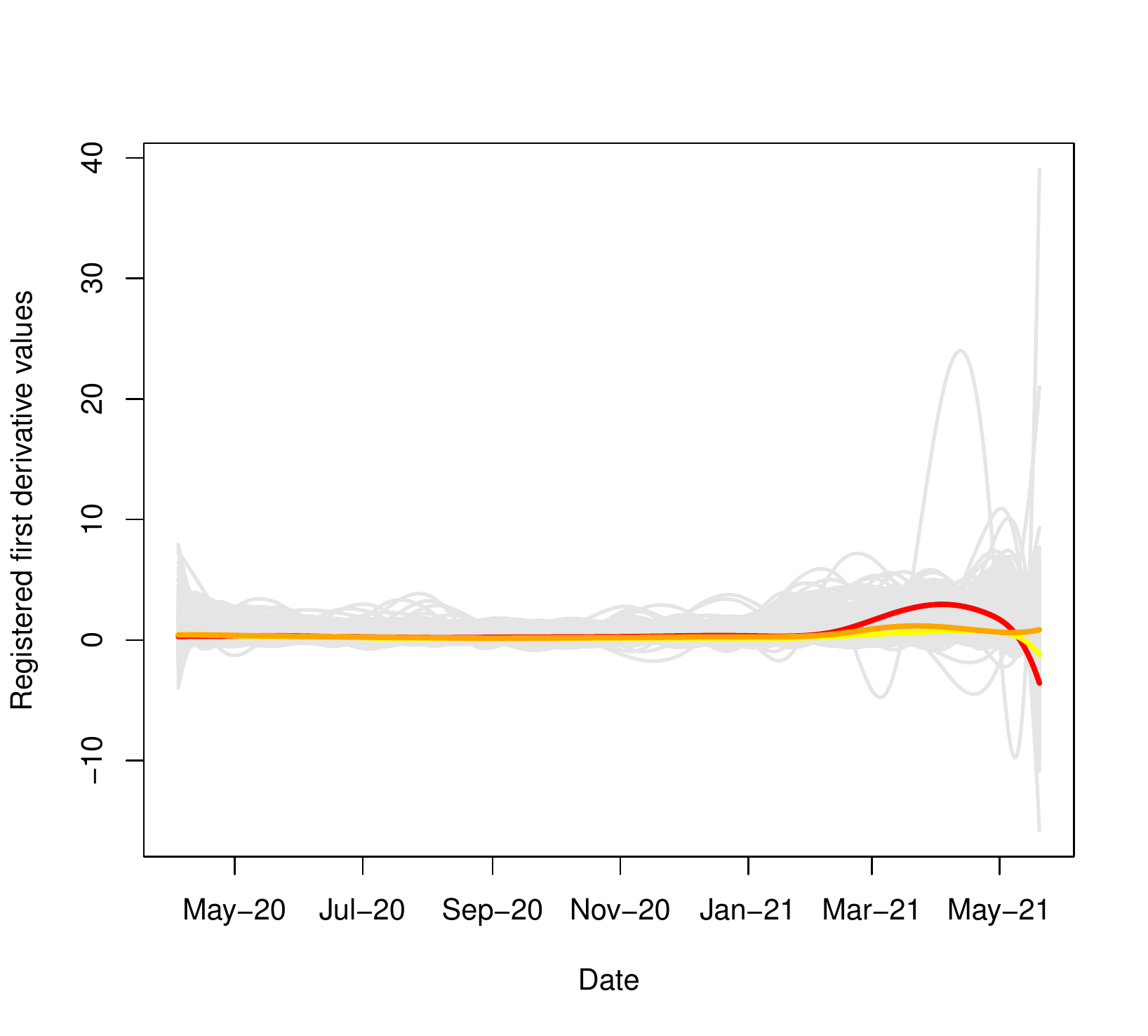}
        }
        \subfigure[]{%
           \label{fig:noncap_2-deriv}
           \includegraphics[width=5.3cm,height=8cm]{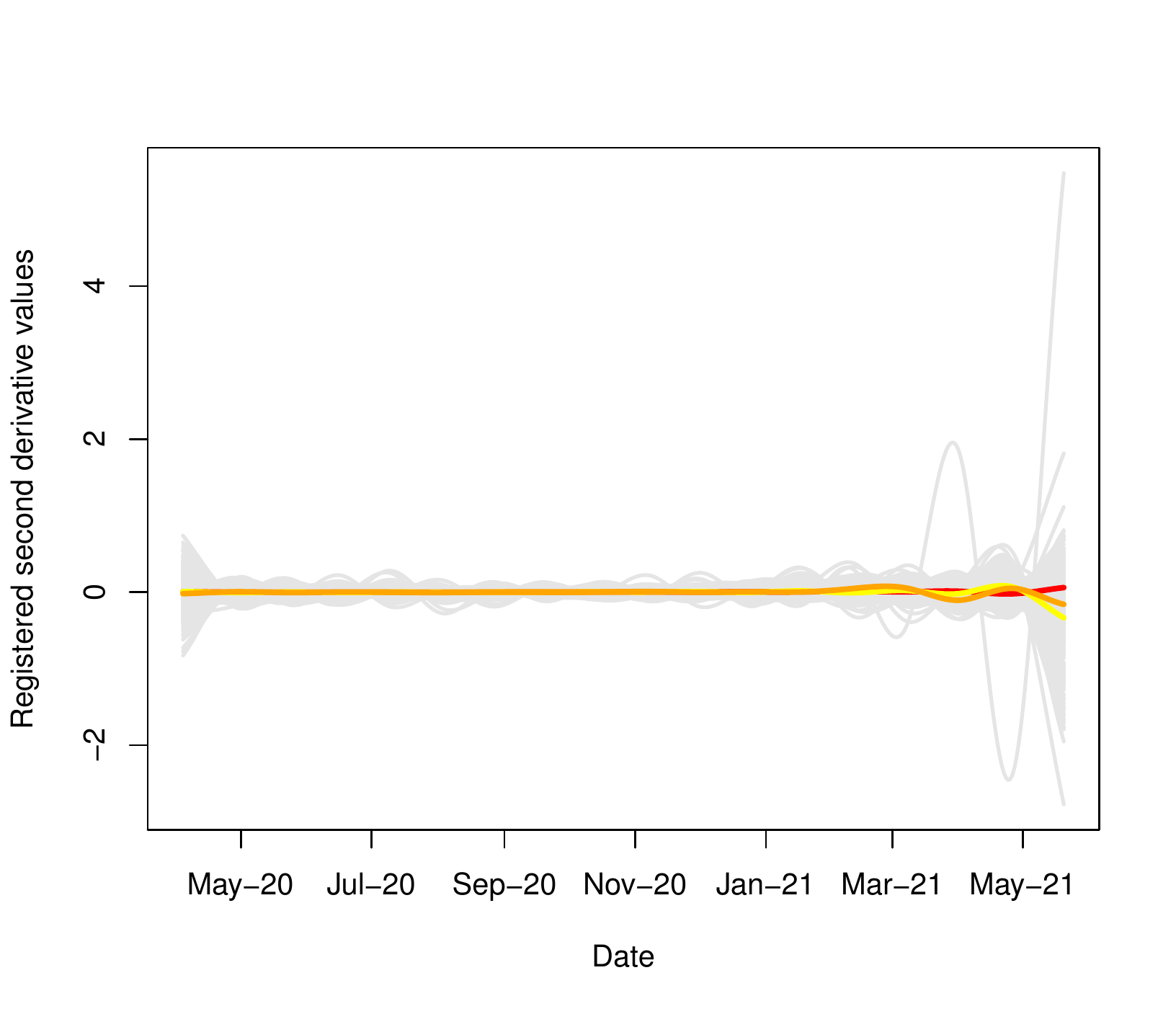}
        }
        \end{center}
    \caption{%
        Estimated functional clusters with K = 3 for non-capital cities: (a) represented curves, (b) first functional derivative and (c) second functional derivative of death curves per 100k inhabitants.}
   \label{figclusternoncapitals}
\end{figure}

Despite the large heterogeneity between all non-capital cities, the procedure is able to identify 3 different groups, as seen in Figure \ref{fig:noncap}. Figure \ref{fig:noncap_1-deriv} shows the result of clustering the velocity of the death count curves. Again, some type of periodic behavior seems to be present. In addition, one of the mean velocity curves related to high alert level appears to increase in February 2021. Figure \ref{fig:noncap_2-deriv} exhibits the outcome of clustering the acceleration (second functional derivative). Here, the periodic behavior is very clear. The spatial location of the estimated functional clusters for the municipalities (non-capital cities) considered by each state are illustrated in Appendix A of the supplementary material. For each figure, the estimated clusters for the three alert levels are displayed using the represented death curves (left column), and the first (middle column) and second functional derivative (right column) of death curves. The results are divided according to the regions that make up the country. Most of the municipalities in the states of the north (Figures 1-2), northeast (Figures 3-5) and south (Figure 9) regions go from being at the low alert level (yellow) to the high level (red) in the functional clusters estimated using the second functional derivative. This means that the acceleration in the cumulative number of deaths in these cities is increasing, which would imply possible new outbreaks with a high number of deaths. Some states in the central-west (Figure 6) and southeast regions (Figures 7-8) present a similar behavior, however it is highlighted that several municipalities specifically in the states of Mato Grosso do Sul (Figures 6(d)-6(f)), Mato Grosso (Figures 6(g)-6(i)) and Belo Horizonte (Figures 8(a)-8(c)) tend to show possible improvements when moving from a high alert level (red) at a low alert level (yellow) due to a slower acceleration than expected in the cumulative number of deaths. 

\section{Function-on Scalar Quantile Regression Model with LASSO Penalty}\label{FOSQR}
In this section we make use of the function-on-scalar quantile regression model described in equation \ref{eq1_1} with LASSO penalty on the functional coefficients. The basic idea is to verify whether it is possible to explain the death count curve per 100k inhabitants at a given level of alert of the functional response by the approximate quantiles using scalar socioeconomic covariates such as: area, altitude, population with access to treated (piped) water, population served by collected municipal solid waste, population with access to electricity, population over 65, economically active population, vulnerable elderly population, illiteracy rate, extreme poverty rate and HDI (human development index). For the estimation of the function-on-scalar quantile regression model we use the represented death curves of the non-capital cities illustrated in Figure \ref{fig:noncap}.  
\begin{figure}[!htbp]
     \begin{center}
         \includegraphics[width=10cm,height=10cm]{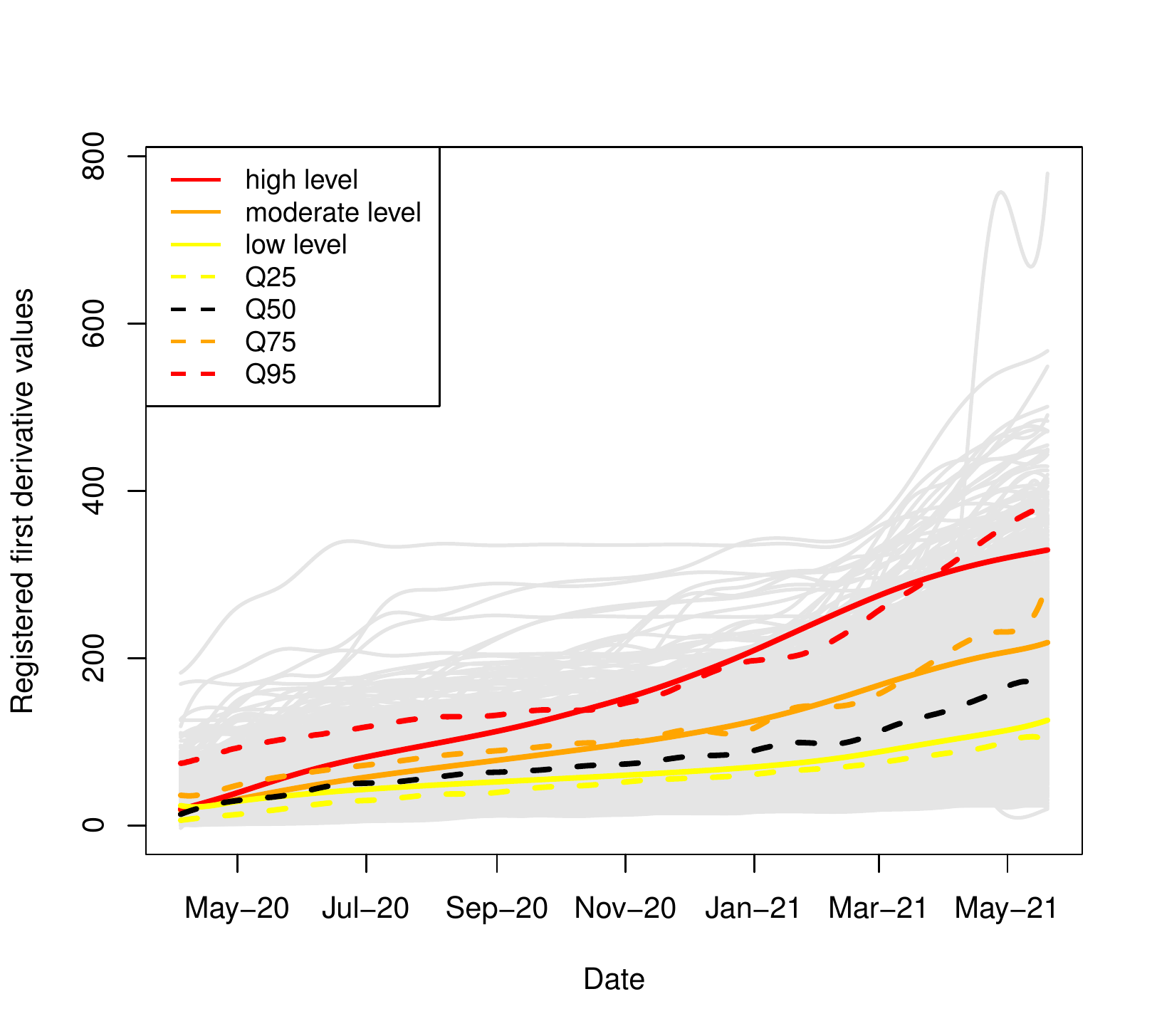}
     \end{center}
    \caption{%
        Approximation of the estimated quantiles of the cumulative number of deaths to the mean curves estimated by the functional clustering method for non-capital cities.} 
   \label{fig:fquantreg}
\end{figure}
The proposed function-on-scalar quantile regression model is:
\begin{eqnarray}\label{fquantmod}
\begin{aligned}
    Q_{\tau}(Y(t)|X)&=\beta_{0,\tau}(t)+\beta_{1,\tau}(t)\text{Area}+\beta_{2,\tau}(t)\text{Elevation}+\beta_{3,\tau}(t)\text{PopPipedWater}\\
&+\beta_{4,\tau}(t)\text{PopSolidWasteColl}+\beta_{5,\tau}(t)\text{PopElecPower}+\beta_{6,\tau}(t)\text{PopOlder65}\\
&+\beta_{7,\tau}(t)\text{EconActPop}+\beta_{8,\tau}(t)\text{VulElderlyPop}+\beta_{9,\tau}(t)\text{IlliteracyRate}\\
&+\beta_{10,\tau}(t)\text{ExtrPoverRate}+\beta_{11,\tau}(t)\text{HDI}
\end{aligned}
\end{eqnarray}
Figure \ref{fig:fquantreg} shows an approximation of the estimated quantile curves using the functional quantile regression model (equation (\ref{fquantmod})) to the estimated mean death curves.. The solid lines indicate the estimated mean curves for each alert level by the functional clustering method and the dashed lines represent the estimated quantile curves at different quantile values given by the function-on-scalar quantile regression model. It can be seen that the estimated quantile curves of 25\% and 75\% (yellow and orange dashed lines) give a better approximation to the estimated mean curves at low and moderate alert levels than the estimated 95\%-quantile curve (red dashed line) for the estimated mean curve at high alert level. The reason for that is the curves of cumulative deaths clustered in the low and moderate alert levels are less heterogeneous than the curves of deaths located in the high alert level where there are atypical curves. However, the approximations of the estimated quantile curves follow the behavior and trends of the estimated mean death curves in most of the observed time. The median curve (black dashed line) is estimated between the mean curves of the low and moderate alert levels, which means that the median curve remained very close to 100 cumulative deaths per 100k inhabitants until March 2021, then exceeded this value and increased more than half of deaths in the following two months.\\

The following figures presented in this section shows the dynamic of the quantile functional regression coefficients. Thus, it is possible to check the conditional effect of the covariates in certain quantiles of the death count curves. 
In fact, for all fixed quantiles these figures exemplify how the quantile coefficients $\beta_{\tau}(t)$ describe the relationship between the functional response variable and the scalar socioeconomic covariates. Large values of $|\beta_{\tau}(t)|$ indicate large impact on the $\tau\text{th}$ quantile of the response variable as function of time.

\begin{figure}[!htbp]
  \centering
  \includegraphics[width=16cm,height=11cm]{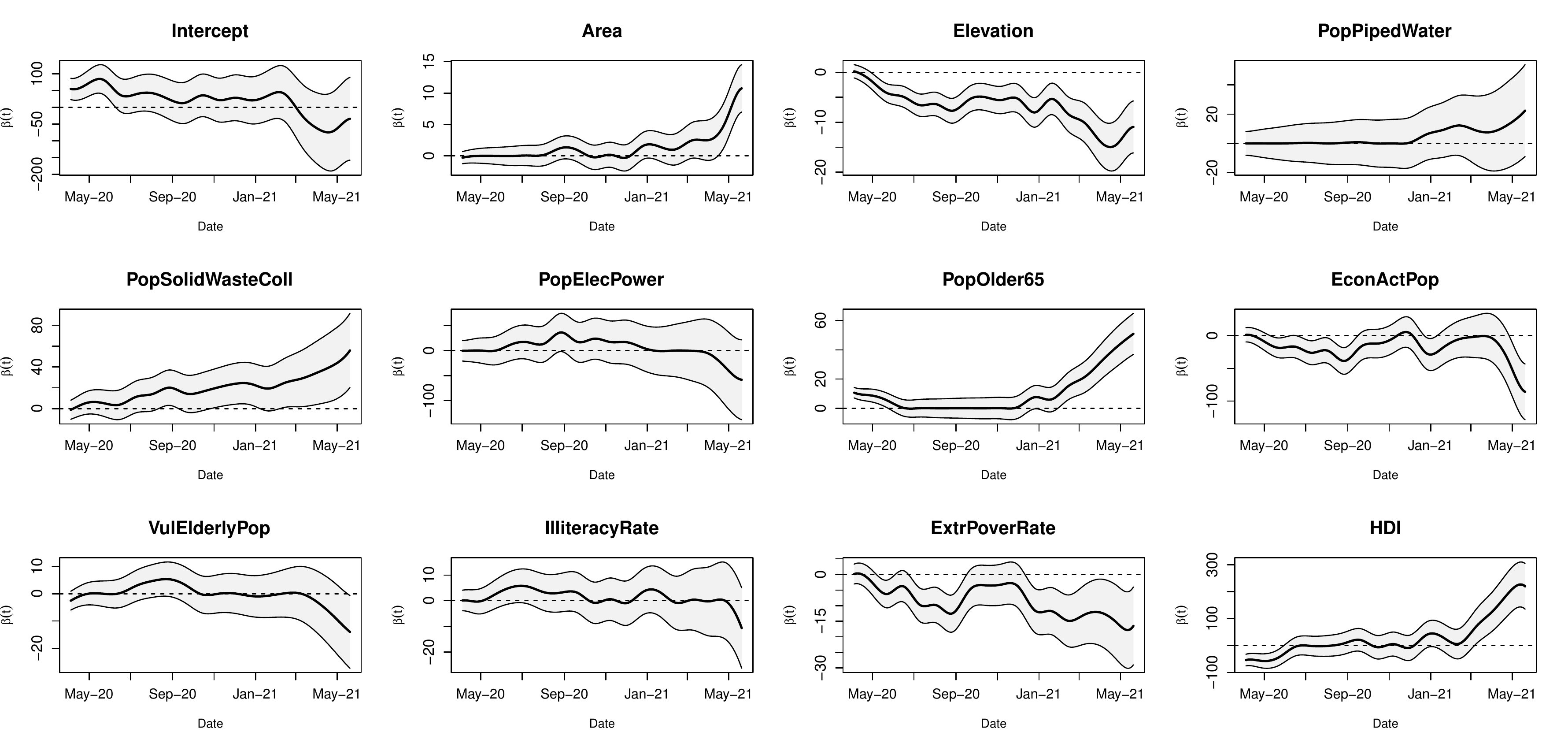}
  \caption{Estimated functional coefficients at 25\% level of cumulative deaths per 100k inhabitants for non-capital cities.}\label{lowlevel}
\end{figure}

One can see, Figure \ref{lowlevel}, the increasing conditional effect of the covariates, population older than 65, population with piped water, HDI and area on the death count curves for the low-level group or functional quantile regression at $25\%$, around the month of February when the second big wave started in Brazil. This means that the cumulative number of deaths for cities at the low alert level tends to increase over time in larger municipalities, with older populations and greater human development.

\begin{figure}[!htbp]
  \centering
  \includegraphics[width=16cm,height=11cm]{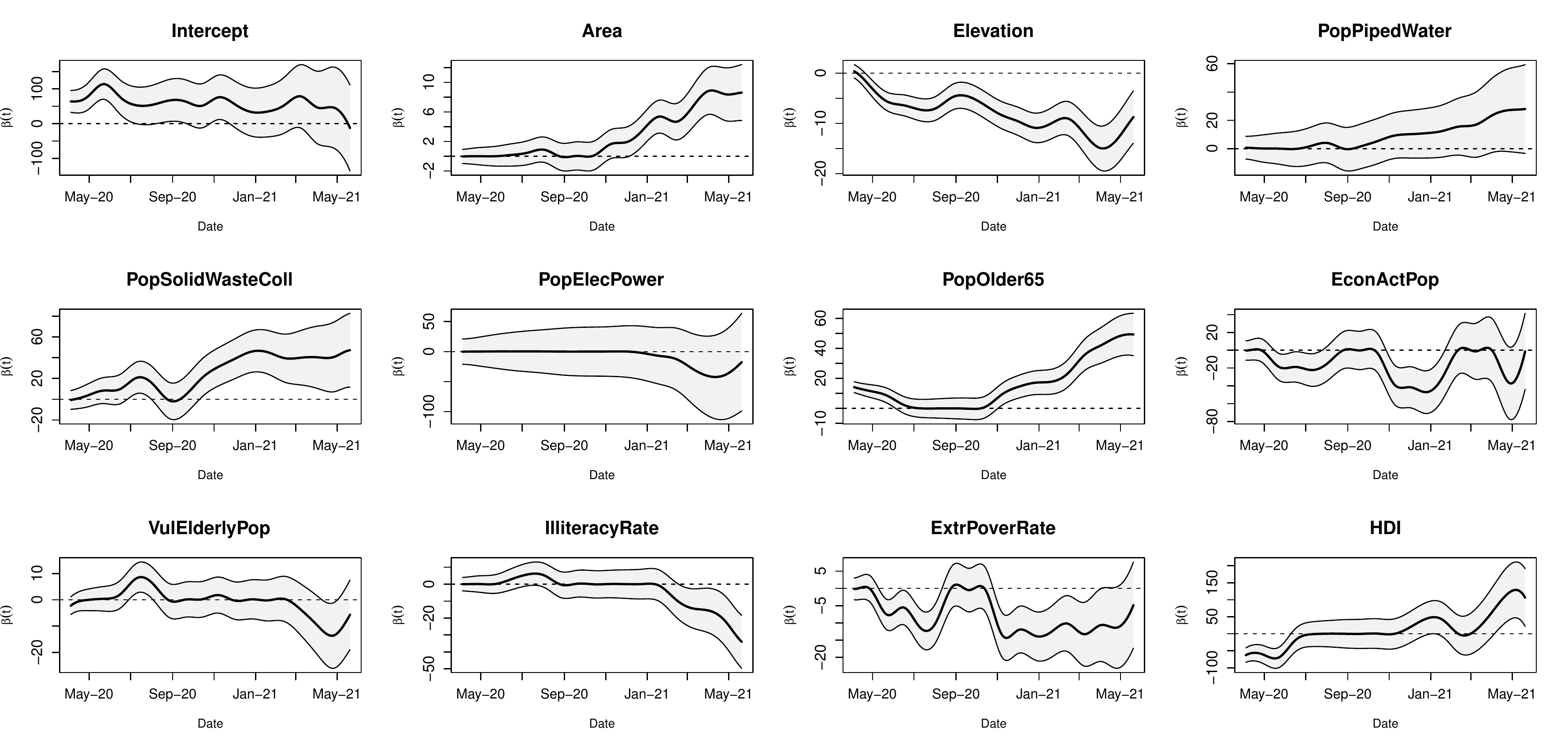}
  \caption{Estimated functional coefficients at 50\% level of cumulative deaths per 100k inhabitants for non-capital cities.}\label{medianlevel}
\end{figure}

Figure \ref{medianlevel} exhibits the dynamic of the functional quantile coefficients at the median level, $50\%$ quantile. It seems that the functional coefficients associated with the scalar socioeconomic covariates, area, population with piped water, population with solid waste collection, population older than 65 and HDI, have a conditional increasing effect on the functional variable response while the others have decreasing conditional effect on the death count curves. The results at this $50\%$ level reveal that large cities with high coverage of solid waste management and available water illustrate a significant increase in the median cumulative number of deaths since September 2020. Also cities with a high human development index, at the end of the time of this analysis, show a positive effect on the median of deaths. The functional effects of the remaining socioeconomic indicators do not have a strong influence on the increase in the median number of deaths over time.

\begin{figure}[!htbp]
  \centering
  \includegraphics[width=16cm,height=11cm]{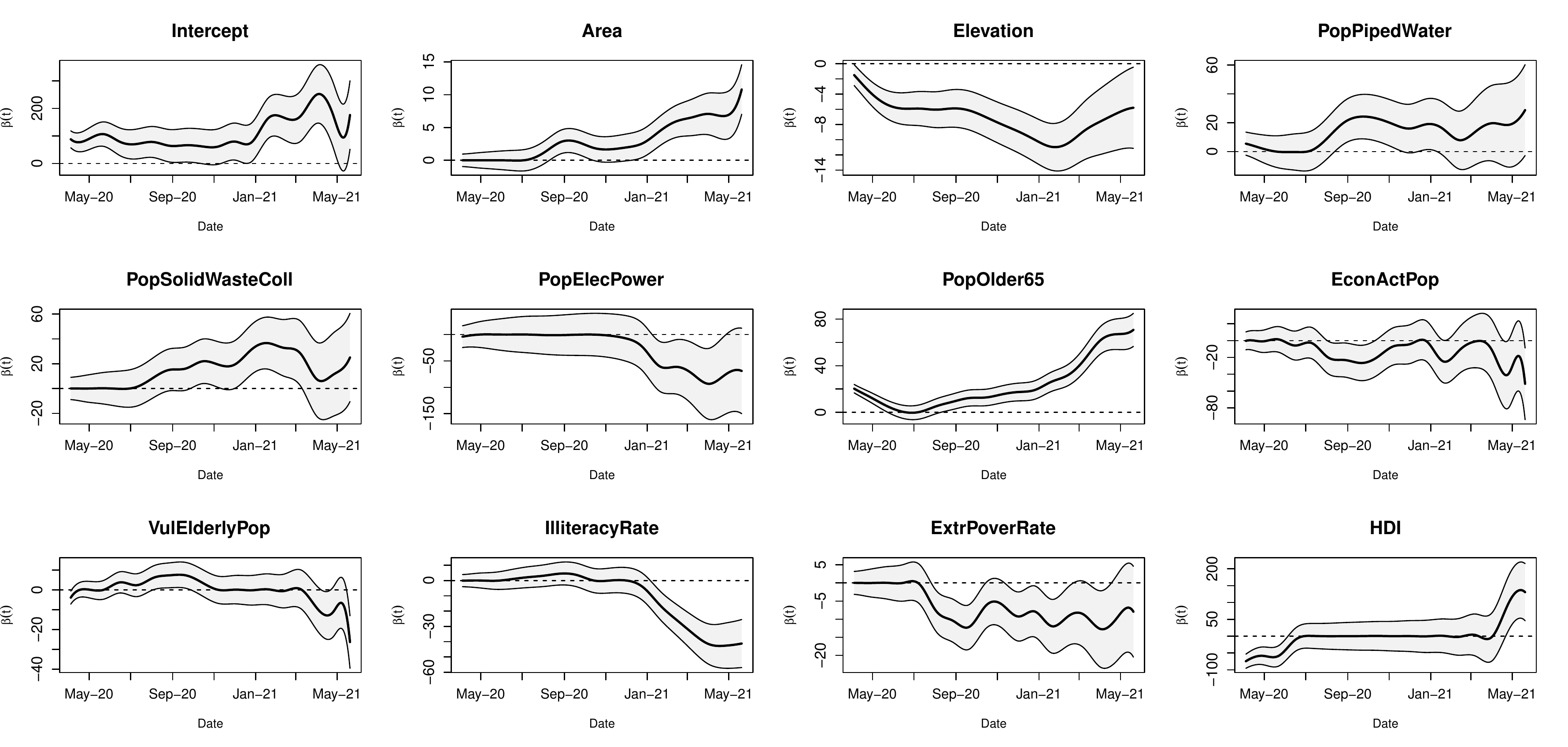}
  \caption{Estimated functional coefficients at $75\%$ level of cumulative deaths per 100k inhabitants for non-capital cities.}\label{moderatelevel}
\end{figure}

Similar to the $50\%$ quantile, Figure \ref{moderatelevel} describes the conditional effect on the functional variable response for the quantile regression level $75\%$ or moderate level group. Particularly, it should be noted that the HDI covariate rapidly increases the conditional effect on the mortality count curves at the beginning of the second big wave. At this moderate level or the $75\%$ quantile, we can say that the estimated functional coefficients continued with the same trends and behaviors in the lower quantiles, although earlier and faster. This means that using higher values in the quantile of the socioeconomic indicators increases (or decreases) more quickly the functional effects on the cumulative number of deaths.

\begin{figure}[!htbp]
  \centering
  \includegraphics[width=16cm,height=11cm]{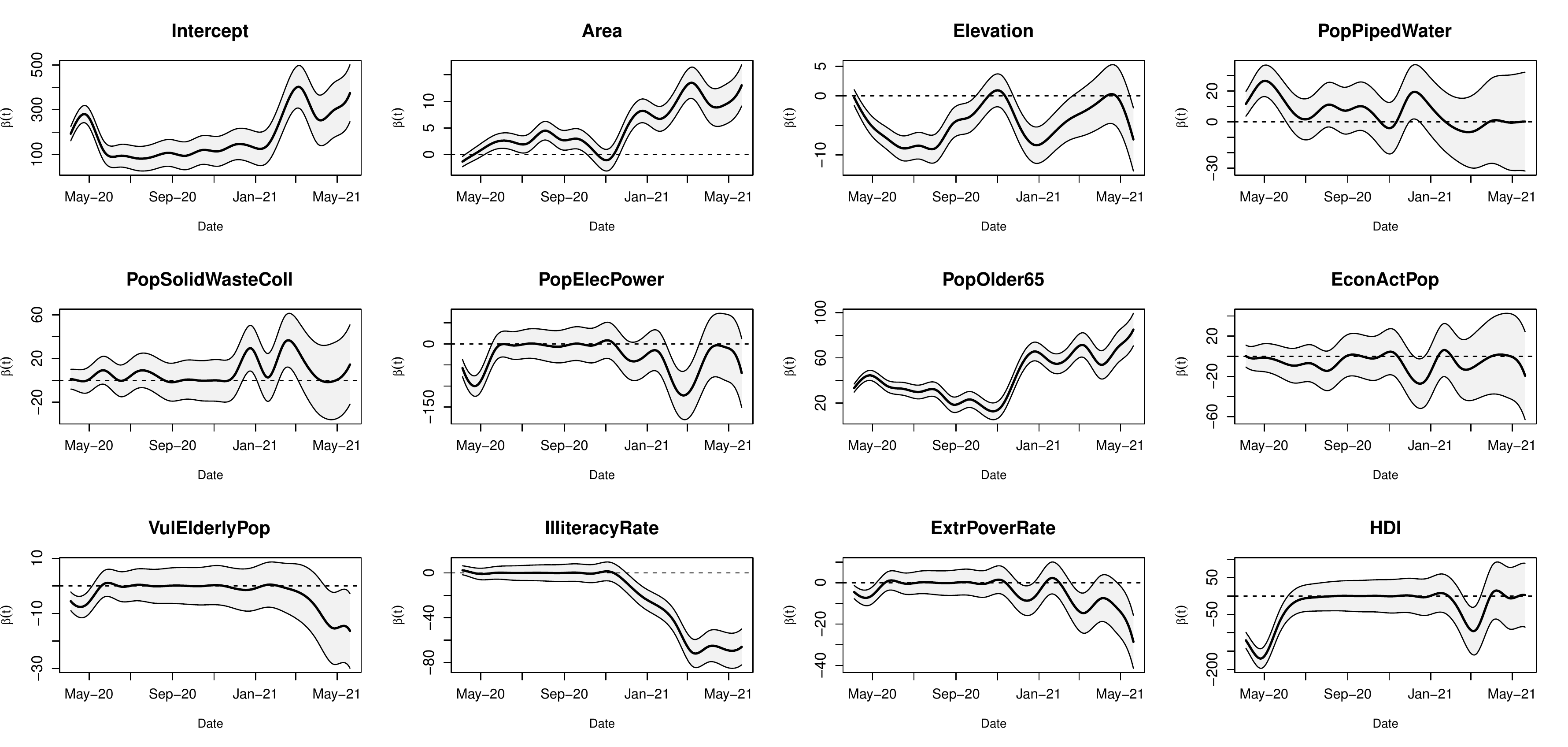}
  \caption{Estimated functional coefficients at 95\% level of cumulative deaths per 100k inhabitants for non-capital cities.}\label{highlevel}
\end{figure}

\begin{figure}[!htbp]
  \centering
  \includegraphics[width=16cm,height=11cm]{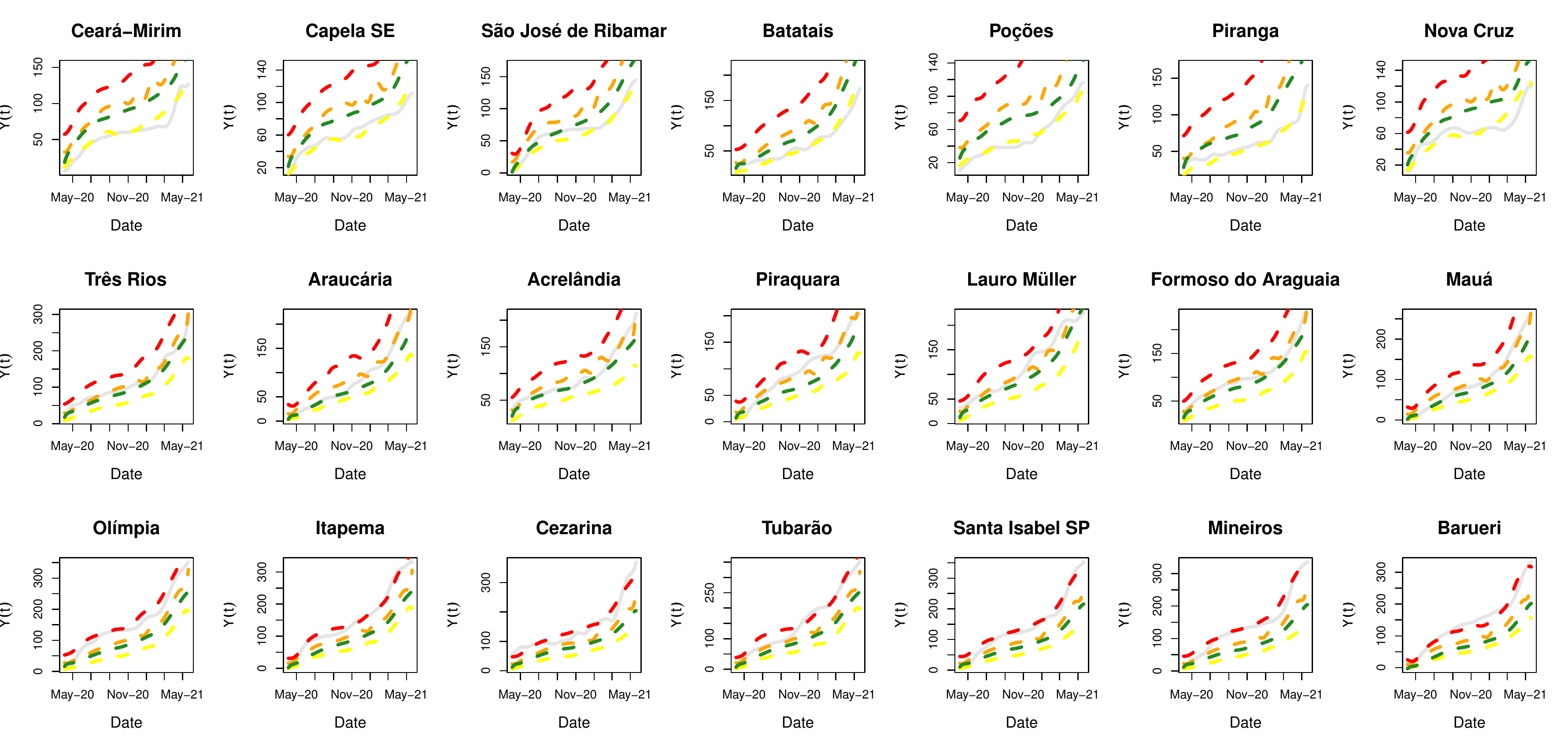}
  \caption{B-splines fitting (gray solid lines), fitted quantile curves for 25\%, 75\% and 95\% (dashed lines, yellow, orange and red) and fitted mean curves (green dashed lines) for random sample of  non-capital cities.}\label{fittedqall}
\end{figure}

For quantile $95\%$, high level, Figure \ref{highlevel} shows the increase of the conditional effect by the covariates, area, population older than 65 while it is noticeable the decrease made by the covariates population with access to electricity, illiteracy rate and extreme poverty rate. These results confirm that the scalar socioeconomic covariates in different quantiles that most impact positively the cumulative number of deaths are the area, population with piped water and solid waste collection and also older than 65 years. The HDI covariate has a positive effect from the start of the second great wave in 2021. Thus, one can say that as the (chosen) quantiles increase their value, the functional effects indicate that the cumulative number of deaths increased earlier and faster in larger cities, with older populations and with greater coverage in environmental sanitation. Also, the negative impacts on the cumulative number of deaths are given by the covariates altitude, population with access to electricity and vulnerable elderly, illiteracy rates and extreme poverty. This implies that the cumulative number of deaths decreased rapidly since the beginning of 2021 in cities with higher elevation, with a high vulnerable elderly population associated with illiteracy and extreme poverty.\\
The visualization of curve fitting of the functional quantile regression model compared to the functional linear model (FLM) in equation (3), can be observed in Figure \ref{fittedqall}. For each randomly chosen city, the represented curve by B-splines (gray solid line), the fitted quantile curves to $25\%$ (yellow dashed line), $75\%$ (orange dashed line) and $95\%$ (red dashed line) and fitted curve by FLM (green dashed line) are displayed. The municipalities in each row belong to the low, moderate and high alert levels estimated by the functional clustering algorithm in the previous section. It can be seen that at each alert level, the proposed functional quantile regression model provides an adequate fit according to the value of the quantile used in the estimated model. That is, cities clustered in a low alert level are properly fitted by the functional regression model in the $25\%$ quantile, cities in the moderate level are appropriately fitted using the $75\%$ quantile in the model and locations belonging to the high level are correctly fitted with the $95\%$ quantile. In the case of the FLM, the fits are poor and do not allow tracking the trajectories of the cumulative number of deaths in such municipalities located at different alert levels.

\begin{figure}[!htbp]
  \centering
  \includegraphics[width=16cm,height=11cm]{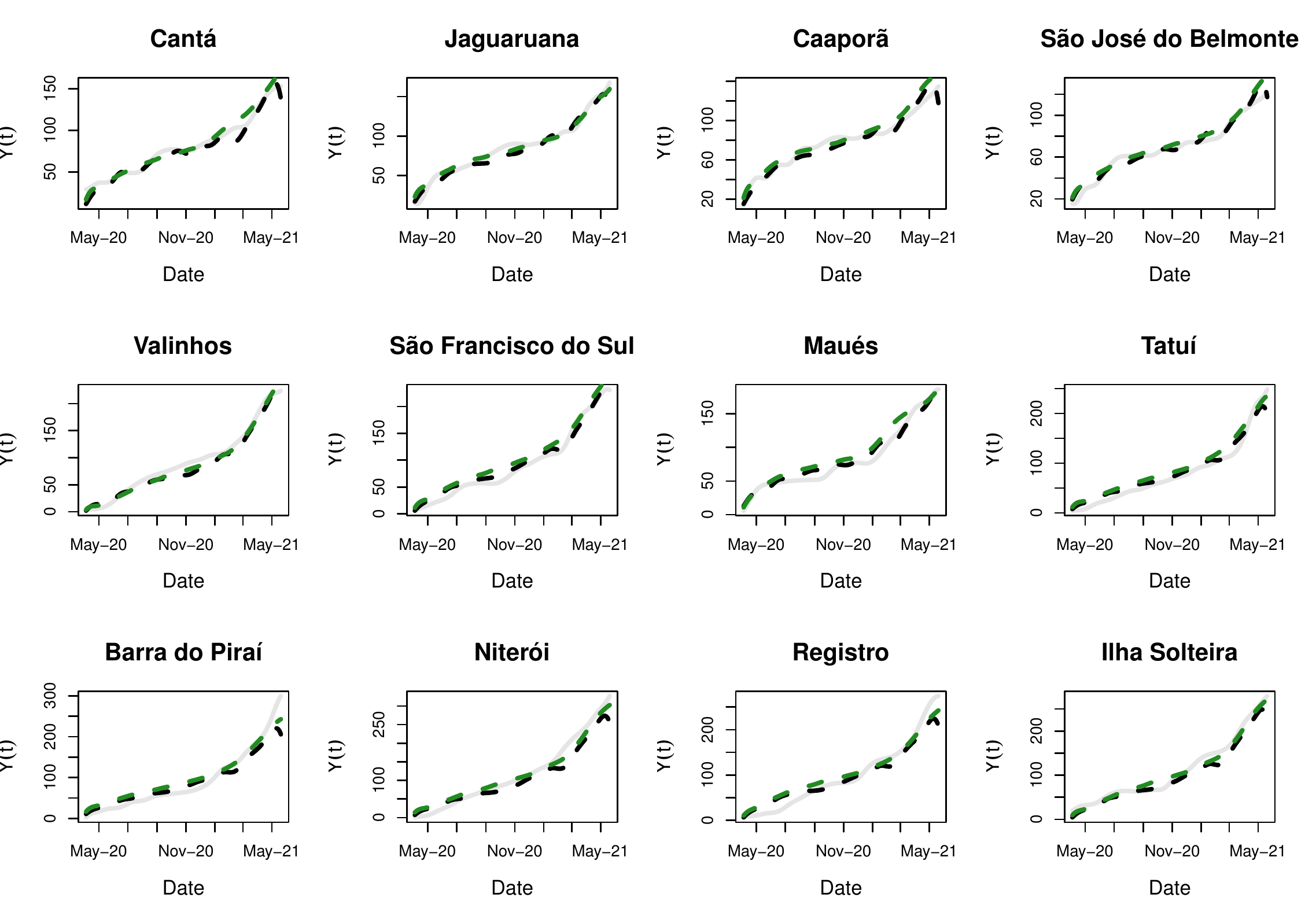}
  \caption{Represented curves (gray solid lines), fitted quantile curves for 50\% (black dashed lines) and fitted mean curves (green dashed lines) for random non-capital cities.}\label{fittedq50s}
\end{figure}

\pagebreak
Figure \ref{fittedq50s} shows the fitted curves for the functional median regression model and the FLM for some represented curves (gray solid lines) of cities. The results show a good fit of the median curve (black dashed line) in most municipalities followed closely by the fitted curves of the FLM (green dashed line). This means that the curve fitting of the FLM improves due to the presence of curves of the municipalities more concentrated in the middle zone of deaths and that are not very distant from the alert levels around them which are less heterogeneous. Therefore, if the death curves are observed further from the middle zone, such as in zones of the alert levels, the FLM is not able to provide a good fit.


\section{Conclusions}\label{conclusions}
In this work, we apply modern statistical methodologies to analyze the death count curves of the Covid-19 pandemic in Brazilian cities. The proposed methodology belongs to the field of Functional Data Analysis which studies data from curves and the relationships between them. Given the irregular nature of the curves, it was necessary to represent them using B-splines basis functions in order to maintain the same dimension for all trajectories. 

Our analysis consists of two different parts. In the first one, we consider an unconditional analysis where we use functional clustering on the dynamics of the represented curves and their functional derivatives. As a result,t hree clusters were estimated as alert levels (low, moderate and high) in the number of deaths over time. The results of the estimated clusters show future critical situations by going from low to high alert levels when death curves are accelerated in most cities. The capital city of Manaus in the state of Amazonas and many cities located in the north and northeast regions are cases where the situation may worsen given their geographic location and limited public health conditions. The dynamic of the curves could be used as public policy and help Health authorities to better understand the different behavior between the clusters. 

The second part is a conditional analysis. We propose a functional quantile regression model in an attempt to verify how scalar socioeconomic variables are able to explain the behavior of the cumulative death curves. The results indicate that for different quantile values, the model has the ability to estimate the functional coefficients associated with the scalar covariates, showing good performance in fitting the death curves for the different alert levels. In comparison with the Functional linear Model, the proposed functional quantile regression model is more suitable for fitting curves at more distant or extreme levels (or quantiles) of the middle zone of the cumulative death curves. We can say that the methodology proposed in this work serves as an alternative strategy for the treatment of curves at any level of functional datasets.

\bibliography{manuscript}
\end{document}


\date{}
\maketitle

\section{Appendix A}
\subsection{North Region}
\begin{figure}[!htbp]
     \begin{center}
        \subfigure[]{%
            \label{ac}
            \includegraphics[width=5.3cm,height=3.5cm]{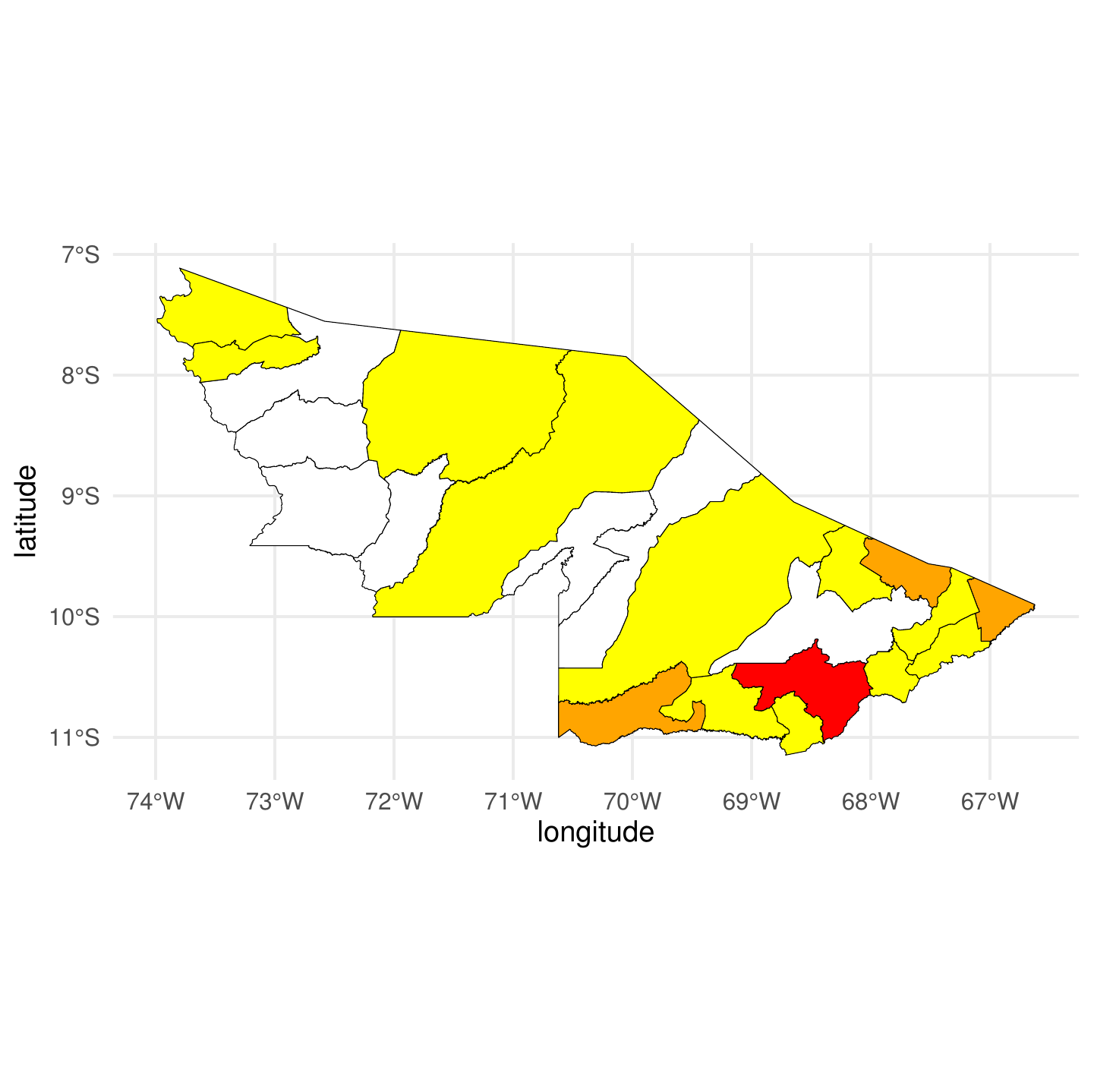}
        }
        \subfigure[]{%
           \label{ac1}
           \includegraphics[width=5.3cm,height=3.5cm]{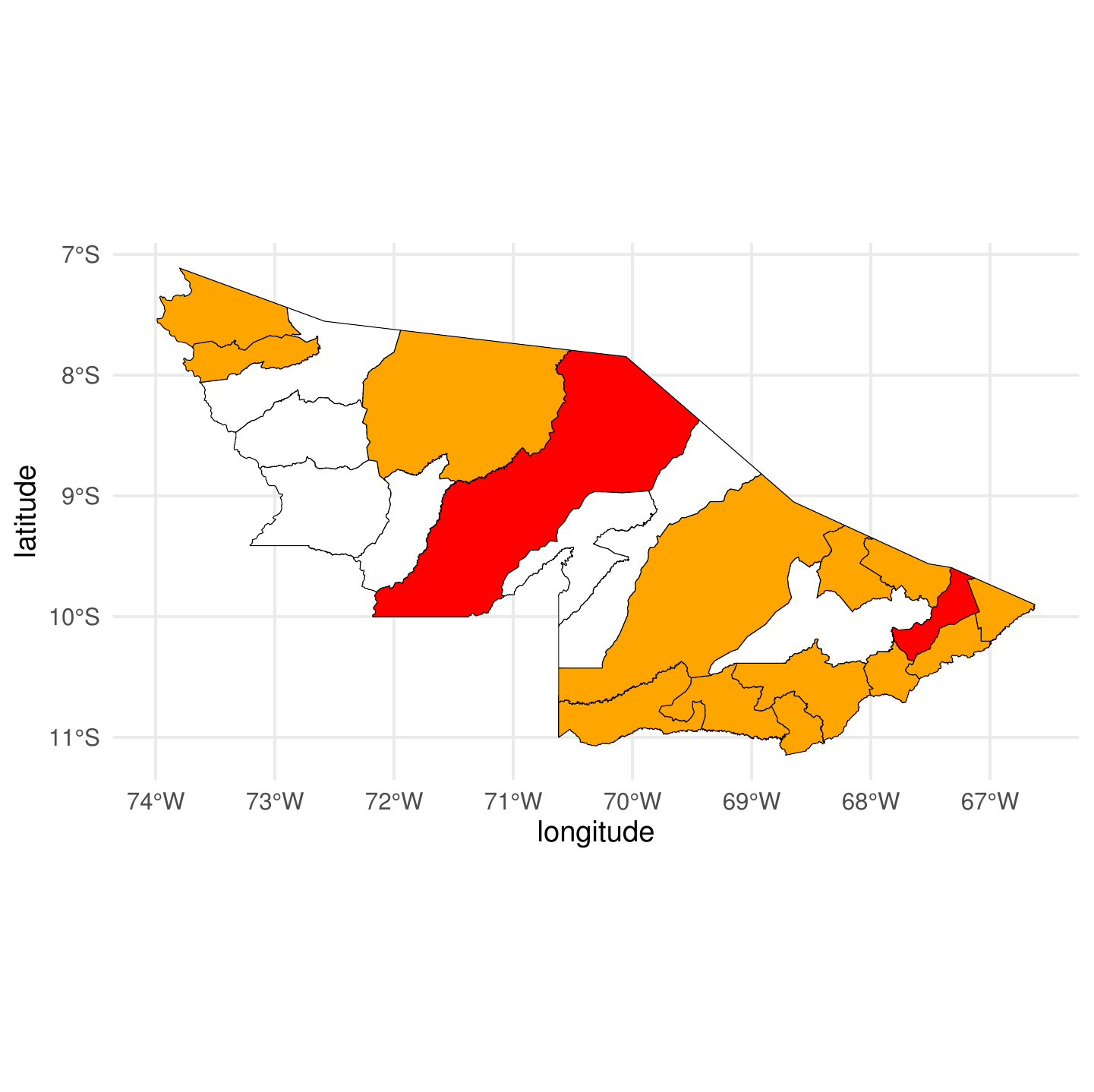}
        }
        \subfigure[]{%
           \label{ac2}
           \includegraphics[width=5.3cm,height=3.5cm]{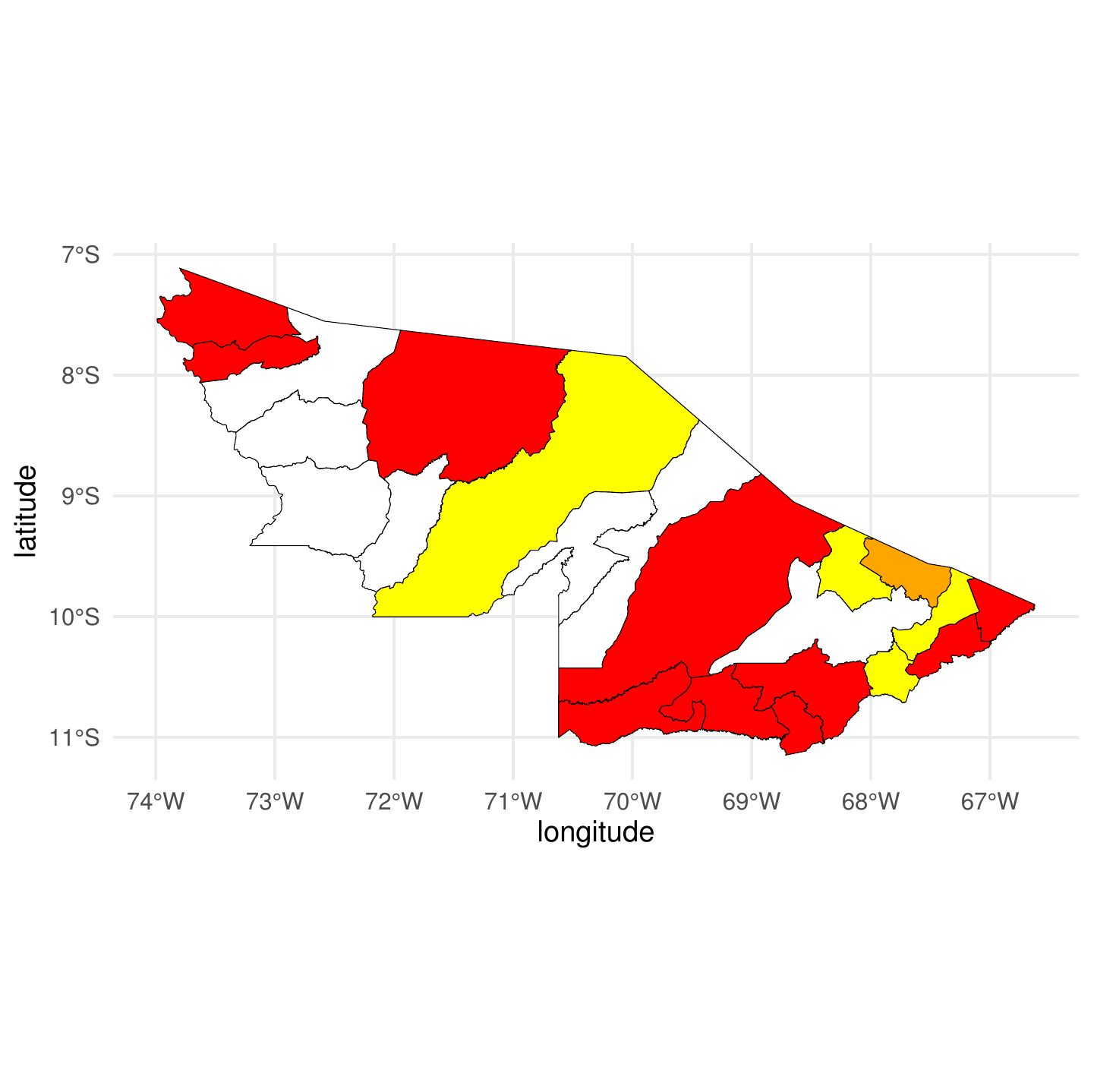}
        }\\
        \subfigure[]{%
            \label{ap}
            \includegraphics[width=5.3cm,height=3.5cm]{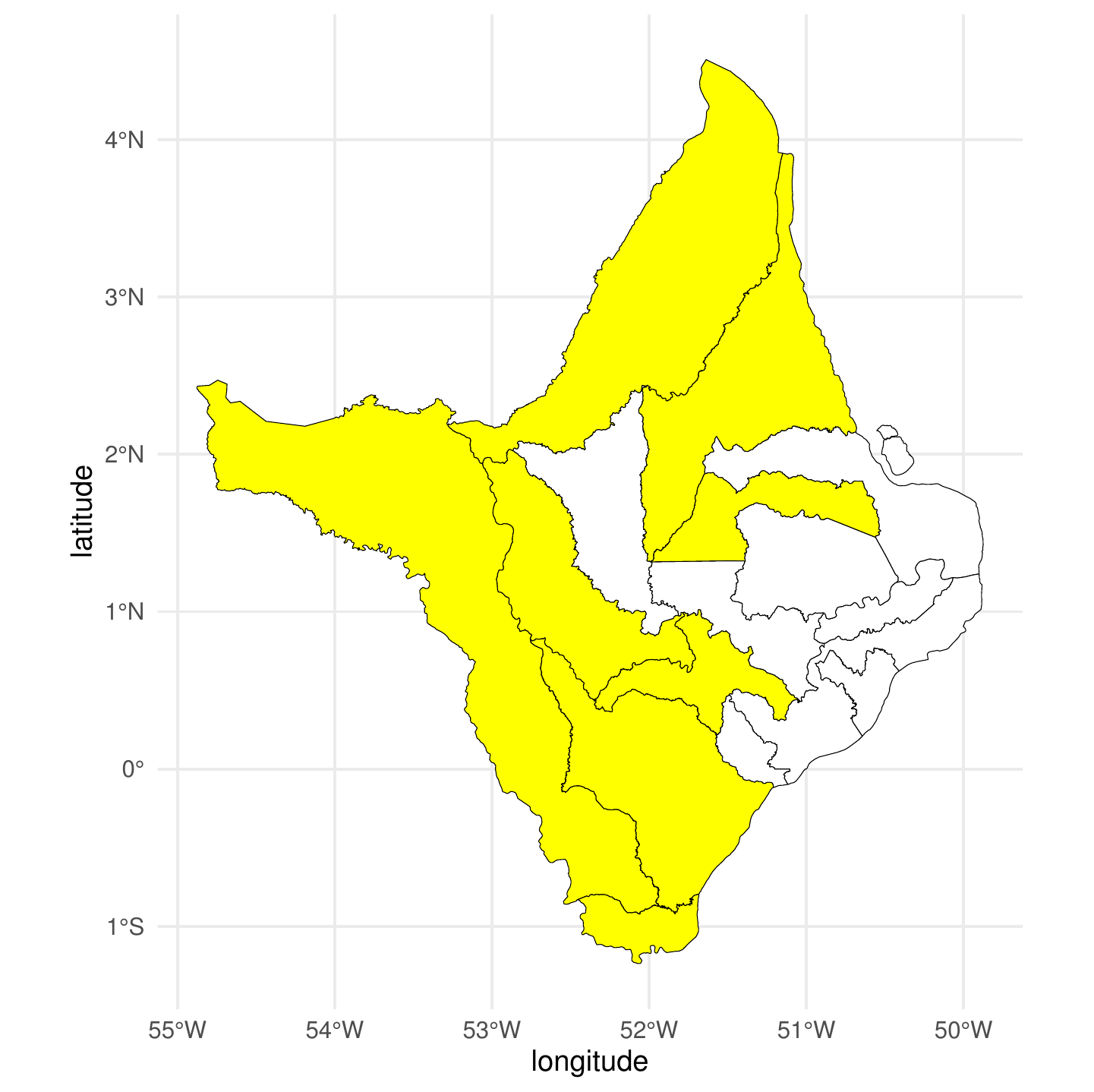}
        }
        \subfigure[]{%
           \label{ap1}
           \includegraphics[width=5.3cm,height=3.5cm]{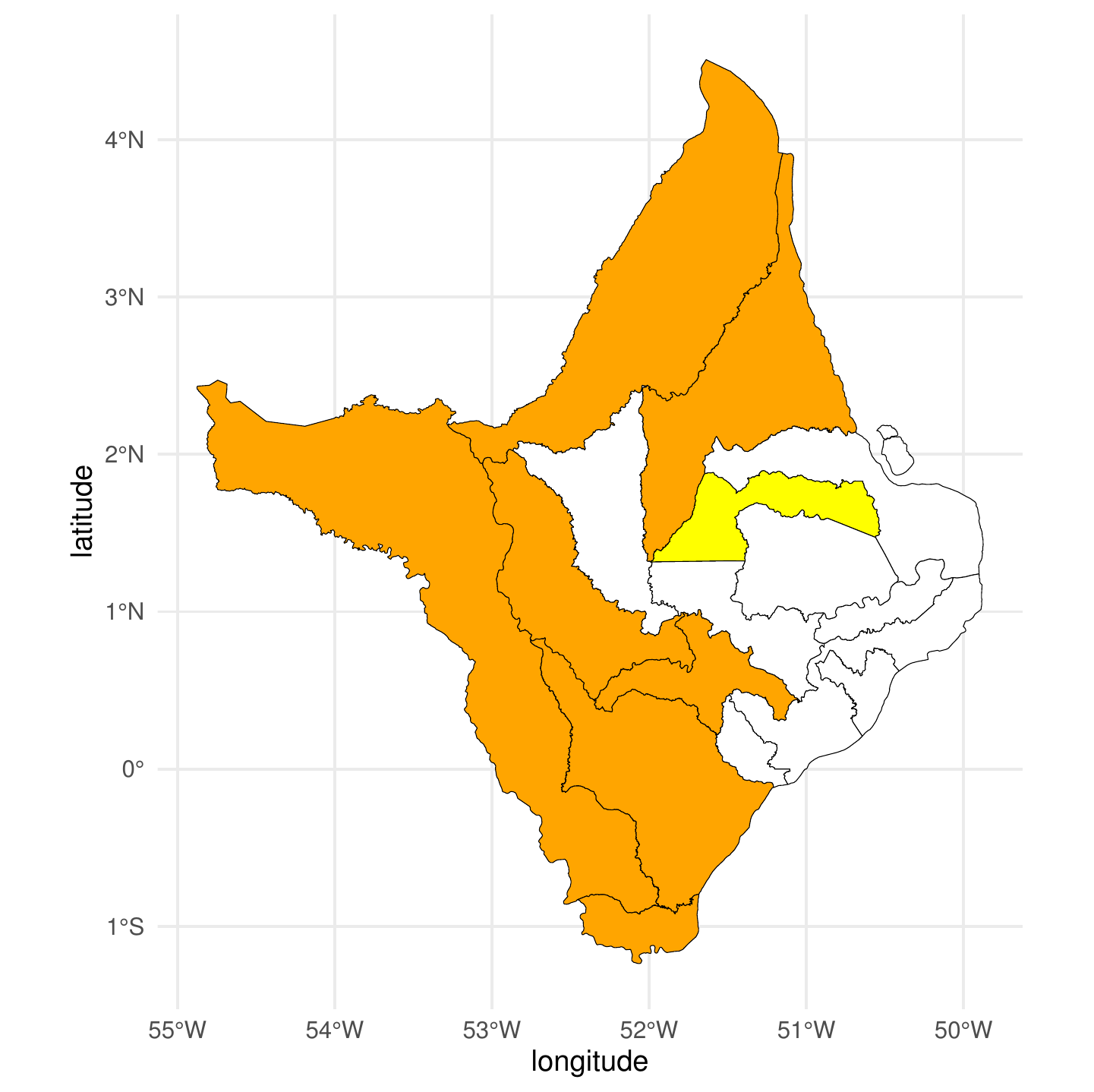}
        }
        \subfigure[]{%
           \label{ap2}
           \includegraphics[width=5.3cm,height=3.5cm]{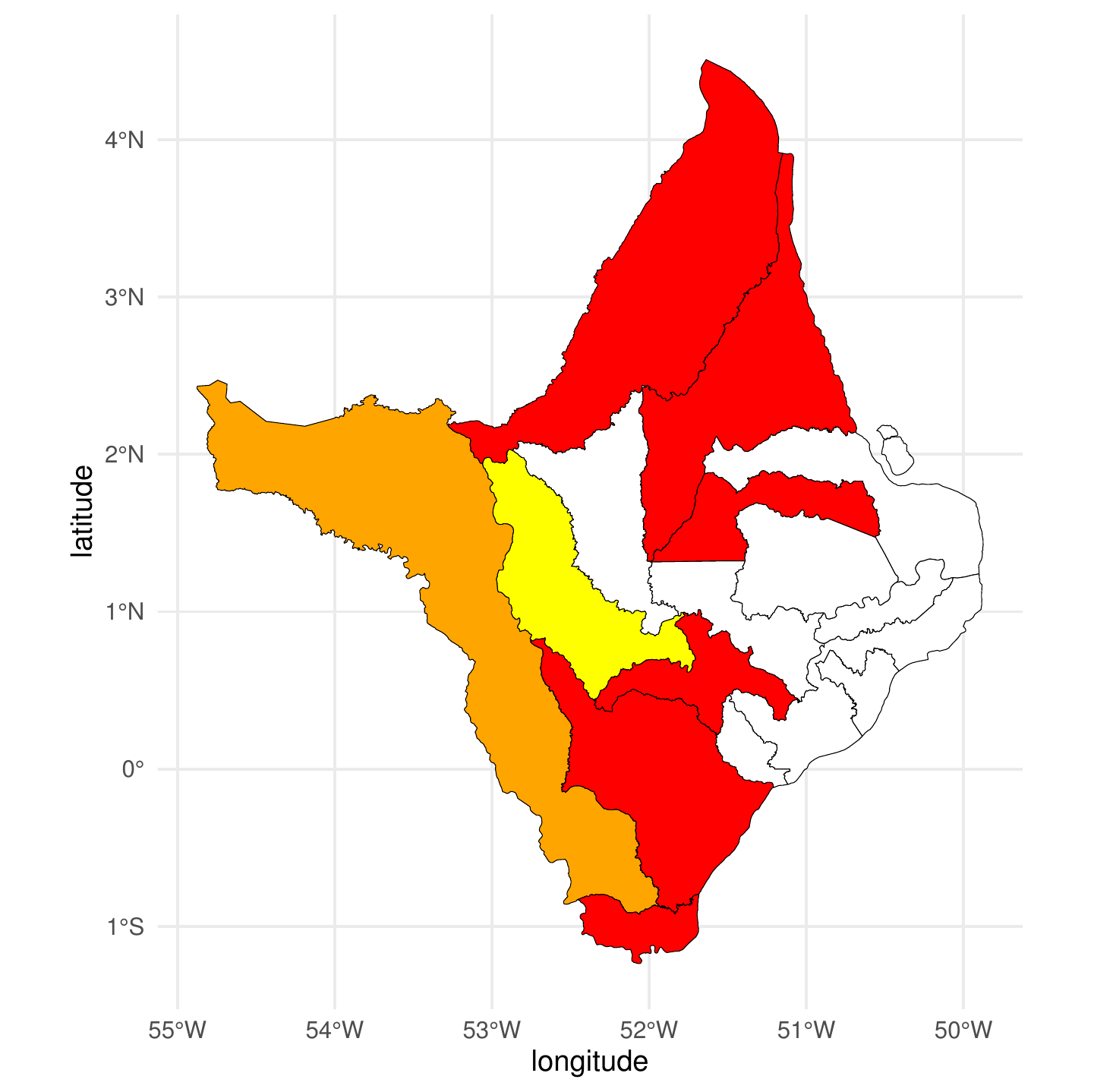}
        }\\
        \subfigure[]{%
            \label{am}
            \includegraphics[width=5.3cm,height=3.5cm]{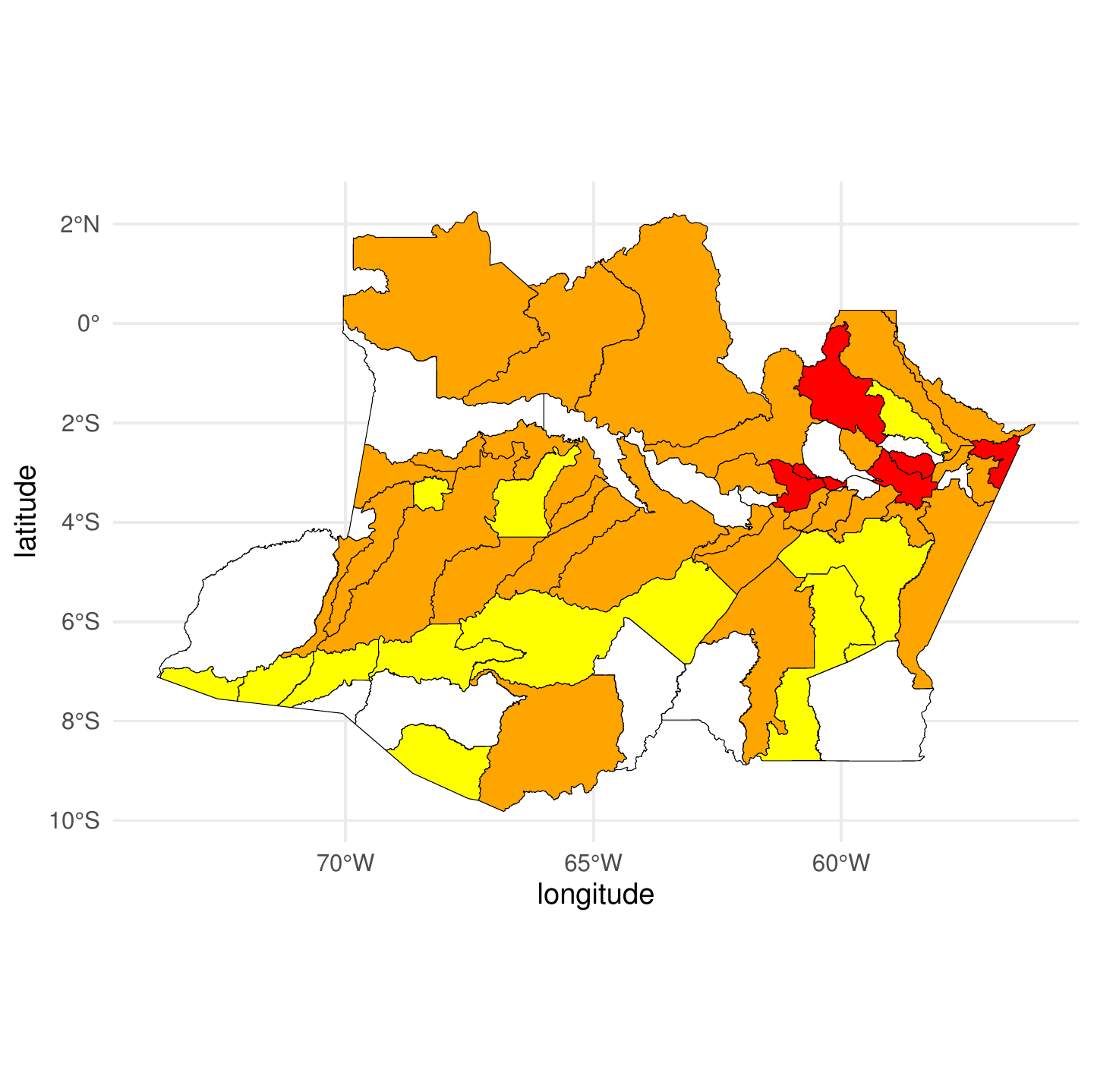}
        }
        \subfigure[]{%
           \label{am1}
           \includegraphics[width=5.3cm,height=3.5cm]{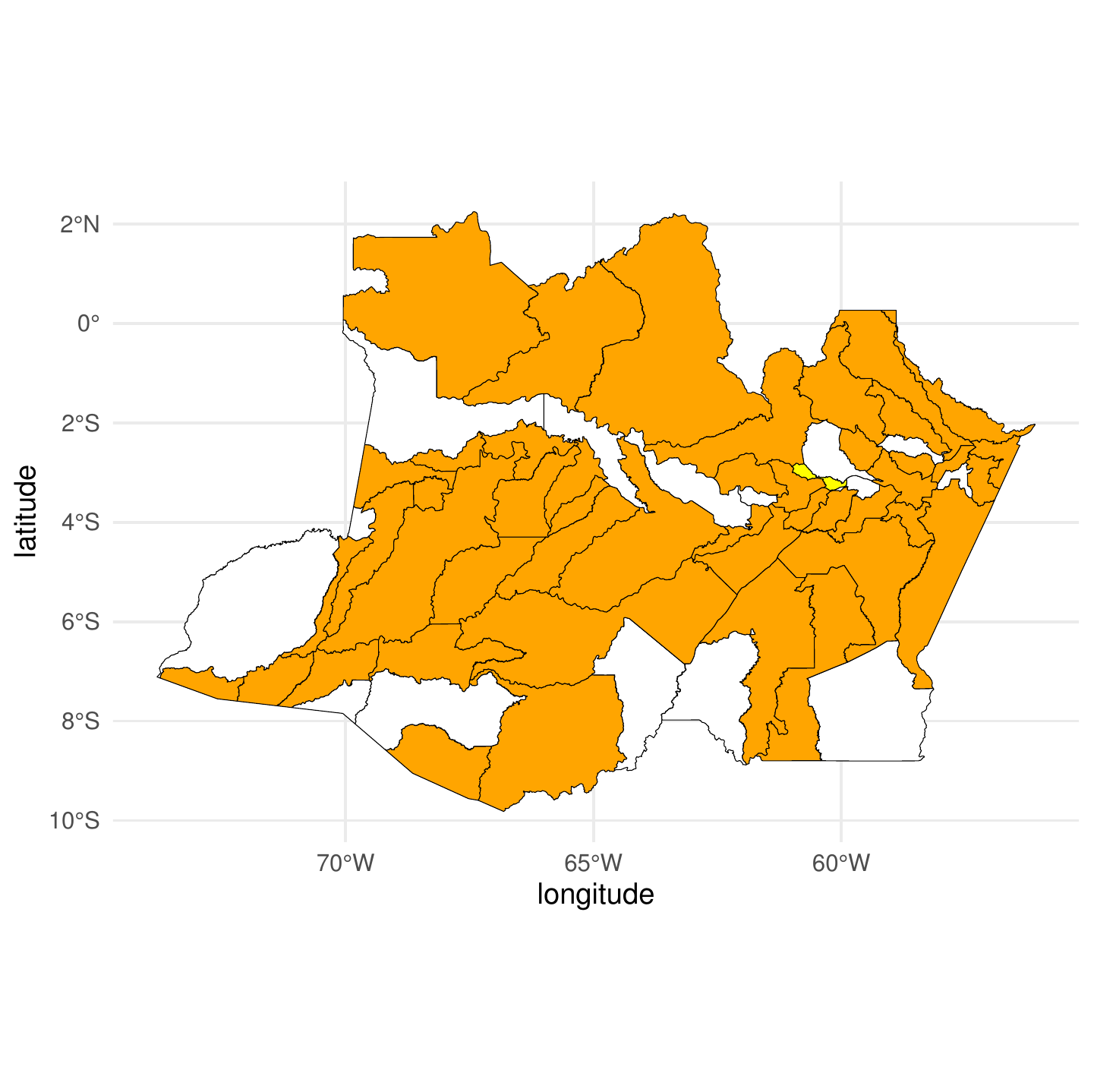}
        }
        \subfigure[]{%
           \label{am2}
           \includegraphics[width=5.3cm,height=3.5cm]{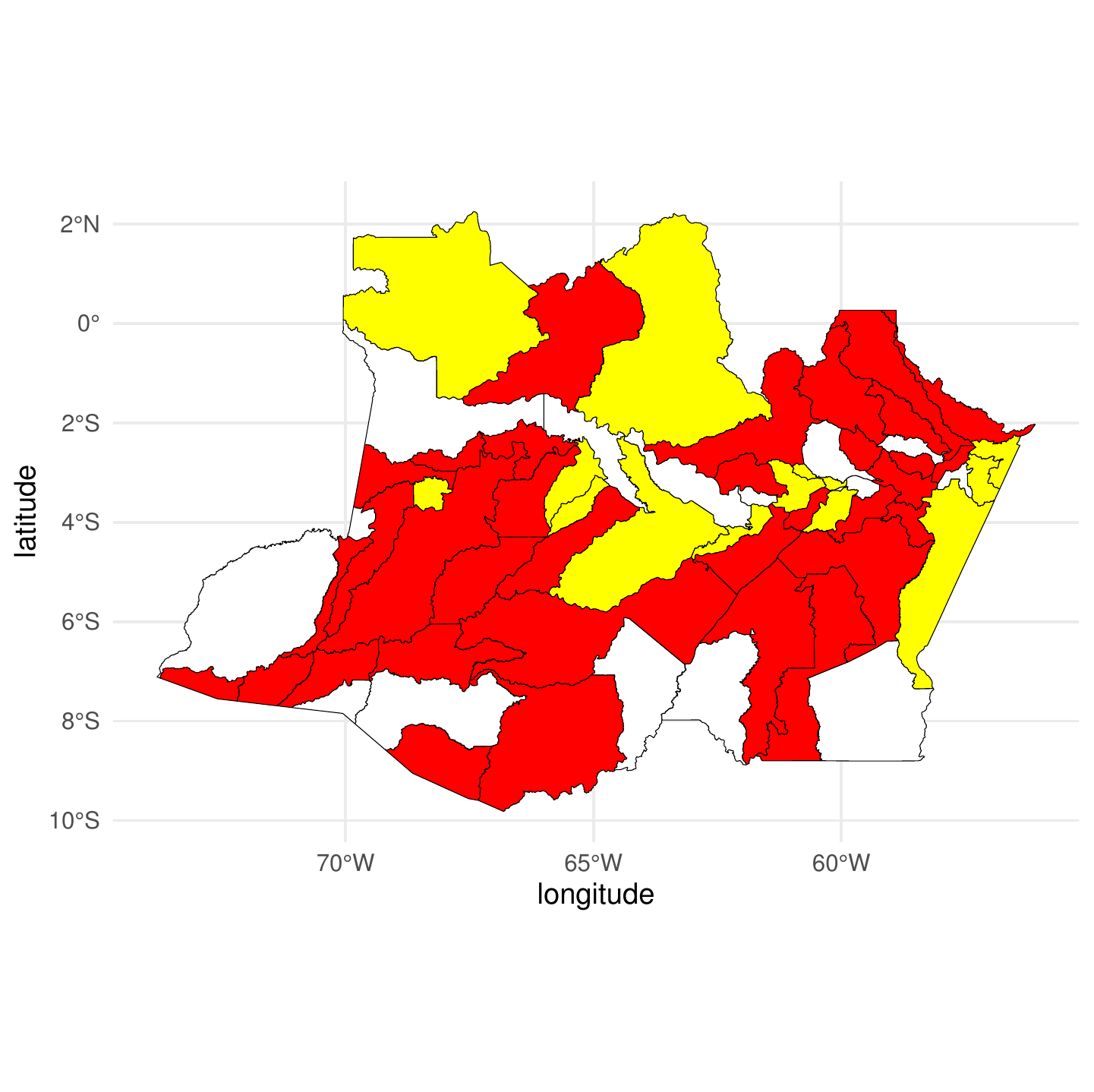}
        }
    \end{center}
    \caption{%
        Spatial location of the functional clustering of the municipalities of the states of Acre (first line), Amap\'{a} (second line) and Amazonas (third line) according to represented death curves (a, d, g), first derivative of death curves (b, e, h) and second derivative of death curves (c, f, i).}
   \label{muni1}
\end{figure}

\begin{figure}[!htbp]
     \begin{center}
        \subfigure[]{%
           \label{pa}
            \includegraphics[width=5.3cm,height=4.5cm]{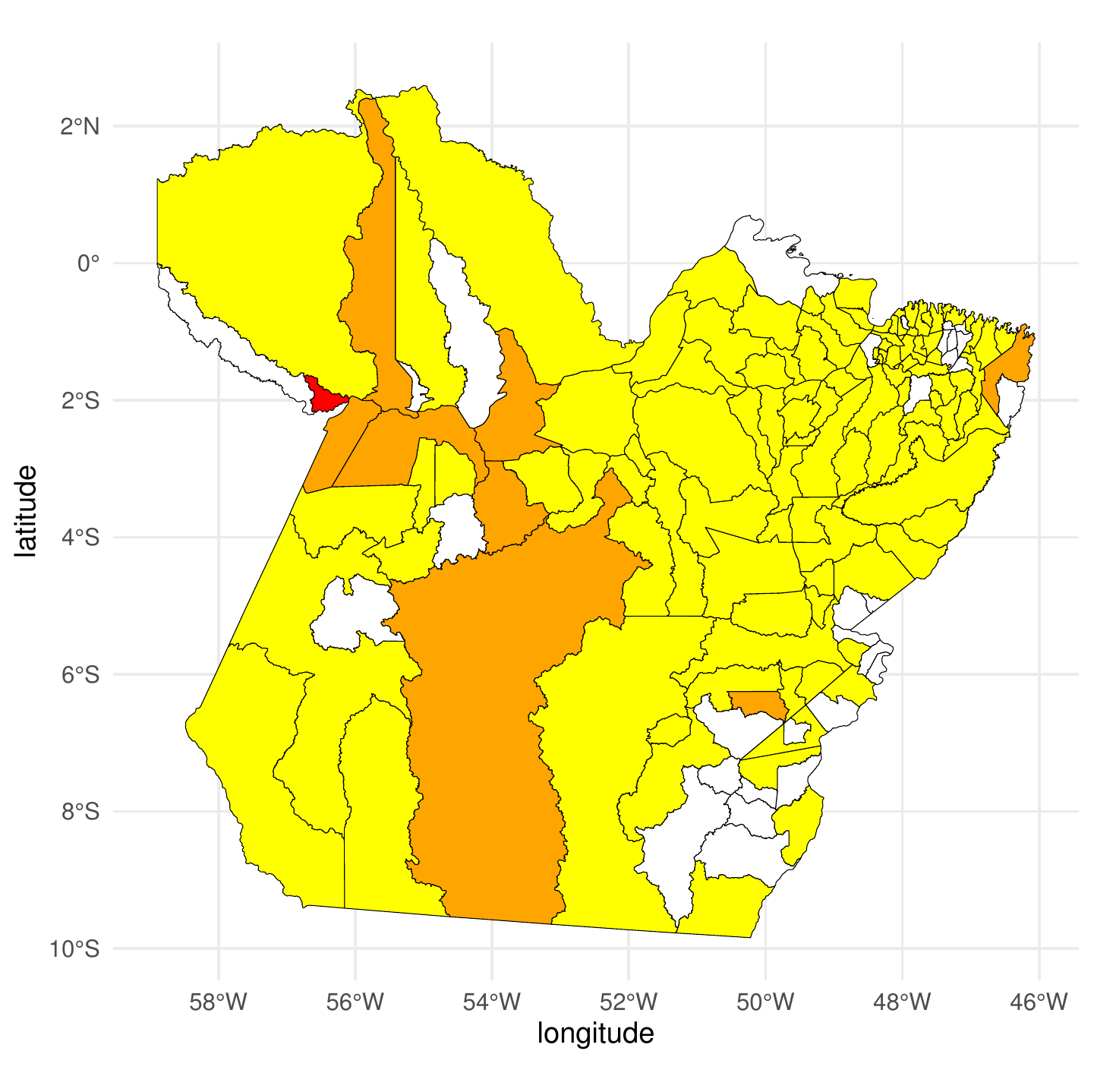}
        }
        \subfigure[]{%
           \label{pa1}
           \includegraphics[width=5.3cm,height=4.5cm]{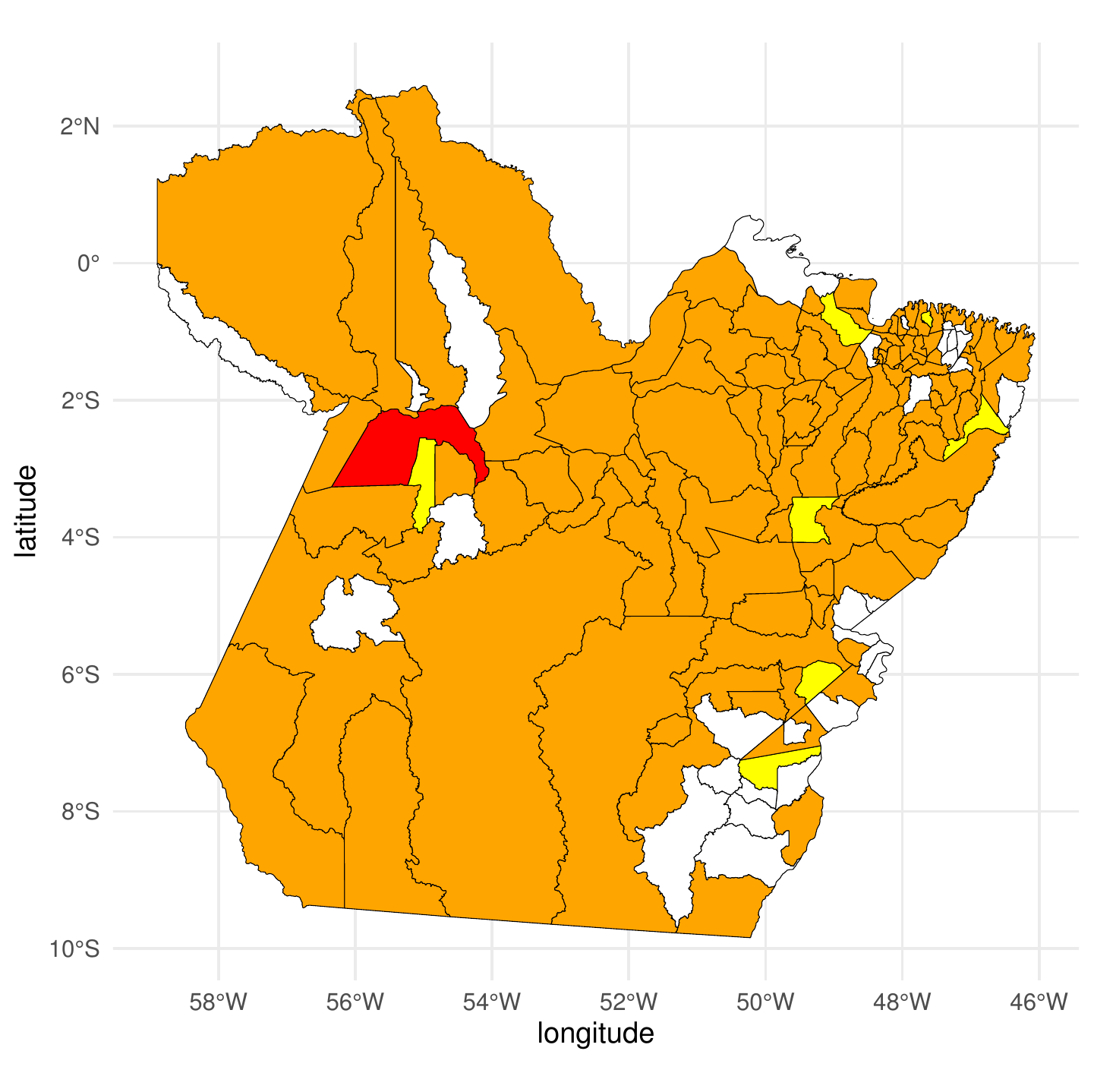}
        }
        \subfigure[]{%
           \label{pa2}
           \includegraphics[width=5.3cm,height=4.5cm]{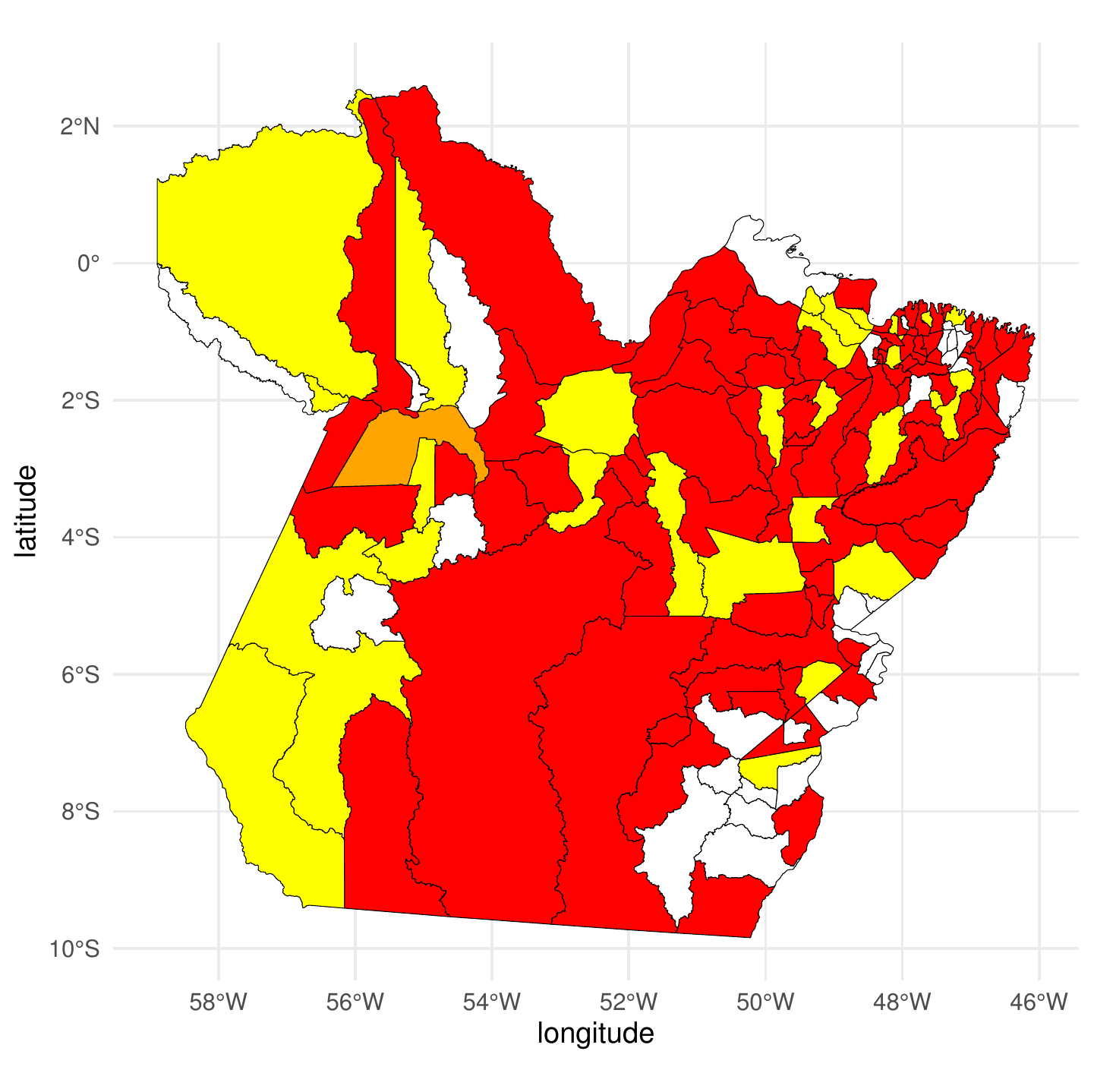}
        }\\
        \subfigure[]{%
            \label{ro}
            \includegraphics[width=5.3cm,height=4.5cm]{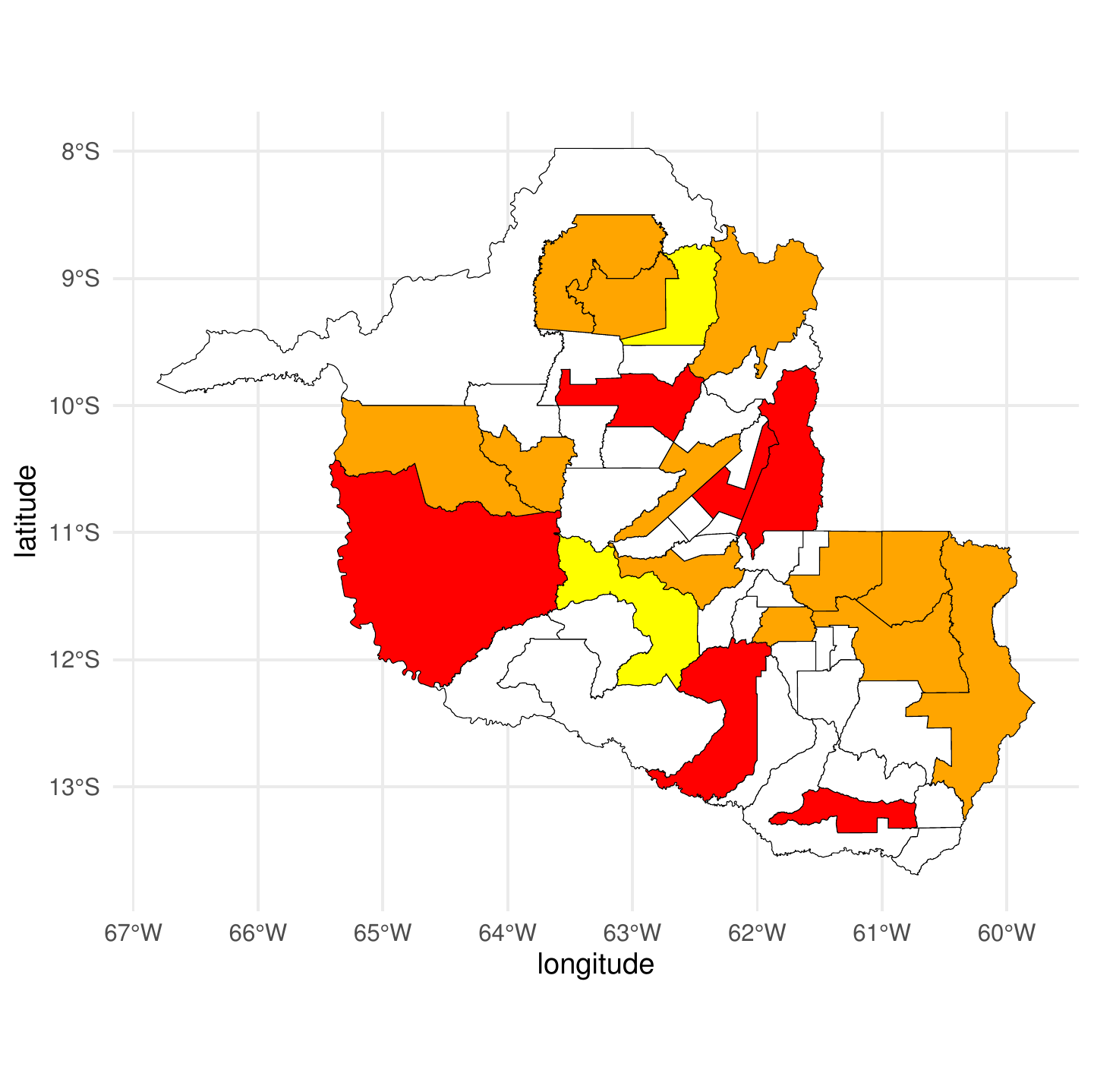}
        }
        \subfigure[]{%
           \label{ro1}
           \includegraphics[width=5.3cm,height=4.5cm]{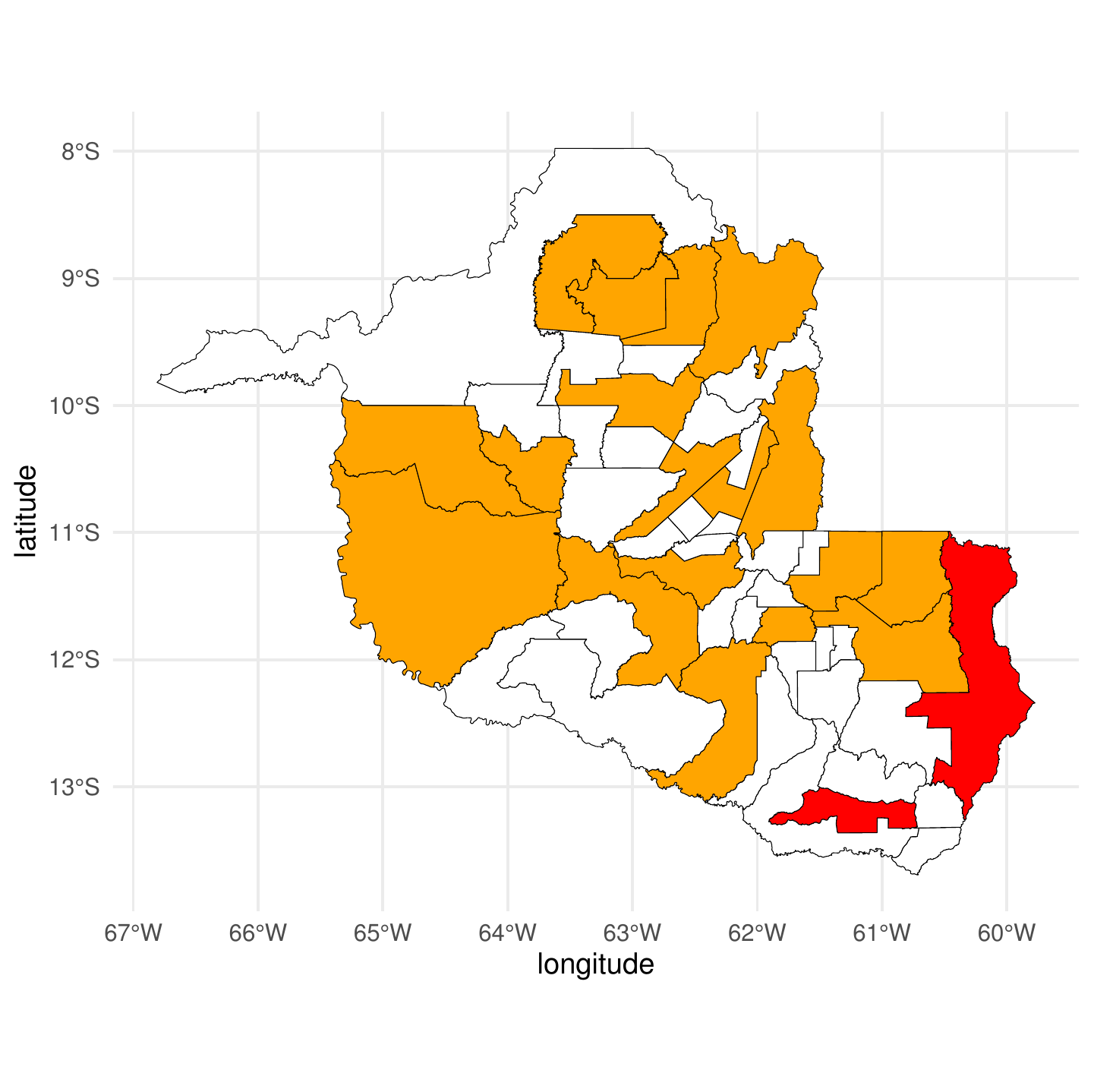}
        }
        \subfigure[]{%
           \label{ro2}
           \includegraphics[width=5.3cm,height=4.5cm]{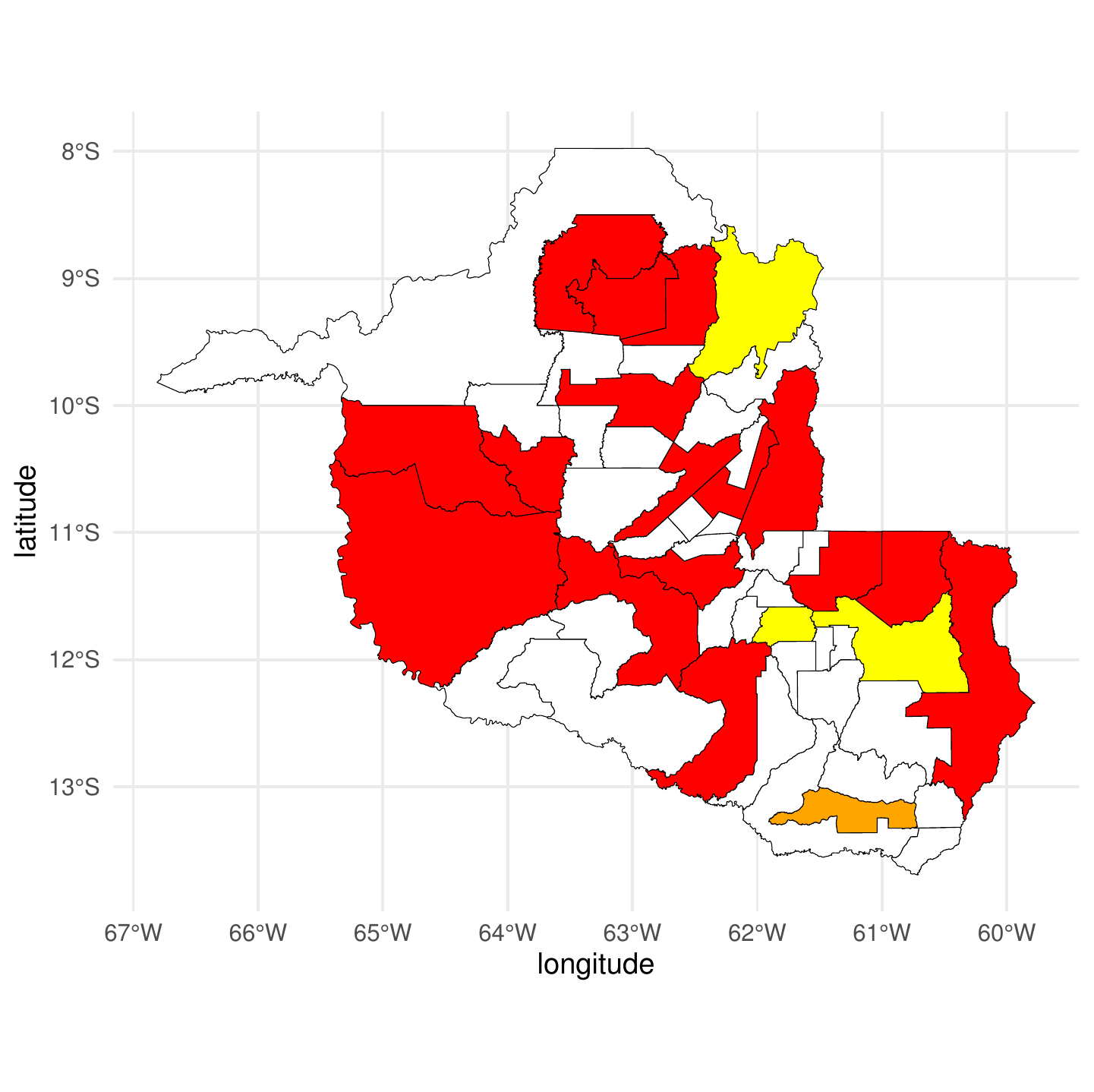}
        }\\
        \subfigure[]{%
            \label{rr}
            \includegraphics[width=5.3cm,height=4.5cm]{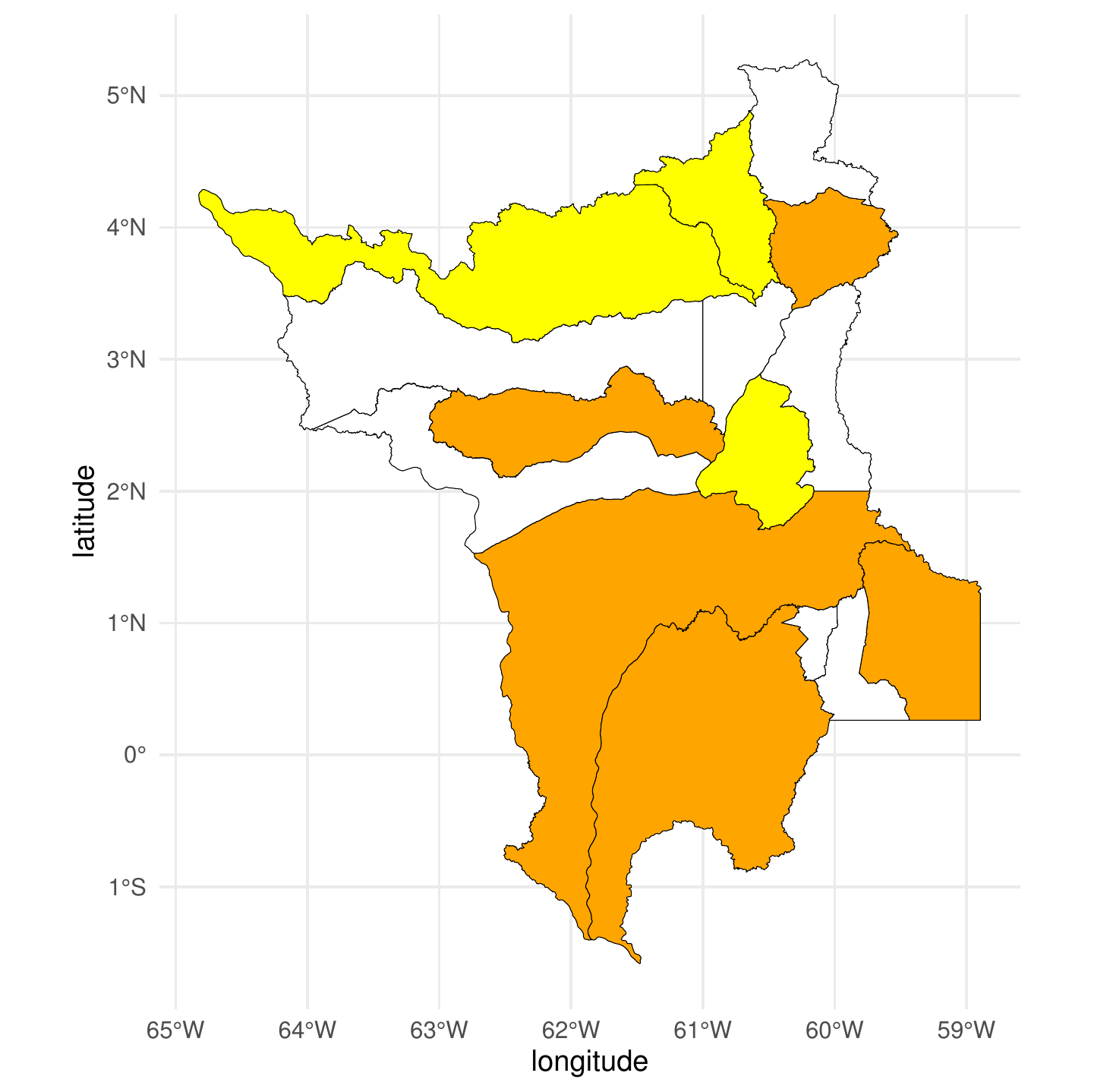}
        }
        \subfigure[]{%
           \label{rr1}
           \includegraphics[width=5.3cm,height=4.5cm]{RRD1.pdf}
        }
        \subfigure[]{%
           \label{rr2}
           \includegraphics[width=5.3cm,height=4.5cm]{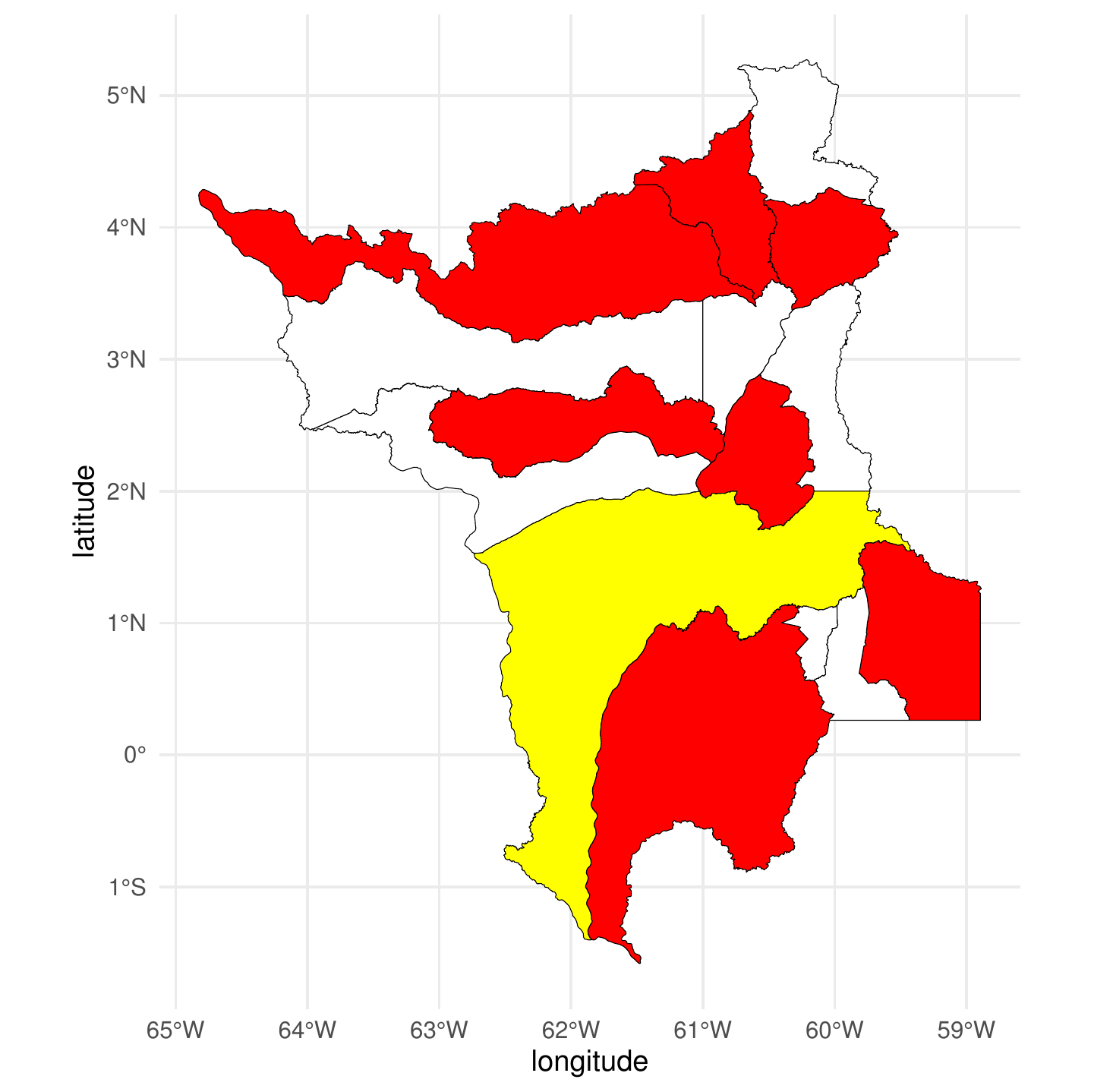}
        }\\
        \subfigure[]{%
            \label{to}
            \includegraphics[width=5.3cm,height=4.5cm]{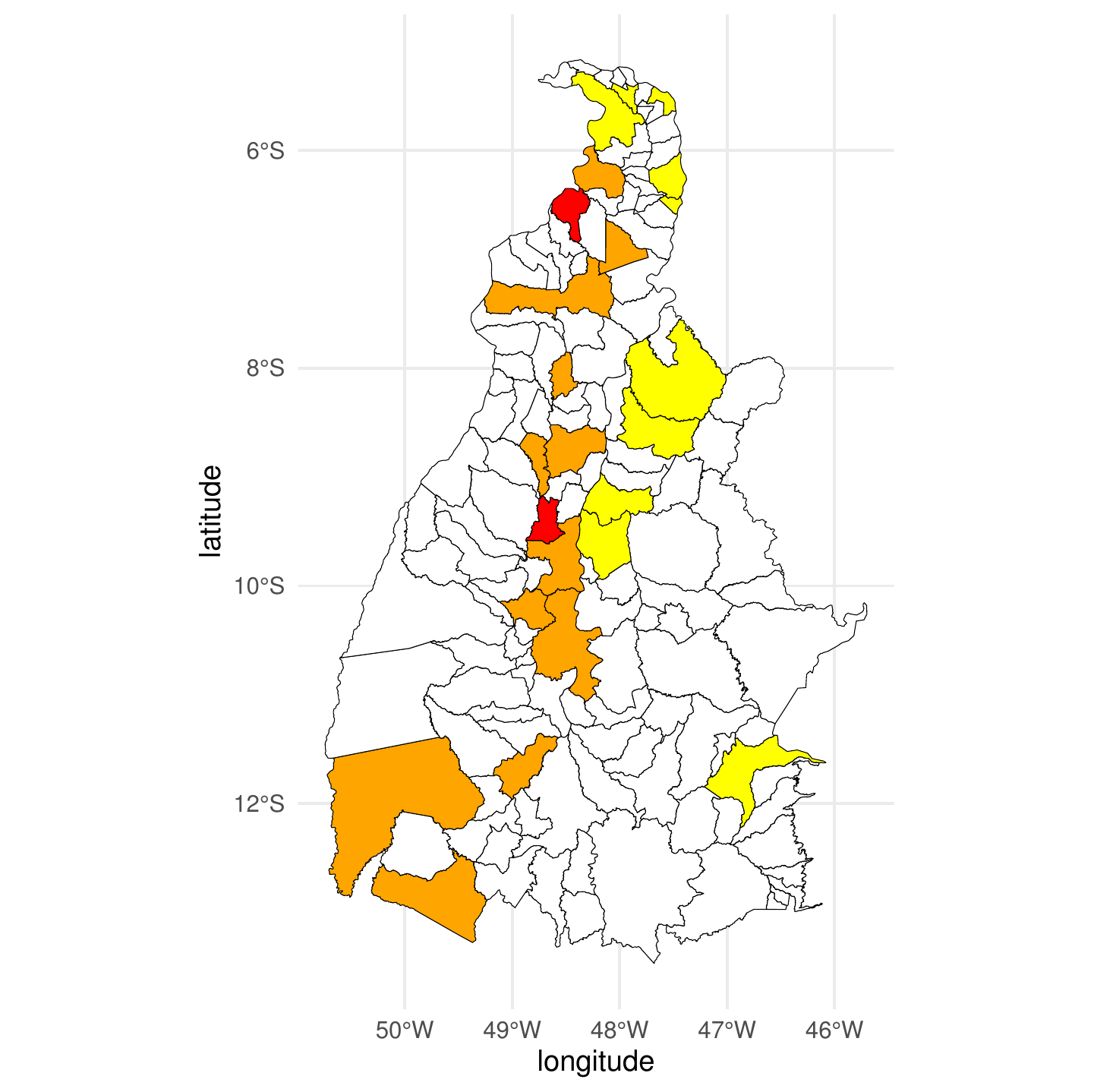}
        }
        \subfigure[]{%
           \label{to1}
           \includegraphics[width=5.3cm,height=4.5cm]{TO.pdf}
        }
        \subfigure[]{%
           \label{to2}
           \includegraphics[width=5.3cm,height=4.5cm]{TO.pdf}
        }
    \end{center}
    \caption{%
        Spatial location of the functional clustering of the municipalities of the states of Par\'{a} (first line), Rond\^{o}nia (second line), Roraima (third line) and Tocantins (fourth line) according to represented death curves (a, d, g, j), first derivative of death curves (b, e, h, k) and second derivative of death curves (c, f, i, l).}
   \label{muni2}
\end{figure}

\newpage
\subsection{Northeast Region}

\begin{figure}[!htbp]
     \begin{center}
        \subfigure[]{%
            \label{al}
            \includegraphics[width=5.3cm,height=6cm]{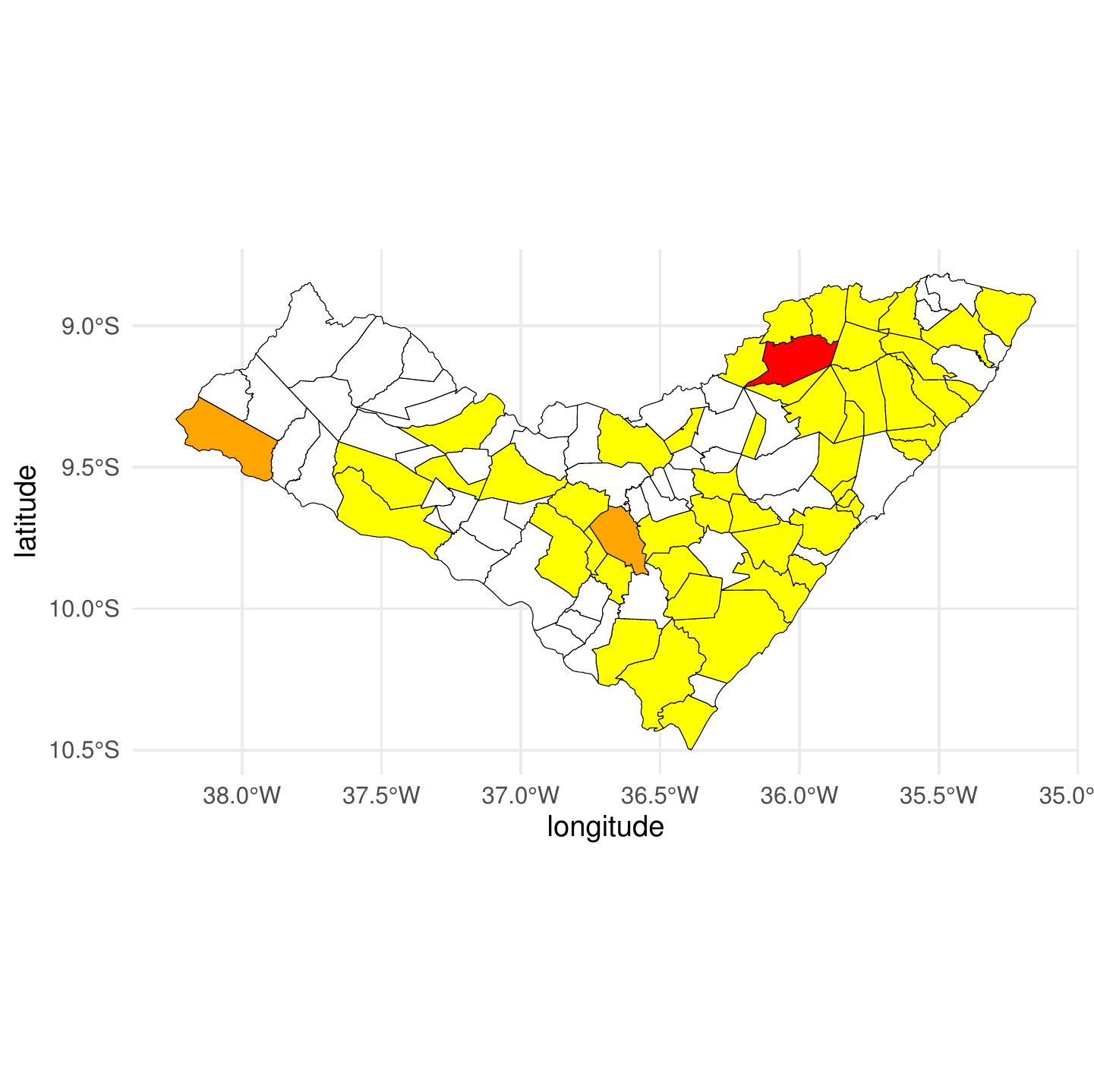}
        }
        \subfigure[]{%
           \label{al1}
           \includegraphics[width=5.3cm,height=6cm]{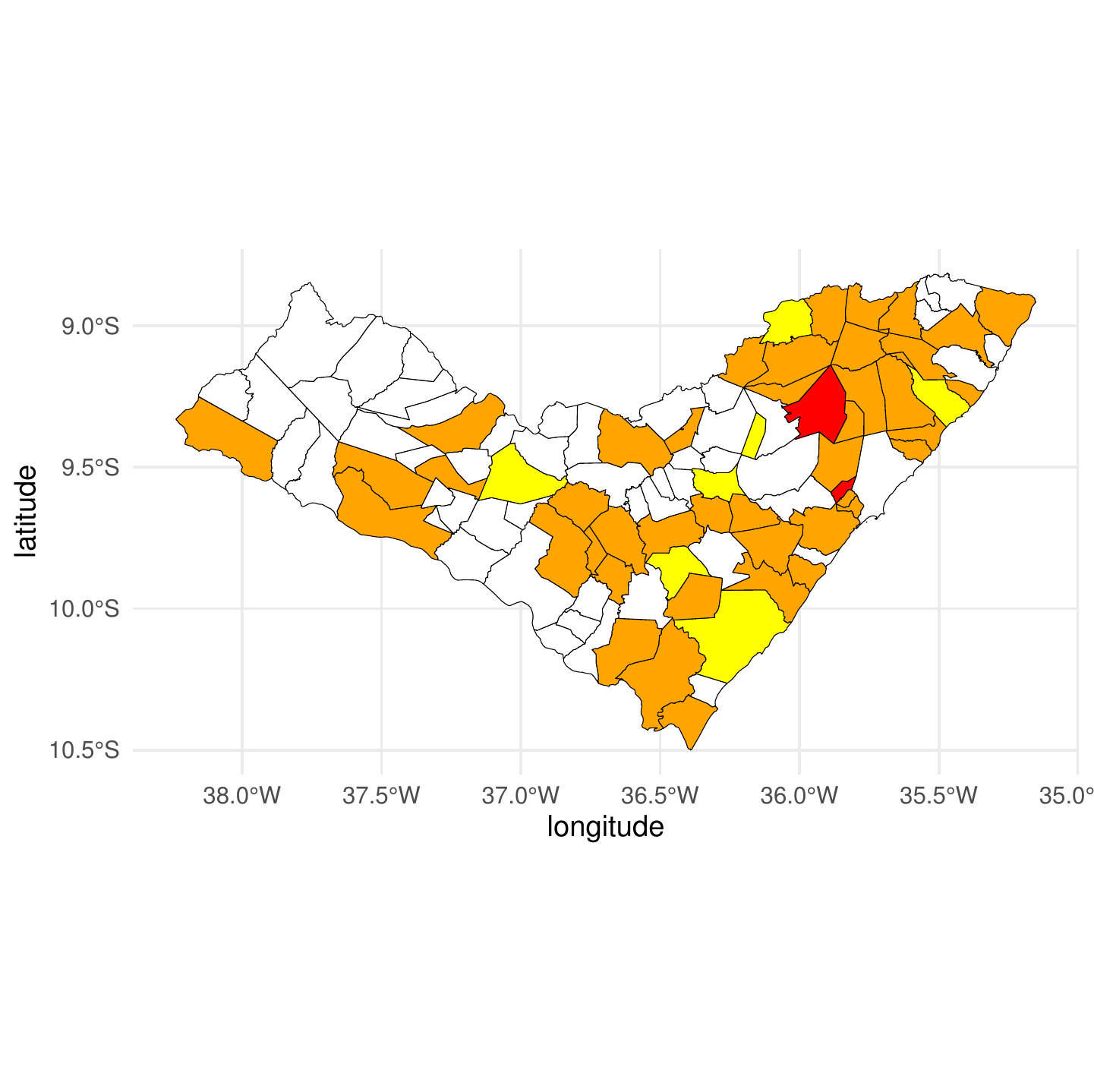}
        }
        \subfigure[]{%
           \label{al2}
           \includegraphics[width=5.3cm,height=6cm]{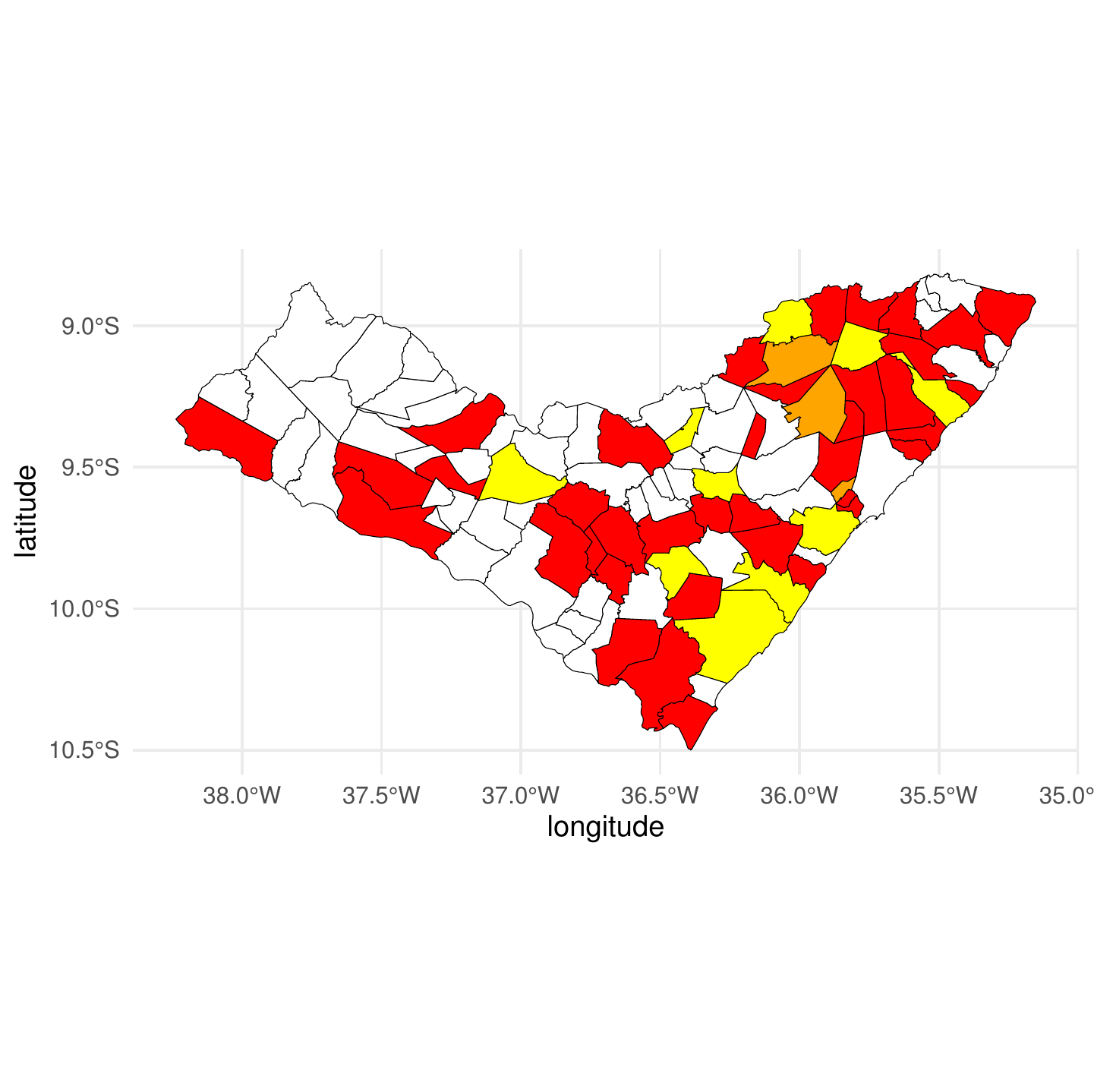}
        }\\
        \subfigure[]{%
            \label{ba}
            \includegraphics[width=5.3cm,height=6cm]{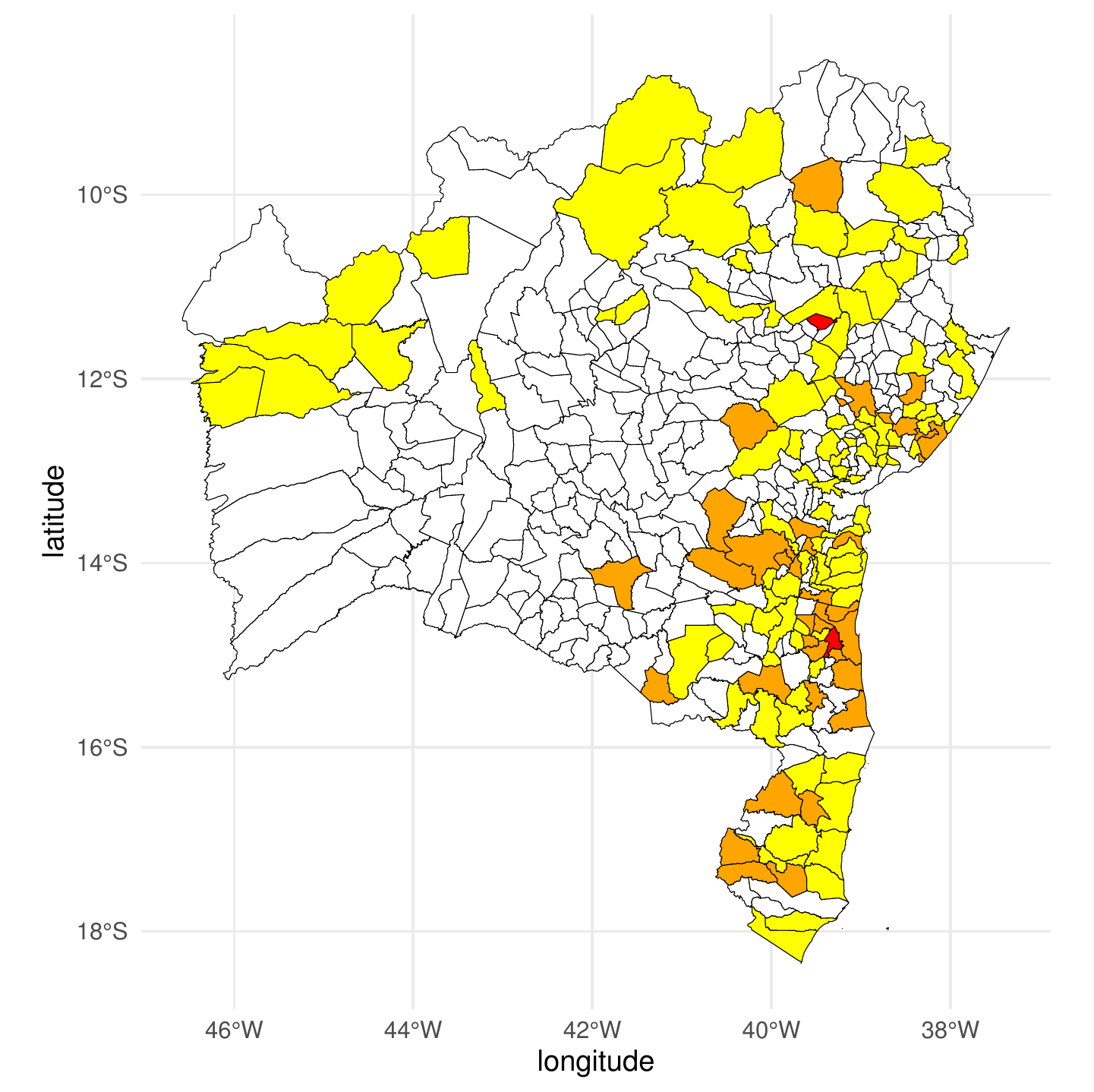}
        }
        \subfigure[]{%
           \label{ba1}
           \includegraphics[width=5.3cm,height=6cm]{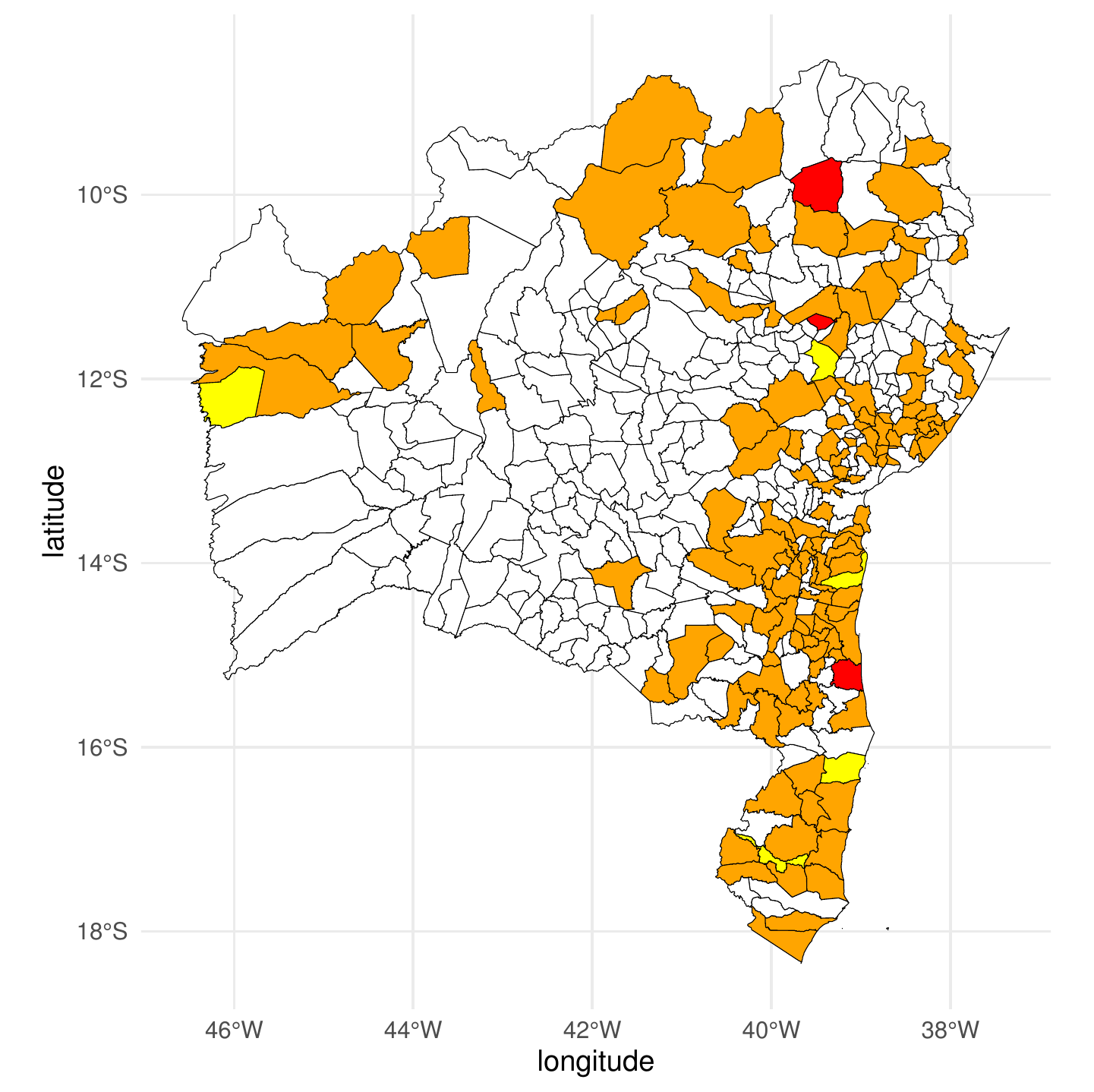}
        }
        \subfigure[]{%
           \label{ba2}
           \includegraphics[width=5.3cm,height=6cm]{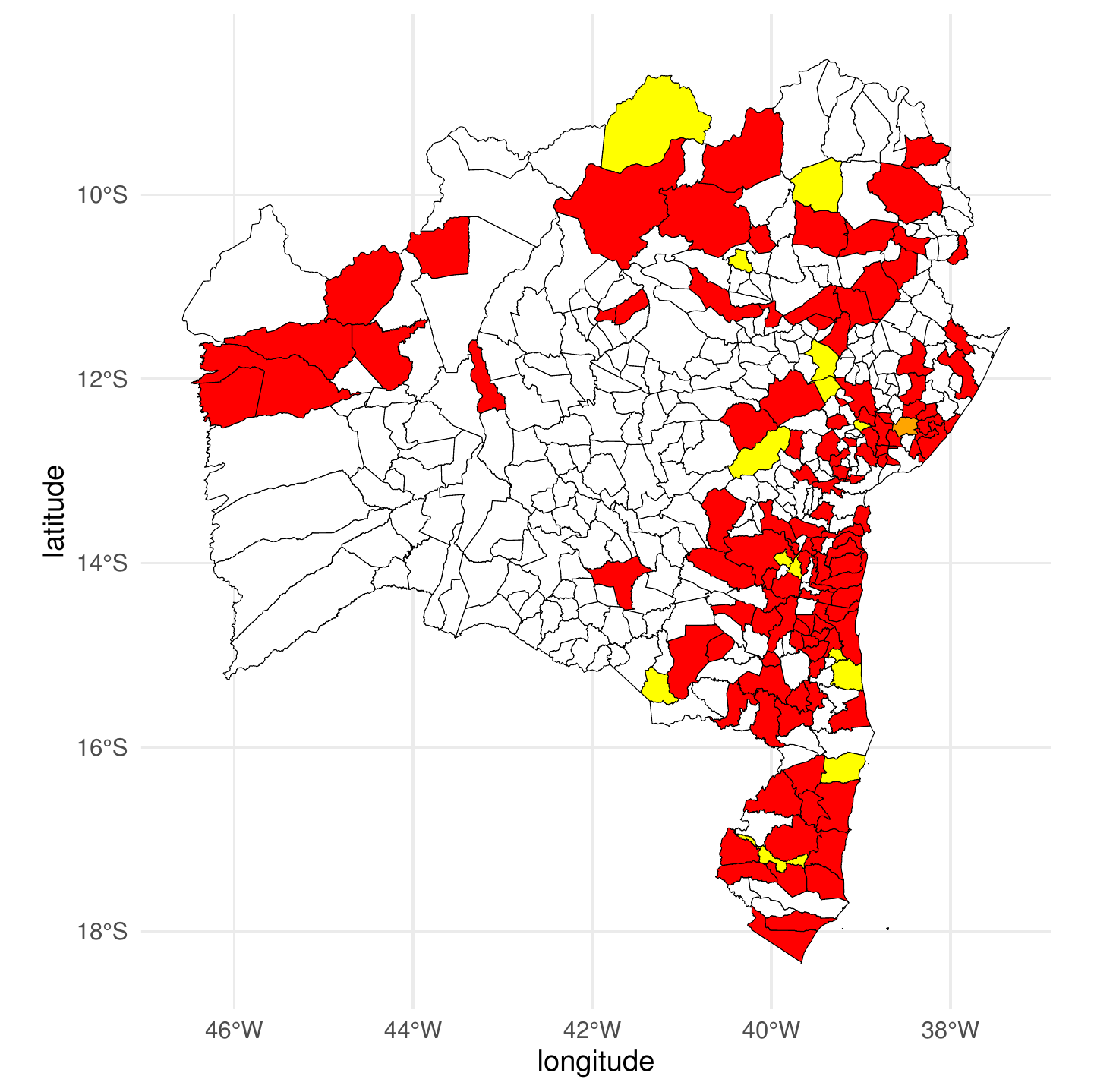}
        }\\
        \subfigure[]{%
            \label{ce}
            \includegraphics[width=5.3cm,height=6cm]{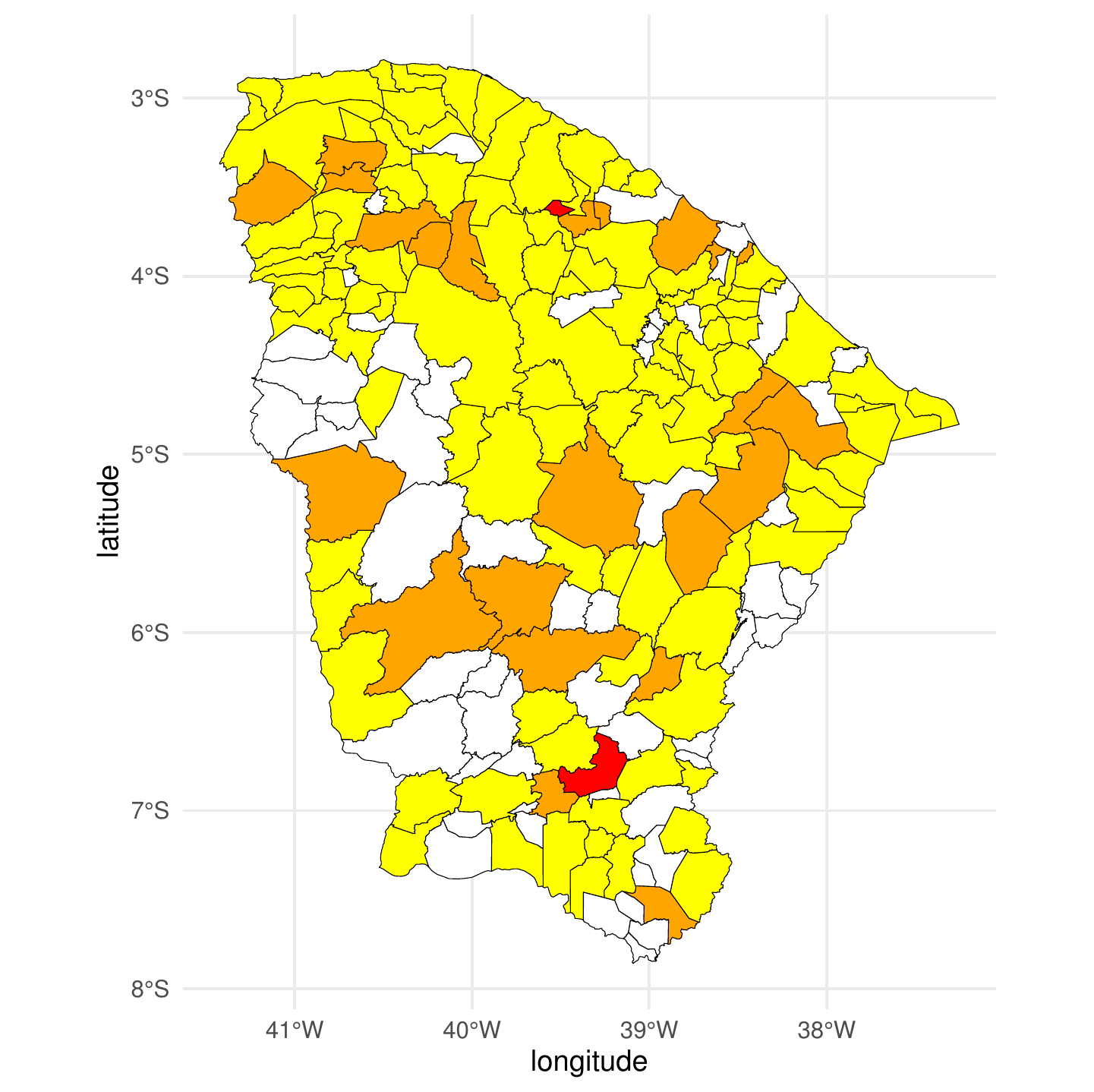}
        }
        \subfigure[]{%
           \label{ce1}
           \includegraphics[width=5.3cm,height=6cm]{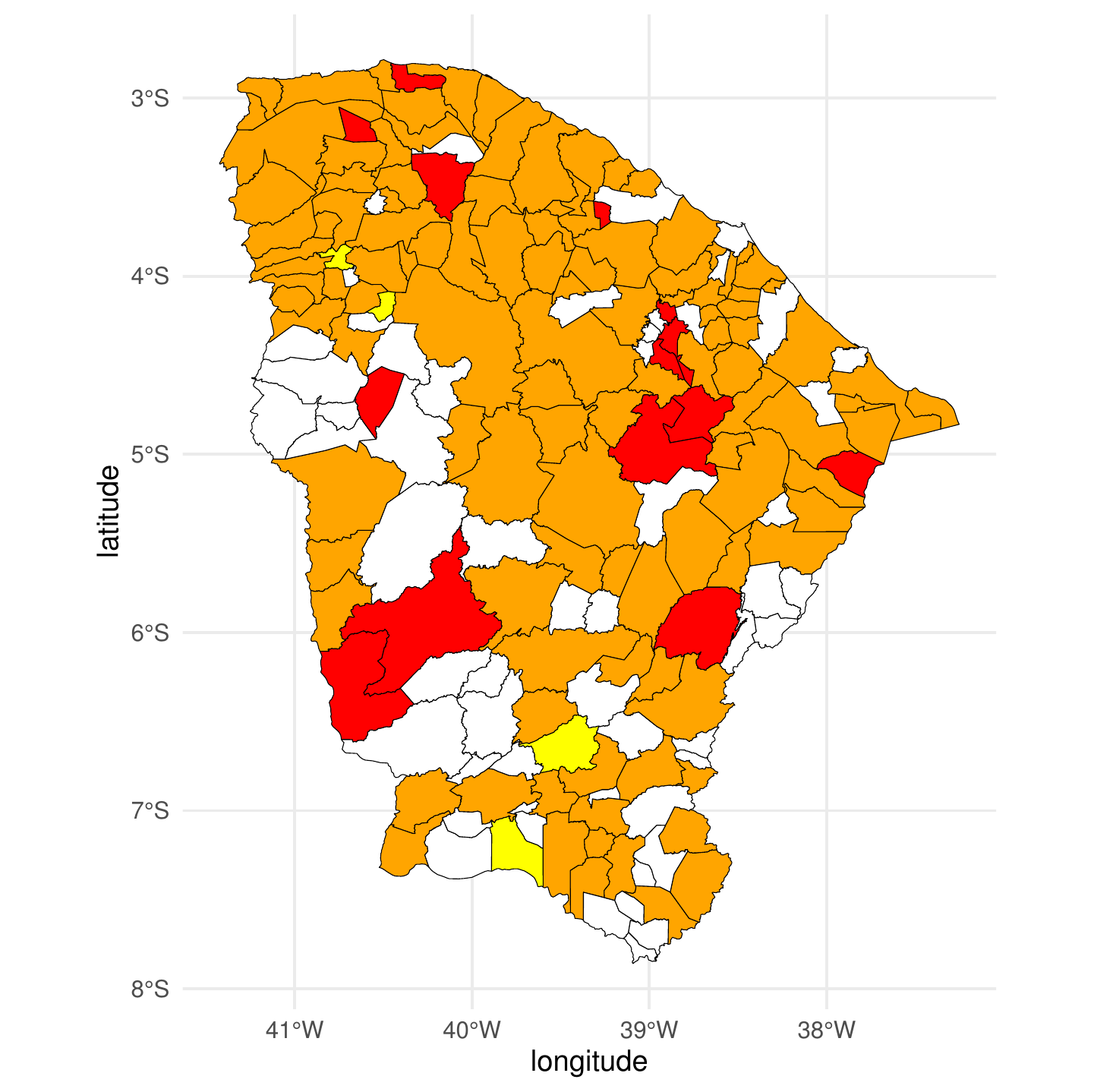}
        }
        \subfigure[]{%
           \label{ce2}
           \includegraphics[width=5.3cm,height=6cm]{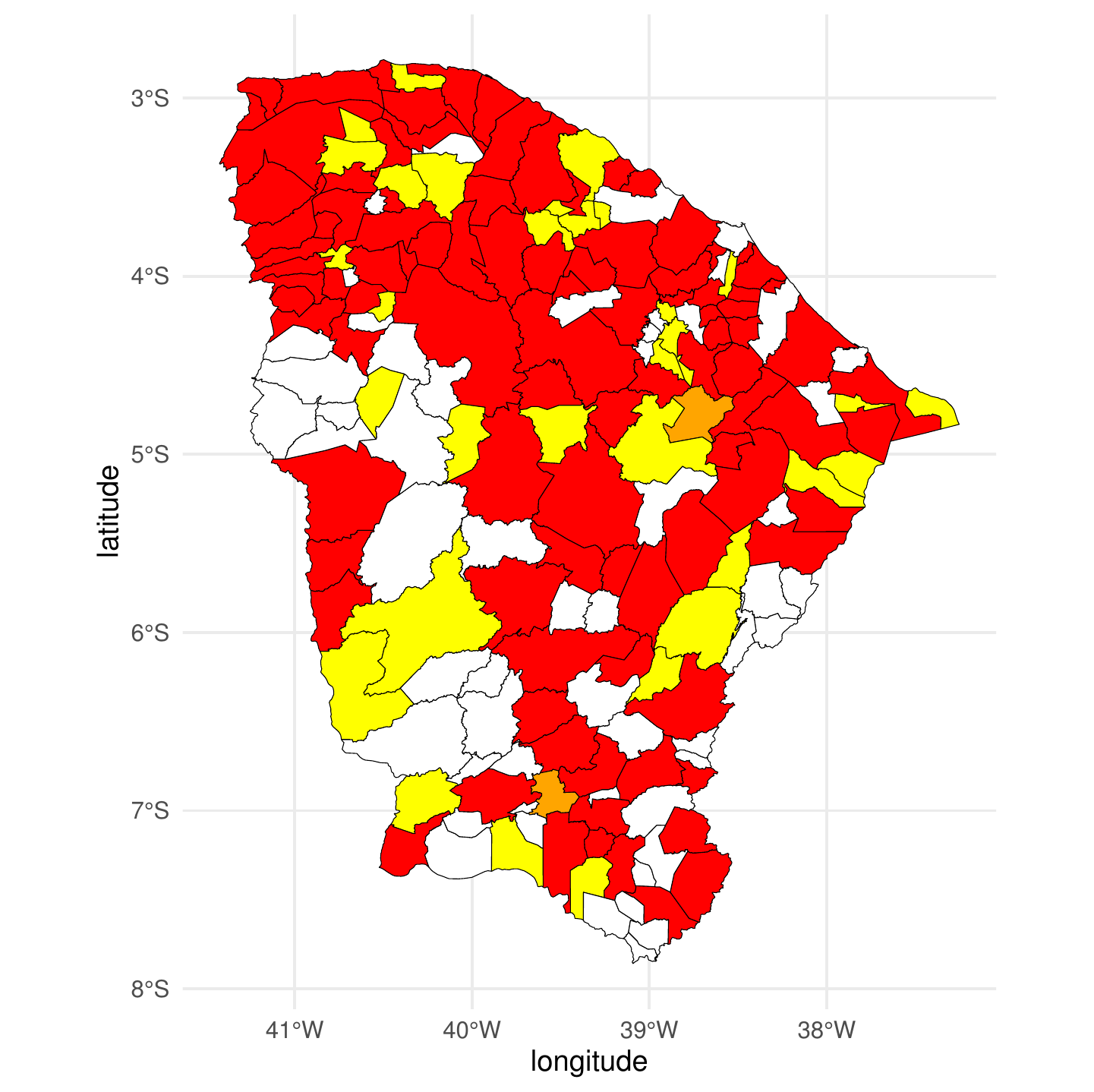}
        }
    \end{center}
    \caption{%
        Spatial location of the functional clustering of the municipalities of the states of Alagoas (first line), Bahia (second line) and Cear\'{a} (third line) according to represented death curves (a, d, g), first derivative of death curves (b, e, h) and second derivative of death curves (c, f, i).}
   \label{muni3}
\end{figure}

\begin{figure}[!htbp]
     \begin{center}
        \subfigure[]{%
            \label{ma}
            \includegraphics[width=5.3cm,height=6cm]{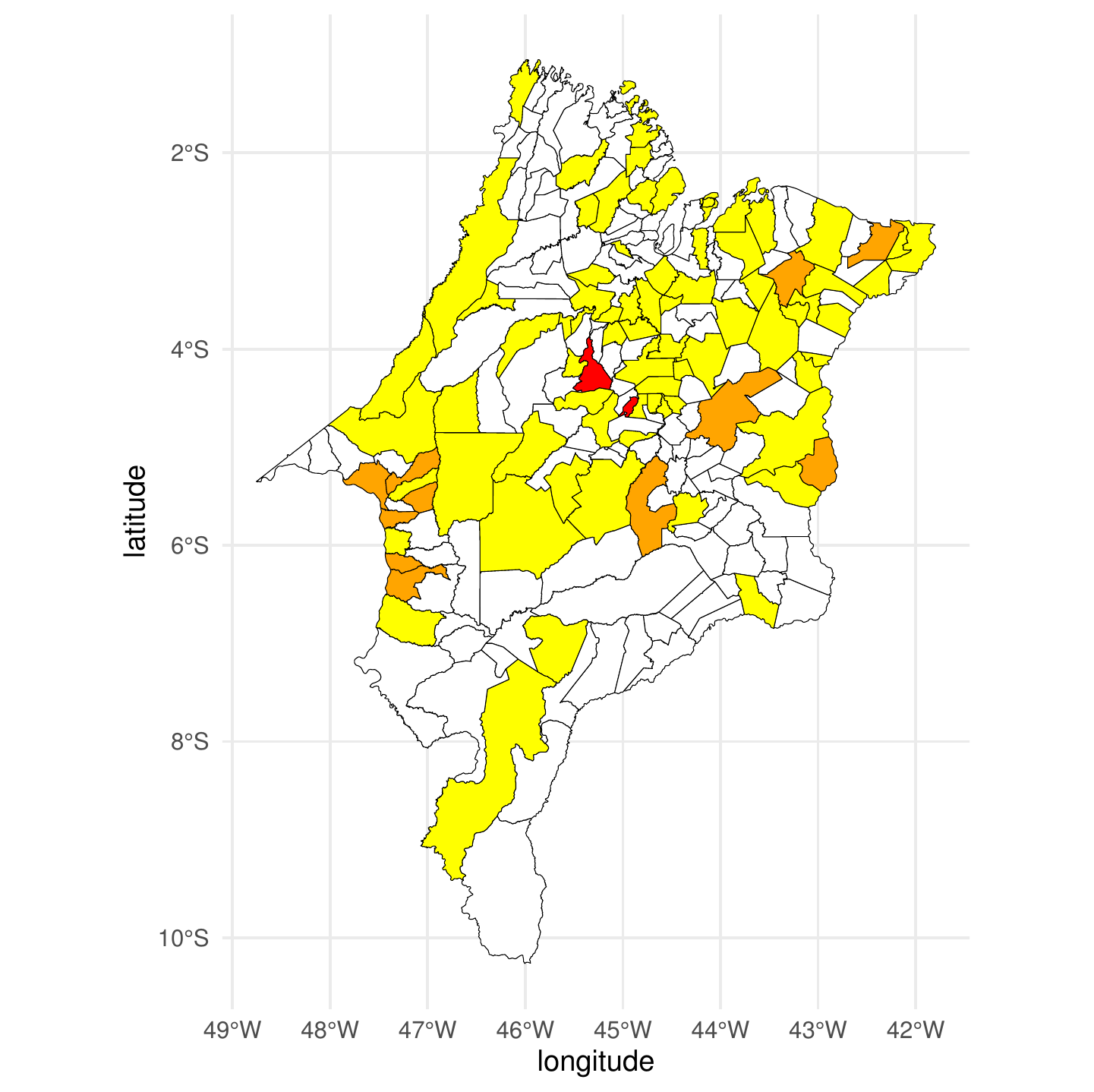}
        }
        \subfigure[]{%
           \label{ma1}
           \includegraphics[width=5.3cm,height=6cm]{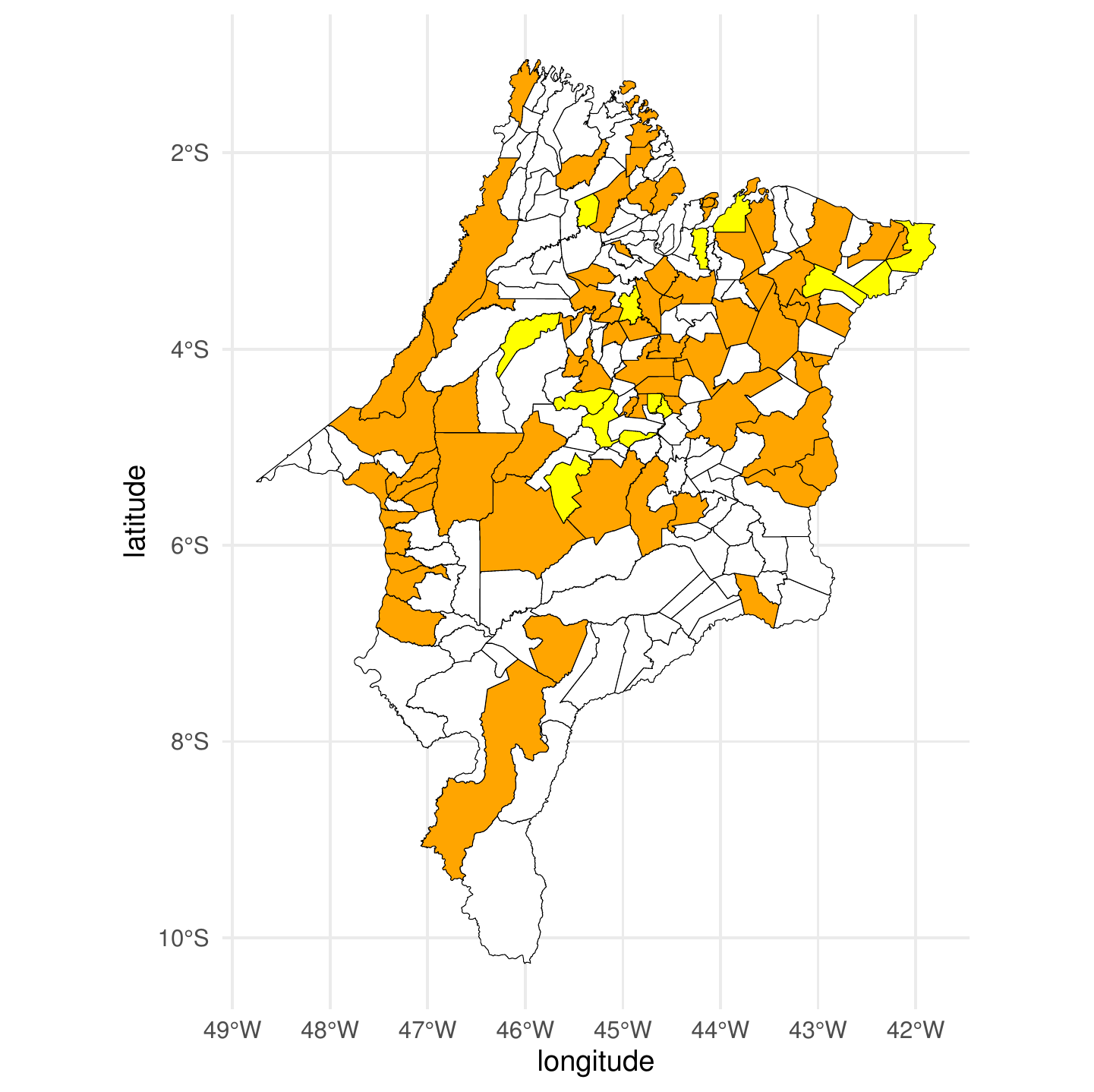}
        }
        \subfigure[]{%
           \label{ma2}
           \includegraphics[width=5.3cm,height=6cm]{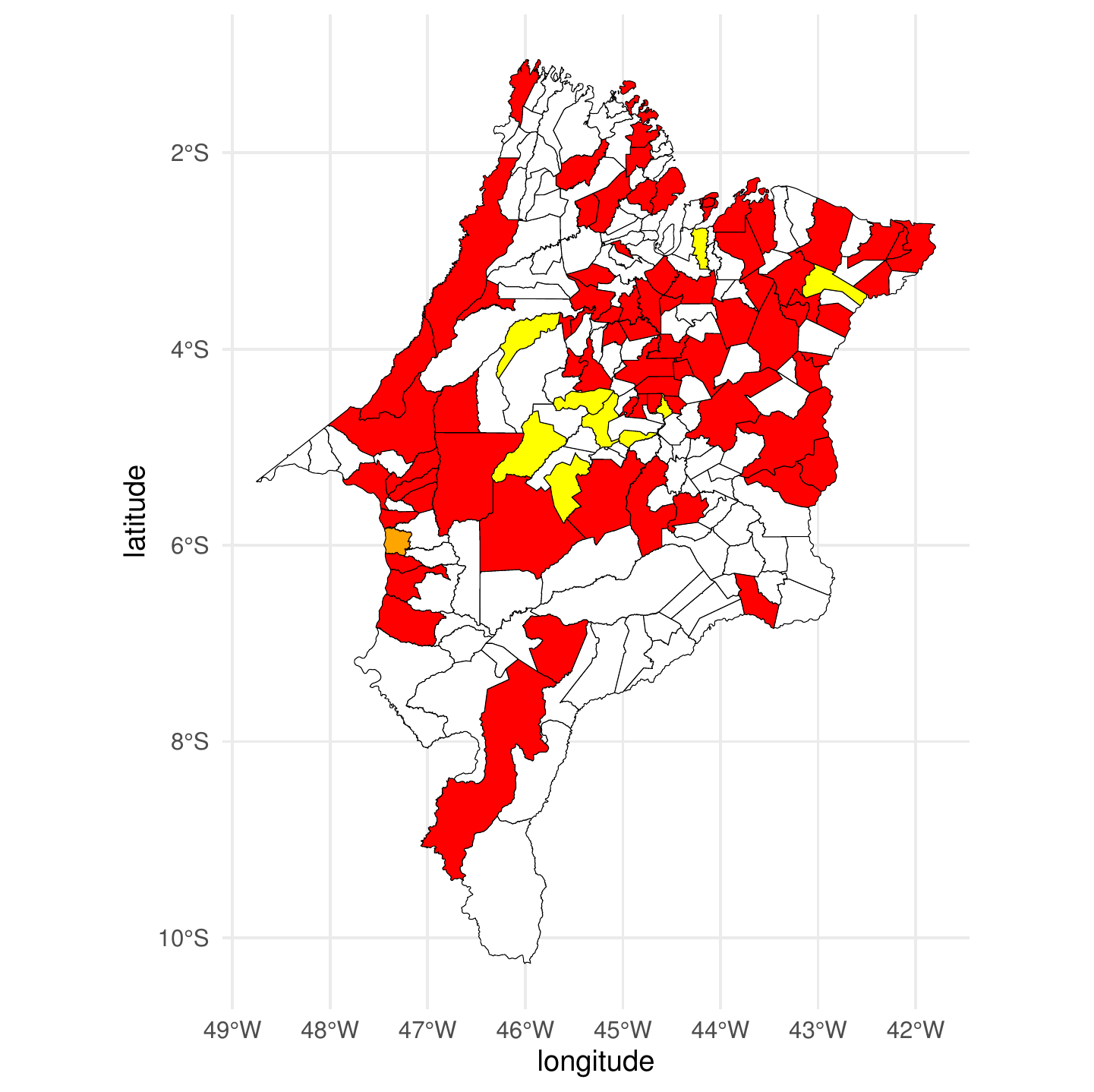}
        }\\
        \subfigure[]{%
            \label{pb}
            \includegraphics[width=5.3cm,height=6cm]{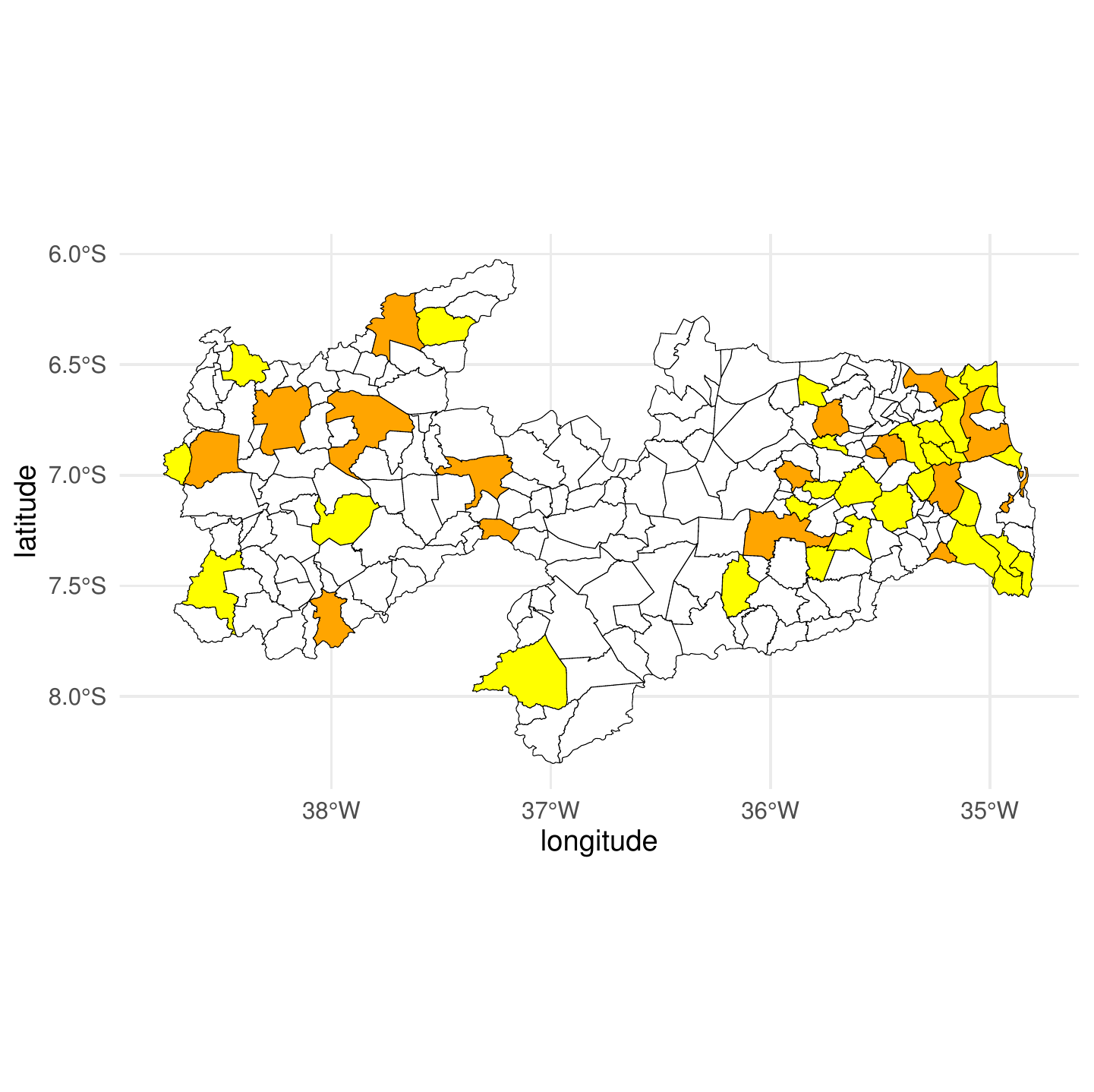}
        }
        \subfigure[]{%
           \label{pb1}
           \includegraphics[width=5.3cm,height=6cm]{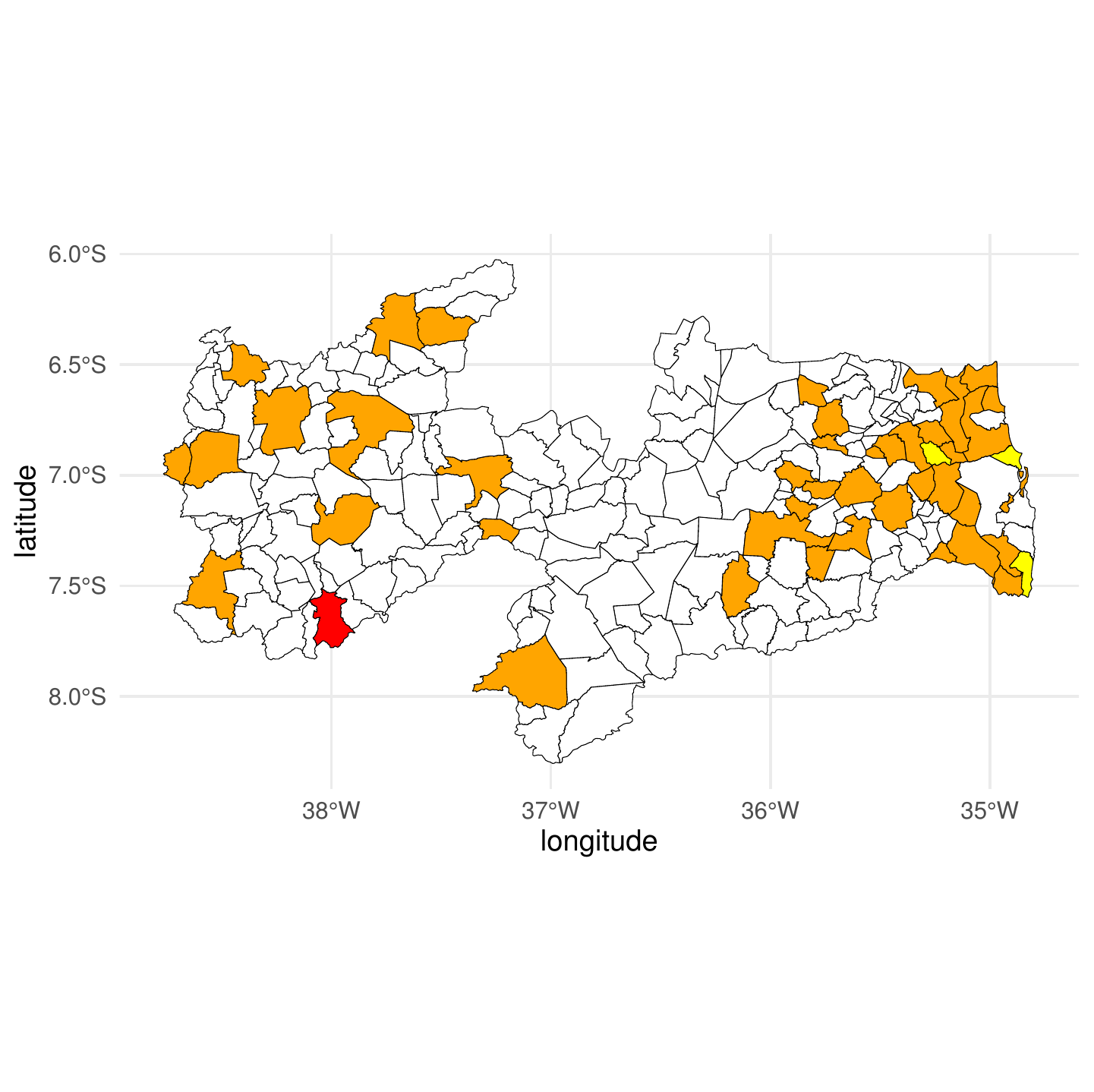}
        }
        \subfigure[]{%
           \label{pb2}
           \includegraphics[width=5.3cm,height=6cm]{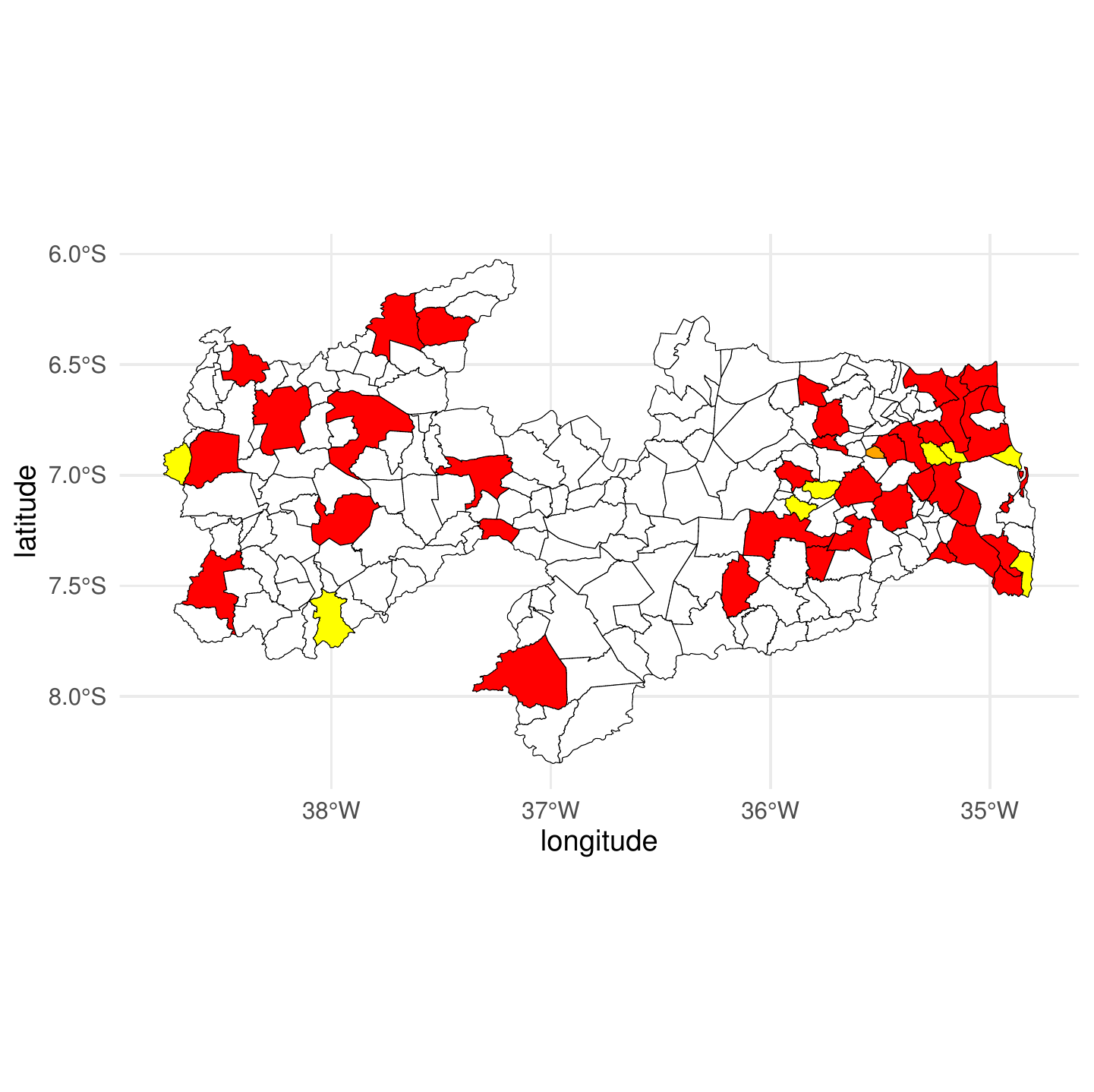}
        }\\
        \subfigure[]{%
            \label{pe}
            \includegraphics[width=5.3cm,height=6cm]{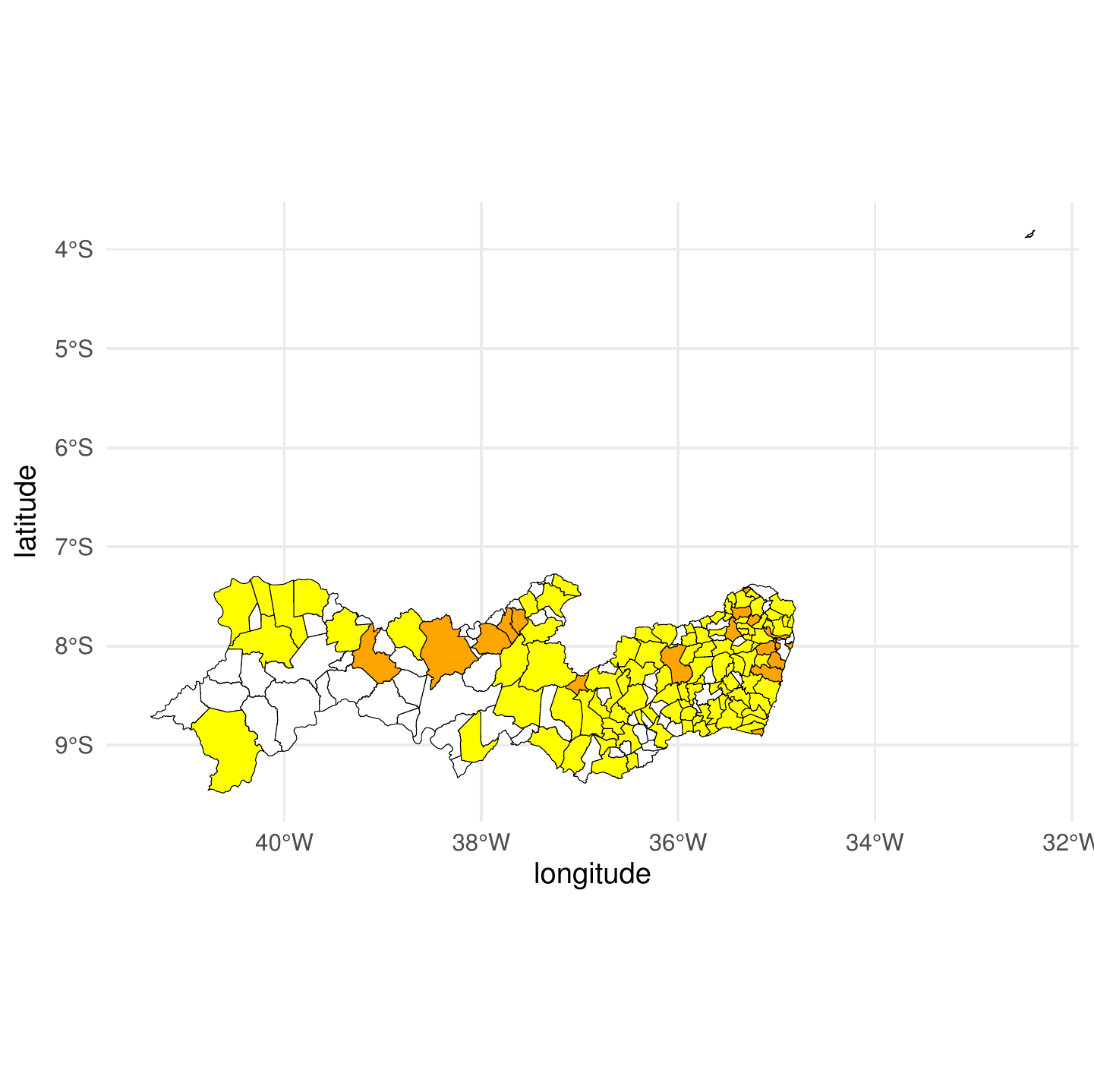}
        }
        \subfigure[]{%
           \label{pe1}
           \includegraphics[width=5.3cm,height=6cm]{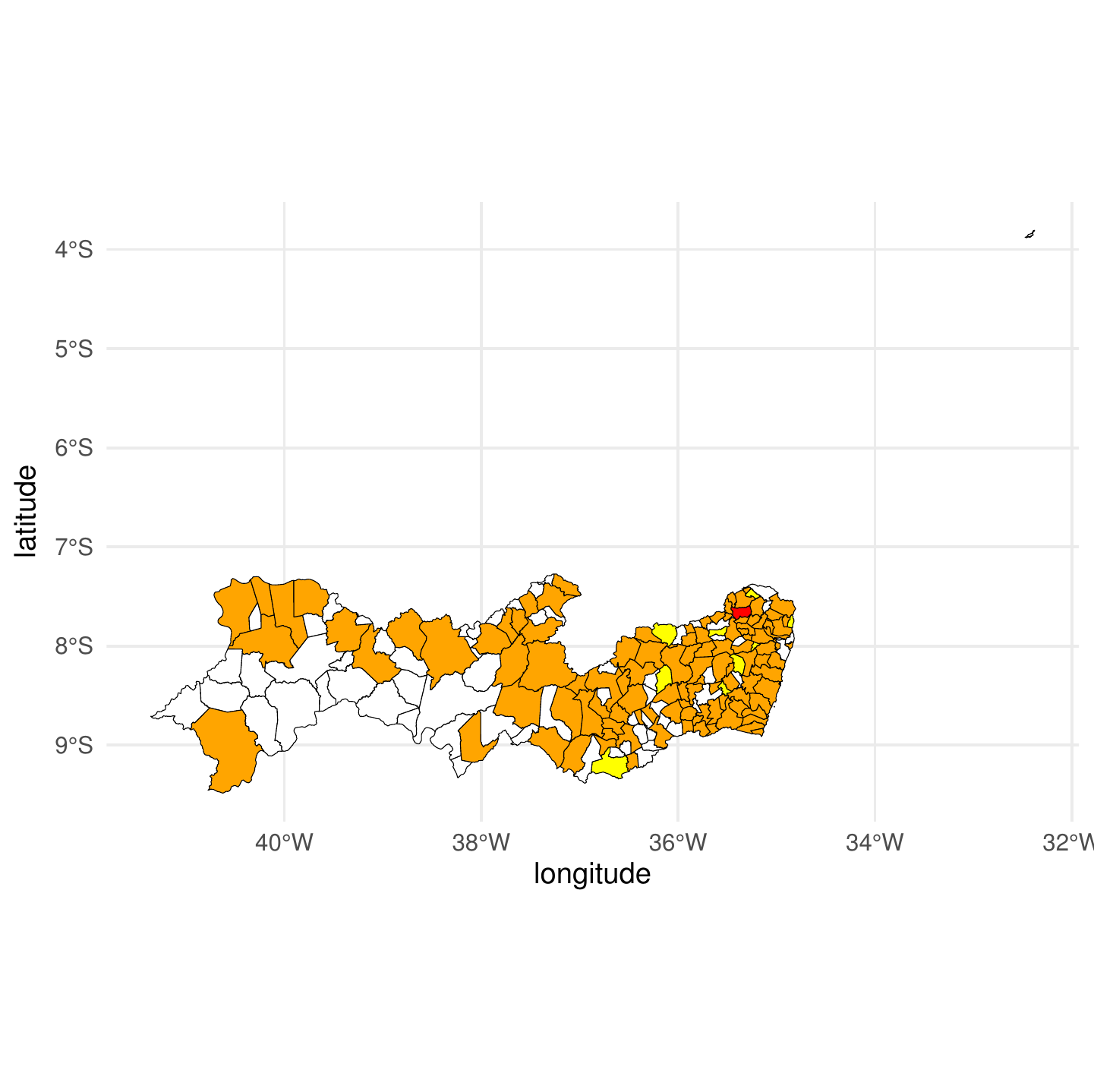}
        }
        \subfigure[]{%
           \label{pe2}
           \includegraphics[width=5.3cm,height=6cm]{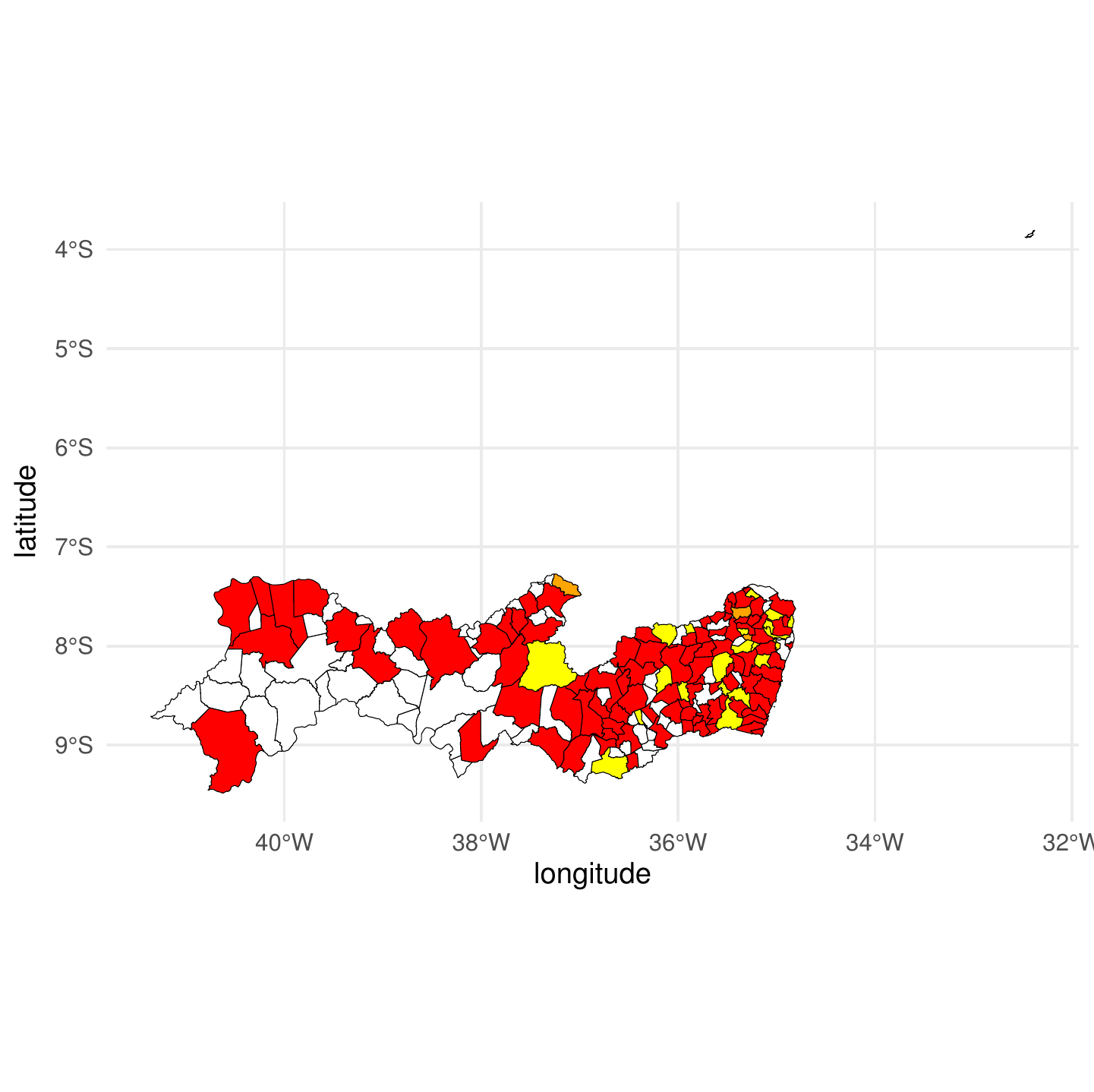}
        }
    \end{center}
    \caption{%
        Spatial location of the functional clustering of the municipalities of the states of Maranh\~{a}o (first line), Para\'{i}ba (second line) and Pernambuco (third line) according to represented death curves (a, d, g), first derivative of death curves (b, e, h) and second derivative of death curves (c, f, i).}
   \label{muni4}
\end{figure}

\begin{figure}[!htbp]
     \begin{center}
        \subfigure[]{%
            \label{pi}
            \includegraphics[width=5.3cm,height=6cm]{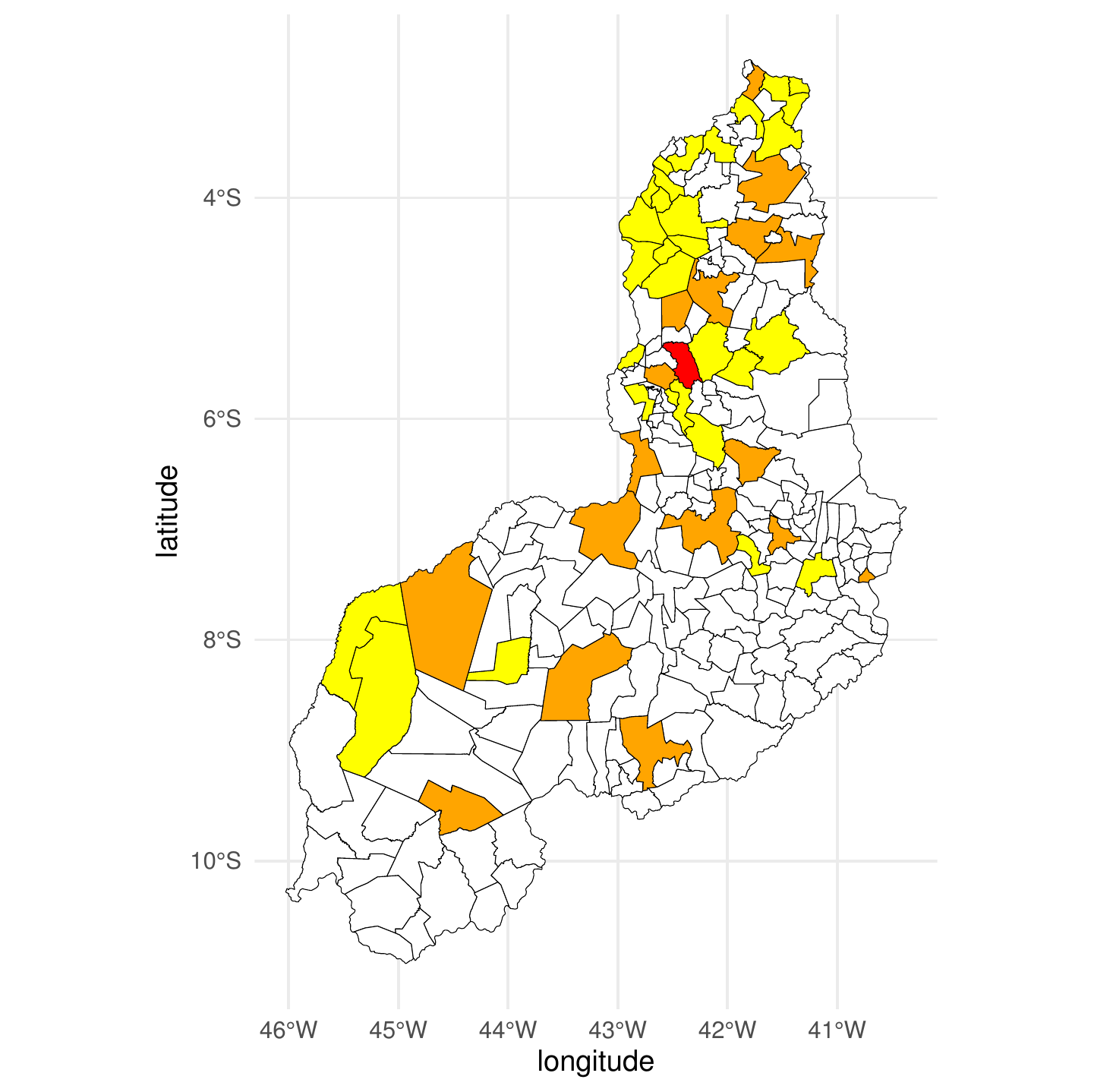}
        }
        \subfigure[]{%
           \label{pi1}
           \includegraphics[width=5.3cm,height=6cm]{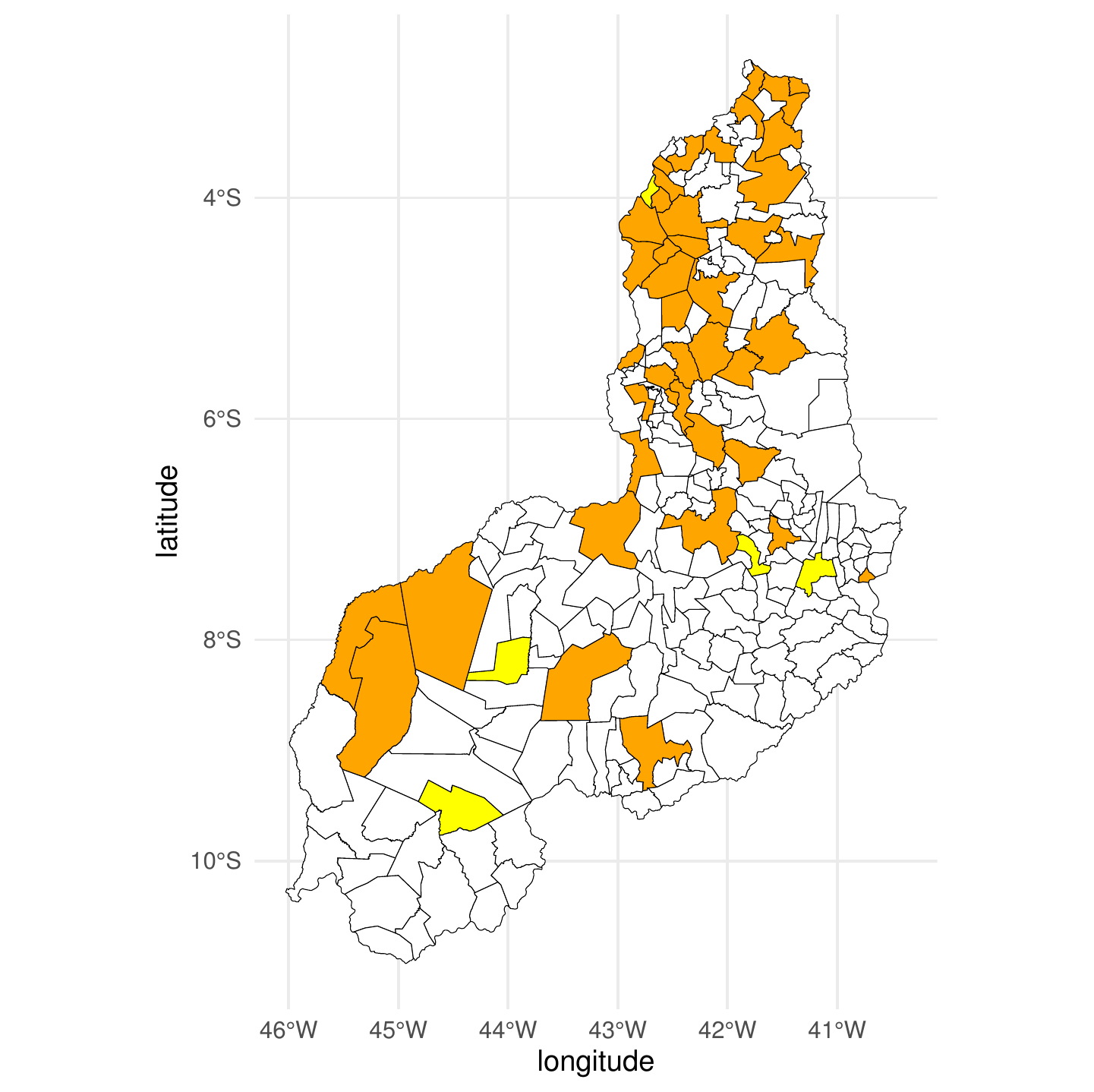}
        }
        \subfigure[]{%
           \label{pi2}
           \includegraphics[width=5.3cm,height=6cm]{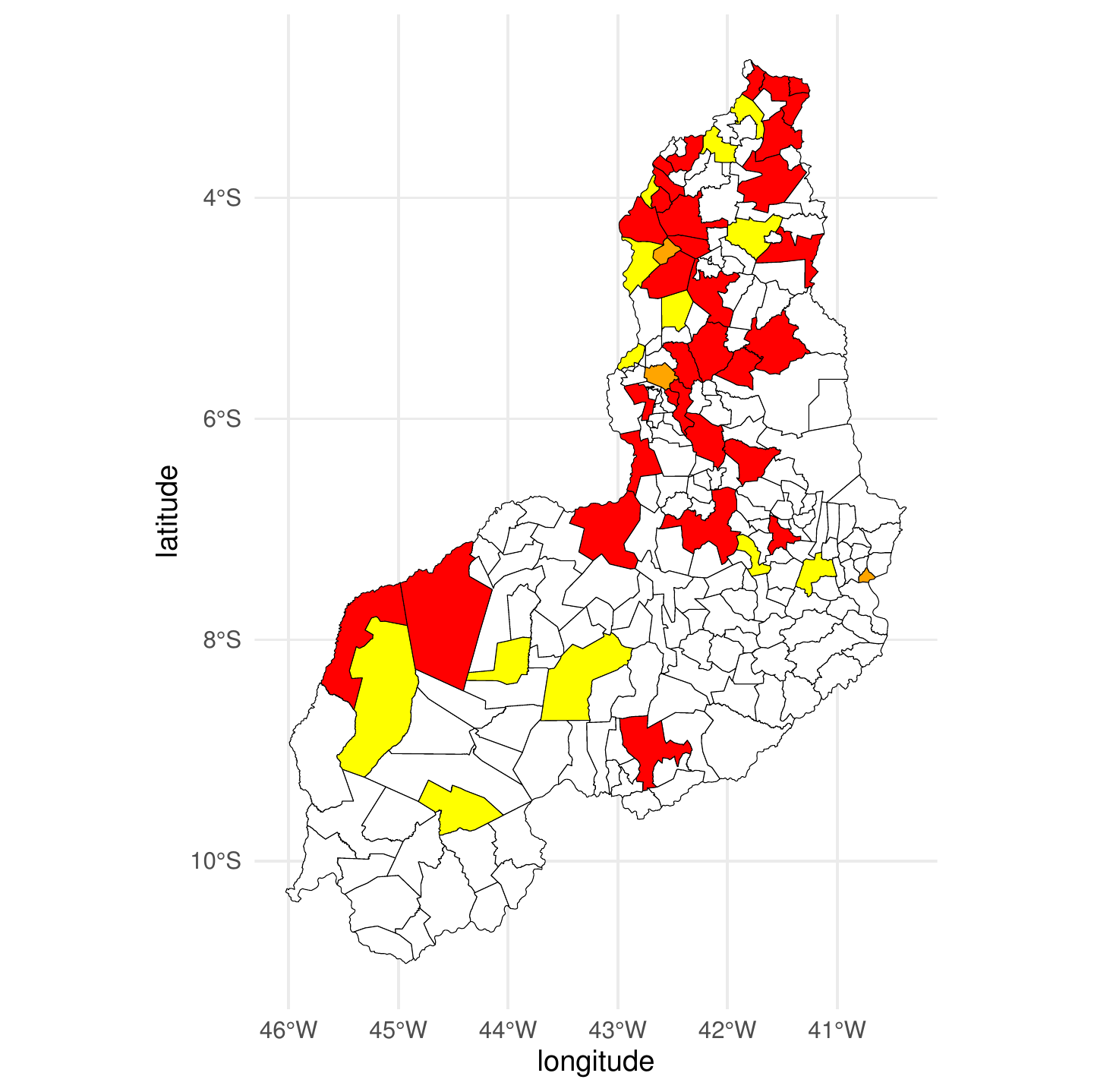}
        }\\
        \subfigure[]{%
            \label{rn}
            \includegraphics[width=5.3cm,height=6cm]{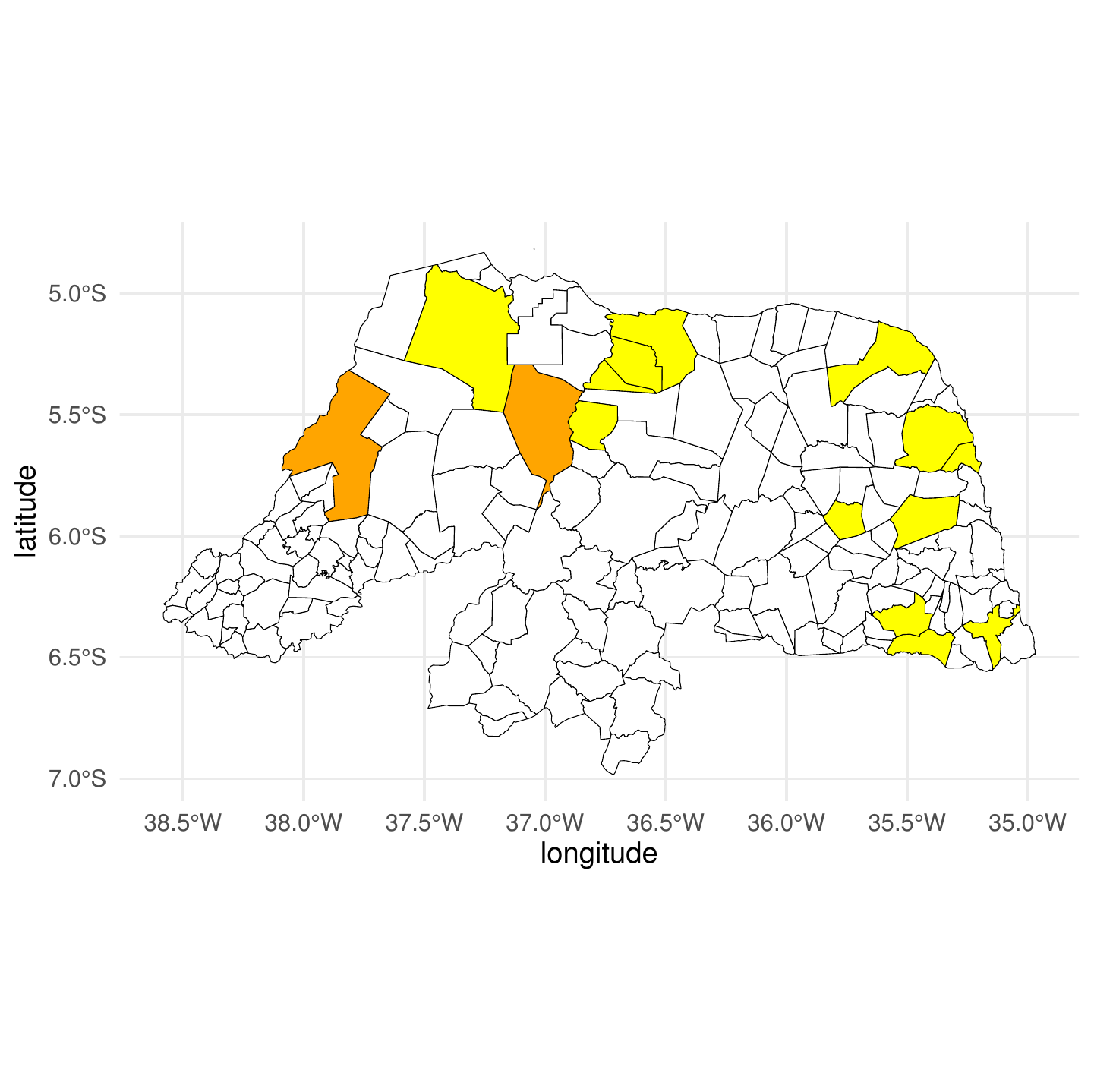}
        }
        \subfigure[]{%
           \label{rn1}
           \includegraphics[width=5.3cm,height=6cm]{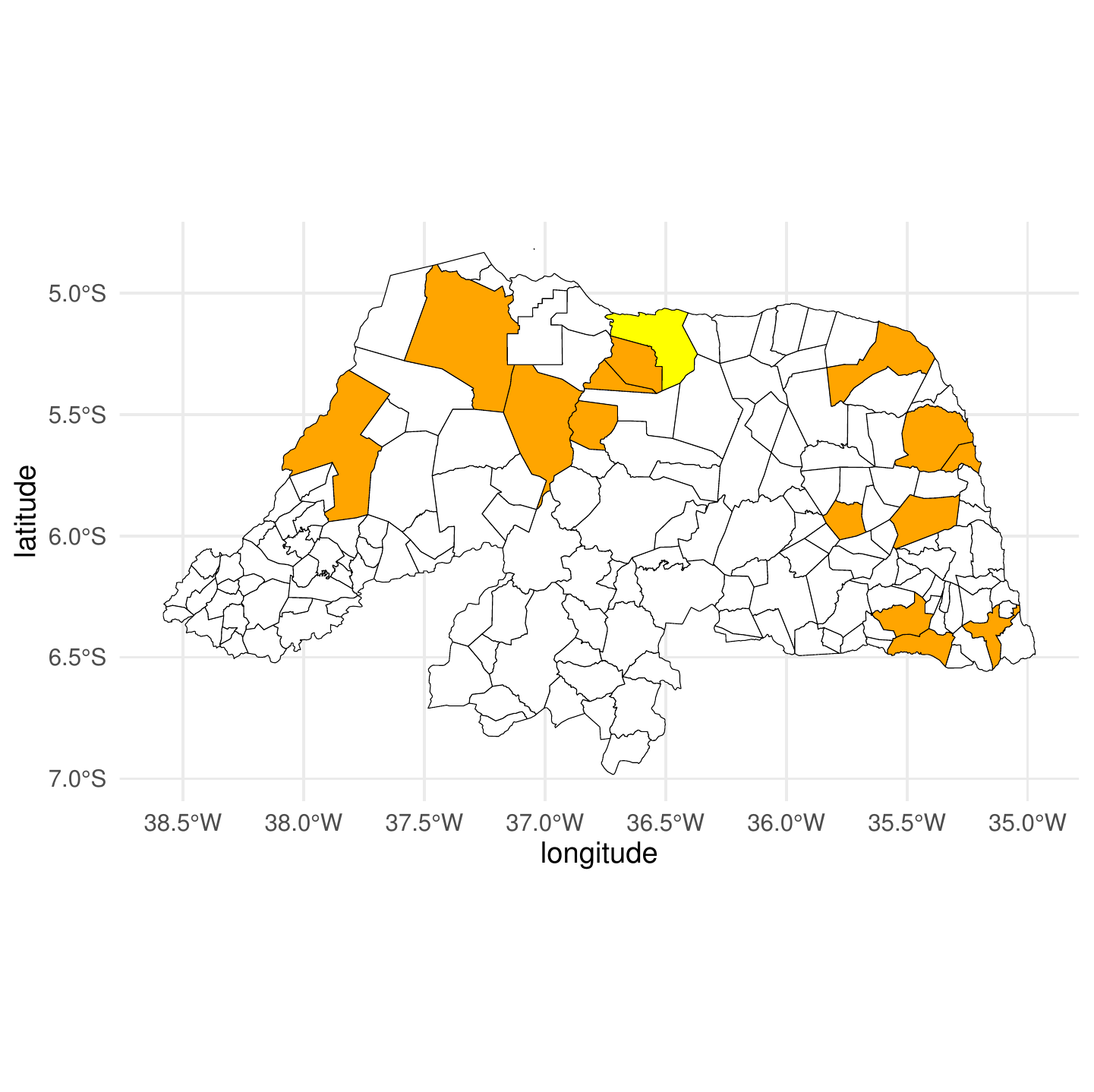}
        }
        \subfigure[]{%
           \label{rn2}
           \includegraphics[width=5.3cm,height=6cm]{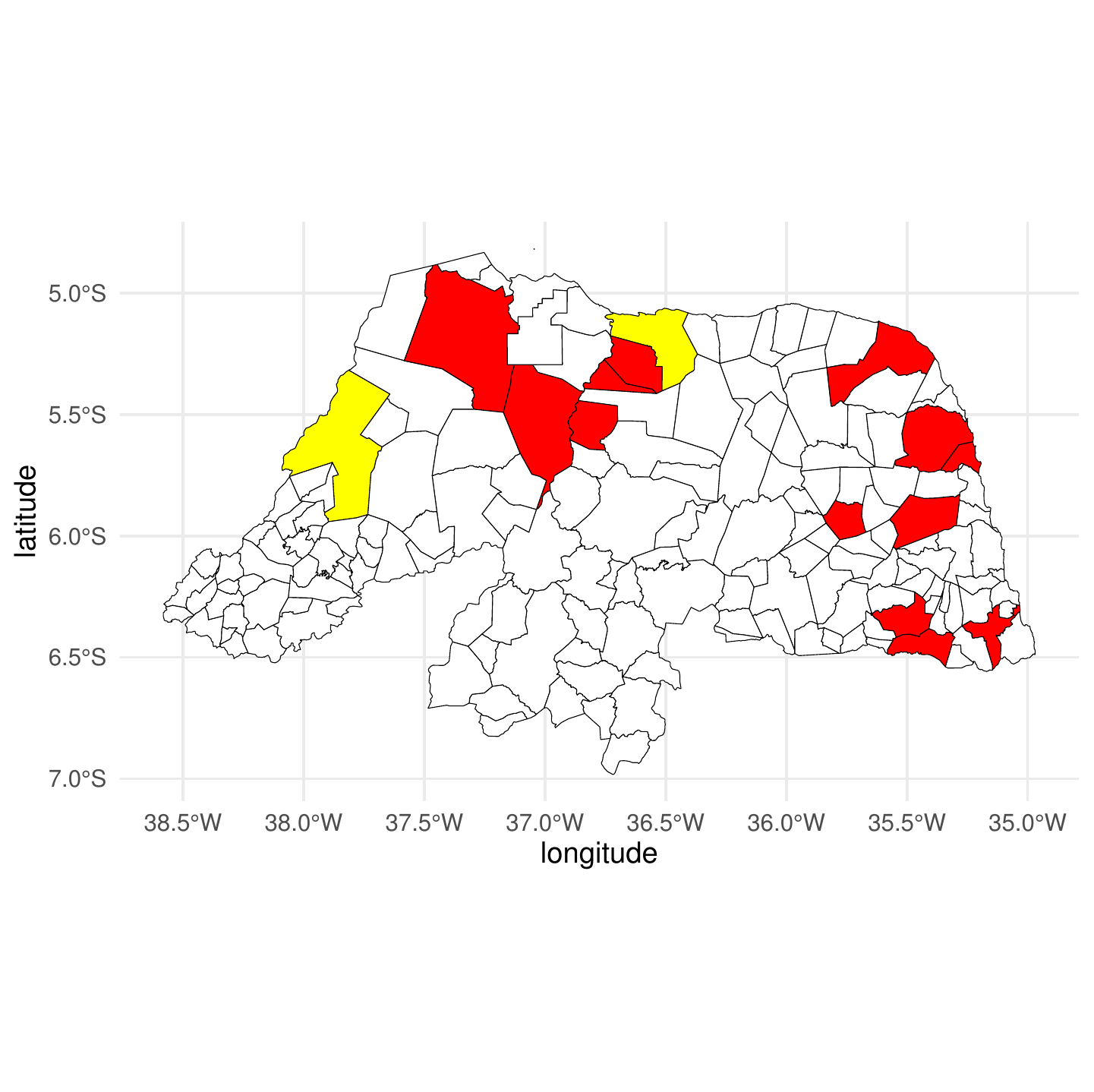}
        }\\
        \subfigure[]{%
            \label{se}
            \includegraphics[width=5.3cm,height=6cm]{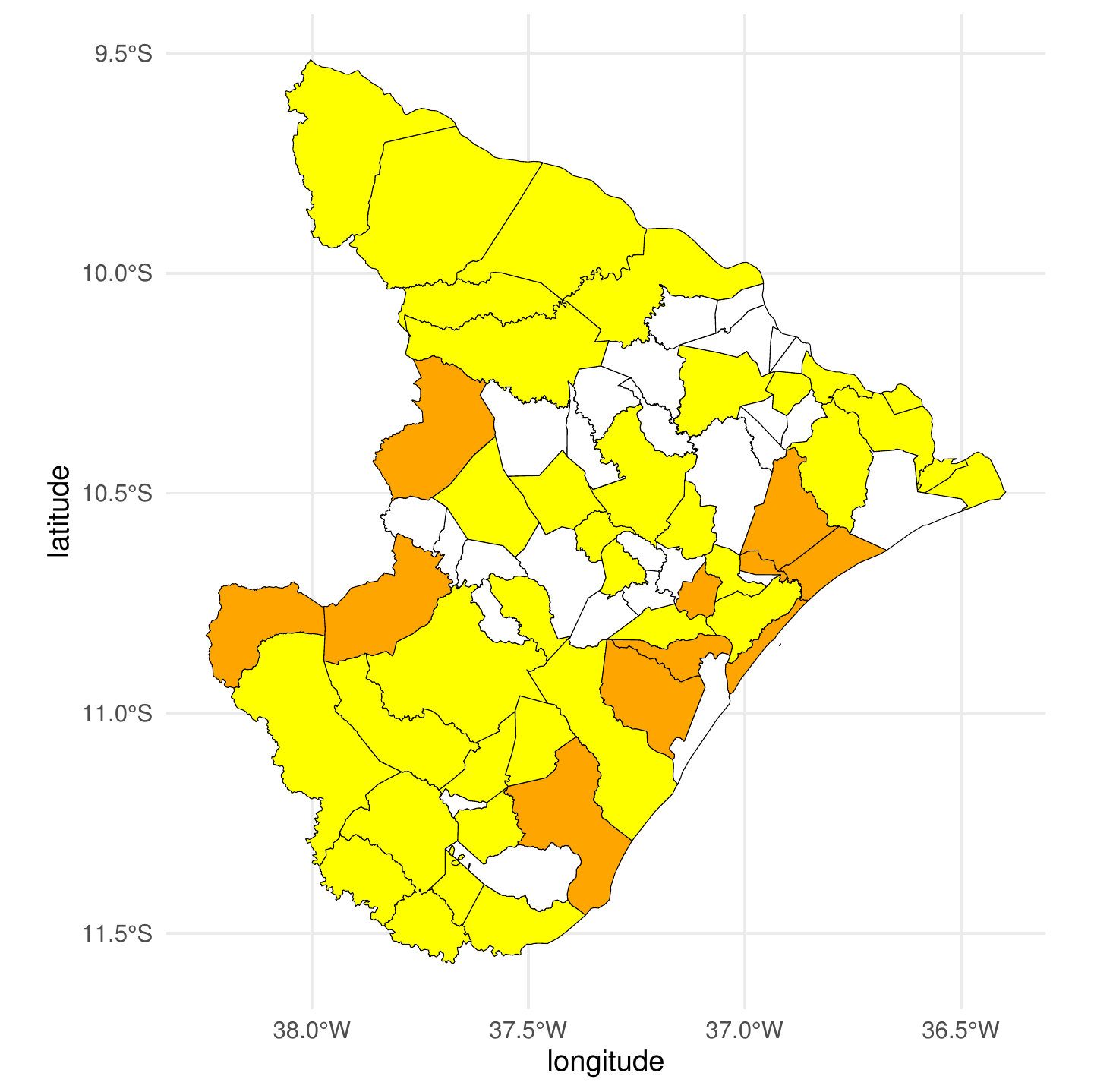}
        }
        \subfigure[]{%
           \label{se1}
           \includegraphics[width=5.3cm,height=6cm]{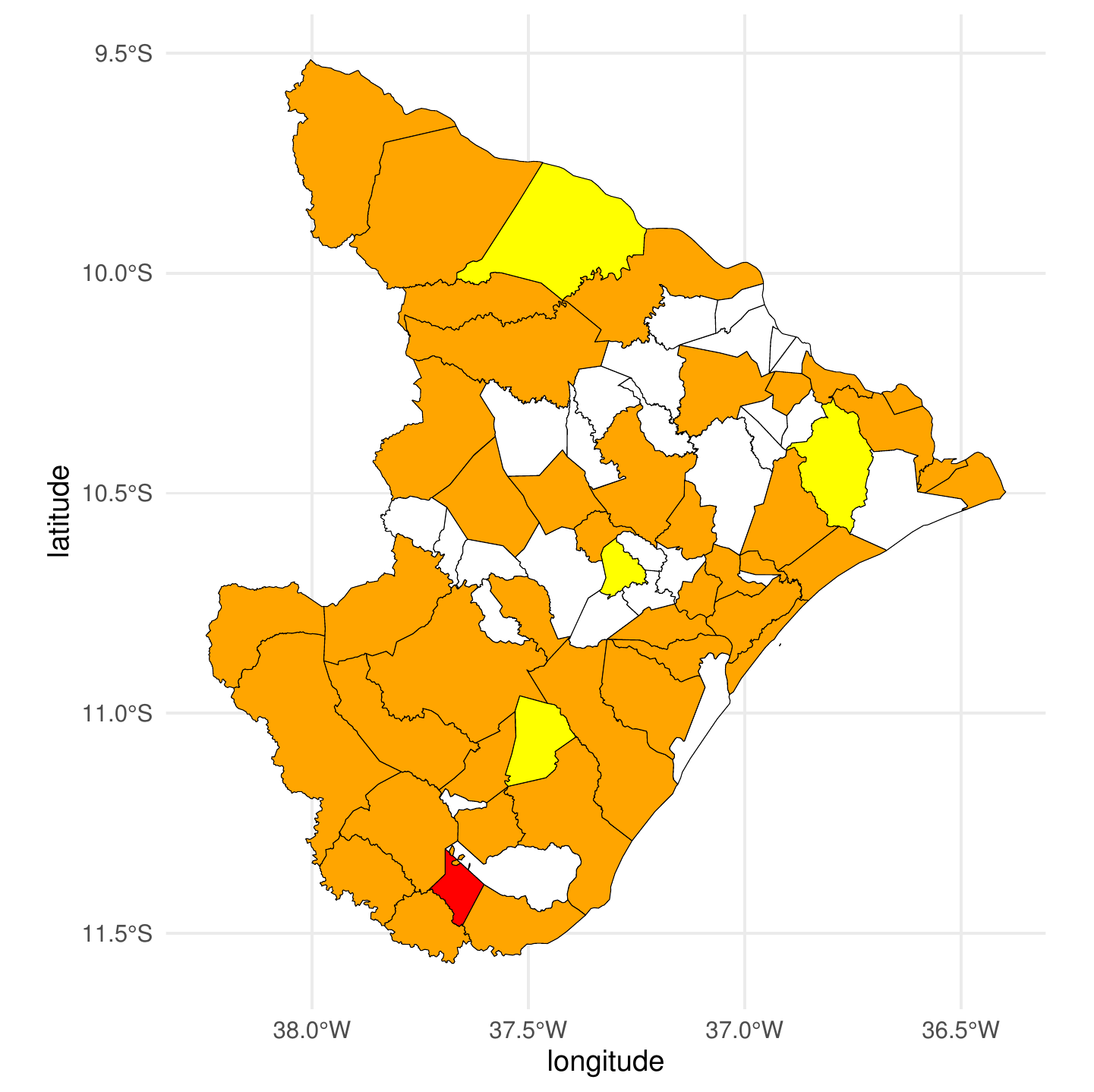}
        }
        \subfigure[]{%
           \label{se2}
           \includegraphics[width=5.3cm,height=6cm]{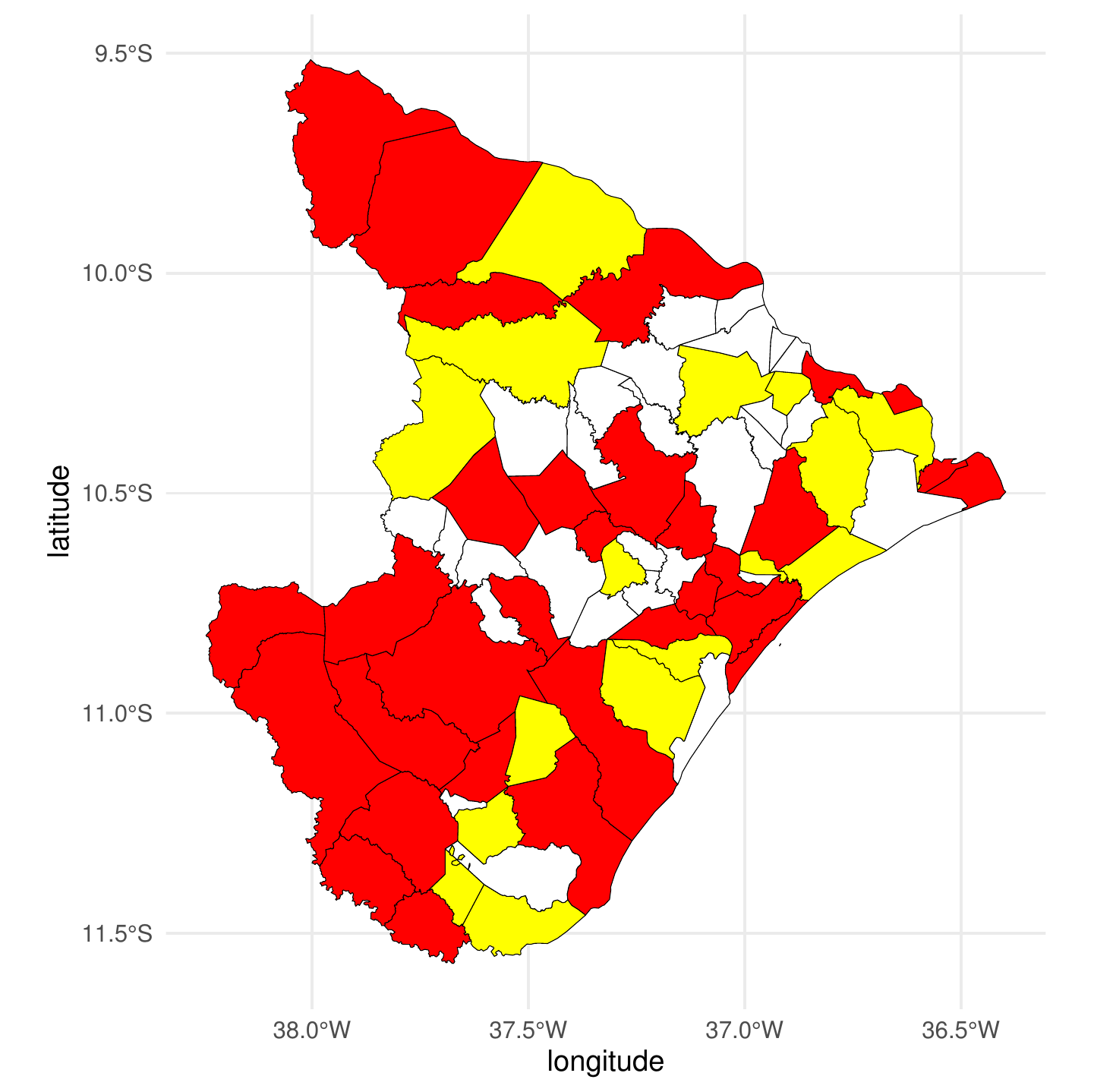}
        }
    \end{center}
    \caption{%
        Spatial location of the functional clustering of the municipalities of the states of Piau\'{i} (first line), Rio Grande do Norte (second line) and Sergipe (third line) according to represented death curves (a, d, g), first derivative of death curves (b, e, h) and second derivative of death curves (c, f, i).}
   \label{muni5}
\end{figure}

\newpage
\subsection{Central-West Region}

\begin{figure}[!htbp]
     \begin{center}
        \subfigure[]{%
            \label{go}
            \includegraphics[width=5.3cm,height=6cm]{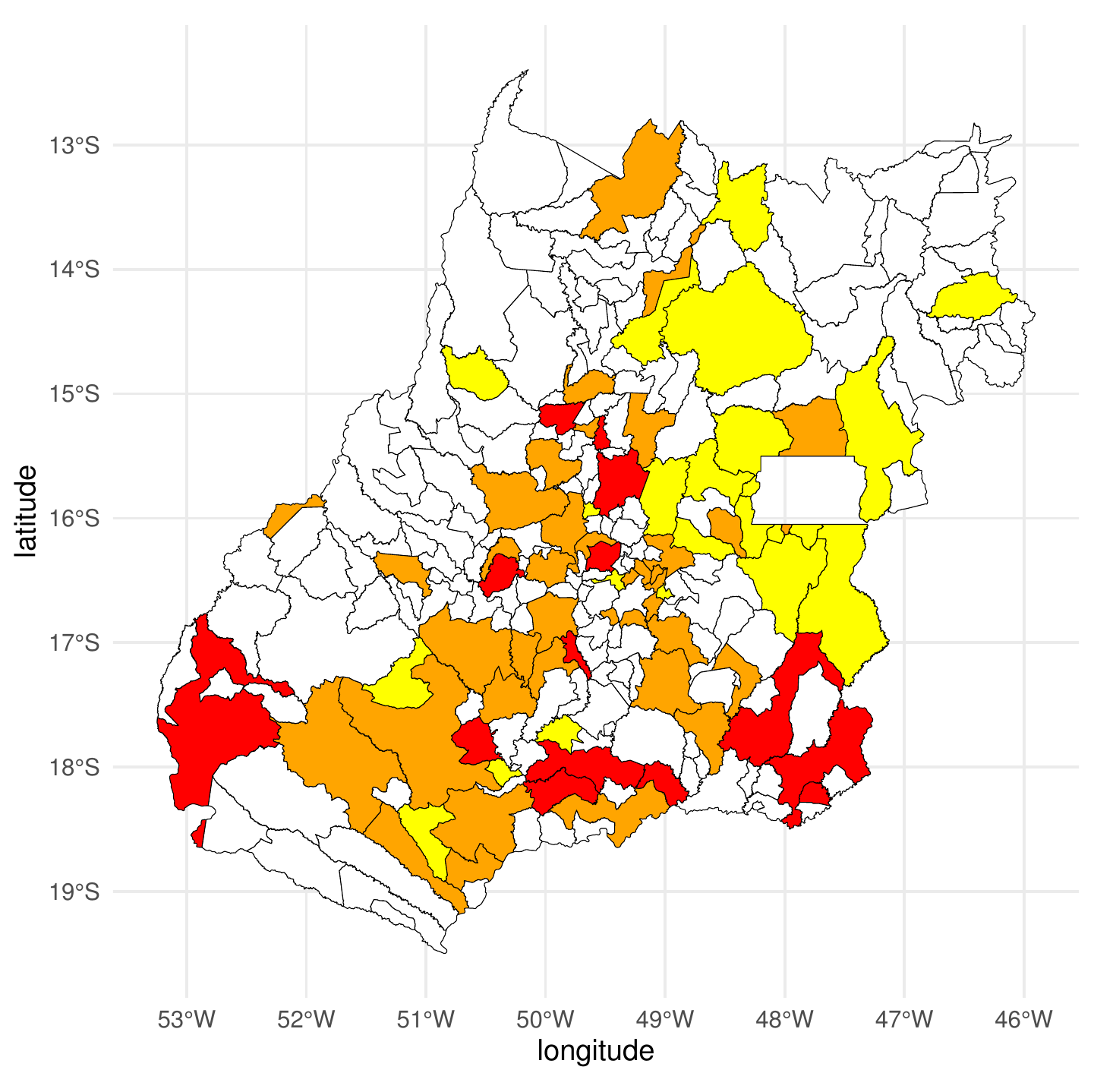}
        }
        \subfigure[]{%
           \label{go1}
           \includegraphics[width=5.3cm,height=6cm]{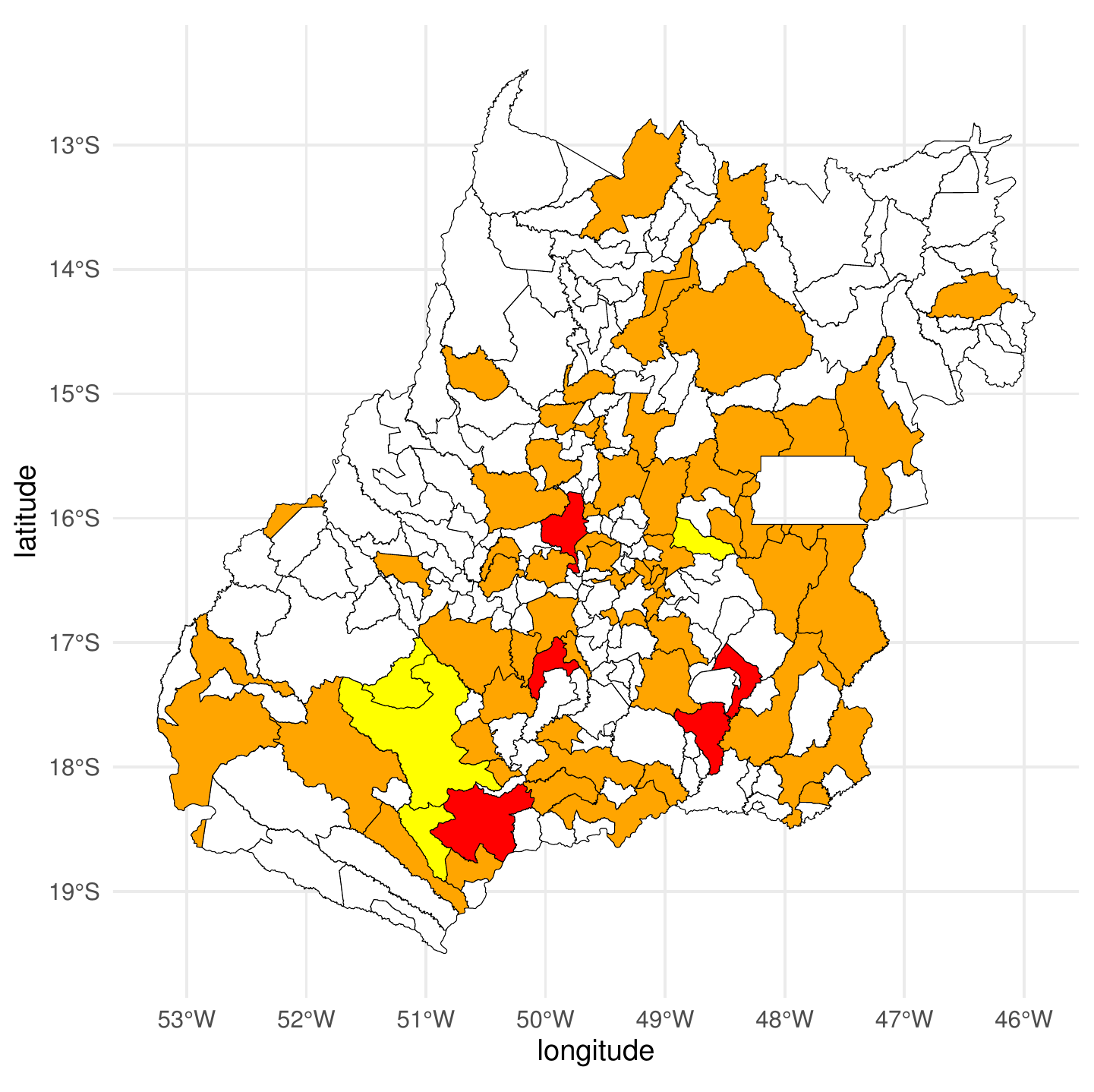}
        }
        \subfigure[]{%
            \label{go2}
           \includegraphics[width=5.3cm,height=6cm]{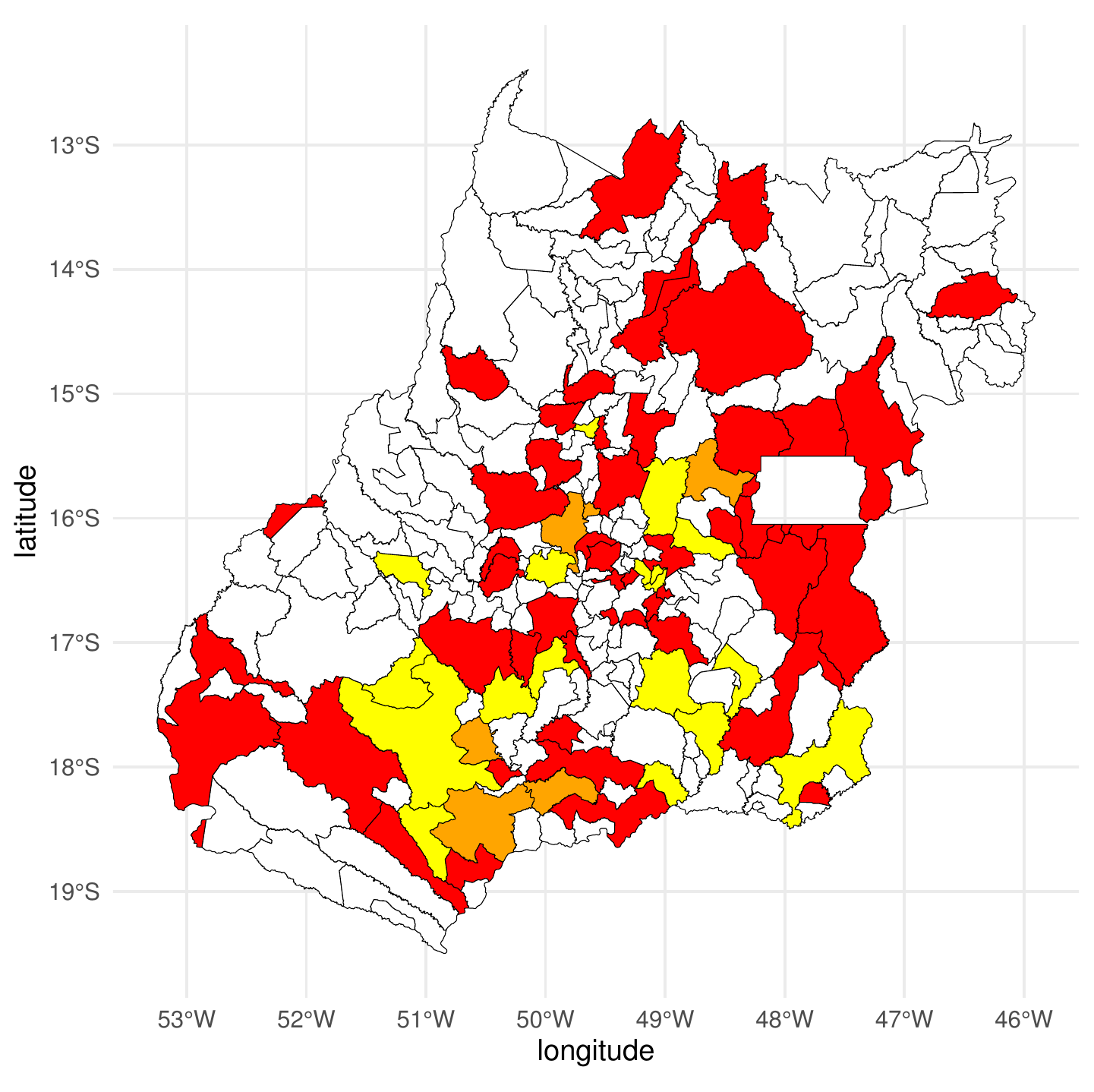}
        }\\
        \subfigure[]{%
            \label{mt}
            \includegraphics[width=5.3cm,height=6cm]{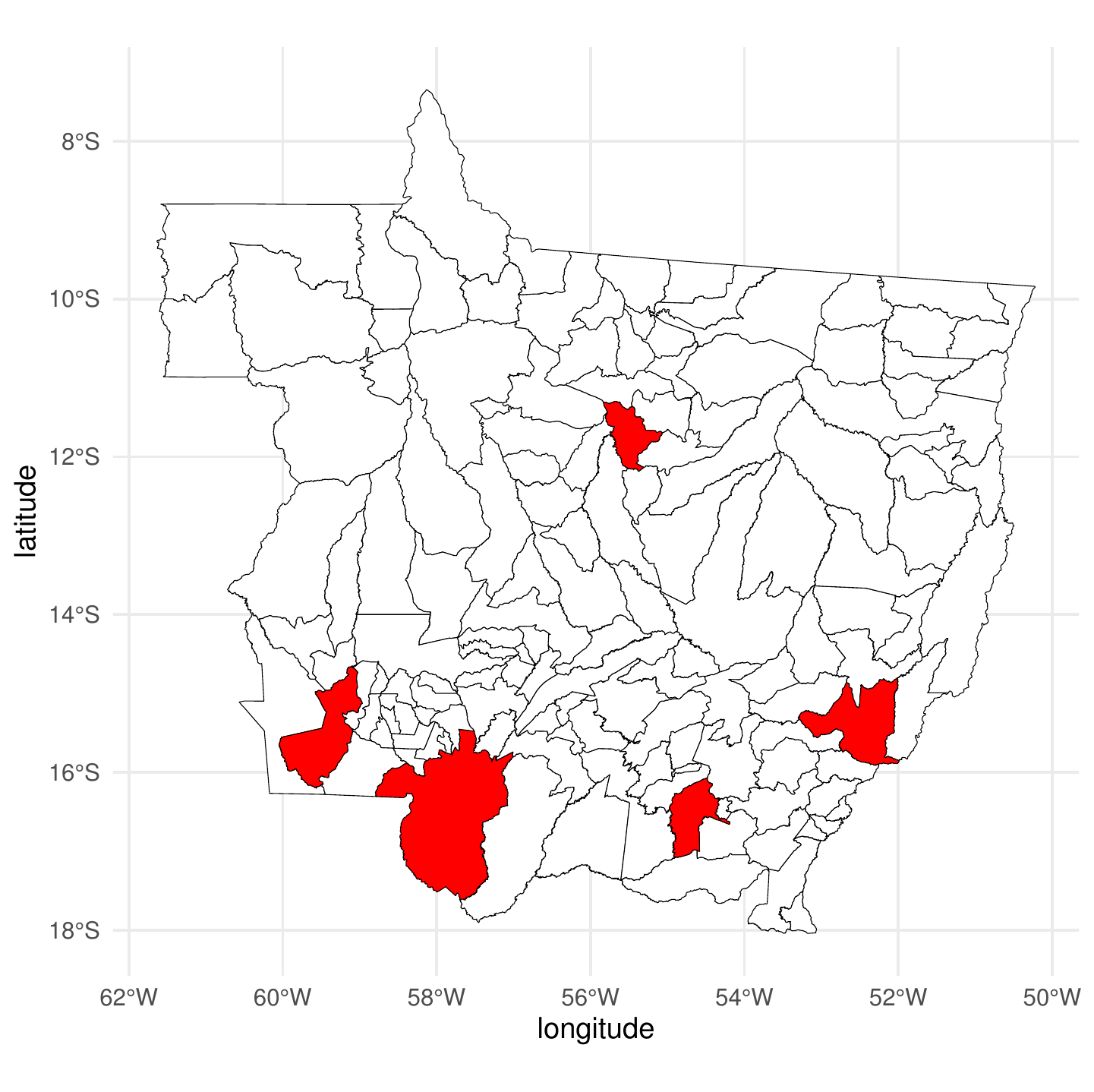}
        }
        \subfigure[]{%
           \label{mt1}
           \includegraphics[width=5.3cm,height=6cm]{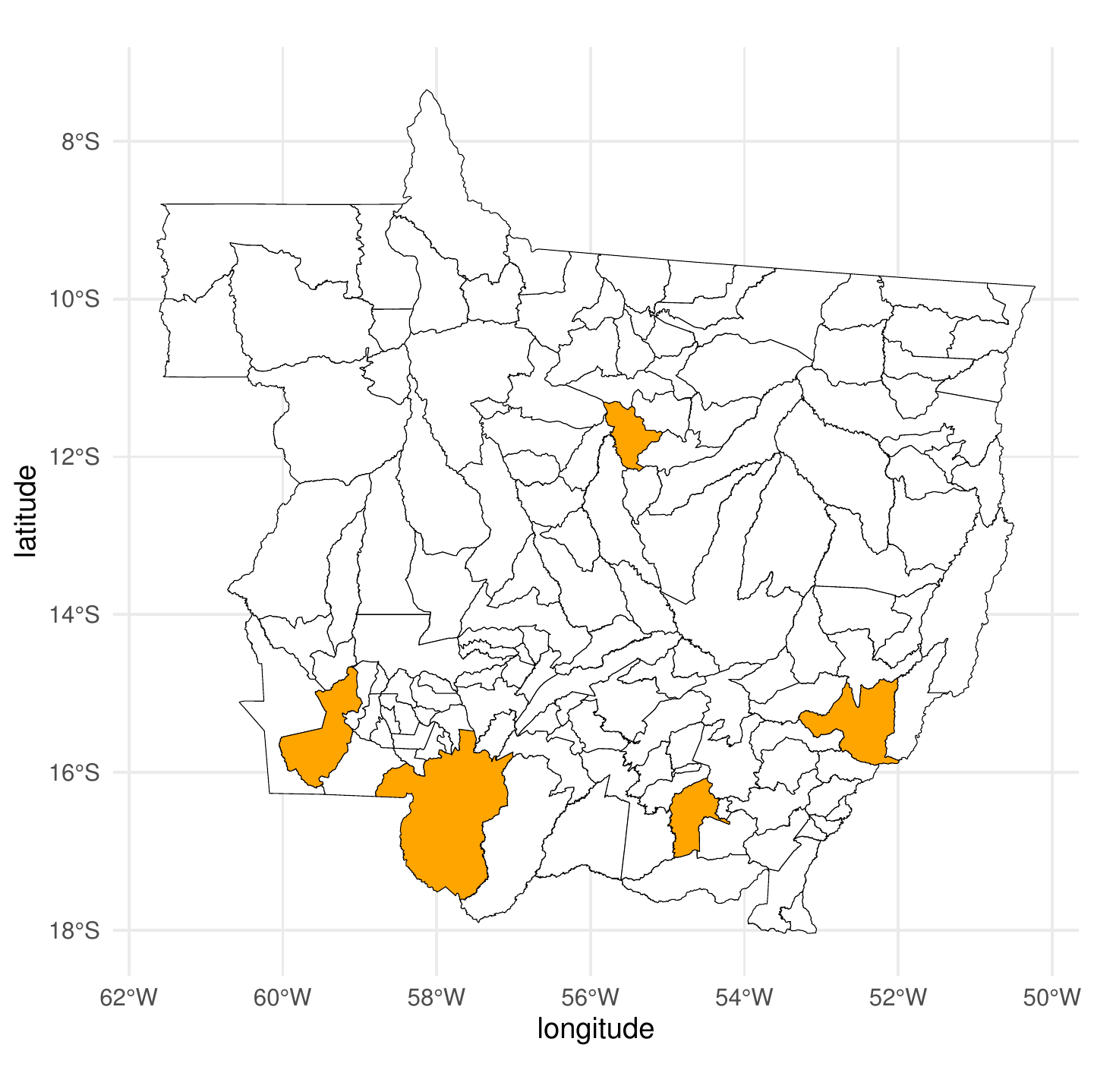}
        }
        \subfigure[]{%
           \label{mt2}
           \includegraphics[width=5.3cm,height=6cm]{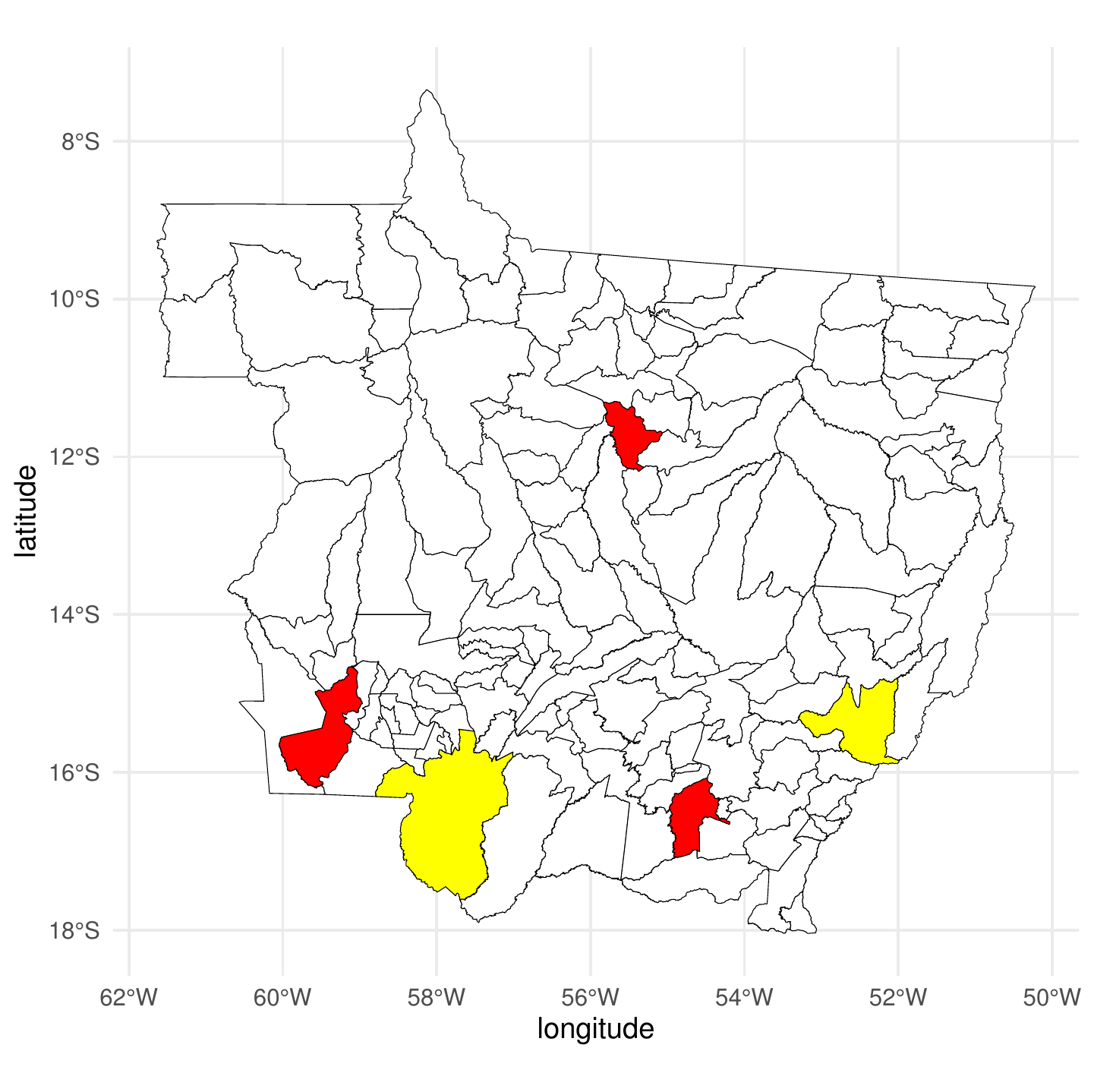}
        }\\
        \subfigure[]{%
            \label{ms}
            \includegraphics[width=5.3cm,height=6cm]{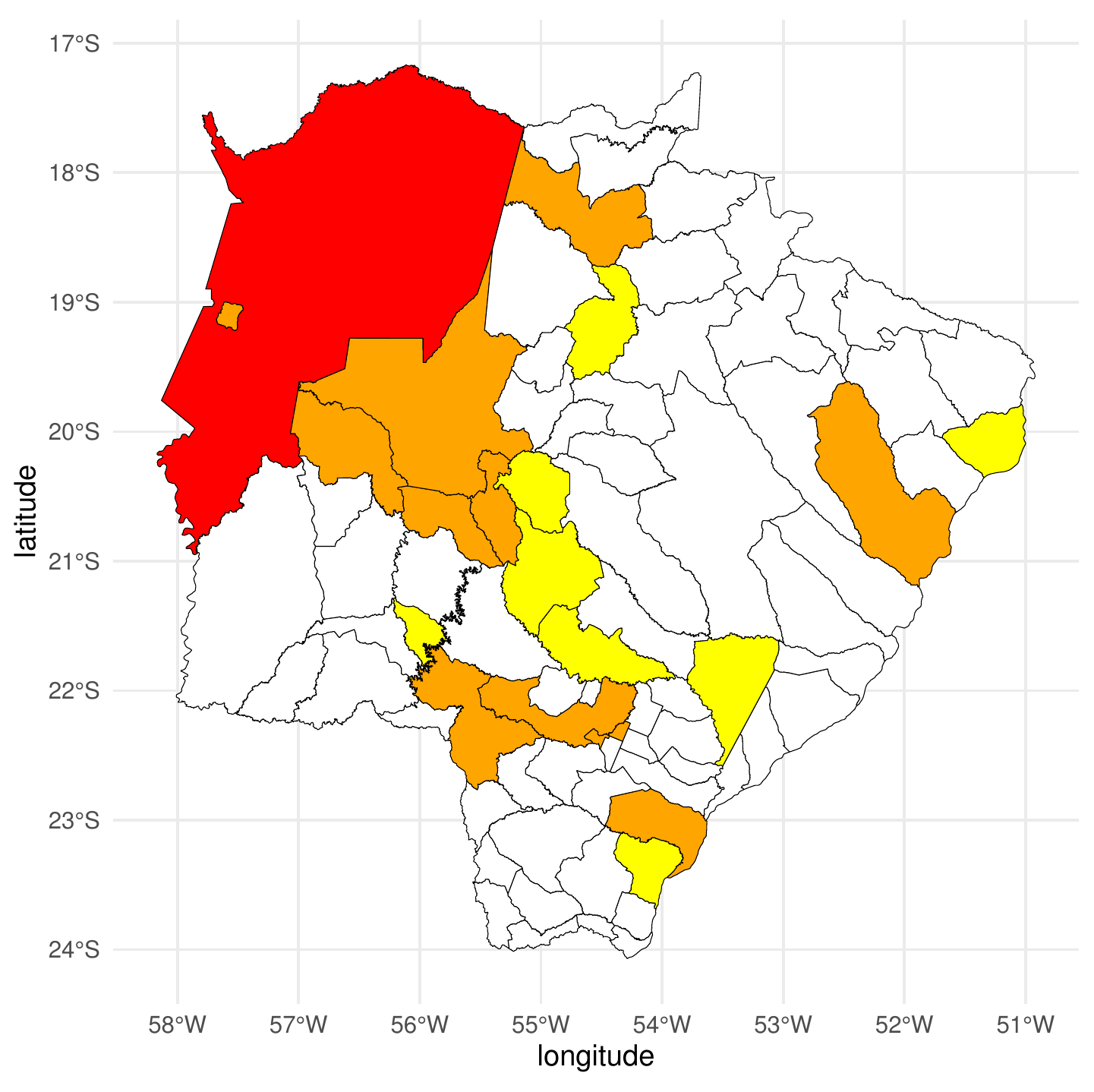}
        }
        \subfigure[]{%
           \label{ms1}
           \includegraphics[width=5.3cm,height=6cm]{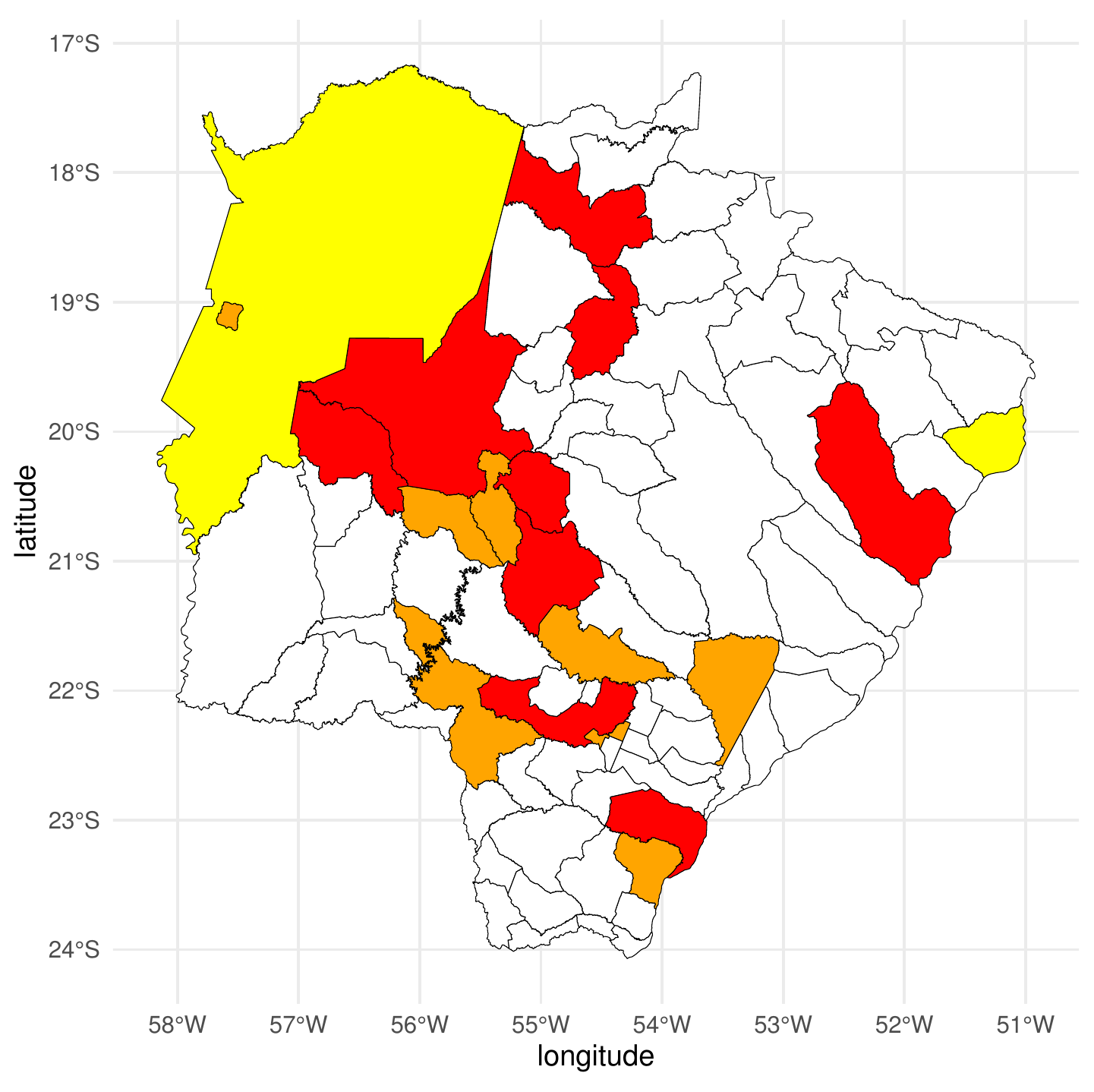}
        }
        \subfigure[]{%
           \label{ms2}
           \includegraphics[width=5.3cm,height=6cm]{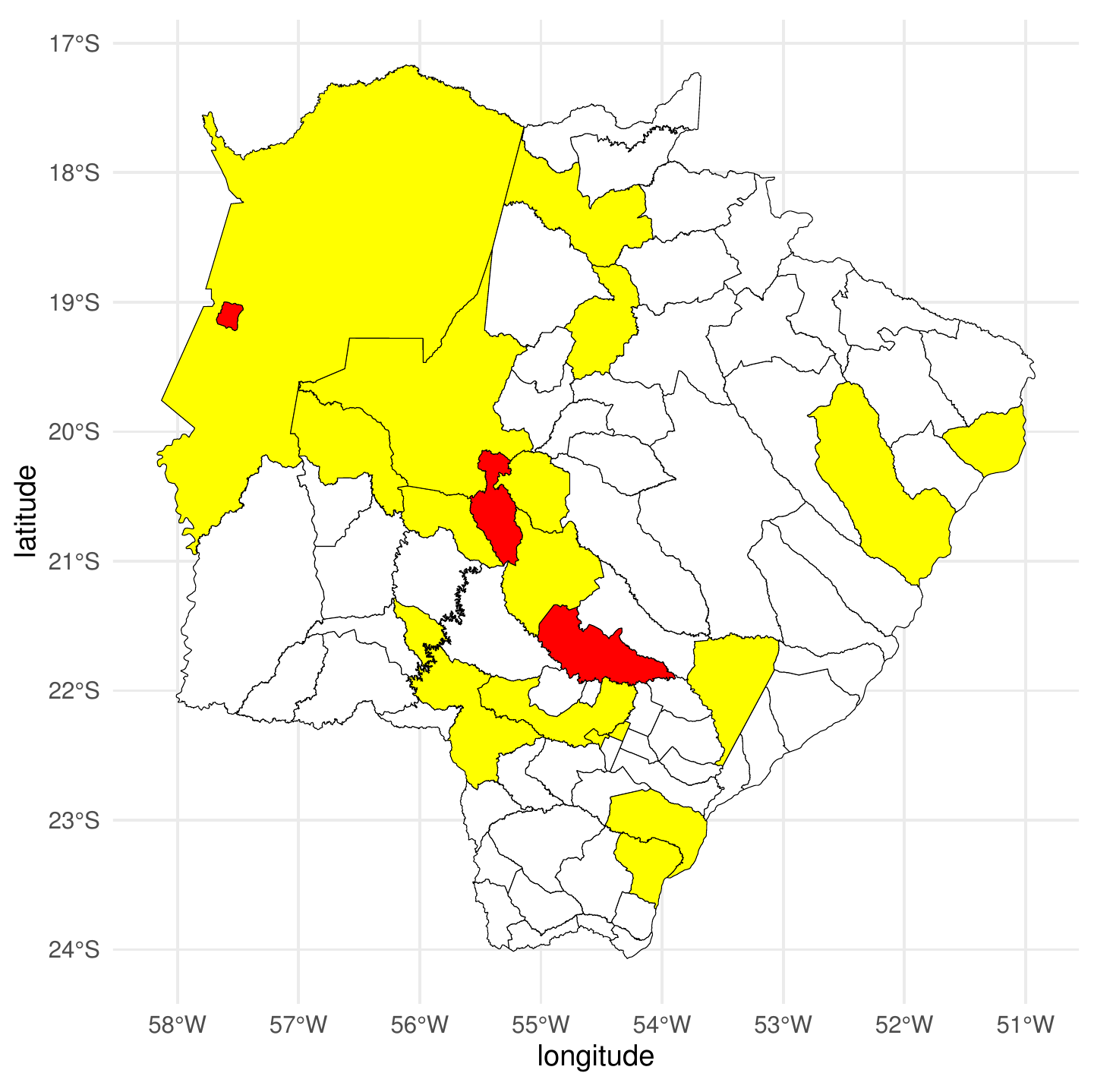}
        }
    \end{center}
    \caption{%
        Spatial location of the functional clustering of the municipalities of the states of Goi\'{a}s (first line), Mato Grosso (second line) and Mato Grosso do Sul (third line) according to represented death curves (a, d, g), first derivative of death curves (b, e, h) and second derivative of death curves (c, f, i).}
   \label{muni6}
\end{figure}

\newpage
\subsection{Southeast Region}

\begin{figure}[!htbp]
     \begin{center}
        \subfigure[]{%
            \label{rj}
            \includegraphics[width=5.3cm,height=6cm]{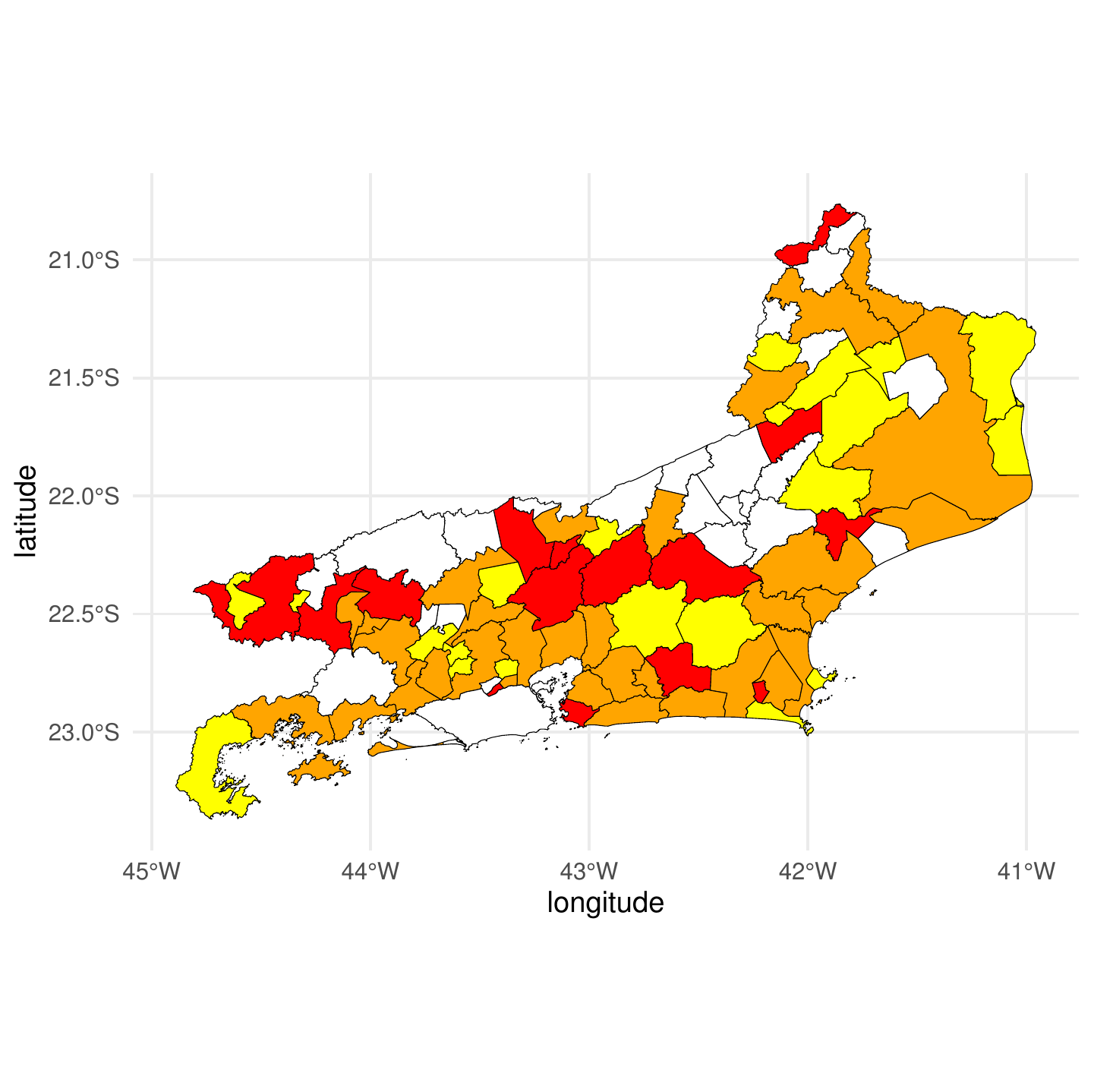}
        }
        \subfigure[]{%
           \label{rj1}
           \includegraphics[width=5.3cm,height=6cm]{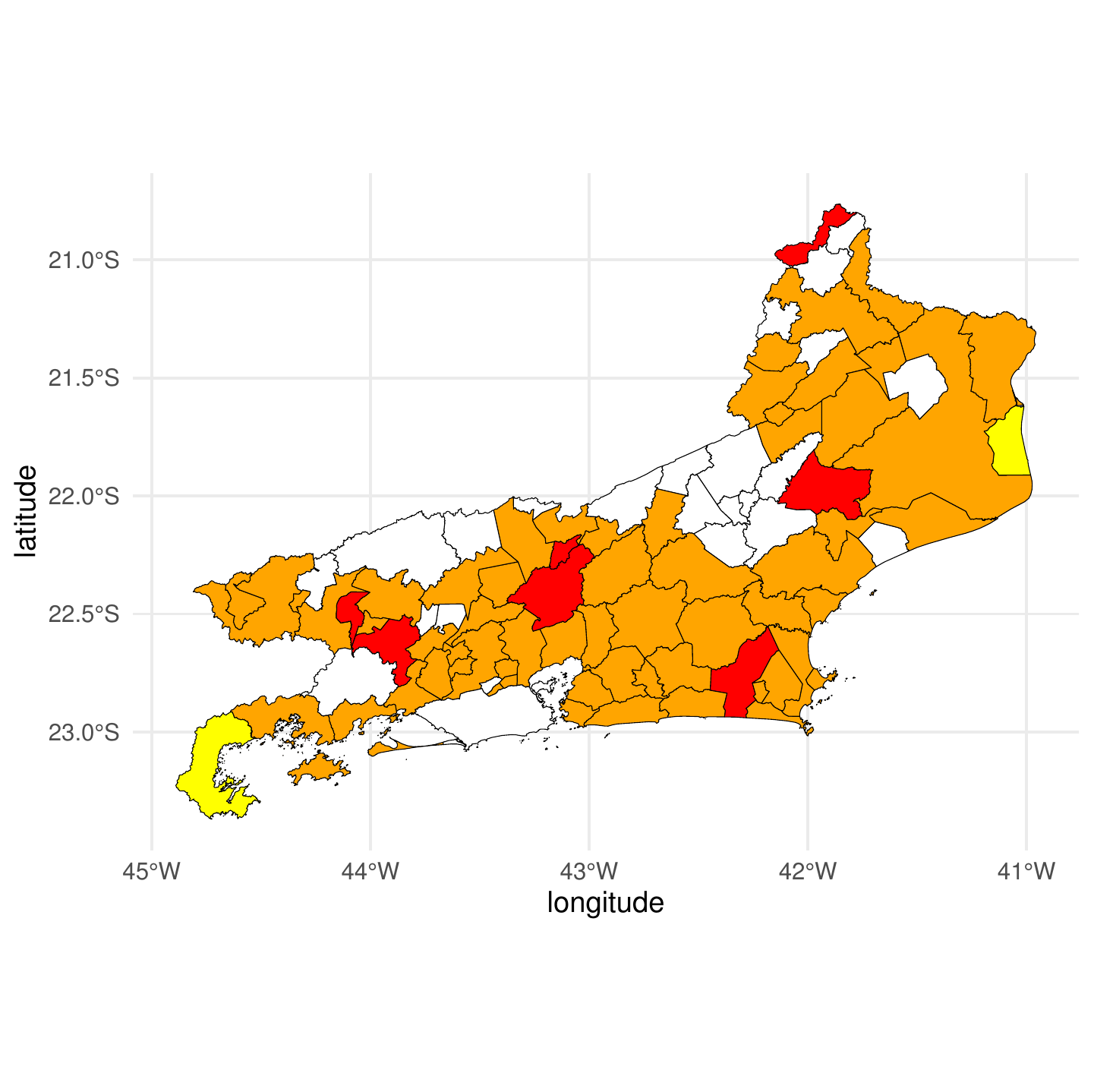}
        }
        \subfigure[]{%
           \label{rj2}
           \includegraphics[width=5.3cm,height=6cm]{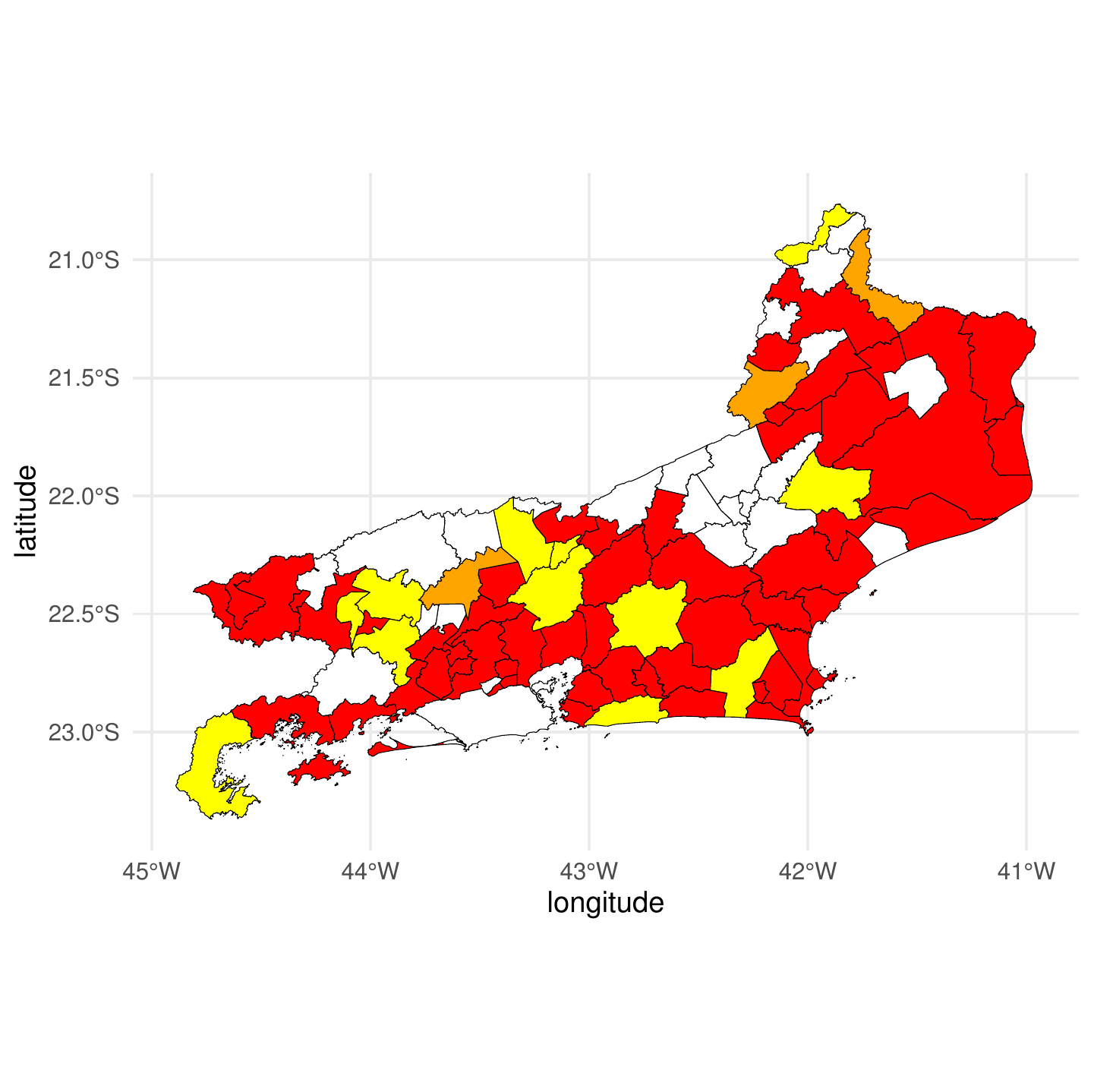}
        }\\
        \subfigure[]{%
            \label{sp}
            \includegraphics[width=5.3cm,height=6cm]{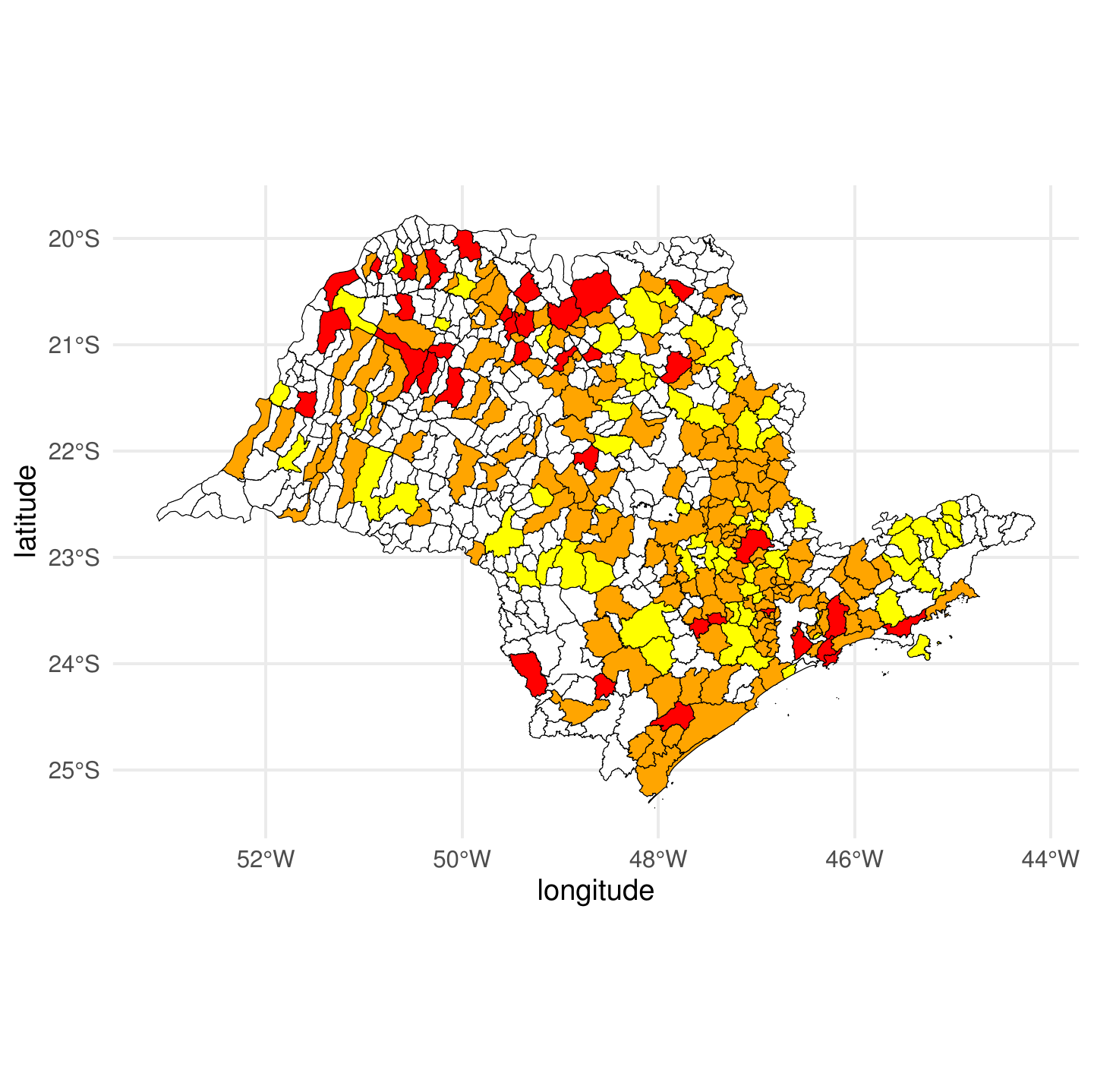}
        }
        \subfigure[]{%
           \label{sp1}
           \includegraphics[width=5.3cm,height=6cm]{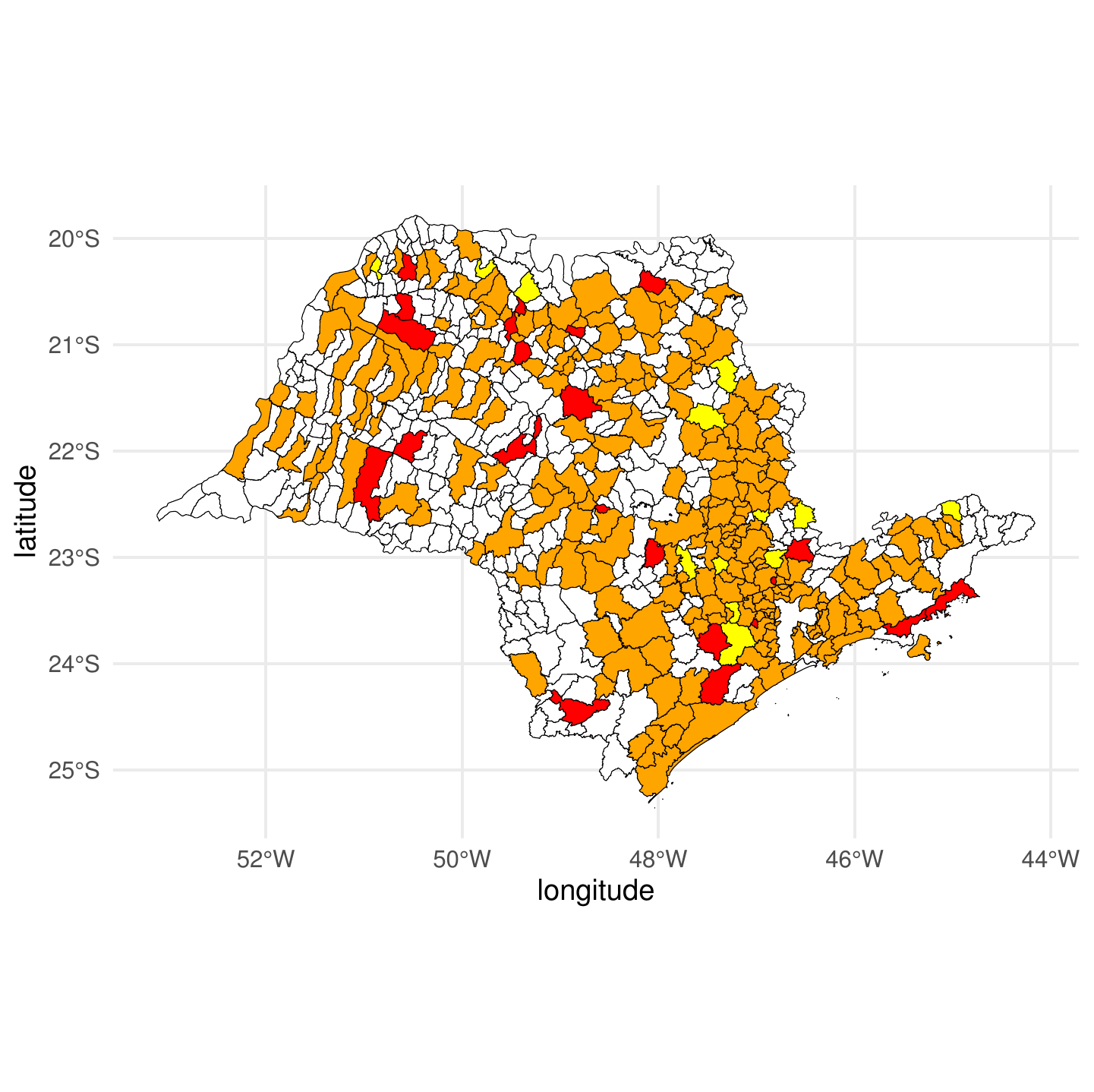}
        }
        \subfigure[]{%
           \label{sp2}
           \includegraphics[width=5.3cm,height=6cm]{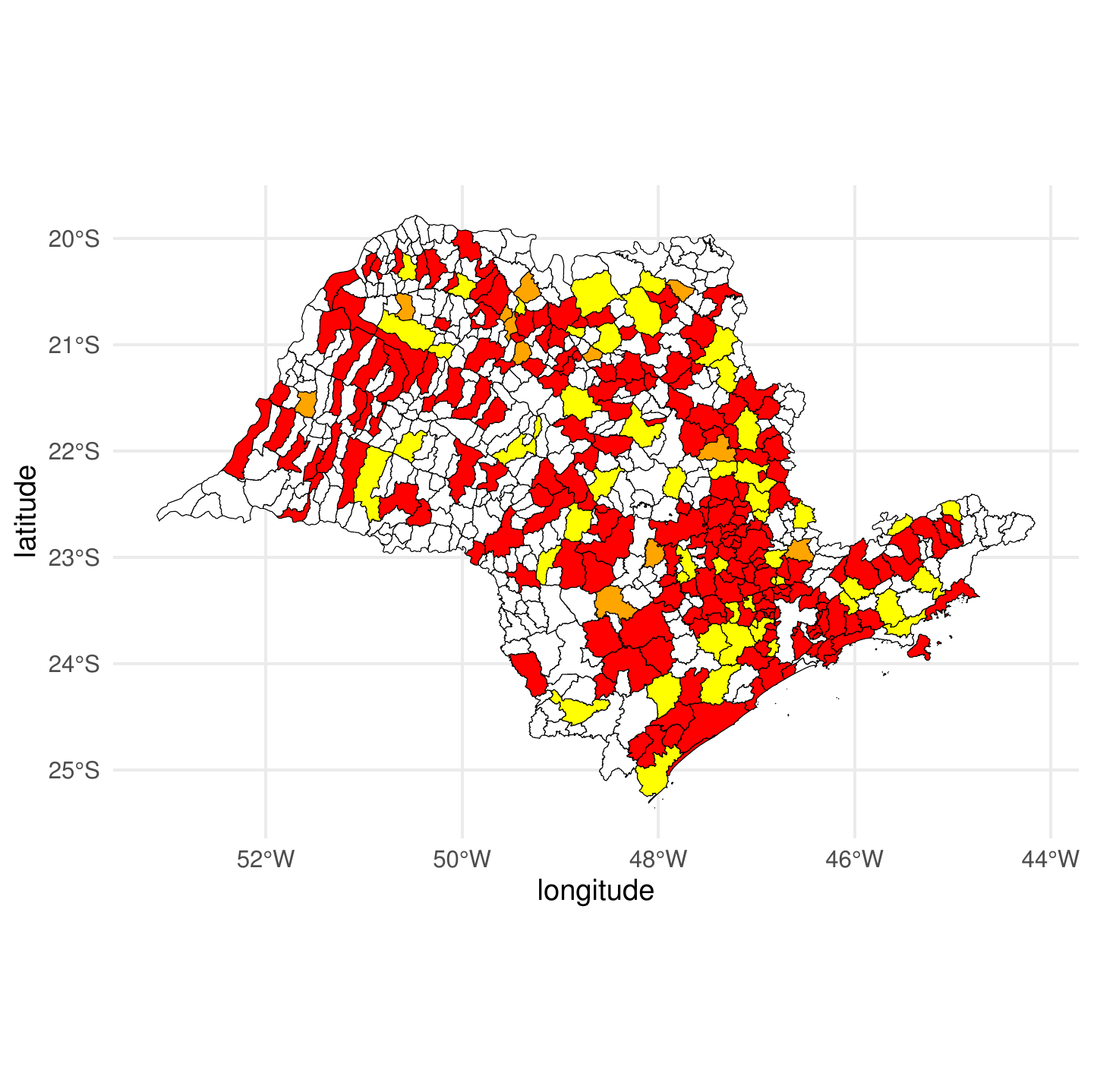}
        }
    \end{center}
    \caption{%
        Spatial location of the functional clustering of the municipalities of the states of Rio de Janeiro (first line) and S\~{a}o Paulo (second line) according to represented death curves (a,d), first derivative of death curves (b,e) and second derivative of death curves (c,f).}
   \label{muni7}
\end{figure}

\begin{figure}[!htbp]
     \begin{center}
        \subfigure[]{%
            \label{mg}
            \includegraphics[width=5.3cm,height=4.5cm]{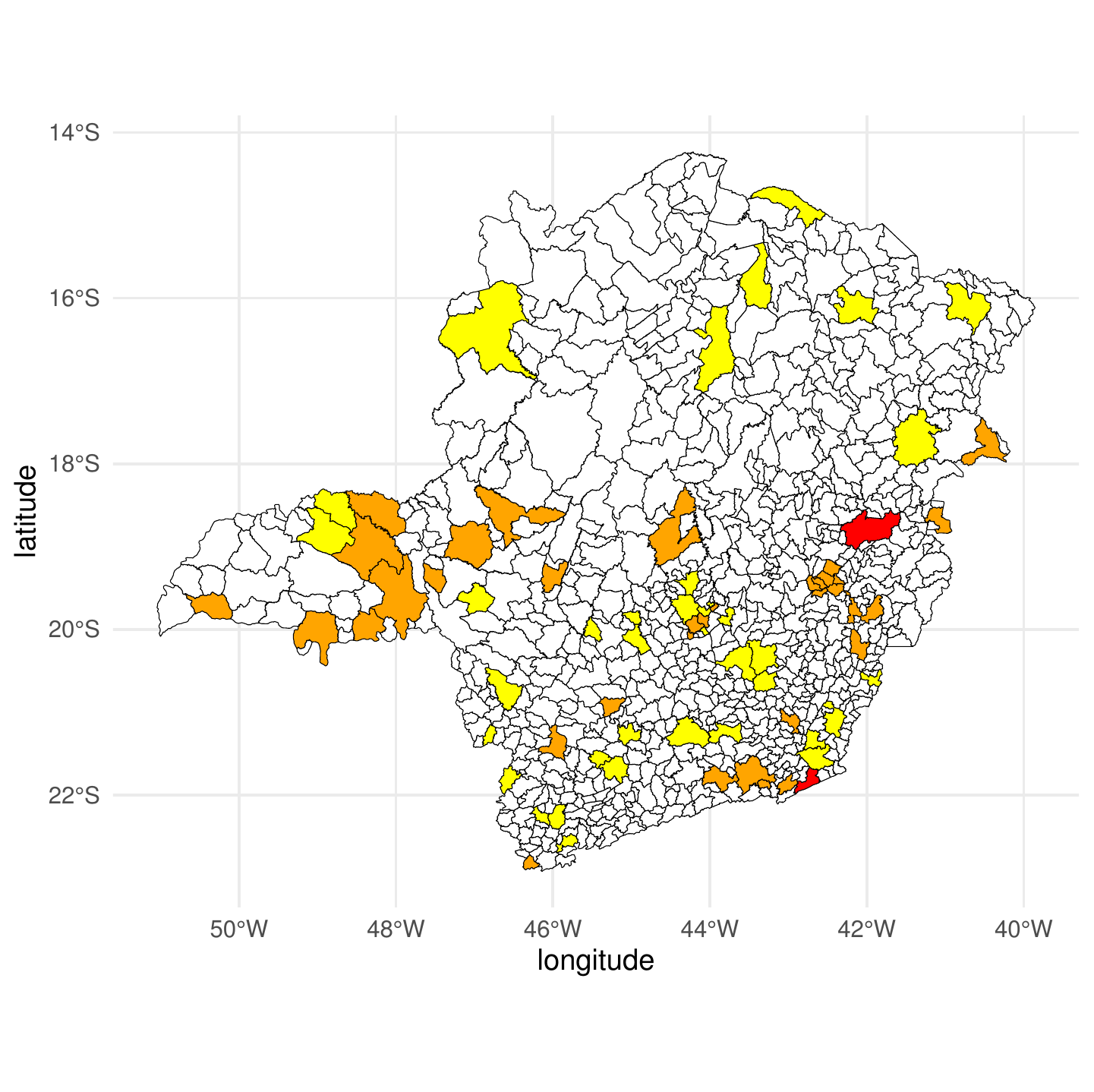}
        }
        \subfigure[]{%
           \label{mg1}
           \includegraphics[width=5.3cm,height=4.5cm]{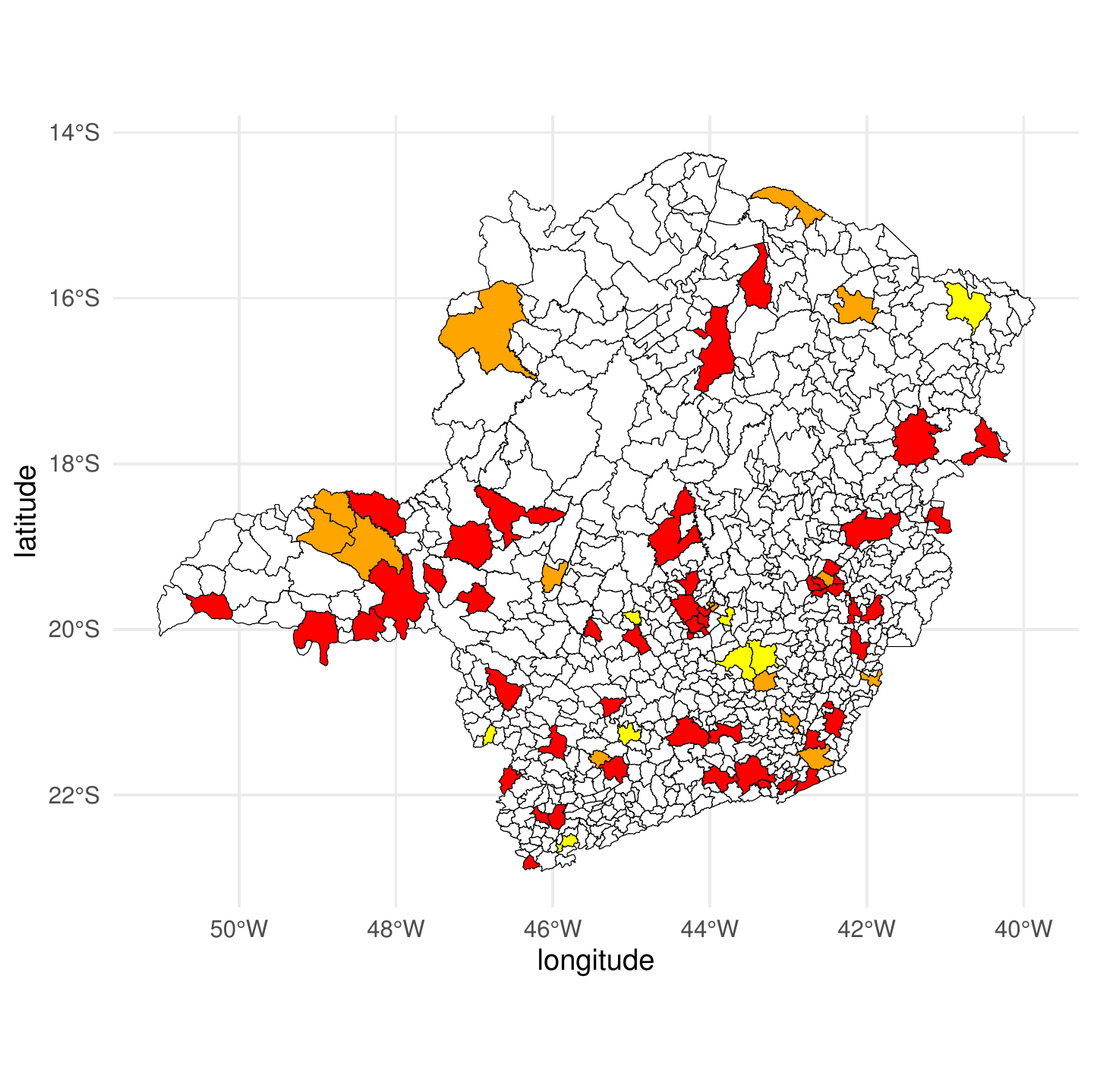}
        }
        \subfigure[]{%
           \label{mg2}
           \includegraphics[width=5.3cm,height=4.5cm]{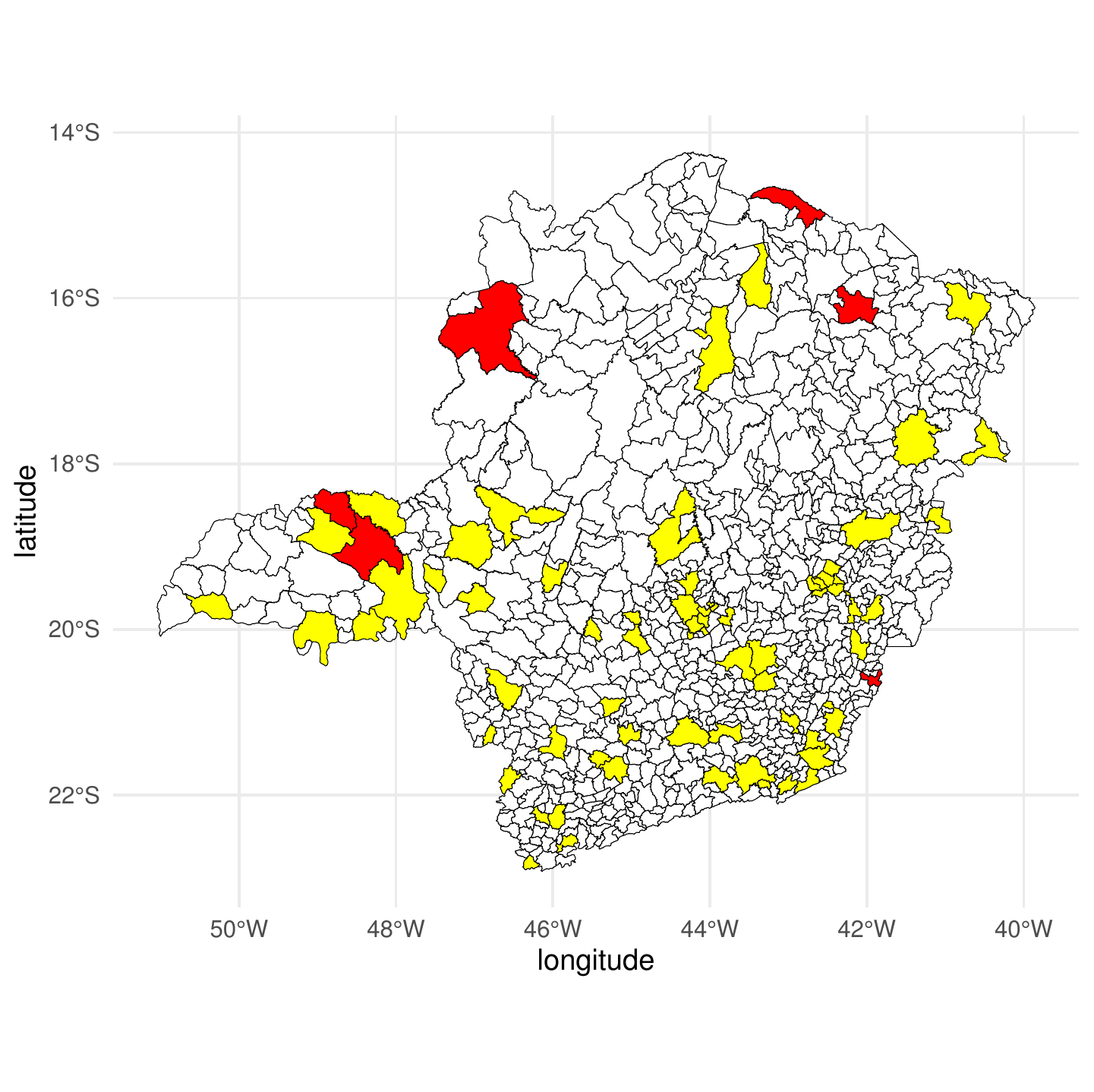}
        }\\
        \subfigure[]{%
            \label{es}
            \includegraphics[width=5.3cm,height=4.5cm]{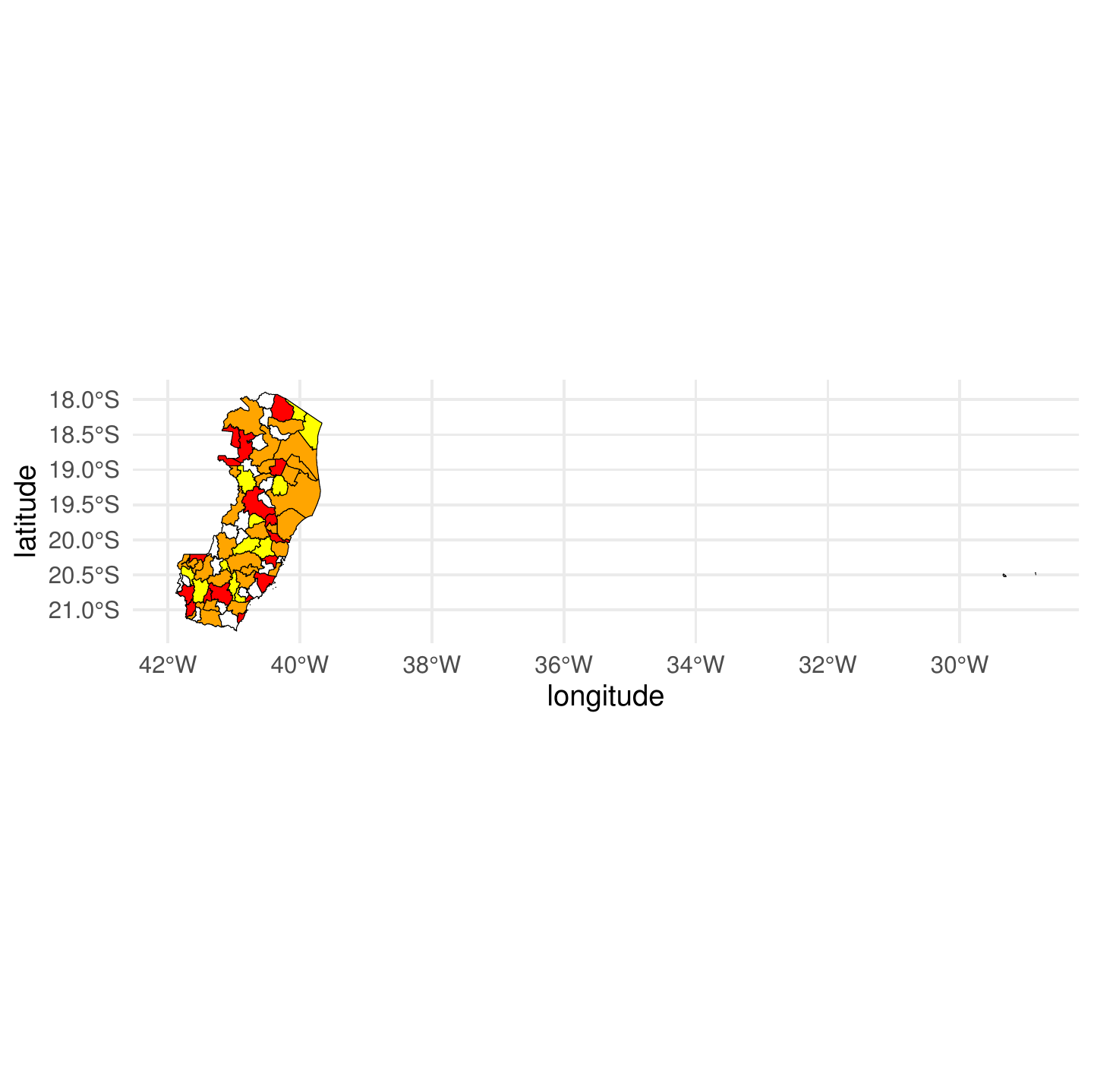}
        }
        \subfigure[]{%
           \label{es1}
           \includegraphics[width=5.3cm,height=4.5cm]{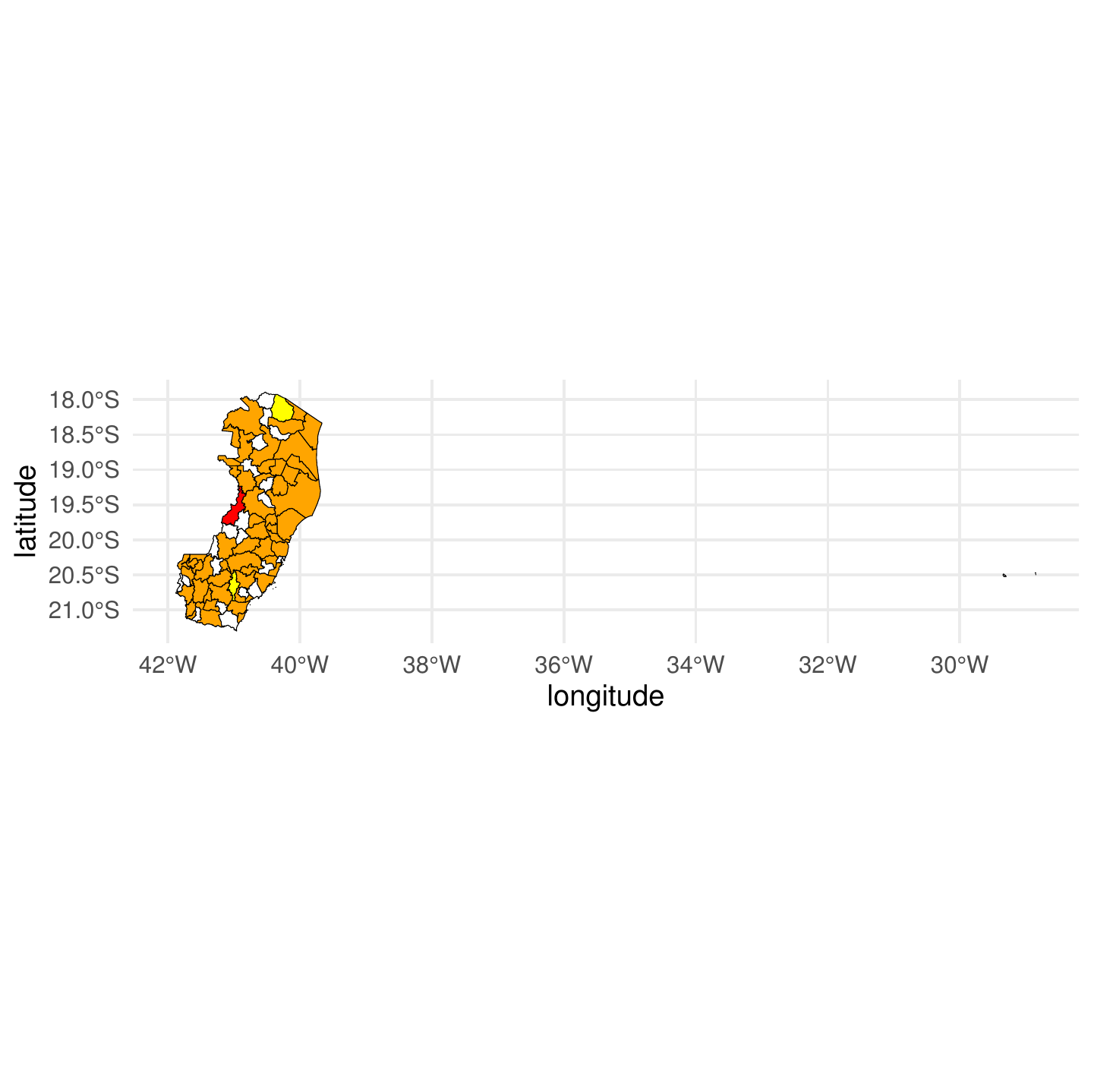}
        }
        \subfigure[]{%
           \label{es2}
           \includegraphics[width=5.3cm,height=4.5cm]{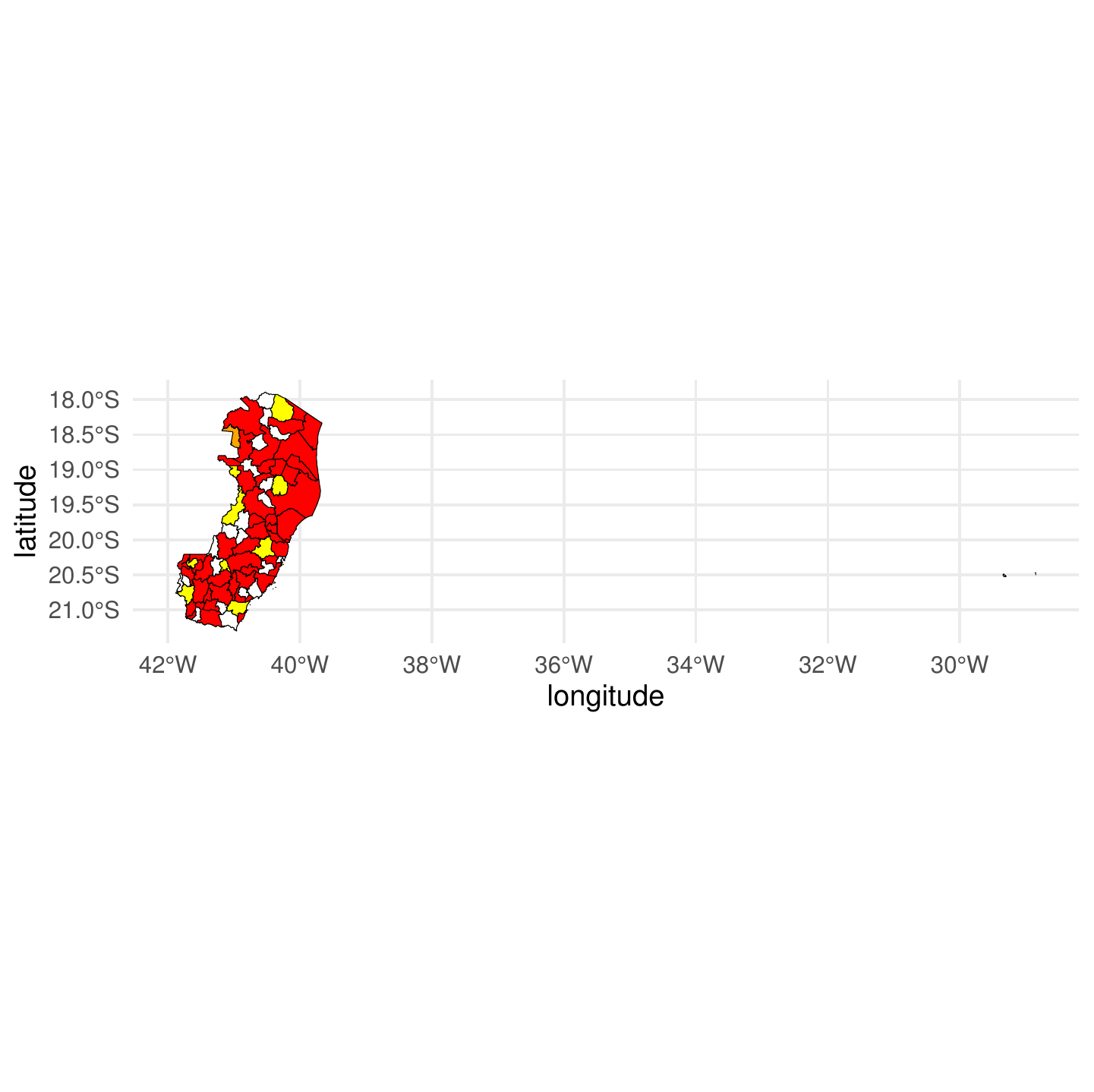}
        }
    \end{center}
    \caption{%
        Spatial location of the functional clustering of the municipalities of the states of Minas Gerais (first line) and Esp\'{i}rito Santo (second line) according to represented death curves (a,d), first derivative of death curves (b,e) and second derivative of death curves (c,f).}
   \label{muni8}
\end{figure}

\newpage
\subsection{South Region}

\begin{figure}[!htbp]
     \begin{center}
        \subfigure[]{%
            \label{pr}
            \includegraphics[width=5.3cm,height=6cm]{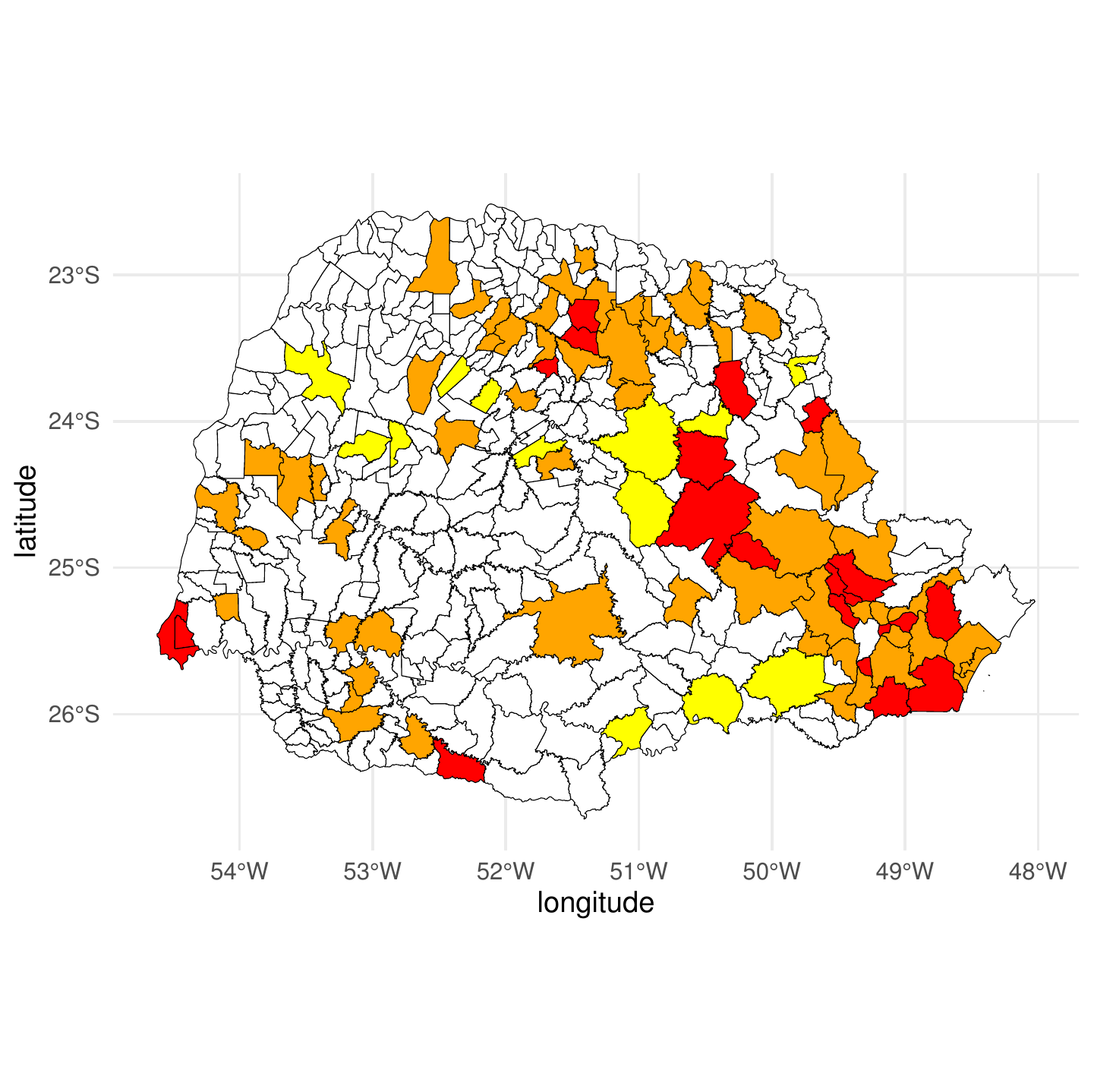}
        }
        \subfigure[]{%
           \label{pr1}
           \includegraphics[width=5.3cm,height=6cm]{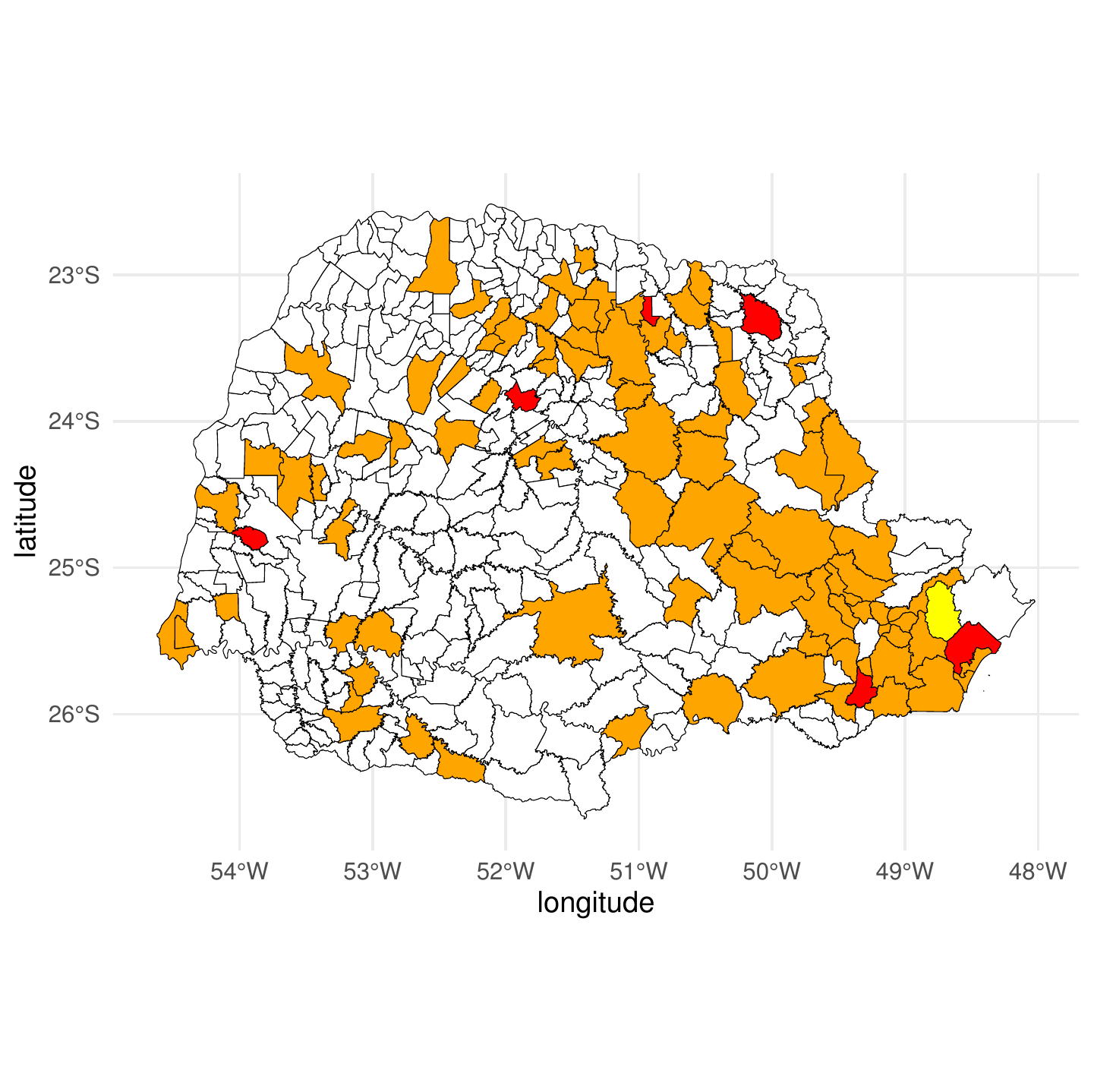}
        }
        \subfigure[]{%
           \label{pr2}
           \includegraphics[width=5.3cm,height=6cm]{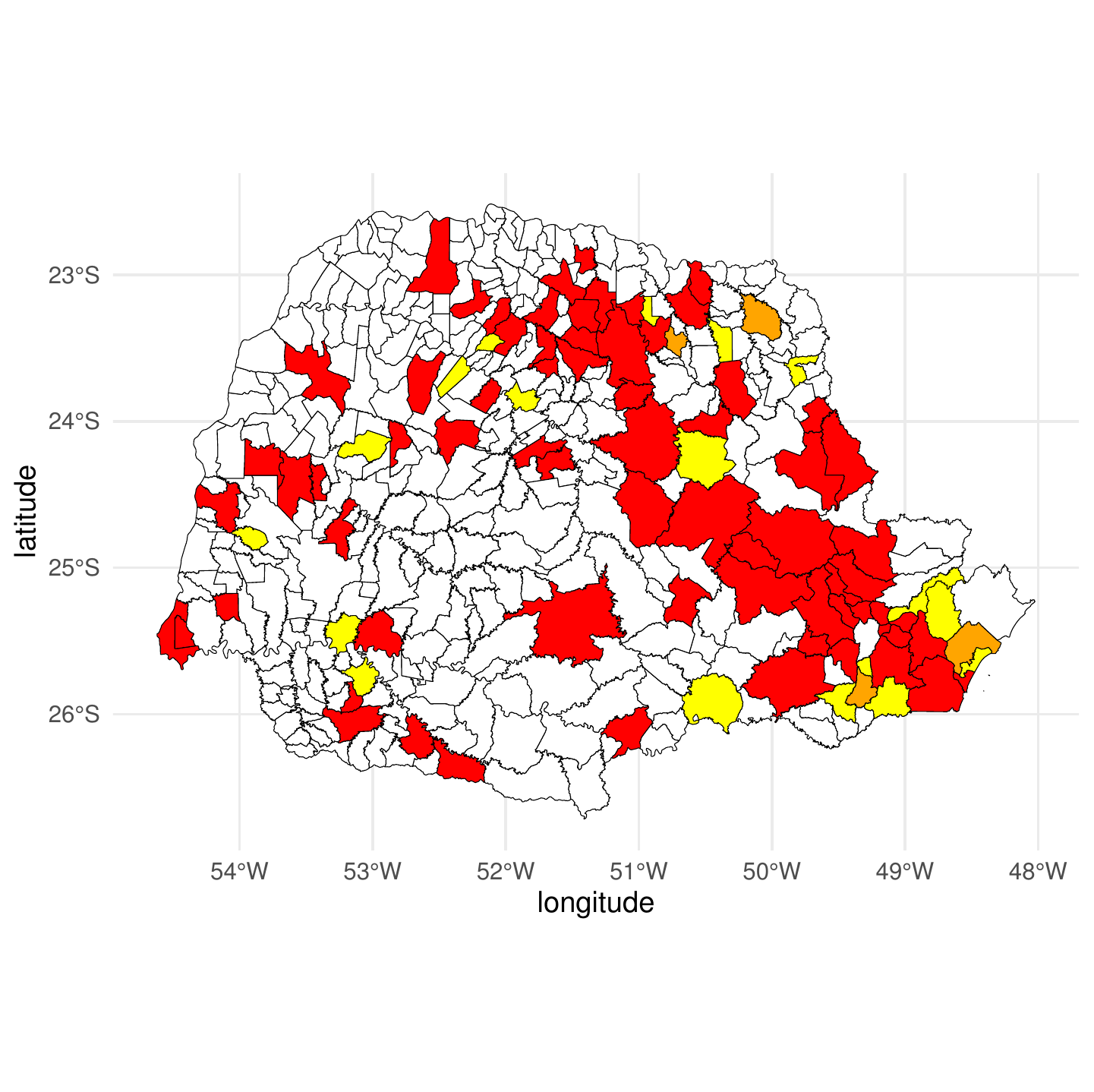}
        }\\
        \subfigure[]{%
            \label{sc}
            \includegraphics[width=5.3cm,height=6cm]{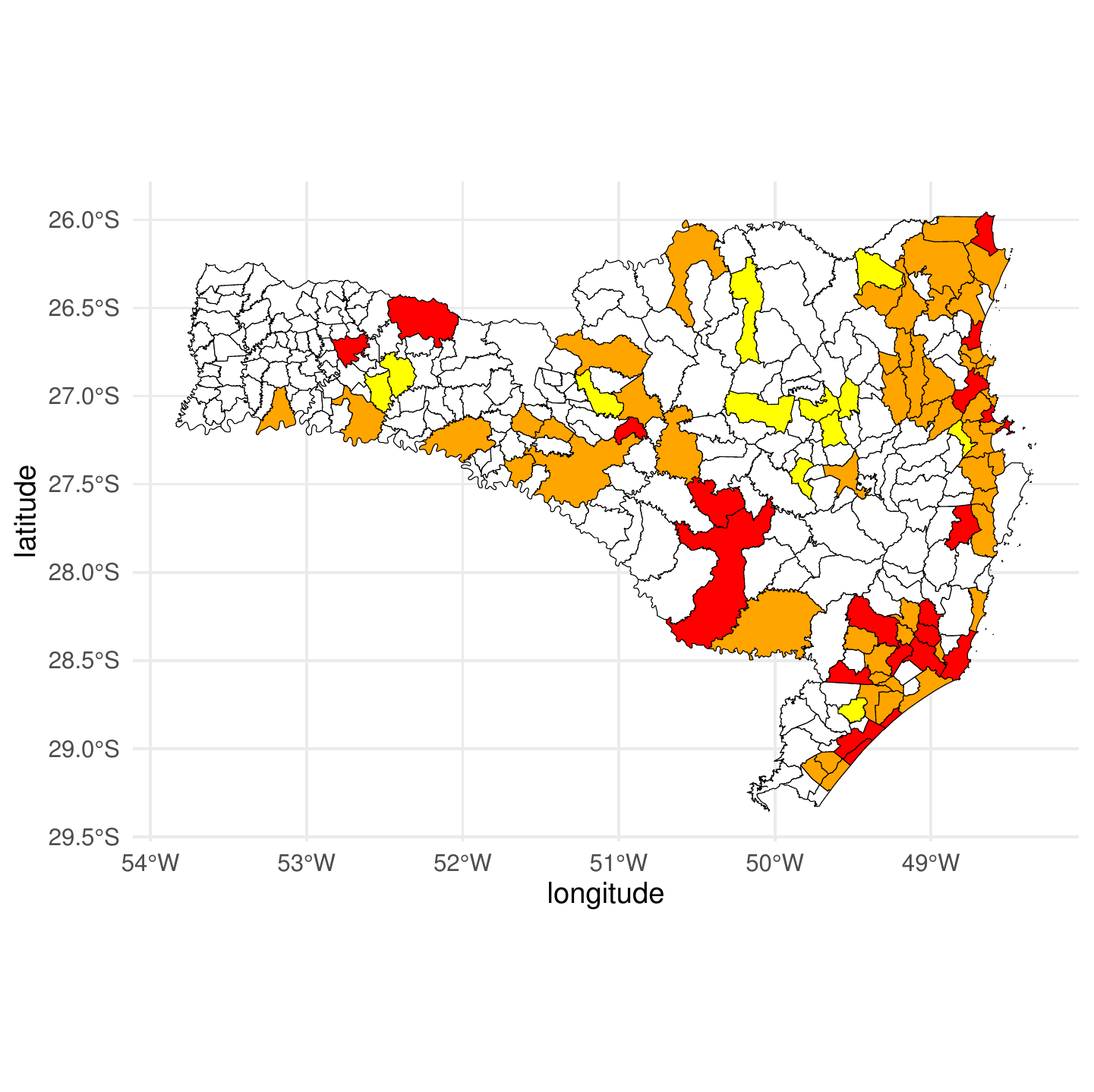}
        }
        \subfigure[]{%
           \label{sc1}
           \includegraphics[width=5.3cm,height=6cm]{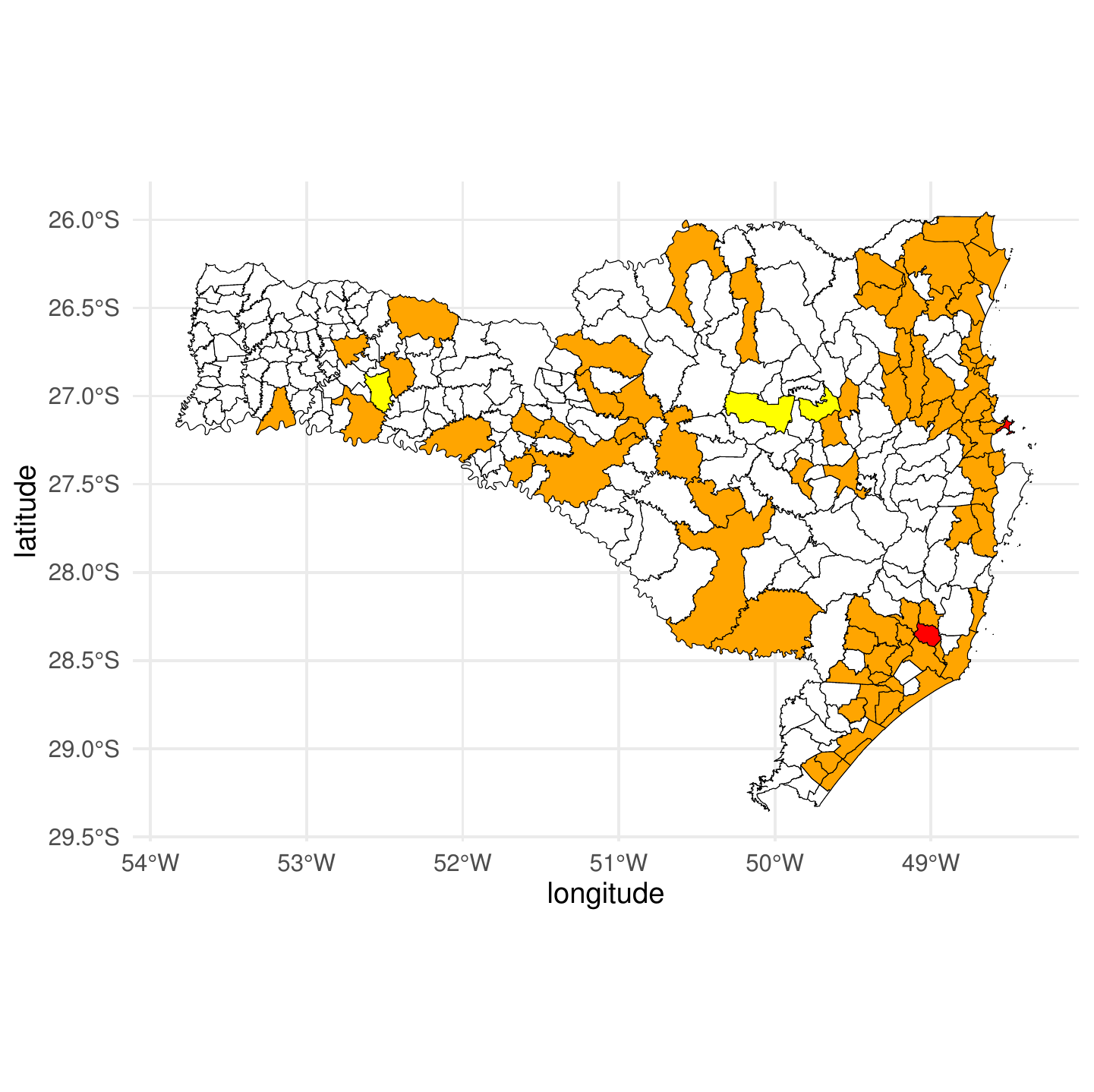}
        }
        \subfigure[]{%
           \label{sc2}
           \includegraphics[width=5.3cm,height=6cm]{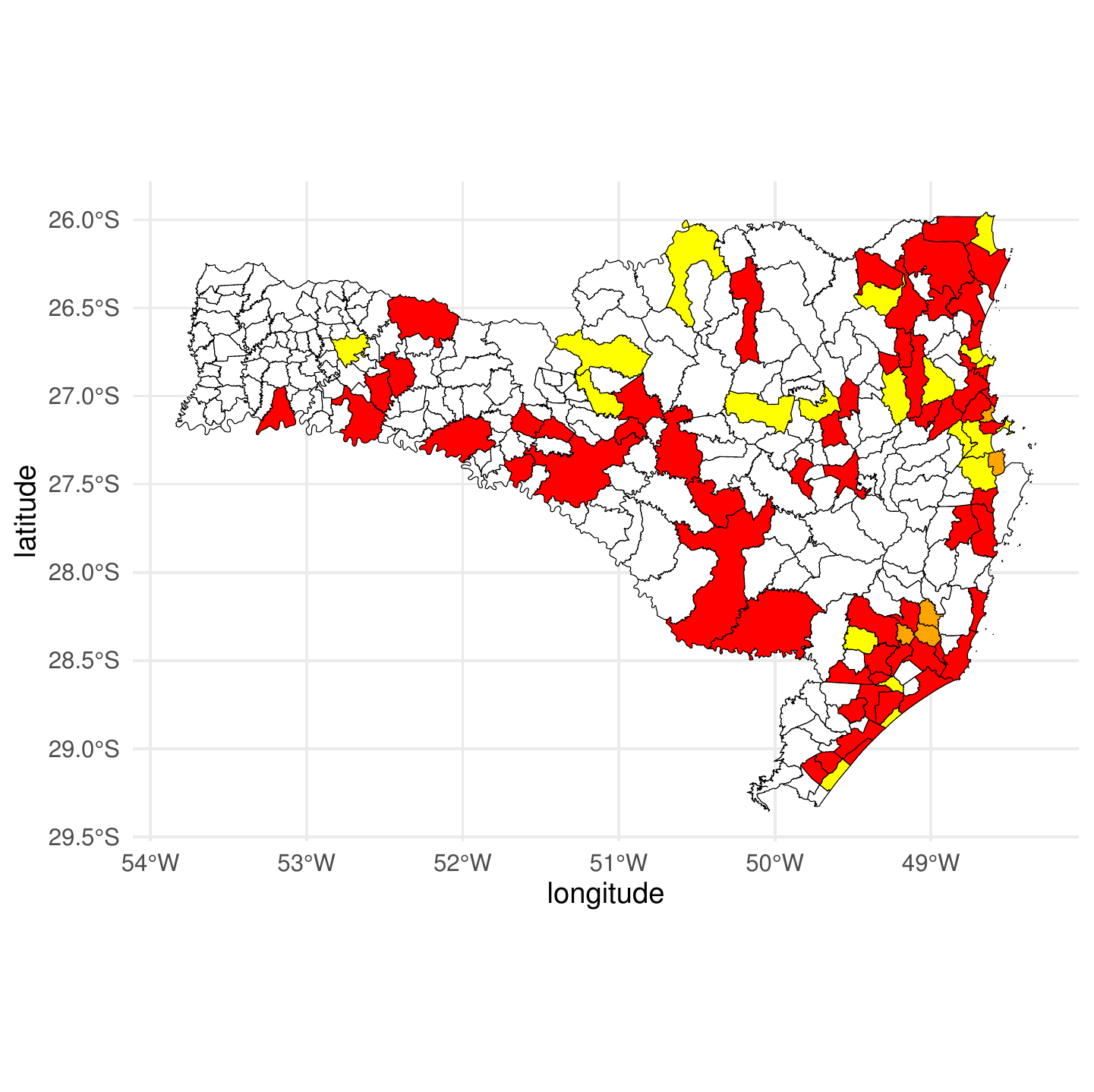}
        }\\
        \subfigure[]{%
            \label{rs}
            \includegraphics[width=5.3cm,height=6cm]{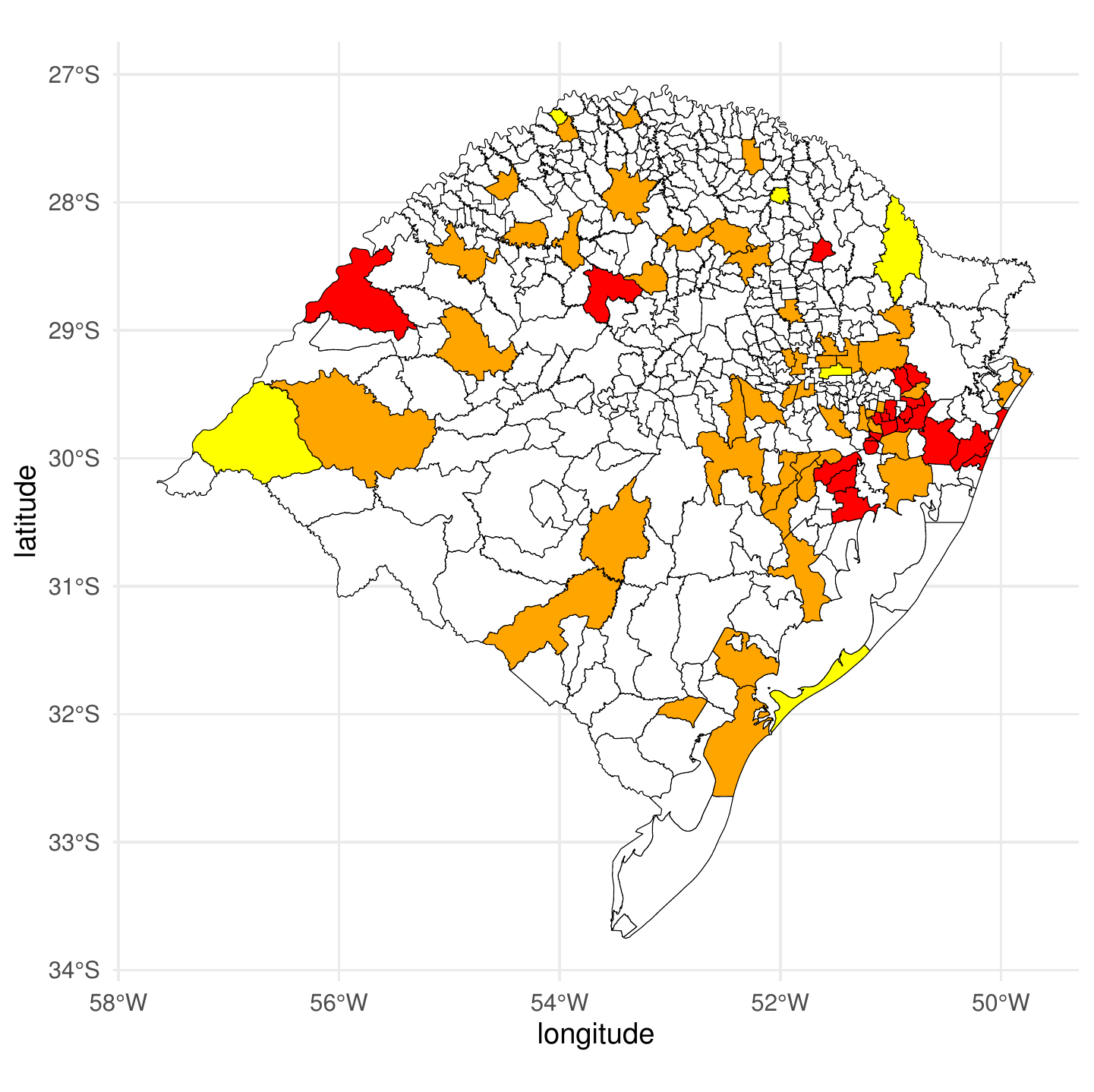}
        }
        \subfigure[]{%
           \label{rs1}
           \includegraphics[width=5.3cm,height=6cm]{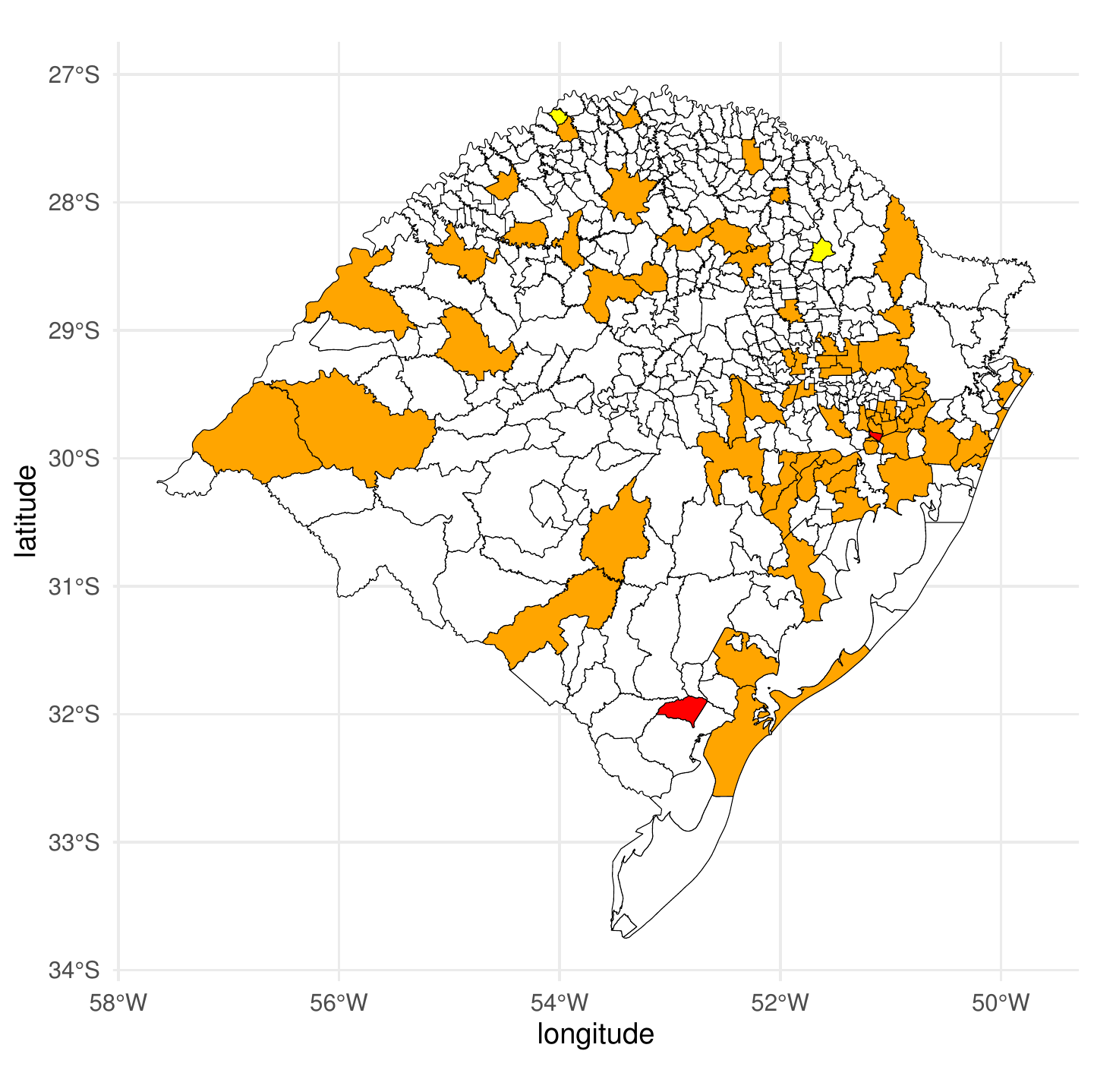}
        }
        \subfigure[]{%
           \label{rs2}
           \includegraphics[width=5.3cm,height=6cm]{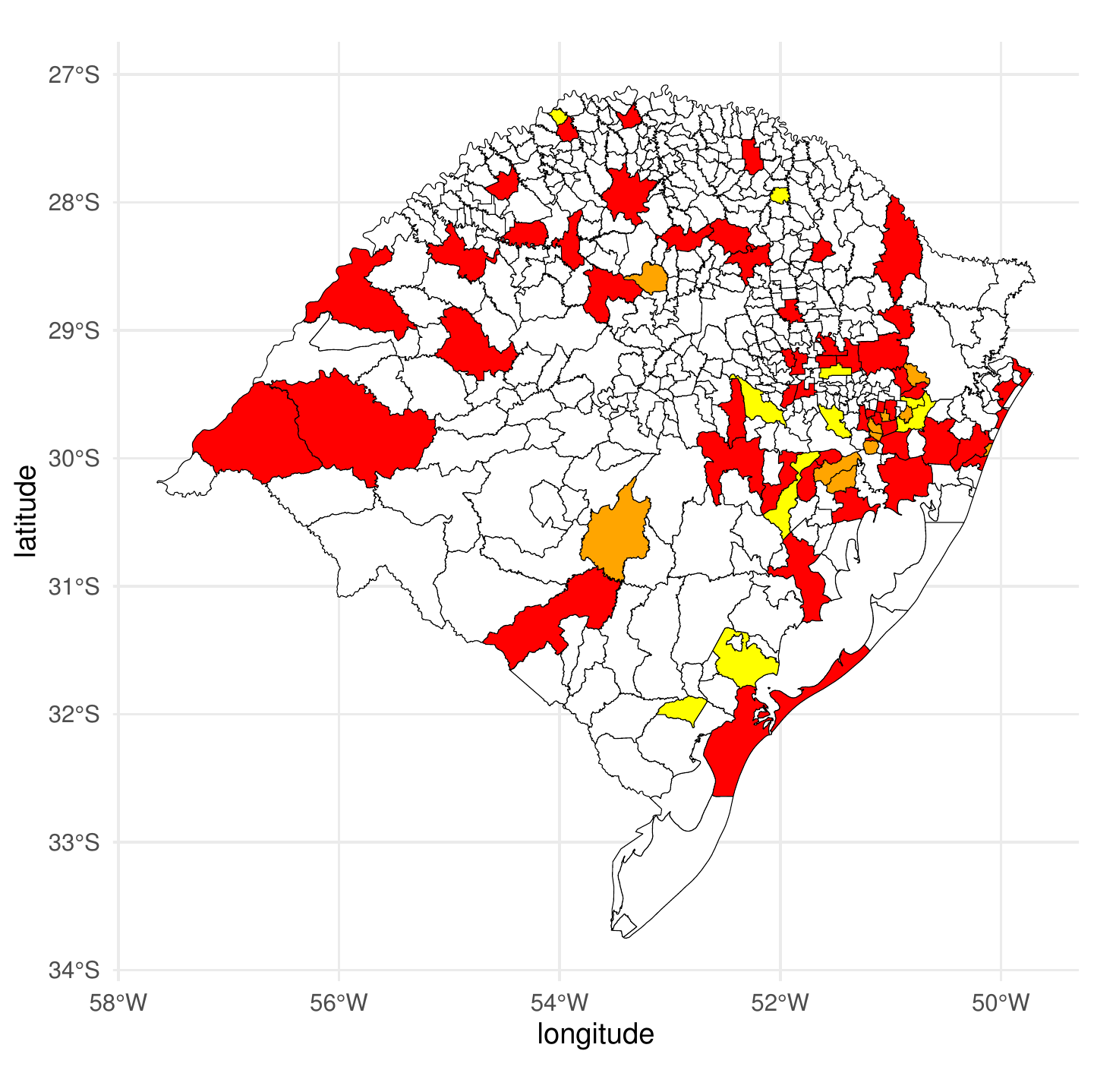}
        }
    \end{center}
    \caption{%
        Spatial location of the functional clustering of the municipalities of the states of Paran\'{a} (first line), Santa Catarina (second line) and Rio Grande do Sul (third line) according to represented death curves (a, d, g), first derivative of death curves (b, e, h) and second derivative of death curves (c, f, i).}
   \label{muni9}
\end{figure}

\pagebreak
